%% file: mader_thesis.tex
\documentclass[a4paper,11pt,titlepage,twoside]{scrreprt}

\usepackage[utf8x]{inputenc}
\usepackage{amsmath, amssymb, graphicx}
\usepackage{hyperref, color,bbm}
\usepackage{slashed,wrapfig}
\usepackage{fancyhdr}

\fancypagestyle{plain}{
\fancyhf{}
\fancyfoot[OR,EL]{\thepage}

}

\fancypagestyle{intropage}{
\fancyhf{}
\fancyfoot[OR,EL]{\thepage}
\fancyhead[OR,EL]{\sffamily{Introduction}}

}

\fancypagestyle{acknpage}{
\fancyhf{}
\fancyhead[C]{\sffamily{Acknowledgements}}
}

\fancypagestyle{titelpage}{
\fancyhf{}

\KOMAoptions{twoside = false}
}

\newcommand{\com}[2]{\ensuremath{\left[ #1 , #2 \right] }}
\newcommand{\acom}[2]{\ensuremath{\left\{ #1 , #2 \right\} }}
\newcommand{\bra}[1]{\ensuremath{\langle #1 \vert}}
\newcommand{\ket}[1]{\ensuremath{\vert #1 \rangle}}

\newcommand{\cp}[3]{\ensuremath{\left( #1\! \times\! #2 \right)^{#3} }}
\newcommand{\cpt}[3]{\ensuremath{\left( #1 \,{\widetilde\times}\, #2 \right)^{#3} }}
\newcommand{\vev}[1]{\ensuremath{\left< #1 \right> }}
\newcommand{\F}[1] { #1_\mathcal{FT} }
\newcommand{\Fvev}[1] {\vev{#1}_\mathcal{FT} }
\newcommand{\var}[2][]{\ensuremath{\frac{\delta #1}{\delta #2}}}

\newcommand{\tr}[2][]{\ensuremath{\mathrm{tr}^{#1}\!\left\{ #2 \right\}}}
\renewcommand{\det}[1]{\ensuremath{\mathrm{det}\left\{ #1 \right\} }}
\renewcommand{\ker}[1]{\ensuremath{\mathrm{ker}\left\{ #1 \right\} }}
\newcommand{\im}[1]{\ensuremath{\mathrm{im}\left\{ #1 \right\} }}

\renewcommand{\log}[1]{\ensuremath{\mathrm{log}\left( #1 \right) }}
\renewcommand{\exp}[1]{\ensuremath{\mathrm{exp}\left( #1 \right) }}

\newcommand{\sa}{\ensuremath{s_{\alpha}}}
\newcommand{\sab}{\ensuremath{\bar s_{\alpha}}}
\newcommand{\se}{\ensuremath{s_{\eps}}}
\newcommand{\seb}{\ensuremath{\bar s_{\eps}}}
\newcommand{\sC}{\ensuremath{s_{C}}}
\newcommand{\sCb}{\ensuremath{\bar s_{C}}}
\newcommand{\set}{\ensuremath{s_{\eta}}}
\newcommand{\setb}{\ensuremath{\bar s_{\eta}}}

\newcommand{\dbar}{\mathchar'26\mkern-12mu d}

\newcommand{\pto}{\stackrel{p^2\rightarrow 0}{\longrightarrow}}
\newcommand{\FP}{\stackrel{\text{FP}}{\longrightarrow}}
\newcommand{\as}{\stackrel{\text{as.}}{\longrightarrow}}

\newcommand{\one}{{\ensuremath{\mathbbm 1 }}}
\newcommand{\complC}{{\ensuremath{\mathbbm C }}}

\newcommand{\GeV}{\ensuremath{\text{GeV}}}
\newcommand{\uone}{\ensuremath{{U(1)}}}
\newcommand{\dzs}{\ensuremath{dz^{\sqrt{}}}}

\renewcommand{\S}{{\ensuremath{\mathcal S}}}
\newcommand{\SYM}{{\ensuremath{\mathcal S_{YM}}}}

\renewcommand{\L}{{\ensuremath{\mathcal L}}}
\newcommand{\Lgf}{\ensuremath{\mathcal L_{gf}}}

\newcommand{\LM}{\ensuremath{\L_M}}
\newcommand{\LYM}{\ensuremath{\L_{YM}}}

\renewcommand{\O}{{\ensuremath{\mathcal O}}}
\newcommand{\M}{{\ensuremath{\mathcal M}}}
\newcommand{\N}{{\ensuremath{\mathcal N}}}
\newcommand{\T}{{\ensuremath{\mathcal T}}}
\newcommand{\V}{{\ensuremath{\mathcal V}}}
\renewcommand{\H}{{\ensuremath{\mathcal H}}}
\newcommand{\D}{\ensuremath{\mathsf{D}}}
\newcommand{\K}{\ensuremath{\mathbf{K}}}

\renewcommand{\P}{\ensuremath{\mathcal{P}}}
\newcommand{\W}{\ensuremath{\mathcal{W}}}

\renewcommand{\c}{\ensuremath{\mathfrak{c}}}

\renewcommand{\k}{\ensuremath{\mathrm{k}}}
\newcommand{\q}{\ensuremath{\mathrm{q}}}
\newcommand{\p}{\ensuremath{\mathrm{p}}}
\newcommand{\f}{\ensuremath{\mathrm F }}

\newcommand{\eps}{\varepsilon}
\newcommand{\vph}{\varphi}
\newcommand{\pp}{\ensuremath{\phi_+}}
\renewcommand{\pm}{\ensuremath{\phi_-}}

\renewcommand{\eqref}[1]{Eq.~(\ref{#1})}
\newcommand{\secref}[1]{Sec.~\ref{#1}}
\newcommand{\appref}[1]{App.~\ref{#1}}
\newcommand{\tabref}[1]{Tab.~\ref{#1}}
\newcommand{\figref}[1]{Fig.~\ref{#1}}

\newcommand{\mathematica}{\ttfamily MATHEMATICA\normalfont}
\newcommand{\form}{\ttfamily FORM\normalfont}

\newcommand{\be}{\begin{equation}}
\newcommand{\ee}{\end{equation}}

\begin{document}

\include{title}
\cleardoublepage
\pagenumbering{roman}
\setcounter{page}{1}

 \tableofcontents

 \linespread{1.2}

\include{1_intro}

 \include{2_LG}

  \include{3_MAG}

 \include{4_ccquartet}

 \include{5_saturation}

 \include{6_conclusio}

\appendix
\allowdisplaybreaks
\include{A1_conventions}

 \include{A2_technicalities}



\end{document}

%% file: title.tex
\begin{titlepage}
\thispagestyle{titelpage}
%

\begin{center}
~

\vfill
\Large{Valentin Mader}

\vfill\vfill\vfill\vfill

{\LARGE\bfseries {Signals of Confinement in the Dyson-Schwinger Equation for the Gauge Boson Propagator}}

\vfill\vfill\vfill

{\bfseries}

\vfill\vfill\vfill
\vfill\vfill\vfill\vfill\vfill\vfill
\vfill\vfill\vfill

\Large{Dissertation}\\ \vfil
\normalsize zur Erlangung des Doktorgrades der Naturwissenschaften\vfill
{Karl-Franzens-Universit\"at Graz}\\
{Institut f\"ur Physik -- FB Theoretische Physik}\\
Betreuer: {Univ.-Prof. Dr. rer. nat. Reinhard Alkofer}\\

\vfill

{Graz, Februar 2014}

\end{center}
\end{titlepage}

%% file: 1_intro.tex
\chapter*{Introduction}
\addcontentsline{toc}{chapter}{Introduction}
\setcounter{page}{1}
\pagenumbering{arabic}

    The last three years, the time this PhD-project has been worked on, were an extraordinary successful period for particle physics. The detection of the Higgs boson by the ATLAS and CMS collaborations at the Large Hadron Collider at CERN \cite{Aad:2012tfa,Chatrchyan:2012ufa} is the climax of a common effort of countless scientist over the last four decades by theoretical and experimental means. All particles of the Standard Model of Particle Physics have now been proven to exist. At the time writing this lines there is no established hint to any theory beyond this highly successful theory.

    Some big questions still need to be answered, however. The Higgs boson cannot answer the big cosmological questions as the unknown origin of the matter/antimatter asymmetry, the unknown nature of Dark Matter and Dark Energy and the unknown unification of the classical theory of gravity with the quantum theories of the elementary particles. Within the particle physics sector, the existence of a Standard Model Higgs brings back attention to the hierarchy problem, the question why the Standard Model features so many distinct scales.

    When looking closer, one problem appears which is overlooked in many major discussions. All theories in the Standard Model of Particle Physics are gauge theories. As such only gauge invariant states can exist in the asymptotic state-space due to Elizur's theorem \cite{Elitzur:1975im}. The success of the Standard Model, however, mostly relies on perturbative calculations which demand gauge-fixing. Perturbatively calculated pole-masses in, e.g., gauge-boson propagators can thus not be the whole truth since these propagators are gauge dependent objects. Only recently it was possible to connect these gauge dependent terms to gauge independent bound-state properties \cite{Maas:2012tj}. Even more mysterious seems the existence of electrically charged asymptotic states in Quantum Electrodynamics (QED) as, e.g., "free"-electrons, which can only exist as non-local states dressed by infinitely many infrared photons, \cite{Bagan:1999jf}.

    Quantum Chromodynamics (QCD), \cite{Fritzsch:1973pi}, the theory of the strong nuclear force in the Standard Model, can be described by perturbative techniques in the high-energy regime due to asymptotic freedom, \cite{Gross:1973id,Politzer:1973fx}. In the last two decades this perturbative calculations have been complemented by non-perturbative computer simulations, e.g. \cite{Durr:2008zz}, and nowadays there is no reasonable doubt that QCD is the theory which correctly describes the formation and interaction of hadrons by the fundamental fields of gluons and quarks. In QCD, however, the problem of asymptotic states is of another quality as in the electroweak sector of the Standard Model described above. While the gluon is perturbatively massless, implying the absence of a Higgs mechanism analogous to the electroweak sector, it is only of finite range as no color-charged object has ever been measured. 
    The absence of color-charged objects from the asymptotic state space is called the \emph{confinement problem} of QCD, \cite{Greensite:2011zz}. The origin of confinement is attributed to the gauge part of QCD only, Yang-Mills theory, \cite{Yang:1954ek}, since there are no asymptotic gluons either.

    In this thesis the quantum equation of motion, the Dyson-Schwinger equation, of the gluon is investigated with respect to signatures of confinement. In classical physics the solution of an equation of motion is the trajectory of the object under consideration. In quantum physics the concept of a definite trajectory is obsolete. The solution of a Dyson-Schwinger equation yields a Green function which encodes propagation and/or interaction of the fields under investigation. If there is some physical truth in the confining property of Yang-Mills theory, it is very likely that signals of it can be found in the propagation of the gluon. The investigation of the Dyson-Schwinger equations of QCD have a long tradition which reach back to the early days of investigating QCD, \cite{Marciano:1977su}. In the last twenty years considerable progress has been gained solving these equations, in particular the Dyson-Schwinger equation for the ghost and gluon propagators in Landau gauge 
    \cite{AlkoferSmekal:2001,Fischer:2006ub,Boucaud:2011ug}. Indeed, in some approximation, a solution was found which is in accordance with the Kugo-Ojima confinement criterion \cite{Kugo:1979gm} and the Gribov-Zwanziger scenario \cite{Vandersickel:2012tz}. In \secref{sec_LG} this solution is reviewed and the approximation is improved. The qualitative results remain unchanged.

    Functional methods, as the Dyson-Schwinger equations are, have one important property which significantly impacts the conclusions one can draw from their solutions. They depend on the gauge. Having a result in one gauge does not necessarily mean there is a corresponding solution in another gauge. It is therefore an inherent interest to perform calculations in functional methods in different gauges to obtain a general picture of the underlying physics from as may points of view as possible. This gauge dependence allows to approach a physical problem from many directions within the same formalism and thus allows for deep insights into the problem under consideration.  In \secref{sec_MAG} the propagator Dyson-Schwinger equations of Yang-Mills theory in the Maximal Abelian gauge are investigated. These equations are rather unknown territory as they have been derived only relatively recent, \cite{Huber:2009wh}. The equations are truncated to a minimal subset which still entails the relevant terms and 
a partial solution is presented.

\thispagestyle{intropage}
    In the second part of this thesis attempts are made to transport the Kugo-Ojima confinement scenario, or specific aspects of it, originally formulated in linear covariant gauge, \cite{Kugo:1979gm}, to other gauges. \secref{sec_cc} treats the quartet mechanism in the generalized covariant gauge, which is an extension of the linear covariant gauge by an additional gauge parameter. In \secref{sec_saturation} the Dyson-Schwinger equation of the gluon propagator in different gauges is investigated. First a corollary of the Kugo-Ojima scenario is extracted, which then can be transported to other gauges. In all gauges investigated similar patterns are discovered. Such patterns cannot be found in the formally equal equation of motion of the gauge bosons of QED and the Abelian Higgs model. This is interpreted as an universal signal of confinement in the quantum equation of motion of the gluon. 

    The following section introduces the reader to the general framework of quantum gauge field theories relevant for this thesis. This includes the gauge-fixing process, renormalization, the derivation of the Dyson-Schwinger equations and aspects of the confinement problem. The main text ends with an outlook in \secref{sec_conclusio}. Conventions, several technical considerations and details of the numerical implementation are shifted to the appendices.

\chapter{Yang-Mills Theory as a Quantum Field Theory\label{sec_intro}}

     Quantum field theory is a framework which combines quantum physics, field theories and special relativity. All theories in the Standard Model of Particle Physics are quantum field theories. As such quantum field theories are treated in numerous work in the literature and all of the material presented in this introductory section is well-established. General aspect of quantum field theories and Yang-Mills theory are worked out along the line of \cite{PascualTarrach,NakanishiOjima:1990,PeskinSchroeder,Muta,Bohm:2001yx,Nair:2005iw}. The confinement problem is treated in \cite{Greensite:2003bk,Alkofer:2006fu,Greensite:2011zz}. 

\section{The Path-Integral Representation\label{sec_YM}}
      Yang-Mills theory is defined by the action $\SYM = \int\!d^4x\,\LYM $ with the Yang-Mills Lagrangian
	  \begin{align} 
	      \LYM & =\frac{1}{4}F_{\mu\nu}^r F_{\mu\nu}^r \label{1_def_LYM}\\
		   & = -\frac{1}{2} \delta^{rs} A_\mu^r\left(\delta_{\mu\nu}\partial^2 - \partial_\mu\partial_\nu \right)A_\nu^s \nonumber\\
		   & \qquad+ (\partial_\mu A_\nu^r) \cp{A_\mu}{A_\nu}{r} + \frac{1}{4} \cp{A_\mu}{A_\nu}{r}\cp{A_\mu}{A_\nu}{r}\,. \label{1_def_LYM2}
	  \end{align}
      The cross-product is defined as $\cp{\phi}{\psi}{a} = g f^{abc}\phi^b\psi^c$. Further details on conventions and definitions are summarized in \appref{secA_con}.  The defining property of Yang-Mills theory is that it is invariant under local gauge transformations  $U(x) = \exp{ig\vartheta^r(x)T^r} \in \mathcal G$ of the gauge-fields $A = A_\mu(x) = A_\mu^r(x)T^r$.
	  \be   A  \rightarrow A^{U}  = U(x)AU^{-1}(x) - \frac{i}{g} \left(\partial_\mu U(x)\right) U^{-1}(x)\,.\label{1_gtA} \ee
      Infinitesimally the gauge field transforms as
	    \be A_\mu^r(x) \rightarrow A_\mu^{r\,\vartheta}(x)  = A_\mu^r(x) + D_\mu^{rs}\vartheta^s(x)\,.\label{1_gtA_inf} \ee
      
      In principle $\mathcal G$ can be any continuous group. In this thesis the groups $SU(N)$ and $U(1)$ are considered exclusively. The generators of the gauge group $T^r$ are considered to be hermitian and traceless, the gauge-parameters $\vartheta^r(x)$ are spacetime dependent real parameters and the gauge-coupling $g$ is a global real constant. The group $U(1)$ is a special case. Since it is Abelian all structure constants vanish and thus the three and four-point vertices in \eqref{1_def_LYM2} disappear.

      The quantum theory corresponding to the classical action $\SYM$ in Euclidean spacetime is given in path-integral representation by the partition function
	  \be Z[J] = \int\![dA] e^{-\SYM + A\cdot J} \,,\label{1_def_PI}\ee
      where $A\cdot J = \sum_{r,\mu} \int\!d^4 x A_\mu^r(x)J_\mu^r(x)$ and $J_\mu^r(x)$ represent sources of the gauge-fields.
      The vacuum expectation value of an operator $\O$ is defined as
	  \be \vev{\O} = \mathcal{N} \int\![dA]\, \O\, e^{-\SYM} \,,\ee
      where $\mathcal N$ is a possibly infinite normalization constant such that $\vev{\one} = 1$. A n-point Green function is the vacuum expectation value of an operator which is a monomial in n fields.

      \subsubsection*{Gauge Orbits and Gauge-Fixing}
      For any gauge configuration $A(x)$ there exist another gauge configuration $A'(x)$ with the same weight factor in the path-integral $e^{-\SYM}$ if $A'(x)$ is just a gauge transformation of the original one. In fact there is an infinite continuous set of such field configurations which only differ by gauge transformations. These gauge configurations build a gauge-orbit 
	  \be O[A] = \left\{A'(x):\,A(x)'=A^U(x) \right\}\,. \label{1_def_orbit}\ee
      The existence of such gauge-orbits is not a problem, if the operator $\O$ is gauge invariant. Then the summation over gauge-equivalent transformations just yields a constant which is absorbed into the normalization. When calculating gauge-variant Green functions, such as, e.g. gluon propagators and vertices, though, these expectation values vanish due to Elizur's theorem \cite{Elitzur:1975im}. Therefore, when calculating such expectation values a proper gauge fixing has to be chosen, which ideally picks one representative out of every gauge-orbit. 

      One might argue the problem of gauge orbits is a problem of the path-integral representation of the quantum theory, which might not show up in other representations. In canonical quantization, however, another problem arises, which is deeply connected to the problem of gauge orbits in the path-integral formulation. The canonical commutation relation for a field $\phi_i$, let $i$ denote any internal index, with the canonical momentum $\pi_j$ of the field $\phi_j$ is given by
	  \be  \com{\phi_i(x)}{\pi_j(y)} = i\delta_{ij}\,\delta^{(4)}(x-y)\,. \label{1_canmom} \ee
      The canonical momentum of the time component of the gluon-field $A_0^r$, however vanishes, $\pi_{A_0^r}  = 0 $. The canonical commutations thus can not be fulfilled for all four components of $A_\mu^r$, and the quantization procedure gets spoiled. In fact the vanishing of a canonical momentum is the sign for a constraint the theory has to obey \cite{Matschull:1996up}.
      
      A standard procedure to deal with the problems considered above in a quantum field theory, named after Faddeev and Popov \cite{Faddeev:1967fc}, is to introduce the following $\one -$operator, 
	  \be \one = \Delta_F \int\![dU] \,\delta\!\left(F^r(A^U)\right) \,. \label{1_def_FP1}\ee
      $F^r(A^U)$ is the gauge-fixing condition, which suppose to vanish at every spacetime point, $\delta\!\left(F^r(A^U)\right) = \prod_{x,r} \delta\!\left(F^r(A^U (x)\right)$. A general expression for the $\delta-$function as a limit of Gaussian functions is used to rewrite the $\delta-$function in \eqref{1_def_FP1} as
	  \be \delta\!\left(F^r(A^U(x))\right) \propto \lim_{\xi\rightarrow 0} \frac{1}{\xi} \exp{-\frac{\left(F^r(A^U(x))\right)^2}{2 \xi}}  \,.\ee
      The Faddeev-Popov determinant in \eqref{1_def_FP1} can be expressed as
	  \be \Delta_F = \det{\M^{rs}(x,y)}\,,\label{1_def_FPdet}\ee
      with the Faddeev-Popov operator
	  \be \M^{rs}(x,y) = \var[F^r(A^U(x))]{\vartheta^s(y)}\Bigg\rvert_{F(A^U)=0} \,. \label{1_def_FPO}\ee
      Following standard literature, e.g. \cite{PeskinSchroeder,Muta}, the absolute value of the determinant in \eqref{1_def_FPdet} is neglected to be able to localize the determinant. In principle this can lead to problems beyond perturbation theory which motivates the Gribov-Zwanziger szenarion below, \cite{Vandersickel:2012tz}.
      
      Following the rules of functional integration a determinant can be expressed as path-integral over fermion fields. Introducing the two scalar Grassmannian fields $c^r$ and $\bar c^r$, the (Faddeev-Popov-)ghost and antighost, allows to further rewrite the determinant
	  \begin{align} 
	      \det{\M^{rs}(x,y)} & = \int[dc][id\bar c]  \exp{i\int\!d^4x\int\!d^4y \,\bar c^r(x) \M^{rs}(x,y)c^s(y)} \label{1_def_FPghosts}	
	  \end{align}
      The Faddeev-Popov gauge-fixing procedure works as follows: Neglecting the source terms in the path-integral \eqref{1_def_PI} and inserting the \one-operator \eqref{1_def_FP1} yields
	   \begin{align}
	      \mathcal N \int\![dA] e^{-\SYM} & = \mathcal N  \int\![dA] \Delta_F \int\![dU] \,\delta\!\left(F^r(A^U)\right) e^{-\SYM}\,.
	      \intertext{Gauge invariance of the action and the measure allows to shift the gauge-fields $A^U$ back to the original gauge-fields $A$}
		    & = \mathcal N  \int\![dA] \Delta_F  \,\delta\!\left(F^r(A)\right) e^{-\SYM} \int\![dU]\,,
	      \intertext{yielding the volume of the group $\int\![dU]$ as an (infinite) constant which can be absorbed in the normalization factor in front of the path-integral. Using the local version of the Faddeev-Popov-determinant and the $\delta-$function, thereby absorbing constants into the normalization,  yields}
		 \mathcal N \int\![dA] e^{-\SYM}   &  = \lim_{\xi\rightarrow 0} \mathcal N'  \int\![dA][dc][id\bar c]  e^{-\SYM -\S_{gf}}\,, \label{1_def_YMgf}
	   \end{align}
      where the gauge-fixing action was defined
	    \be \S_{gf} =  \int d^4x \frac{\left(F^r(A(x))\right)^2}{2 \xi} - i\int\!d^4x \int\!d^4y\, \bar c^r(x) \M^{rs}(x,y)c^s(y) \,. \label{1_def_Sgf} \ee
      Landau gauge is defined by the gauge-fixing condition
	  \be F^r(A^U) = -\partial_\mu A^{r\,U}_\mu = 0\,,\label{1_def_LG}\ee
      which yields the Faddeev-Popov operator $\M^{rs}(x,y) =  -\partial_\mu D^{rs}_\mu \delta(x-y) $ and the gauge-fixing action 
	\be \S_{gf}^{lc} = \int d^4x \left\{ \frac{\left(\partial_\mu A_\mu^r(x)\right)^2}{2 \xi} + i  \bar c^r(x) \partial_\mu D_\mu^{rs} c^s(x) \right\}\,. \label{1_def_Sgf_LG} \ee
      Landau-gauge is obtained for the limes $\xi \rightarrow 0$ of $\S_{gf}^{lc}$ which has to be performed at the very end of a calculation.

    In this work the convention of \cite{Kugo:1979gm} is applied and real ghosts and antighosts,
	  \be (c^r)^\dag = c^r \qquad \text{, and, }\qquad (\bar c^r)^{\dag} = \bar c^r \,.  \label{1_ghostherm}\ee
    are used. With this hermiticity assignment to the ghosts the gauge fixing  Lagrangian is hermitian. Ghost and antighost carry canonical dimension $1$ and ghostnumber $+1$ and $-1$, respectively. For some different hermiticity assignment with complex ghost fields consider \cite{Baulieu:1981sb}. The relation between these two hermiticity assignments is resolved in appendix A of \cite{AlkoferSmekal:2001}.

      \subsubsection*{Renormalization}
      When multiplicative renormalizability is preserved, renormalized fields can be introduced by means of multiplicative renormalization constants. In Yang-Mills theory in the linear covariant gauge, \eqref{1_def_YMgf}, the following terms are subject to renormalization, e.g. \cite{Muta},
	 \begin{align}
	      A_{0\,\mu}^s &= \sqrt{Z_3}\,A_{R\,\mu}^s \,,& g_0 &= Z_g\, g_R \,,  & \xi_0  &= Z_\xi \,\xi_R\,,\label{1_def_rc} \\
	      c_0^s  &= \sqrt{\tilde Z_3}\,c_{R}^s\,, & \bar c_0^s  &= \sqrt{\tilde Z_3} \,\bar c_{R}^s\,. \nonumber
	 \end{align}
      The possibly infinite bare terms, denoted by then index $0$ on the lhs of the equations, are split into their finite renormalized counterparts, denoted with index $R$, and the possibly infinite renormalization constants $Z_i$ on the rhs.  In \eqref{1_def_rc} there is no direct reference to the vertex renormalization constants but only the gauge coupling itself is renormalized. The vertex renormalization constants are defined as
	  \be Z_1 = Z_g Z_3^{3/2}\,,\qquad Z_4 = Z_g^2 Z_3^2\,,\qquad\text{and, }\qquad  \tilde Z_1 = Z_g \tilde Z_3 Z_3^{1/2}\,. \label{1_def_Z1Z4}\ee
     The equivalence of $Z_g$ in \eqref{1_def_Z1Z4} is a consequence of the unbroken gauge symmetry. \eqref{1_def_rc} features four independent renormalization constants. This set can and will be restricted by further constraints in the course of the thesis.
  
     Power counting would allow for an additional term in the Yang-Mills Lagrangian, a gluon-mass term $\propto m_A^2 A_\mu^a A_\mu^a$. However, this term is not invariant under gauge transformations \eqref{1_gtA} and thus it is forbidden by symmetry. In practical calculations this term might still be relevant, though. A clumsy choice of the regularization scheme can introduce quadratic divergences, which for dimensional reasons can only be removed by a mass-counter term. The gluon mass term and the corresponding counter term thus are apparent in the renormalized Yang-Mills Lagrangian. Gauge-invariance, however, demands the renormalized mass to vanish. The renormalization condition
	  \be Z_m m_{R\,A}^2 = 0 \label{1_rc_Zm} \ee
    has to be imposed onto all Green functions of the theory. In practical calculations this means that the quadratic divergences have to be removed completely. So called gauge-invariant regularization schemes, e.g. dimensional regularization, fulfill this condition implicitly and quadratic divergences cancel to any order in perturbation theory \cite{Muta}. In the following the indices $0$ and $R$ will be dropped, as it will be clear from the context if renormalized or bare quantities are used.

\section{BRST Symmetry\label{sec_BRST}}
      By construction the action $\S_{lc} = \SYM + \S_{gf}^{lc}$ is not invariant under local gauge transformations. However, up to equations of motion, it isl invariant under BRST-transformations, named after Becchi, Rouet, Stora and Tyutin \cite{Becchi:1975nq,Tyutin:1975qk},
	    \begin{align}
		s A_\mu^r &= (D_\mu c)^r\, & sc^r &= -\frac{1}{2} \cp{c}{c}{r}\, \label{1_def_BRST} \\
		s\bar c^r &= b^r\, & sb^r & = 0 \,. \nonumber
	    \end{align}
      In \eqref{1_def_BRST} a new field was introduced, the Nakanishi-Lautrup (NL) field $b^r$. It is the BRST-transformation of the antighost and serves as a Lagrange multiplier for the gauge-fixing condition. One characteristic property of the BRST transformations is their nilpotency $s^2=0$. The corresponding BRST symmetry can be used for proofs of fundamental properties of quantum gauge theories such as, e.g., renormalizability \cite{Becchi:1975nq} of the theory and unitarity of the S-Matrix and a construction of a physical Hilbert-space \cite{Kugo:1979gm}.
      
      BRST symmetry also allows for a generalization of the gauge-fixing procedure beyond the path-integral method. Demanding that any gauge-fixing Lagrangian should be invariant under BRST-transformations and exploiting the nilpotency property it can be shown that any gauge-fixing Lagrangian can be written as, \cite{Kugo:1981hm},
	    \be \Lgf = -i s(\bar c^r F^r(A_\mu,c,\bar c,b,\psi))\,. \label{1_genGF}\ee
      The function $F^r$ is the gauge-fixing condition and can in principle be any polynomial with ghost-number $0$ in the fields $A_\mu^r, c^r, \bar c^r,b^r$ and possible matter fields $\psi$. The linear covariant gauge is obtained by the gauge condition 
	    \be F^r = \partial_\mu A_\mu^r + i \frac{\xi}{2} b^r \ee
      which yields the gauge-fixing Lagrangian
	    \begin{align} 
	    	\L^{lc}_{gf} &= \frac{\xi}{2} b^2 - i b^r \partial_\mu A_\mu^r + i \bar c^r (\partial_\mu D_\mu c)^r\,.\label{1_def_Lgf_lc}
		\intertext{Integrating out the NL-field by its equation of motion yields}
		 \L^{lc}_{gf} & =  \frac{1}{2\xi} (\partial_\mu A_\mu)^2 + i \bar c^r (\partial_\mu D_\mu c)^r\,, \label{1_def_SgfKU2}
	    \end{align}
      i.e. exactly the same gauge-fixing Lagrangian as obtain via the Faddeev-Popov method \eqref{1_def_Sgf_LG}. Beside its function as a Lagrange multiplier for the gauge-fixing condition, the NL field also serves as a canonical momentum for the $A_0^r-$field, such that the canonical equal-time commutation relations for the $A_\mu^r-$fields can be fulfilled.

      The gauge-fixed Lagrangian $\Lgf^{lc}$ is not only invariant under the BRST transformation \eqref{1_def_BRST}, but also under anti-BRST transformations \cite{Ojima:1980da,Baulieu:1981sb}
	    \begin{align}
		\bar s A_\mu^r &= (D_\mu \bar c)^r\,, & \bar s \bar c^r &= -\frac{1}{2} \cp{\bar c}{\bar c}{r} \,, \label{1_def_aBRST}\\
		\bar s c^r &= -b^r - \cp{\bar c}{c}{r} \,,&  \bar s b^r & = -\cp{\bar c}{b}{r} \,.\nonumber
	    \end{align}
      The anti-BRST transformations are nilpotent and connected to the original BRST transformations by anti-commutivity, \cite{Baulieu:1981sb,ThierryMieg:1985yv},
	    \be  \bar s^2 = \acom{s}{\bar s} = 0\,. \ee
      In terms of field-transformations the anti-commutivity condition reads
	    \be s\bar c^r + \bar s c^r + \cp{\bar c}{c}{r} = 0\,. \ee
      The role of this additional symmetry is still under debate, since it does not provide any new physical insights. It can be used for the same purposes as the BRST-transformations, but does not add any additional information \cite{NakanishiOjima:1990}. Still it can be used to generate a natural justification for the background field method \cite{Binosi:2013cea} or to restrict the allowed terms in the Lagrangian. Neglecting topological terms, the most general Lagrangian, which fulfills the conditions, \cite{ThierryMieg:1985yv},
	  \begin{enumerate} 
	  \item[  i,] dimension 4
	  \item[ ii,] Lorentz-invariance and global color invariance
	  \item[iii,] BRST and anti-BRST invariance
	  \item[ iv,] ghost number $0$,
	  \end{enumerate}
      is given by
  	  \be \L = \LYM + \frac{i}{2}  s\bar s \left(  A_\mu^r A_\mu^r - \lambda_1\, \bar c^r c^r\right)) + \lambda_2\, s(\bar c s \bar c)\,. \label{1_eq_gengauge}\ee
      This is quite restrictive since one already has for the  Landau gauge limit of $\L_{gf}^{lc}$
	  \be \Lgf^{LG} = -is\left(\bar c^r \,\partial_\mu A_\mu^r \right) = i \bar s \left(c^r \,\partial_\mu A_\mu^r \right)  = \frac{i}{2} s\bar s\left(A_\mu^r A_\mu^r \right)\,. \label{1_eq_lincov_ssb}\ee

      In the literature there is some discussion if BRST-symmetry is realized in nature. Restricting the functional integration to the first Gribov-region, which is one possible way of defining a nonperturbative gauge-fixing procedure and what will be presented in \secref{sec_gz}, leads to a spontaneous breaking of BRST-symmetry \cite{Maggiore:1993wq,Vandersickel:2012tz,Dudal:2012sb,Capri:2013naa}. The possible breaking BRST symmetry agrees well with the fact, that BRST symmetry cannot be trivially realized on the lattice \cite{Neuberger:1986vv} since all correlation functions yield $\frac 0 0$ in a BRST--gauge-fixed theory. The root of this problem lies in the fact that the gauge-fixing partition function multiplied to the theory is a topological field theory which calculates the Euler characteristics of the gauge group  \cite{Baulieu:1996rp,Baulieu:1996kb}. Unfortunately this topological number vanishes. A possible solution of this problem is to define non-polynomial gauge-fixing conditions 
      \cite{vonSmekal:2007ns,vonSmekal:2008es,Serreau:2012cg,Serreau:2013ila}. Moreover there are arguments that such a soft breaking  of BRST symmetry renders the S-Matrix gauge-dependent and therefore is inconsistent \cite{Lavrov:2011wb}.
      
      Despite this subtleties in most of this work BRST-symmetry is assumed not to be broken spontaneously. This can be justified since, with one exception in \secref{sec_saturation}, the restriction to the Gribov horizon is not implemented in this thesis but the Faddeev-Popov gauge-fixing action is used. The spontaneous breaking of BRST symmetry of such a theory in the continuum, at least, has not been shown, yet.

\section{Dyson-Schwinger Equations and Ward Identities \label{sec_dse}}

      The Dyson-Schwinger equations \cite{Dyson:1949,Schwinger:1951aa,Schwinger:1951ab} are the equation of motions for the Green functions of a quantum field theory. They provide relations which are true independent on the specific regularization and renormalization scheme. They are an infinite coupled set of integral equations which define all Green functions of the theory. A full solution of all Dyson-Schwinger equations is equivalent to solving the theory. 

      On the way of solving the theory, if this is possible at all, a plausible approach is to take into account only a finite subset of this equations and model the neglected n-point functions. Such truncations are unavoidable for many practical calculations using Dyson-Schwinger equations. Nevertheless these equations provide an ab-initio approach to investigate a quantum field theory by non-perturbative means. In particular they provide insights into the detailed structure of the theory.

      Ward identities are relations among Green functions quite akin to the Dyson-Schwinger equations. The follow from symmetries of the quantum theory. The derivation is based on the same footing and the general Ward identity just drops out on the way to the Dyson-Schwinger equations. It will turn out to be helpful in particular in the second half of this thesis. A special kind of Ward identities are Slavnov-Taylor identities, of which one is investigated in the following subsection.
      
      For the derivation of the Dyson-Schwinger equations Schwinger's variational approach is used in the following, \cite{Schwinger:1951xk}, which is the standard derivation found in many textbooks, e.g. \cite{Rivers:1987,Bohm:2001yx,Nair:2005iw}.  To start, consider any quantum field theory with field content $\phi = \{\phi_i\}$,  whose defining partition function in Euclidean space is given by
	  \be Z[J] = \int\![d\phi] e^{-\S + \S_J}  \ee
      with the classical action $\S$ and the source terms $\S_J = \phi_i \cdot J_i$. The dot product represents a scalar product in all internal spaces as, e.g. flavor, color and position space. Any operator $\O$ shall be given as a polynomial in the fields $\phi_i$ and its vacuum expectation value in presence of the sources $J$ is defined as
	   \be \vev{\O}_J = \mathcal{N}\int\![d\phi]\,\O\, e^{-\S + \S_J} \,,\label{1_def_OJ}\ee
      with some normalization constant $\mathcal N$. Since $\vev{\O}_J$ only depends on the sources and not on the fields, it is left invariant by any variation in the fields. Excluding anomalies one finds for a general variation in the fields $\delta$,
	   \be \vev{\delta \O}_J = \vev{\O \,\delta\left(\S-\S_J\right)}_J\,.  \label{1_def_WI_s}\ee
      Aboves equation yields two important relations. One the one hand, setting the sources to $0$ yields the general Ward identity
	   \be \vev{\delta \O} = \vev{\O \,\delta(\S)}  \label{1_def_WI}\,,\ee
      where the notation $\vev \O = \vev{\O}_{J=0}$ was used. One the other hand, since both, the operator $\O$ and the action $\S$ are polynomial in the fields, the variation $\delta$ can be expressed as a derivative. Taking the operator to be $\one$ one finds the functional form of the Dyson-Schwinger equations,
	   \be 0  = \vev{\var[(\S-\S_J)]{\phi_i}}_J = \left( \var[\S]{\phi_i}\Biggr\rvert_{\phi_k = \var{J_k}} - J_i \right) Z[J]  \label{1_def_DSE_1}\ee
      where in the second step all fields $\phi_k$ are replaced by derivatives with respect to the corresponding sources $J_k$. Varying \eqref{1_def_DSE_1} with respect to further sources and setting all sources to $0$ at the end yields the Dyson-Schwinger equation for the corresponding Green function. As an example,  the Dyson-Schwinger equation for the two-point function of $\phi_j$ and $\phi_i$, is derived by taking the derivative of \eqref{1_def_DSE_1} with respect to the source $J_j$. Setting all sources to $0$ yields
	  \be \delta_{ij} = \vev{\var[\S]{\phi_i}\,\phi_j}\,. \label{1_DSE_prop}\ee
      This equation will be heavily used in various forms in this thesis with $\phi$ being different gauge-boson and ghost fields in various theories.  In \secref{sec_cc} and \secref{sec_saturation} the form \eqref{1_DSE_prop} will be used. In \secref{sec_LG} and \secref{sec_MAG} the one-particle irreducible (1PI) versions of these equations are investigated.

      To derive the Dyson-Schwinger equation for the connected and 1PI Green functions the corresponding generating functionals for the connected Green functions
	  \be W[J] = \log{Z[J]} \label{1_def_W}\ee
      and the 1PI Green functions
	  \be \Gamma[\Phi] = -W[J] + \Phi\cdot J \label{1_def_Gamma}\,\ee
      are introduced. The effective action $\Gamma$ depends on the "classical fields" $\Phi_i = \vev{\phi_i}_J$. These two generating functionals are related via the equations
	  \be \var[ {\Gamma[\Phi]} ] {\Phi_i} = J_i\,, \quad\text{and, }\quad \var[W]{J_i} = \Phi_i \,. \ee
      Using the identities
	  \be \var{J_i} = \var[\Phi_j]{J_i} \var{\Phi_j} = \var[^2 {W[J]}]{{ J_i\delta J_j}} \var{\Phi_j}  \ee
      and $  f\left(\var{\psi}\right) \exp{F(\psi)} = \exp{F(\psi)} \left( \var[F(\psi)]{\psi} + \var{\psi} \right)$, 
      one finds for the functional Dyson-Schwinger equations for the connected and 1PI Green functions
      \begin{align}
       0 & = \var[S]{\phi_i}\Bigg\rvert_{\phi_k = \var[W]{J_k} + \var{J_k}} + J_i\,, \label{1_dse_DSEW} \\
       0  & = \var[S]{\phi_i}\Bigg\rvert_{\phi_k = \Phi_k + \sum_j\var[^2 {W[J]}]{{ J_k\delta J_j}} \var{\Phi_j}} +  \var[ {\Gamma[\Phi]} ] {\Phi_i} \,. \label{1_def_DSEG}
      \end{align}
      
      One issue in practical calculations of Dyson-Schwinger equations is their renormalization. Generally, if the quantum field theory exists and is finite, so are the Dyson-Schwinger equations. All apparent divergences are absorbed in the appropriate renormalization constants. A finite number of renormalization conditions are then imposed onto a finite set of Green functions. When performing a truncation and/or choosing an inappropriate regularization, one might interfere with this "self-renormalizing" structure. 

      E.g., in the numerical solution of a truncated set of Dyson-Schwinger equations as performed in this thesis, one usually regularizes the theory with a hard cutoff. In Yang-Mills theory in four spacetime dimensions this introduces quadratic divergences. Within the calculation the root of these divergences does not matter, i.e. truncation or cutoff induced. They have to be removed completely as enforced by the renormalization condition \eqref{1_rc_Zm}. A study where dimensional regularization has been used in a Dyson-Schwinger equation is \cite{Schreiber:1998ht}.

    \subsubsection{Slavnov-Taylor Identity for the Longitudinal Gluon Propagator}
      One quite famous consequence of the general Ward-Identity \eqref{1_def_WI}, which also has some relevance in the course of this thesis, is the non-renormalization of the longitudinal gluon propagator in linear covariant gauge. Ward identities with respect to unbroken BRST symmetry are called Slavnov-Taylor identities \cite{Taylor:1971ff,Slavnov:1972fg}. Consider Yang-Mills theory in linear covariant gauge, \eqref{1_def_Lgf_lc}. The following identity holds if BRST symmetry is not broken,
	  \be 0 = s\vev{i\partial_\mu A_{\mu}^r(x)\,\bar c^s(y) } = i \vev{\left(\partial_\mu D_\mu c \right)^r(x)\,\bar c^s(y)} + i\vev{\partial_\mu A^r_{\mu}(x)\,b^s(y)}\,. \label{1_def_sti} \ee
      The Dyson-Schwinger equation of the anitghost two-point function determines the first term in \eqref{1_def_sti} which immediately leads to 
	  \be i\vev{\partial_\mu A^r_{\mu}(x)\,b^s(y)} = -\delta(x-y)\delta^{rs}\,. \label{1_dA_b_STI}\ee
      Integrating out the NL-field and taking the Fourier-transform yields
	  \be p_\nu \Fvev{A^r_{\mu}(x)\,A^s_\nu(y)} = \xi \delta^{rs}\,\frac{p_\mu}{p^2} \label{1_long_gl_STI}\,.\ee
      \eqref{1_long_gl_STI} is the Slavnov-Taylor identity for the longitudinal gluon-propagator. It is analogous to the Ward-Takahashi identity in QED and it states that in Landau gauge the gluon propagator is transverse. Introducing renormalized quantities, \eqref{1_long_gl_STI} can be rewritten to
	    \be p_\nu \Fvev{A^r_{R\,\mu}(x)\,A^s_{R\,\nu}(y)} = \frac{Z_\xi}{Z_3}\,\xi_R \delta^{rs}\,\frac{p_\mu}{p^2} \label{1_long_gl_STI_ren}\,.\ee
      Since both sides of the equation are finite one finds
	    \be  \frac{Z_\xi}{Z_3} = \text{finite} \equiv 1\,. \label{1_STI_lg}\ee
      The longitudinal gluon propagator in general covariant gauge does not renormalize. It is stable under radiative corrections to any order in perturbation theory, \cite{PascualTarrach}.

\section{Aspects of Confinement\label{sec_conf}}

      Physics is a quantitative natural science. A physical phenomenon usually is considered to be "understood", if experimental data and theoretical calculations agree within the boundaries on precision of both approaches. However, in addition, physical research always has the ambition to explain \emph{how} specific phenomena occur, i.e. to uncover underlying mechanisms and trace them back to some few basic assumptions. The confinement problem is of the latter kind. While computer simulations promise to reproduce the measured hadron spectrum within the next few years, the mechanisms which lead to the characteristic properties of the spectrum still remain unclear. 

      The experimental result on color confinement can be summarized in the statement that \emph{"no particles with fractional electric charge have ever been seen"}, \cite{Perl:2004qc}. As quarks do have fractional electric charges, this finding expresses the confinement of \emph{quarks} \cite{Beringer:1900zz}. \emph{Color} confinement is a stronger idea as it needs quarks and gluons to be confined. A bound on the cross-sections of free-gluon productions has been presented in \cite{HidalgoDuque:2011je}. It was found that gluon-confinement is not as established. Still, for the course of this thesis color confinement as an experimental fact is assumed. Over thirty years after the introduction of the fundamental theory of the strong interaction this experimental finding is not yet explained by theoretical calculations in an satisfying manner.
      
      One reason why confinement is not fully explained, yet, is that the question "What is confinement?" is not agreed upon \cite{Greensite:2011zz}. Different definitions of confinement yield different order parameter. The absence of color-charged asymptotic fields is also true in gauge theories coupled to a Higgs-field \cite{'tHooft:1979bi,Frohlich:1981yi}. Even it can be shown that the confining and Higgs phases are analytically connected \cite{Fradkin:1978dv,Lang:1981qg,Caudy:2007sf,Capri:2012ah,Capri:2013oja}, which raises the question if there is a physical difference between the Higgs and confining phase at all. This point is addressed in \secref{sec_saturation}. In the current section some definitions of confinement which are relevant for this thesis are presented. A more complete list of confinement criteria and a detailed discussion of their (dis-)advantages can be found in \cite{Greensite:2011zz}.

      If there is some physical reality in the confinement criteria presented below, they cannot be independent but must be related somehow. A remarkable investigation related the low-energy behavior of the ghost propagator in Coulomb gauge as expected from the Gribov-Zwanziger scenario to the color-electric permeability of the vacuum \cite{Reinhardt:2008ek}. The relation between the Gribov-Zwanziger and Kugo-Ojima scenarios is discussed in \cite{Dudal:2009xh}. Moreover in \cite{Suzuki:1983cg,Hata:1992np} it was found that a simple connection of the Kugo-Ojima criterion and the Dual-Superconductor criterion by means of the Maximal Abelian Gauge is not possible. In \secref{sec_saturation} a possible unification of these criteria in terms of the gluon Dyson-Schwinger equation is investigated. Not touched here, but definitively worth mentioning, are the studies \cite{Braun:2007bx,Fister:2013bh} which present calculations of the expectation value of the Polyakov loop at finite temperature by means of functional 
methods.

\subsection{The Kugo-Ojima Scenario\label{sec_KO}}
      In \cite{Kugo:1979gm} Kugo and Ojima develop the covariant operator formalism for quantum gauge theories, which is an alternative approach to the path integral. Their central result concerning the confinement of quarks and gluons is that any physical asymptotic state of a quantum gauge theory in general covariant gauge is essentially colorless \emph{even in the absence of a Higgs-phase},
	  \be Q^r_{color} \ket{\Psi_{phys}} = 0\,.  \label{1_KO}\ee
      Since physically here (and henceforth) means \emph{measurable}, i.e. an object is considered to be physical if it can be identified in an experiment, \eqref{1_KO} just states, that in any experiment only colorless objects can be found. However, as we can measure an electron by ionizing an atom, \eqref{1_KO} can not be true for QED. Thus in QED \eqref{1_KO} must fail which is very likely connected to the fact the the electric charge in QED is not well-defined in the sense explained below.

      To get meaningful information from \eqref{1_KO} one needs to know how to construct a charge-operator $Q_{color}^r$ and what states shall be considered to be "physical", i.e. one needs a definition of the Hilbert space $\mathcal H = \left\{ \ket{\Psi_{phys}} \right\}$. The Kugo-Ojima scenario acts in the linear covariant gauge of Yang-Mills theory which was introduced above,
	  \be \L = \LYM + \frac{\xi}{2}\,b^2 - i b^r \partial_\mu A_\mu^r + i \bar c^r \partial_\mu D_\mu^{rs} c^s\,. \label{1_KO_lag}\ee
      In this section the canonical quantization procedure will be used. All fields in \eqref{1_KO_lag} are considered to be Heisenberg operators which fulfill the canonical equal time commutation relations in Euclidean spacetime, see \appref{app_currents} for details.

\subsubsection*{Color Charge}
      In addition to the BRST invariance discussed previously, the theory \eqref{1_KO_lag} is invariant under global color rotations 
	  \be \delta_\vartheta \phi^r = \cp{\phi}{\vartheta}{r} \label{1_KO_gc}\ee
      for all fields $\phi_i \in \{A_\mu,c,\bar c,b\}$. Using the equation of motion of the gluon, the Noether current of global color symmetry can be rewritten as, \cite{Kugo:1979gm},
	  \be J_\mu^r = - \partial_\nu F_{\nu\mu}^r + i s(D_\mu\bar c)^r \equiv \mathcal G^r _\mu+ \mathcal N^r_\mu \,. \label{1_KO_current}\ee
      The corresponding charge is the global color charge $Q_{color}^r$ which reads
	  \be  Q^r_{color} = \int\!d^3x\left( - \partial_\nu F_{\nu0}^r + i s(D_0\bar c)^r \right) \,. \label{1_KO_Qc}\ee
      The global color charge $Q^r_{color}$ is well defined if the integral \eqref{1_KO_Qc} exists. This statement is equivalent to the fact that the current $J_\mu^r$ does not couple to any massless state,
	  \be \bra{\Omega} J_\mu^r \ket{\Psi(p^2=0)} = 0\,. \ee
      Perturbatively, the antighost is massless. For $Q_{color}^r$ to be well-defined, non-perturbative effects have to kick in and for the asymptotic state of the antighost $\bar c^r \as \bar \gamma^r$ there must be a function $u(p^2)$ such that
	  \be (D_\mu \bar c)^r \pto \left(\one + u (p^2) \right) \partial_\mu \bar \gamma^r\,.  \label{1_def_u}\ee
      The Kugo-Ojima criterion now states that the charge $Q_{color}^r$ is well defined, and thus \eqref{1_KO} holds, if the current $\mathcal G^r_\mu$ does not couple to massless modes, i.e. there is no long-range gluon propagation, and the function $u(p^2)$ fulfills
	  \be u(p^2)\pto - 1 \,. \label{1_KO_crit} \ee
      The individual existence of both contributions of the global color charge \eqref{1_KO_Qc} excludes the spontanous breaking of some global symmetry, i.e. a Higgs-mechanism.

\subsubsection*{Physical Hilbert Space}  
      Having a criterion for the color charge in \eqref{1_KO} to be well-defined one still needs a definition of physical states. The problem that arises is, that in Lorentz covariant theories zero- and negative-norm states arise. The full vector space of states is denoted by $\V$, that part of $\V$ which shall contain only non-negative norm states by $\V_{phys}$, the space of zero-norm states by $\V_0$ and the positive definite Hilbert space by $\H$. The physical Hilbert space can be defined as the quotient space
	  \be \H = \overline{\V_{phys}/\V_0} \ee
      if the following conditions are fulfilled,
	\begin{enumerate}
	  \item[  i,] the Hamiltonian is hermitian, $H^\dag = H$,
	  \item[ ii,] $\V_{phys}$ is invariant under time evolution, 
	  \item[iii,] $\V_{phys}$ does not contain negative-norm states.
	\end{enumerate}
      While the first condition is fulfilled by construction of the Hamiltonian and the second one can be ensured via the definition of $\V_{phys}$ by means of a time-independent subsidiary condition, the main difficulty lies in proving that the as such defined $\V_{phys}$ fulfills condition three.
      
      In QED the Gupta-Bleuler mechanism \cite{Gupta:1949rh,Bleuler:1950cy} ensures the cancelation between the timelike and longitudinal degrees of freedom of the photon. The corresponding subsidiary condition reads
	  \be \partial_\mu A_\mu^+\ket{\Psi_{phys}} = 0 \label{1_KO_GP} \,,\ee
      which can be generalized to
	  \be Q_B \ket{\Psi_{phys}} = 0 , \label{1_KO_subs} \ee
      for general quantum gauge theories in the covariant gauge \cite{Kugo:1979gm}. The operator $Q_B$ is the Noether charge of the global BRST symmetry, \eqref{1_def_BRST}, 
	  \be  Q_B = \int\!d^3x\left((D_\mu c)^r\left(F_{0\mu}^r - i b^r\delta_{0\mu} \right) - \frac{i}{2}\cp{c}{c}{r}\partial_0 \bar c^r - i b^r (D_0 c)^r  \right)\,. \label{1_def_QB} \ee
      An important assumption in the Kugo-Ojima construction is that this BRST charge is unbroken and thus annihilates the vacuum,
	    \be Q_B\ket \Omega = 0 \qquad \Longleftrightarrow \qquad \vev{\com{Q_B}{\O}} = 0 \quad ,\forall \O\,. \ee
      The nilpotency property of the BRST transformations is transported to the BRST charge, $Q_B^2 = 0$. A valid vector space $\V_{phys}$ is defined as that set of states, which do not carry BRST charge,
	    \be \V_{phys} = \left\{\ket{\psi}: Q_B\ket{\psi}=0 \right\}\,.\ee 
      All negative norm contributions cancel by the so-called quartet mechanism described below. Taking this result already for granted one can define the physical, positively definite, Hilbert space of the quantum gauge theory by the cohomology construction
	  \be \H = \ker{Q_B}\!/\, \im {Q_B} \,.\label{1_KO_HS}\ee
      This means that the Hilbert space consists of equivalence classes of states, where equivalence between two states is given, if they only differ by a BRST exact term,
	  \be \H = \left\{ \left[\ket{\psi}\right] : s\ket{\psi}=0\,\text{ and } \,\ket{\psi'}\sim\ket{\psi} \Leftrightarrow \ket{\psi'} - \ket{\psi} = s\left(\ket{\chi}\right);\,\forall \ket{\chi} \in \V \right\} \,.\label{1_KO_HS_states}\ee

\subsubsection*{Quartet Mechanism}
      The last missing piece in the construction of the physical Hilbert space is the proof that negative norm states cancel if the subspace $\V_{phys}$ is defined as in \eqref{1_KO_subs}. It is given in \cite{Kugo:1979gm} and will be sketched here, only. Quantum gauge theories in the linear covariant gauge as defined in \eqref{1_KO_lag} are invariant under a global rescaling of the ghost-fields, $\delta_\lambda c^r = \lambda c^r$ and $\delta_\lambda \bar c^r = - \lambda \bar c^r$ for some complex parameter $\lambda \in \complC$. The corresponding Noether-current yields the conserved charge
	  \be Q_c = \int\!d^3x\,\Bigl(i\bar c^r\,(D_0c)^r - i (\partial_0\bar c)^r c^r\Bigr)\,,\ee
      which is proportional to the generator of the ghost-number
	  \be \com{iQ_c}{c^r} = c^r \qquad \text{ and } \qquad \com{iQ_c}{\bar c^r} = -\bar c^r \,.\ee

      The ghost-number generator and the BRST-charge generate a closed (graded) algebra
	  \be \acom{Q_B}{Q_B}  = 0\,,\qquad \com{iQ_c}{Q_B}  = Q_B\,, \qquad\com{iQ_c}{iQ_c} = 0\,.\nonumber
	   \label{1_KO_algebra}\ee
      Since both generators of the algebra \eqref{1_KO_algebra} are conserved charges of the theory, all states can be classified using the transformation properties under the corresponding transformations. The nilpotency of the BRST charge allows for a very simple classification,
	\begin{enumerate}
	 \item[  i,] physical particles, $Q_B\ket \psi = Q_c \ket \psi = 0$ 
	 \item[ ii,] singlet pairs of FP-conjugate states, $Q_B\ket \psi = 0$, $Q_c \ket \psi \neq 0$
	 \item[iii,] quartets, $Q_B\ket \psi \neq 0$, $Q_c \ket \psi \neq 0$.
	\end{enumerate}

      While for the physical particles positivity is assumed to define a consistent theory and the singlet pairs are neglected due to violation of the spin-statistics theorem, the last class of representations deserves more discussion. Take an operator $\chi$ with ghost-number $n$, which can be assumed to be even without loss of generality, $\com{iQ_c}{\chi} = n \chi$ and non vanishing BRST-variation $\gamma = \com{Q_B}{\chi} \neq 0$ which has ghost-number $n+1$. There exist unique operators $\beta$ with ghost-number $-n$ and $\bar \gamma$ with ghost-number $-(n+1)$ such that
	  \be \one  = \vev{\bar \gamma\, \gamma} = \vev{\beta\,\chi} \,. \ee
      The unity operator $\one$ in the equation above may carry factors of $p^2$ for dimensional reasons. Using that BRST symmetry is unbroken and the definition of $\bar \gamma$ and $\beta$
	  \be  \vev{\com{Q_B}{\bar \gamma\, \chi}}= 0 =\vev{\bar \gamma\, \gamma} - \vev{\beta\,\chi}  \label{1_KO_sti} \ee
      one finds that the operators $\bar \gamma$ and $\beta$ are related via BRST-transformations 
	  \be \beta = \acom{Q_B}{\bar \gamma}\,. \ee
      
      A set of operators $\{\chi,\beta,\gamma,\bar \gamma \}$, of which $\chi$ and $\beta$ are bosonic and $\gamma$ and $\bar\gamma$ are fermionic, is called a quartet if
      \begin{enumerate}
       \item[ i,] they are metric partners,
	  \be \vev{\chi\,\beta} = \vev{\bar \gamma\, \gamma} = \one \label{1_KO_MP}\,,\qquad \text{and,}\ee
	\item[ii,] they are related by the BRST transformations
	  \be \gamma = \com{Q_c}{\chi} \qquad \text{ and } \qquad \beta = \acom{Q_B}{\bar \gamma} \label{1_KO_quart}\,.\ee
      \end{enumerate}
      
      It can be shown that all contributions of a quartet to transition matrix-elements of states within the vector space $\V_{phys}$ as defined in \eqref{1_KO_subs} cancel, \cite{Kugo:1979gm}. A necessary condition for the existence of such quartets is an unbroken BRST symmetry for \eqref{1_KO_sti} to hold.

      One important example of such a quartet is the so-called "elementary quartet" which contains the longitudinal gluon, the NL field and ghost and antighost contributions. Identifying the asymptotic states
	    \begin{align}
		 A_\mu^r & \as  \partial_\mu \chi^r\,, &  (D_\mu c)^r & \as  \partial_\mu \gamma^r \,,\label{1_KO_elq}\\
		b^r & \as \beta^r\,, &    \bar c^r & \as \bar \gamma^r\,, \nonumber
	    \end{align}
      one can verify  the BRST transformation rules \eqref{1_KO_quart} for the asymptotic states. Using the Dyson-Schwinger equation of the antighost one finds that indeed \eqref{1_KO_elq} form a quartet. Their contributions do cancel in any physical S-matrix element. Due to dimensional reasons, the Fourier transform of the $\one-$operator picks up a factor of $p^2$. It reads
	  \be \F{\one} \rightarrow -i \frac{\delta^{rs}}{p^2} \,.\ee
      The quartet features two massless correlation functions,
	    \be \Fvev{\beta^s(y) \,\chi^r(x)} = \Fvev{\bar \gamma^s(y)\, \gamma^r(x)} = -i\frac{\delta^{rs}}{p^2}\,. \label{1_KO_mp}\ee
      Note the relation of the elementary quartet to the Slavnov-Taylor identity for the longitudinal gluon propagator, e.g. \eqref{1_dA_b_STI}.

      As explained above the quartet mechanism provides a proper generalization of the Gupta-Bleuler mechanism in QED. Indeed \eqref{1_KO_elq} and \eqref{1_KO_mp} can readily transported to QED in linear covariant gauge by replacing $D_\mu\rightarrow \partial_\mu$. However, as it stands, it does not explain the absence of the transverse gluons and quarks from the asymptotic state space. These particles form bound-states, which are non-perturbative and thus not treatable in this simple framework.  Only recently the appropriate Bethe-Salpeter equations for the transverse-gluon--ghost and quark-ghost bound states were derived and the IR-leading diagrams could be identified \cite{Alkofer:2011pe}.  In \secref{sec_cc} a generalization of the quartet mechanism beyond the linear covariant gauge is discussed.

\subsubsection*{Ghost Enhancement}
      The subsequent considerations follow \cite{smekalpriv}. In \eqref{1_KO_elq} and \eqref{1_def_u} an asymptotic state was assigned to the antighost and the covariant derivatives of the ghost and antighost, but not to the ghost-field itself. Consider the definitions of the asymptotic fields, \eqref{1_KO_elq}. The asymptotic field of the ghost is assumed to be proportional to the one of its covariant derivative and is denoted by $[c^r]^{as} = \lambda \gamma^r$ for some $\lambda \in \complC$. The Dyson-Schwinger equation for the ghost propagator reads
	  \be \delta^{rs}\delta(x-y) = \vev{\left(-i\partial_\mu D_\mu\bar c - i\cp{\bar c}{\partial_\mu A_\mu}{} \right)^r(x)\,\,c^s(y)}\,, \ee
      which yields in the asymptotic limit
	  \be -i\delta^{rs} = p^2 \left(\one +u(p^2) \right)\,\lambda\, \Fvev{ \bar\gamma^r(x) \,\gamma^s(y)} -\lambda\, \Fvev{\left[ \cp{\bar c}{\partial_\mu A_\mu}{r} (x)\right]^{as}\,\,\gamma^s(y)} \,.\label{1_KO_DSE_as}\ee
      Imposing the Landau gauge condition $\partial_\mu A_\mu^r=0$ and using \eqref{1_KO_mp} yields
	  \be c^r \as \frac{1}{1+u(p^2)} \,\gamma^r\,. \ee
      The asymptotic field of the ghost pics up a factor, which diverges for vanishing momentum if the Kugo-Ojima criterion \eqref{1_KO_crit} holds. This divergence exactly cancels the vanishing of the asymptotic field of $(D_\mu\bar c)^r$ such that the unity on the left-hand side of \eqref{1_KO_DSE_as} is preserved. When considering the infrared asymptotics of the ghost propagator,
	  \be \Fvev{\bar c^r \,c^s} \as \frac{-i}{1+u(p^2)} \frac{\delta^{rs}}{p^2} \,,\label{1_KO_GP_scaling}\ee
      one finds that, if the Kugo-Ojima criterion is fulfilled, i.e. $u(p^2)\pto -1$, then the ghost propagator, for vanishing momentum, is more divergent that a massless pole in Landau gauge.

      For general values $\xi$ there are two possibilities. First if there is no bound state of the longitudinal gluon and the antighost, then exactly the same argumentation applies as above and the asymptotic field of the ghost is infrared enhanced. However, if there is such a bound state and the last term in \eqref{1_KO_DSE_as} does not vanish the asymptotic field of the ghost becomes infrared-finite and one has
	    \be -i\delta^{rs} = -\lambda \Fvev{\left[ \cp{\bar c}{\partial_\mu A_\mu}{r} (x)\right]^{as}\,\,\gamma^s(y)} \,.\ee
      In this case the ghost propagator resembles the one of a massless field,
	    \be \Fvev{\bar c^r \,c^s} \as -i\lambda\, \frac{\delta^{rs}}{p^2} \,.\label{1_KO_GP_mass}\ee

      When calculating the ghost propagator by means of Dyson-Schwinger equations, Exact Renormalization group equations, Lattice gauge theory and other methods, the "scaling solution"  \eqref{1_KO_GP_scaling} and the "decoupling solution" \eqref{1_KO_GP_mass} were found. In \secref{sec_LG_DSEs} this dichotomy of solutions will be treated in some detail. From the point of view of the Kugo-Ojima criterion the statement is clear: If the Kugo-Ojima criterion is fulfilled, in Landau gauge there is only the scaling solution. For any $\xi > 0$ there are two possible solutions.

\subsection{The Dual-Superconductor Picture\label{sec_dsc}}

      \begin{figure}[t]
	  \centering
	  \includegraphics[width=.65\textwidth]{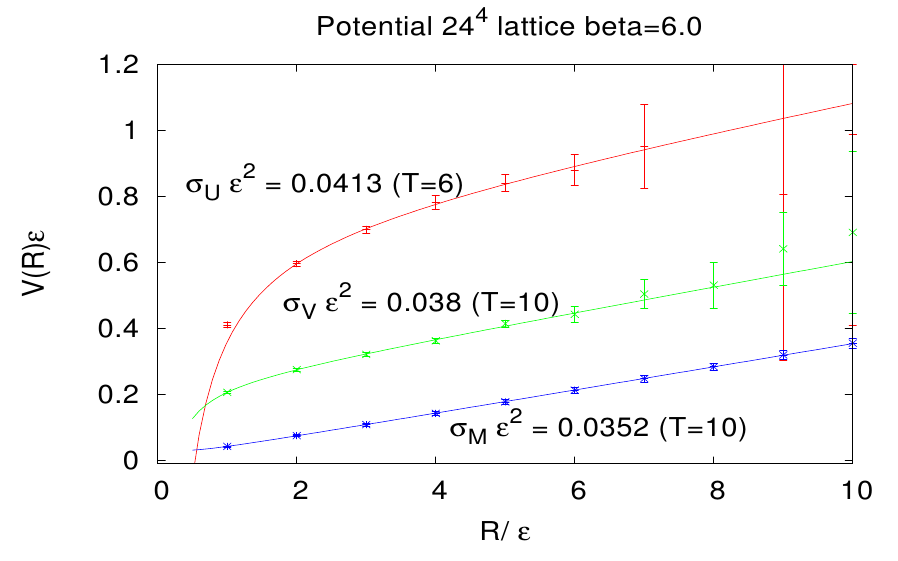}
	  \caption{The linear rising potential as obtained in the minimal option of the Cho-Faddeev-Niemi decomposition of $SU(3)$--Yang-Mills theory with lattice spacing $\varepsilon$. The quark-antiquark potential as obtained from the full Wilson-loop (red), the resrticted contributions (green) and the chromo-magnetic monopoles only (blue). The monopoles contribute $\approx 85\%$ of the string-tension. Re-printed from \cite{Kondo:2010pt} with permission of the authors. \label{fig:1_linrisepot}}
      \end{figure}

      The Dual-Superconductor picture of the Yang-Mills vacuum \cite{Mandelstam1976245,'tHooft:1995fi,Bohm:2001yx,Nair:2005iw} is an intriguing, but more heuristic and less quantitative idea of understanding confinement as a dynamical characteristic of the theory. Consider an ordinary Type-II-superconductor. If in the superconducting phase one applies a weak magnetic field it gets repelled completely by the Meissner-effect. Cranking up the magnetic field will destroy the superconductivity at some critical magnetic field strength. Type-II-superconducters have an additional phase, where the magnetic field pushes through the supercondutor while most of the material remains in the superconducting phase. The magnetic field pierces the material in small spots called Abrikosov vortices, which are small flux-tubes of magnetic field. The magnetic field is squeezed into this flux tubes by the supercurrent of the Cooper-pairs, i.e. condensed electric monopoles. 
      The idea of the dual-superconductor picture is that the Yang-Mills vacuum is a dual color-superconductor, i.e. the magnetic field is represented by the color-electric field and the electric monopoles become color-magnetic monopoles. The Abrikosov vortex is the color-flux inside a meson. Baryons are introduced accordingly.

      The analogy between Abrikosov-vortices in ordinary Type-II-superconductors and hadrons in the Yang-Mills vacuum can be pushed quite far. First, the magnetic-flux can be seen to be ``confined'' into the Abricosov vortices by the supercurrent. Also if one would introduce hypothetical magnetic monopoles at the beginning and end of the Abrikosov vortices, one would have a linear rising potential between these monopoles, exactly as one has for quarks \cite{Bohm:2001yx}. One immediate conclusion from the idea of dual-superconductivity is that the "Abelian parts" of the gauge-fields should dominate the low-energy-regime \cite{Ezawa:1982bf}.

      Being an heuristic idea of quark-confinement the dual-superconductor picture still attracted some attention and many studies on this field have been carried out. Using lattice simulations one can directly search for the chromo-magnetic monopoles in the vacuum, what is, however, technically demanding, e.g.\cite{DelDebbio:1994sx,PhysRevD.66.011503,Bonati:2011jv}. Another possibility to investigate if confinement is caused by the condensation of chromo-magnetic monopoles is to use the Cho-Faddeev-Niemi decomposition \cite{Cho:1979nv,Cho:1980nx,Faddeev:1998eq} and extract the string-tension, which stems from different parts of the gauge-fields \cite{Kondo:2010pt}, see \figref{fig:1_linrisepot}. It is found that chromo-magnetic monopoles contribute $\approx 85\%$ of the $SU(3)$--string tension.

       The dual-superconductor picture motivates the use of the Maximal Abelian gauge (MAG), \cite{Kronfeld:1987vd,Kronfeld:1987ri}, which will be introduced in \secref{sec_MAG}. The idea behind this gauge is to maximize the effect of the "Abelian" gluons and investigate the "Hypothesis of Abelian dominance". Abelian dominance could be confirmed in the gluon propagators both in an analytical investigation of the IR asymptotics of Dyson-Schwinger and exact Renormalization group equation \cite{Huber:PhD,Huber:2011fw}, and in lattice gauge theory studies \cite{Amemiya:1998jz,Gongyo:2012jb,Gongyo:2013sha}.

\subsection{The Gribov-Zwanziger Scenario \label{sec_gz}}
      The Faddeev-Popov method of gauge-fixing relies on the fact that the gauge-fixing condition, $F^r(A^U)$, picks exactly one representative out of every gauge-orbit \eqref{1_def_orbit}. Unfortunately, for non-Abelian gauge theories this is not the case. Consider a gauge field $A_\mu^r(x)$ which fulfills the Landau gauge condition $\partial_\mu A_\mu^r(x) = 0$ and a gauge transformation $\vartheta^r_{0}(x)$ which is a non-trivial zero mode of the Faddeev-Popov operator
	  \be \left(-\partial_\mu D_\mu\right)^r\vartheta^r_0(x) = 0 \,. \ee
      The gauge transformation of $A_\mu^r(x)$ then also fulfills the Landau-gauge condition
	  \be  \partial_\mu A_\mu^{r\,U_0} = \partial_\mu A_\mu^r + \partial_\mu D_\mu^{rs}\vartheta^s_0(x) = \partial_\mu A_\mu^r = 0 \,. \ee
      If a non-trivial zero-mode of the Faddeev-Popov operator exists, then the Landau gauge-fixing condition is not unique. This fact was first observed by Gribov \cite{Gribov:1977wm} and later put on  a more profound basis by Singer \cite{Singer:1978dk}. The gauge configuration $A_\mu^{r\,U_0}(x)$ is called a Gribov-copy of $A_\mu^r(x)$. Examples of Gribov copies can even be constructed explicitly, \cite{Henyey:1978qd}.

      Gribov proposed to restrict the functional integration to these configurations, whose Faddeev-Popov operator has positive eigenvalues, only. This region is called the first Gribov-region. Its boundary, the (first) Gribov-horizon, is characterized by the vanishing of the smallest eigenvalue of the Faddeev-Popov operator. Restricting the path integral to the first Gribov-region amounts in a drastic change of the propagators in the theory compared to perturbative calculations. While the ghost propagator gets enhanced in the infrared \cite{Gribov:1977wm},
	  \be D^{rs}(p^2) =\frac{1}{1-\sigma(p^2)} \,\frac{ \delta^{rs}}{p^2} \stackrel{p^2\rightarrow 0}{\approx} \frac{ \delta^{rs}}{p^4}\,,\label{1_gz_ghp}\ee
      the gluon propagator gets infrared suppressed,
	  \be D^{rs}_{\mu\nu}(p^2) = \delta^{rs}\left( \delta_{\mu\nu} - \frac{p_\mu p_\nu}{p^2} \right) \frac{g^2 p^2}{p^4 + \lambda^4} \stackrel{p^2\rightarrow 0}{\approx} 0\,,\label{1_gz_glp}\ee
      which is generally interpreted as a signature of confinement.

      Later on Zwanziger localized the restriction to the first Gribov-region which results in additional terms and additional fields in the quantized action \cite{Zwanziger:1988jt,Zwanziger:1989mf,Zwanziger:1992qr}, a pedagogical review can be found in \cite{Vandersickel:2012tz}. It can be shown that this additional terms in the action can be written in an BRST exact manner. This BRST symmetry, however, is spontaneously broken by the vacuum \cite{Maggiore:1993wq,Dudal:2012sb,Vandersickel:2012tz}. 
      
      To localize the \emph{"horizon-condition"}, i.e. the restriction of the path-integral to the first Gribov region, one introduces the additional fields $\{ \phi^{rs}_\mu, \phi^{rs}_\mu, \omega^{rs}_\mu,$ $ \bar \omega^{rs}_\mu \}$ of which $\phi$ and $\bar\phi$ are bosonic and $\omega$ and $\bar\omega$ are fermionic and where $r$ and $s$ are independent color indices. As usual $\mu$ is a Lorentz index. These new fields are related via the BRST transformations
	  \be s\phi^{rs}_\mu =  \omega^{rs}_\mu \,,\quad s \omega^{rs}_\mu =  0\,,\quad  s\bar \omega^{rs}_\mu =  \bar\phi^{rs}_\mu \,\quad\text{ and }\quad s \bar\phi^{rs}_\mu =  0\,.\ee
      The horizon condition is implemented by the additional Lagrangian
	   \begin{multline} \L_{hc} = s(\partial_\mu \bar \omega_\mu^{rs}D_\mu^{rt}\phi^{ts}_\mu)
		= \partial_\mu\bar \phi^{rs} D_\mu^{rt}\phi^{ts}_\mu - \partial_\mu \bar \omega_\mu^{rs}D_\mu^{rt}\omega^{ts}_\mu -  g f^{rtu}\partial_\mu \bar \omega_\mu^{rs} (D_\mu c)^t\phi_\mu^{us} \,\end{multline}
      which completes the Faddeev-Popov Lagrangian beyond perturbation theory,
	    \be \Lgf^{GZ} =  \L^{LG}_{gf} + \L_{hc}\,. \ee
     As in any case of a spontaneously broken theory it is advantageous to express the Lagrangian in terms of fluctuations about the vacuum of the broken theory. Therefore new fields are introduced,
      \begin{subequations}\label{1_gz_shift}
	    \begin{align}
		    \phi^{rs}_{\mu}(x) & = \vph^{rs}_{\mu}(x) - \gamma^{1/2} x_\mu \delta^{rs} \, ,\\
		    \bar \phi^{rs}_{\mu}(x) & = \bar\vph^{rs}_{\mu}(x) + \gamma^{1/2} x_\mu 
		    \delta^{rs} \, , \\
		    b^{r}(x) & = \hat b^{r}(x) + i \gamma^{1/2}x_\mu \tr[r]{\bar \varphi_\mu}(x) \, ,\\
		    \bar c^r(x) & = \hat{ \bar c}^r(x) + i \gamma^{1/2}x_\mu \tr[r]{\bar \omega_\mu}(x) \,,
	    \end{align}
      \end{subequations}
      where $  \tr[r]{\Psi_\mu} = g f^{rst} \Psi^{st}_{\mu}$. The new vacuum is then defined by $\vev{\vph^{rs}_{\mu}}= \vev{\bar\vph^{rs}_{\mu}}=0$. After the shift \eqref{1_gz_shift} one gets the Gribov-Zwanziger Lagrangian in Landau gauge, \cite{Vandersickel:2012tz},
      \begin{align}
      \L^{GZ}_{gf}& =  i \partial_\mu \hat b^r A_\mu^r - i (\partial_\mu \hat{\bar c}^r)(D_\mu c)^r +\partial_\mu\bar \vph^{rs} D_\mu^{rt}\vph^{ts}_\mu - \partial_\mu \bar \omega_\mu^{rs}D_\mu^{rt}\omega^{ts}_\mu  \label{1_gz_lag}\\
	      & -  g f^{rtu}\partial_\mu \bar \omega_\mu^{rs} (D_\mu c)^t\vph_\mu^{us}  + \gamma^{1/2} \left( D_\mu^{rs} (\vph^{sr}_{\mu} - \bar\vph^{sr}_{\mu}) - g f^{rtu}(D_\mu c)^t \bar \omega_\mu^{ur} \right) - \gamma d N_c  \,.\nonumber
      \end{align}
      The Lagrangian $\L^{GZ}_{gf}$ incorporates confinement already at tree-level as it reproduces the Gribov propagators \eqref{1_gz_ghp} and \eqref{1_gz_glp} in first order perturbation theory \cite{Vandersickel:2012tz}.      

      Although the shift \eqref{1_gz_shift} explicitly depends on the spacetime variable $x_\mu$ the resulting Lagrangian does not. This "Maggiore-Schaden" construction \cite{Maggiore:1993wq} can be made more stringent by localizing it by adding a function with small compact support \cite{Vandersickel:2012tz}. The BRST transformations of the shifted fields read
	  \begin{subequations}\label{1_gz_shiftBRST}
		\be sA_\mu^r  = (D_\mu c)^r \, ,\quad sc^r =-\frac 1 2 \cp{c}{c}{r}\,, \quad s \hat{\bar c}^r  = \hat b^r\,,\quad s\hat b^a  = 0 \, ,\ee
		\be s\vph^{rs}_\mu  = \omega^{rs}_\mu \, , \quad s\omega^{rs}_{\mu}  = 0 \, , \quad s \bar \omega ^{rs}_\mu(x)  = \bar \vph^{rs}_\mu(x) + \gamma^{1/2}\delta^{rs}x_\mu \, , \quad s\bar\vph^{rs}_\mu  = 0 \, .\ee
	  \end{subequations}

      The transformations \eqref{1_gz_shift} are scaled by the Gribov parameter $\gamma$ with mass-dimension $4$. It is a measure for the "strength" of the spontaneous breakdown of the BRST symmetry:
	  \be \vev{\bar \phi^{rs}_\mu} = \vev{s\bar \omega^{rs}_\mu} = \vev{\bar\vph^{rs}_{\mu}(x) + \gamma^{1/2} x_\mu \delta^{rs}} = \gamma^{1/2} x_\mu \delta^{rs}\,.\ee
      The Gribov parameter extremizes  the quantum effective action $\Gamma$,
	  \be \frac{\partial \Gamma}{\partial \gamma} = 0\,. \ee
      It is interesting to investigate in how far the new gauge-fixing terms change the behavior of the correlation functions. Perturbative calculations to one- and two-loop order in three \cite{Gracey:2010df} and four \cite{Ford:2009ar,Gracey:2009zz} spacetime dimensions show no sign of qualitative changes in the IR behavior of gluon and ghost propagators. Even more it is found that the Dyson-Schwinger equations and their asymptotic solution in the infrared stay unaffected by the introduction of the new fields \cite{Zwanziger:2001kw,Huber:2009tx,Huber:2010cq}.

%% file: 2_LG.tex
\chapter{Landau Gauge \label{sec_LG}}
      Due to its simple structure, Landau gauge is very much suited for investigations of the Dyson-Schwinger equations of QCD. By far most of the studies in the Dyson-Schwinger approach have been performed in this gauge. In this thesis the coupled system of the propagator Dyson-Schwinger equations of Yang-Mills theory in this gauge is investigated. The \emph{"one-loop-only-truncation"}, which was developed in \cite{Fischer:2002hna,Fischer:2003zc} is presented in some detail, as it is well-known and the basic concepts and technical steps needed in later sections are presented by means of this model calculation. Subsequently the truncation is improved by adding the sunset diagram to the calculation. The full inclusion of a two-loop term into a self-consistent Dyson-Schwinger calculation is the main technical advance of this thesis.

      The coupled system of propagator Dyson-Schwinger equations of Yang-Mills theory possesses two kinds of solution, called "scaling" and "decoupling", which are partially also found in other non-perturbative approaches. After introducing the equations, the status of the discussion on this ambiguity is summarized.

\section{Dyson-Schwinger Equations and Yang-Mills Propagators\label{sec_LG_DSEs}}
      In the last section the Lagrangian of Yang-Mills theory in linear covariant was derived. Introducing appropriate renormalization constants into \eqref{1_def_LYM2} and \eqref{1_def_SgfKU2} yields
	  \begin{multline} \L_{YM} + \Lgf^{lc} = -\frac{Z_3}{2} A_\mu^r\left(\delta_{\mu\nu}\partial^2 - (1- \frac{1}{Z_\xi\,\xi})\partial_\mu\partial_\nu \right)A_\nu^r  + Z_1\, (\partial_\mu A_\nu^r) \cp{A_\mu}{A_\nu}{r} \\
	  + \frac{Z_4}{4} \cp{A_\mu}{A_\nu}{r}\cp{A_\mu}{A_\nu}{r}  + \tilde Z_3\, i \bar c^r \,\partial^2 c^r - \tilde Z_1\, i \left(\partial_\mu\bar c\right)^r \cp{A_\mu}{c}{r} \,,\label{2_def_LG}\end{multline}
      where Landau gauge corresponds to the limit $\xi \rightarrow 0$. The propagators of the gluon $D_{\mu\nu}^{ab}(p)$ and ghost $D^{ab}(p)$ are parametrized as
      \be  D_{\mu\nu}^{ab}(p) = \delta^{ab} \,\frac{Z(p^2)}{p^2}\,\left( \delta_{\mu\nu} - \frac{p_\mu p_\nu}{p^2} \right) \,,\qquad\text{and,}\qquad D^{ab}(p) = i \frac{G(p^2)}{p^2}\,\delta^{ab} \label{LG_def_props} \,.\ee
 
    \begin{figure}[ht]
	\centering
	\includegraphics[width = .75\textwidth]{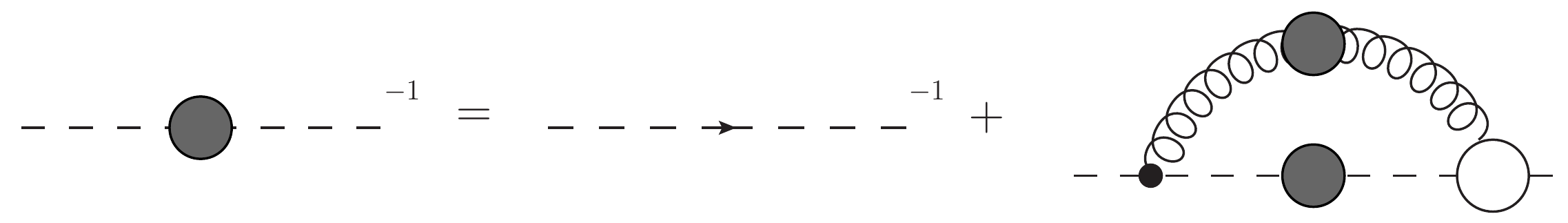}
	\caption{\label{fig_LG_ghost}The Dyson-Schwinger equation for the ghost propagator, \eqref{2_LG_ghDyson-Schwinger equation_1}. Grey and white blobs denote fully dressed propagators and vertices, respectively.}
      \end{figure}
    
      For explicit calculations it is advantageous to use the 1-PI formulation of the Dyson-Schwinger equations as given in \eqref{1_def_DSEG}. Taking derivatives with respect to the corresponding fields in \eqref{1_def_DSEG} and using the action \eqref{2_def_LG} yield the Dyson-Schwinger equations for the gluon and ghost propagators, \cite{AlkoferSmekal:2001,Fischer:2003zc},
      \begin{subequations}\label{2_DSE}
       \begin{align}
	D^{ab\,-1} & = \tilde Z_3\left( D^{0\,ab}\right)^{-1} \label{2_LG_ghDyson-Schwinger equation_1}\\
	  &- \tilde Z_1 g^2N_c \int\!\dbar^4 \q \Gamma_\mu^{0\,acd}(\p,\p_2,\p_1)D^{ce}(\p_1)D_{\mu\nu}^{df}(\p_2)\Gamma_{\nu}^{efb}(\p_1,-\p_2,\p)  \,, \nonumber \\
	 D_{\mu\nu}^{ab\,-1} & =Z_3 \left( D_{\mu\nu}^{0\,ab}\right)^{-1}\label{2_LG_glDyson-Schwinger equation_1} \\
	  & -\tilde Z_1 \int\!\dbar^4 \q\, \Gamma_\mu^{0\,acd}(\p,\p_2,\p_1) D^{de}(\p_1)\Gamma^{bfe}(-\p,-\p_2,\p_1)D^{fc}(-\p_2) \nonumber\\
	  & + Z_1\frac{1}{2} \int\!\dbar^4\q\,\Gamma_{\mu\alpha\beta}^{0\,acd}(\p,-\p_1,-\p_2)D_{\beta\gamma}^{de}(\p_2)D_{\alpha\delta}^{cf}(\q)\Gamma_{\nu\gamma\delta}^{bef}(-\p,\p_2,\p_1) \nonumber \\
	  & + \frac{1}{2} Z_4 \int\!\dbar^4 \q\, \Gamma_{\mu\nu\alpha\beta}^{0\,abcd}D_{\alpha\beta}^{cd}(\q) \nonumber\\ & - \frac{1}{6} Z_4 \int\!\dbar^4 \q_1\!\!\int\!\dbar^4 \q_2\, \Gamma_{\mu\rho\sigma\tau}^{0\,arst}D_{\alpha\sigma}^{cs}(\p_1) D_{\beta\rho}^{dr}(\p_2) D_{\gamma\tau}^{et}(\p_3) \Gamma_{\nu\alpha\beta\gamma}^{bcde} \nonumber \\
	  & + \frac{1}{2} Z_4 \int\!\dbar^4 \q_1\!\!\int\!\dbar^4 \q_2\, \Gamma_{\mu\rho\sigma\tau}^{0\,arst}     \Gamma_{\alpha\nu\beta}^{cbd}(\p_1,-\p,-\p_4)\Gamma_{\gamma\delta\epsilon}^{efg}(\p_2,\p_4,\p_3) \nonumber \\ & \hspace{73mm} \times\,\, D_{\alpha\rho}^{cr}(\p_1) D_{\sigma\gamma}^{se}(\p_2) D_{\tau\epsilon}^{tg}(\p_3)D_{\beta\delta}^{df}(\p_4) \,, \nonumber 
      \end{align}
      \end{subequations}
      where the bare n-point functions are denoted by the superscript $0$. 

      The solutions of \eqref{2_DSE} are the fully dressed ghost and gluon propagators. They contain all information about the propagation of the Yang-Mills fields in Landau gauge. They have been investigated from the beginning of non-perturbative investigations in QCD \cite{Smit:1974je,Eichten:1974et,Marciano:1977su}. Since then, approximation and truncation schemes improved steadily, from a very simple ansatz for the gluon loop, \cite{Mandelstam:1979xd}, to a more advanced technique of subtracting quadratic divergences,\cite{Brown:1988bn}, the discovery of ghost dominance and a solution of the coupled equations \cite{vonSmekal:1997is,vonSmekal:1997vx,Atkinson:1997tu,Fischer:2002hna} to a first approximate inclusion of two-loop terms, \cite{Bloch:2003yu}. Since the mid of the 1990's a simple but powerful technique was used to gain information from this equations in the infrared. Assuming power laws for the propagator dressing functions in the infrared
	  \be Z(p^2) \propto (p^2)^{\delta_A} \qquad \text{, and,} \qquad G(p^2) \propto (p^2)^{\delta_c} \ee
      one searches for \emph{self-consistent solutions}. With improving truncations such a solution could be found, which uniquely relates the infrared-coefficients of the gluon and ghost propagators with the scaling relation \cite{Zwanziger:2001kw,Lerche:2002ep,Fischer:2002hna}
	  \be \frac{1}{2}\delta_A = -\delta_c \equiv \kappa = \frac{93-\sqrt{1201}}{98} \approx 0.595 \,. \label{2_scalRel}\ee
      This solution is nowadays called the scaling solution of the Yang-Mills propagators and its characteristic is the definite relation between the ghost and gluon infrared exponents. In accordance with the Kugo-Ojima and Gribov-Zwanziger scenarios, the scaling solution yields a ghost propagator which is more divergent than a massless pole and an infrared vanishing gluon propagator. The corresponding running coupling $\alpha_{s}(p^2)$ gains an infrared fixed point $\alpha_s(0) > 0$. 

    \begin{figure}[t]
	\centering
	\includegraphics[width = \textwidth]{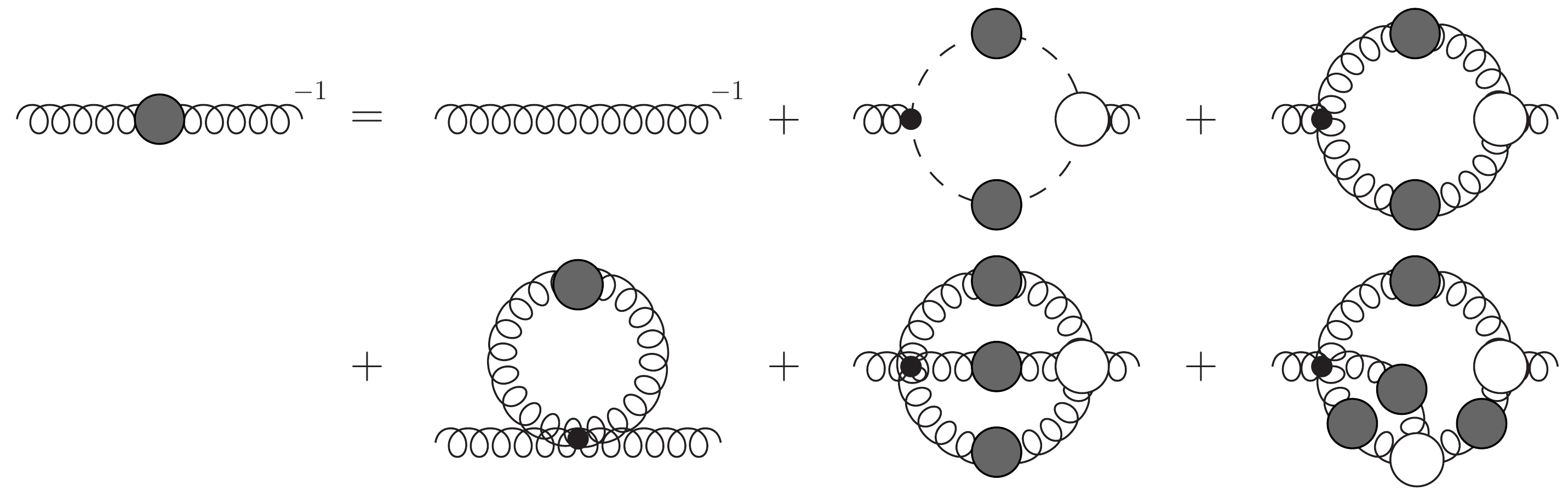}
	\caption{\label{fig_LG_glue}The Dyson-Schwinger equation for the gluon propagator, \eqref{2_LG_glDyson-Schwinger equation_1}. Grey and white blobs denote fully dressed propagators and vertices, respectively.}
      \end{figure}

      Both, Dyson-Schwinger equations and exact renormalization group equations, define the Green functions of a quantum field theory. They form two complementary sets of infinite coupled equations. A comparison between the infrared asymptotics of the two sets of equations cumulated in a proof, that, if the scaling solution exists, it is unique \cite{Fischer:2006vf,Fischer:2009tn}. For any proper vertex Green function with $n$ ghost and $m$ gluon legs, $\Gamma^{(2n,m)} = Z^{(2n,m)} \mathbf T$, where the tensor structure $\mathbf T$ carries the canonical dimension, in uniform scaling, the infrared asymptotics of the dressing function $Z^{(2n,m)}$ can be given in terms of the infrared exponent $\kappa$
	  \be  Z^{(2n,m)}(\lambda p_i) \propto \lambda^{(n-m)\kappa} Z^{(2n,m)}(p_i) \,.\label{2_generalSR}\ee

      \subsubsection*{Scaling vs. Decoupling}
      Since the discovery of the scaling solution in the Dyson-Schwinger equations, it has been tried to verify this solution also by means of other non-perturbative techniques. While it was found in Exact renormalization group equations \cite{Pawlowski:2003hq} and stochastic quantization \cite{LlanesEstrada:2012my}, it could not be found by means of Lattice gauge theory in four spacetime dimensions. Instead a family of solutions for the Yang-Mills propagators was found where the ghost propagator possesses a simple massless pole and the gluon propagator becomes finite in the infrared, e.g. \cite{Maas:2011se,Oliveira:2012eh,Sternbeck:2012mf,Bogolubsky:2013rla} and references therein. 
      The corresponding running coupling vanishes in the infrared, $\alpha_s(0) = 0$. This solutions have also been found for the Dyson-Schwinger equations, \cite{Aguilar:2008xm,Boucaud:2008ji,Fischer:2008uz,Boucaud:2011ug}, a refinement of the Gribov-Zwanziger action \cite{Dudal:2008sp} and stochastic quantization \cite{LlanesEstrada:2012my} and are called the "decoupling" solutions or sometimes also the "massive" solutions. The term "massive" seems unfortunate, however. First of all, a pole mass of the gluon is ruled out by perturbation theory \cite{Beringer:1900zz}. Second, a screening mass for the gluon defined as
	  \be  m^2_{screening} = \lim_{p^2\rightarrow 0} \frac{p^2}{Z(p^2)} \,. \ee
      does not discriminate between scaling and decoupling solutions, because this screening mass is non-vanishing for both solutions. For the scaling solution, however, it is infinite. Moreover it is questionable if an as such defined mass is gauge independent.

      For phenomenology it is irrelevant which solution might be the correct one, \cite{Blank:2010pa}, which is intuitively clear if one considers that the difference of the two solutions becomes manifest only far beyond the confinement scale $\Lambda_{QCD}$. It also turns out that the family of decoupling solutions depends on one parameter and possesses the scaling solution as a limit. This parameter can be interpreted as an additional renormalization of the ghost dressing function \cite{Fischer:2008uz}, some additional gauge-fixing parameter \cite{Maas:2009se,Maas:2013vd} or the strong running coupling at some renormalization scale \cite{RodriguezQuintero:2010wy,RodriguezQuintero:2011vp,Boucaud:2011ug}. 
      On the lattice one finds different decoupling solutions if gauge-fixing is performed with respect to the lowest eigenvalue of the Faddeev-Popov operator \cite{Sternbeck:2012mf}. 

      From physics point of view, there are arguments for both solutions. The decoupling solutions seem to be more general since it is found by more different approaches and minimizes the effective action \cite{LlanesEstrada:2012my}. Moreover in a perturbative renormalization group analysis one finds that only the decoupling solutions corresponds to an infrared-stable fixed-point \cite{Weber:2011nw}. In contrast in the strong coupling limit of lattice gauge theory one finds reminiscent of the  scaling solution, however with a strong debate about the interpretation of the data \cite{Sternbeck:2008mv,Cucchieri:2009zt}. 

      In perturbation theory massless gauge-bosons lead to severe divergences in the infrared. The scaling solution orders this infrared divergences of Yang-Mills theory in a specific self-consistent manner for all Green functions, \eqref{2_generalSR}, while for the decoupling solution such an ordering is not known. For confinement, beside the gluon propagator, one especially interesting Green function is the four-quark vertex as it should encode the linear rising potential between heavy quarks. This linear behavior corresponds to a $1/k^4$-singularity  in momentum space \cite{Gromes:1981cb}. While the corresponding $1/k^4$-singularity is found for the scaling solution \cite{Alkofer:2008tt}, the decoupling solution needs an infrared diverging coupling constant $g$ to obtain the $1/k^4$ behavior \cite{Vento:2012wp}, or alternatively, a diverging sum of infinitely many Green functions. 
      
      Moreover, all arguments on scaling and decoupling solutions depend on the dimensionality: while there is a multitude of solutions in $d=3,4$ spacetime dimensions, in $d=2$ spacetime dimensions only the scaling solution exists and is found both in the continuum and on the lattice \cite{Cucchieri:2012cb,Huber:2012zj,Zwanziger:2012xg}.
      
      To summarize: At the time writing this thesis, the deep-infrared behavior of the gluon and ghost propagators stays inconclusive. Neither the scaling, nor the decoupling solutions have been ruled out. No unambiguous interpretation of the interpolation parameter has been given. It is even questionable if there is a physical, i.e. measurable, difference between the solutions, as it is hard to imagine an experiment which probes QCD far below the confinement scale. Still, the solution of this puzzle promises deep insights into the theoretical structure of the low-energy regime of Landau-gauge Yang-Mills theory in particular and the non-perturbative definition of gauge-fixed non-Abelian quantum gauge theories in general.

      In this thesis, whenever calculating dressing functions of gluon and ghost propagators, scaling type of solutions will be used. This should not be interpreted as a bias but has been chosen for convenience and internal consistency. All numerical calculations could have been done with decoupling type of solutions equivalently.

\section{Review: The One-Loop-Only Truncation \label{sec_2_olo}}
    The solution of the Landau-gauge--Yang-Mills propagator Dyson-Schwinger equations in \emph{one-loop-only truncation} as developed in \cite{Fischer:2002hna,Fischer:2003zc} is reviewed in this section. This calculations serves as a role model for the inclusion of the sunset diagram in the following subsection and the calculations in the Maximal Abelian gauge in \secref{sec_MAG}. Thus the methods and techniques are explained in some detail. 

    \subsection{Construction and Vertex Models}
    The one-loop-only truncation was developed in \cite{Fischer:2002hna,Fischer:2003zc}. Its construction principle is to exclude all diagrams which contain four-gluon interactions. The missing vertices are the three-point functions, the ghost-gluon and three-gluon vertices. The basic structure of vertex models here and in the complete thesis is to take the tree-level structure of the vertex, furnished with a scalar dressing function. The dressing function usually consist of some product of the propagator dressing functions to account for back-coupling effects in the iterated solution process. The asymptotics of the vertex dressing function is constrained by infrared analysis of the Dyson-Schwinger equations and perturbation theory. This kind of vertex modeling has been proven to be very successful in Dyson-Schwinger calculations. However, a certain dependence on the structure of the dressing-functions cannot be denied, \cite{Pennington:2011xs,Huber:2012kd,Blum:2014gna}.
    
     In the perturbative regime, the ghost-gluon vertex in Landau gauge is constrained by Taylor's non-renormalization theorem, \cite{Taylor:1971ff}, 
    \be \tilde Z_1 = 1 \,.\label{2_Taylor}\ee
    Assuming this non-renormalization to be true also in the infrared, what is confirmed by recent dynamical calculations \cite{Huber:2012kd}, one can model the ghost-gluon vertex by its tree-level structure only. The three-gluon vertex is modeled as 
	\begin{multline} \Gamma_{\mu\nu\sigma}^{abc}(q,k,p) \approx D_{3\Gamma}(q,k,p)\Gamma_{\mu\nu\sigma}^{0\,abc}(q,k,p) \\= \frac{1}{Z_1}\,\left(Z(q^2) Z(k^2) \right)^\alpha \left(G(q^2) G(k^2) \right)^\beta \Gamma_{\mu\nu\sigma}^{0 \,abc}(\q,\k,\p)\label{2_def_3g}\end{multline}
    with $k^2 = (\p-\q)^2$. The parameters $\alpha$ and $\beta$ can be fixed to the ultraviolet and infrared behavior. Assuming the scaling relation \eqref{2_scalRel} holds and assuming an infrared constant three-gluon vertex one immediately finds $\beta - 2\alpha =0$. The high-energy behavior will be fixed later. Here the vertex model of \cite{Fischer:2002hna,Fischer:2003zc} was chosen, which is sufficient for our purposes.  For a more elaborate bose-symmetric vertex construction consider \cite{Huber:2012kd}. Only recently two groups reported on self-consistently back-coupled calculations of the three gluon vertex \cite{Blum:2014gna,Eichmann:2014xya}.

    To simplify the numerical solution it is advantageous to project the truncated gluon-equation with some Lorentz-tensor to gain a scalar equation. Here some subtlety arises. In Landau gauge the gluon-propagator is transverse and features one dressing function as defined in \eqref{LG_def_props}. However, in a solution of the truncated gluon-Dyson-Schwinger equation, quadratic divergences arise, which manifest itself in the $\delta_{\mu\nu}$ part of the propagator. Thus a longitudinal contribution to the gluon-propagator arises, which properly has to be subtracted. Brown and Pennington therefore developed a projector, which exactly subtracts all contributions from the $\delta_{\mu\nu}$ part in the truncated Dyson-Schwinger equation and thus cancels all quadratic divergences, \cite{Brown:1988bn}. This projector can be generalized \cite{Fischer:2002hna,Fischer:2003zc}. In $d$ spacetime dimensions it reads
    \be \mathcal P_{\mu\nu}(p,\zeta,d) = \frac{1}{d-1} \left( \delta_{\mu\nu} - \zeta \frac{p_\mu p_\nu}{p^2} \right) \label{LG_def_BP}\,.\ee
    One finds $ \delta_{\mu\nu}  \mathcal P_{\mu\nu}(p,\zeta,d) = \frac{d-\zeta}{d-1}$ and thus all quadratic divergences are subtracted for $\zeta=d$. However one has no control of possible over-subtractions. If the truncation of the gluon Dyson-Schwinger equation would respect transversality the calculation would be independent on the parameter $\zeta$, as must any physically relevant result. However, the simple truncation here does not respect transversality and thus is not independent on this parameter. Still one can use the dependence on this parameter to gain more insights into the details of the calculation as has been done in \cite{Fischer:2002hna,Fischer:2003zc}. The color structure is traced out via projecting both sides of the equations with $\frac{1}{N_c}\delta^{ab}$.

    The Landau gauge propagator Dyson-Schwinger equations in one-loop only truncation, projected with the generalized Brown-Pennington projector, read, \cite{Fischer:2002hna,Fischer:2003zc},
    \begin{subequations}\label{2_olo}
    \begin{align}
    \frac{1}{G(p^2)} = \tilde Z_3 &- g^2 N_c \int\!\dbar^4 q \,\frac{1-z_q^2}{k^4}\,G(q^2)Z(k^2) \label{2_LG_ghDyson-Schwinger equation}\\
    \frac{1}{Z(p^2)} =  Z_3 &+ \frac{g^2 N_c}{p^2\,3} \int\!\dbar^4 q \,\frac{q^2(1-\zeta\,z_q^2) - (1-\zeta) p\, q\, z_q}{q^2\,k^2}\,G(q^2)G(k^2) \label{2_LG_glDyson-Schwinger equation}\\ 
    & \,+\frac{g^2 N_c}{p^2\,3}\, \int\!\dbar^4 q\,\K_{gl}\,\left(Z(q^2) Z(k^2) \right)^{(1+\alpha)} \left(G(q^2) G(k^2) \right)^{2\alpha} \nonumber
    \end{align}
    \end{subequations}
    with  the kernel of the gluon-loop
    \begin{align}
	\K_{gl} &= \frac{1}{q^2} \Biggl[\frac{\zeta-5}{2}+ \frac{1}{k^2}\Biggl(q^2 \left(-\frac{9}{2} + 5\zeta \,z_q^2\right) + p^2 \left(\frac{\zeta-10}{2} + \frac{9+\zeta}{2} \,z_q^2\right) \label{LG_def_kerQ}\\ 
	&\hspace{5mm} -  p\, q \,z_q\,\left(2+4\zeta+\zeta z_q^2\right) \Biggr) +\frac{1}{k^4}\Biggl(q^4 \left(\frac{2-\zeta}{2} + \zeta z_q^2\right) + p^4 \left(\frac{1}{2} +  z_q^2\right) \nonumber   \\
	&\hspace{5mm}+ p^2q^2\,\frac{z_q^2}{2}(\zeta - 1)- 3\, p^2q^2 + p\,q \, z_q \left( q^2 \left(1-\zeta z_q^2 \right) + p^2 \left(1- z_q^2 \right) \right)\Biggr) \Biggr]\nonumber \,.
    \end{align}
    Before solving the coupled set of equations \eqref{2_olo} it is beneficial to investigate their high and low energy asymptotics. In the ultraviolet regime they should match resummed perturbation theory since the dressing functions $Z(p^2)$ and $G(p^2)$ are directly related to the corresponding wave function renormalization constants. In the infrared regime a power-law ansatz leads to the scaling solution \eqref{2_scalRel}.

    \subsection{UV-Analysis \label{sec_2_UV_olo}}
    Due to asymptotic freedom the high-energy regime of QCD and Yang-Mills theory can be approached perturbatively and the ultraviolet behavior of the QCD Green-functions in standard gauges is well known, e.g. \cite{Altarelli:1981ax,Muta}.  To fix the conventions, we define the renormalization functions $\beta(g)$ and $\gamma_i(g)$ as
    \begin{align}
      \beta(g) & = \frac{\partial\, g_R}{\partial\, \log{\mu}} = -\beta_0 \,g^3 - \beta_1 \,g^5 + \dots \\
      \gamma_i(g) & = \frac{\partial\, Z_i}{\partial\, \log{\mu}} =  \gamma_{i0} \,g^2 + \gamma_{i1} \,g^4 + \dots \,.\label{2_def_gi}
    \end{align}
    While the QCD $\beta-$function is gauge-invariant and its first expansion coefficient for $N_f-$fermion flavors is given by the famous $\beta_0 =\frac{1}{(4\pi)^2} \left(\frac{11 N_c}{3} - \frac{2}{3} N_f\right)$, the $\gamma-$functions do depend on the choice of gauge. Summing up the leading logarithmic contributions one finds that a two-point function $G_2(p^2;g,\mu)$ behaves in contrast to its tree-level counter-part $G^{(0)}_2(p^2;g_0,\Lambda)$ as, \cite{Collins:1984},
    \be G_2(p^2;g,\mu)\Big\rvert_{leading \,log} = \left(1+\beta_0 g^2 \,\log{\frac{p^2}{\mu^2}} \right)^{ \frac{\gamma_{i0}} {2\beta_0}}G_2^{(0)}(p^2;g_0,\Lambda)\,. \label{2_resummedPT}\ee

    The solutions of the Dyson-Schwinger equations are the fully dressed Green functions. To resemble resummed perturbation theory to lowest order, the propagator dressing functions $d_i \in \{Z(p^2),G(p^2)\}$ have to behave according to \eqref{2_resummedPT}
    \be d_i(p^2)  \stackrel{p^2\gg 1}{\longrightarrow} \left(1+\beta_0 g^2\, \log{\frac{p^2}{\mu^2}} \right)^{\gamma_i} \label{2_def_gend} \ee
    with the renormalized coupling $g$, the renormalization scale $\mu$ and the the anomalous dimension of the field $i$, $\gamma_i = \frac{\gamma_{i0}}{2\beta_0}$. In Landau gauge Yang-Mills theory the values for the anomalous dimensions are , \cite{Muta},
      \be  \gamma_A = -\frac{13}{22}\,,\qquad \text{and,}\qquad \gamma_c = -\frac{9}{44} \,. \label{2_gAgc}\ee
    
    The anomalous dimensions of the field determine the behavior of the renormalization constants when the renormalization scale is changed. Relations as the non-renormalization theorem \eqref{2_Taylor} thus also constrain the anomalous dimensions of the theory. Using the identity $\tilde Z_1 = Z_g Z_3^{\frac{1}{2}} \tilde Z_3$ and applying the logarithmic derivative on both sides of \eqref{2_Taylor} yields the relation
	  \be 0 =- \frac{1}{2}-\frac{\gamma_A}{2} - \gamma_c \,. \label{2_Taylor_gi}\ee
    The ultraviolet behavior of the three-gluon vertex is determined by the anomalous dimension of its renormalization constant $Z_1$. Combining the definition of $Z_1$, \eqref{1_def_Z1Z4}, with the non-renormalization theorem \eqref{2_Taylor} yields the identity $Z_1 = \frac{Z_3}{\tilde Z_3}$ and thus the corresponding anomalous dimension is given by $\gamma_{3A} = \gamma_c-\gamma_A = \tfrac{17}{44}$. Taking into account the anomalous dimension of the factor $Z_1^{-1}$ in \eqref{2_def_3g} one then finds for the vertex parameter $\alpha = -\gamma_{3A}$ which completely fixes the vertex model.

    When investigating the ultraviolet-behavior of a set of truncated Dyson-Schwinger equations as \eqref{2_olo} one performs several approximations to gain insight into the high-energy behavior of the equations. Demanding self-consistency of the equation then yields constraints for the anomalous dimensions of the fields. First one assumes all momenta to be large and thus the dressing functions are described by their perturbative behavior \eqref{2_def_gend}. As the logarithms vary slowly, it is safe to neglect the dependence of the dressing functions on the angular variables and substitute
	  \be d_i(k^2) \stackrel{p^2,q^2\gg 1}{\longrightarrow} d_i(q^2) \,,\ee
    which allows to pull the dressing functions outside of the angular integrals. Second, for any finite external momentum $p^2$, the main contribution to the integral stems from the interval of integration $[p^2,\Lambda^2]$, as the cutoff can be arbitrarily large and the integrals increase with the cutoff. Using the Euclidean measure \eqref{A_measure_ol} and performing the angular integrals one finds
    \begin{subequations}\label{2_olo_ymax}
	\begin{align}
	\frac{1}{G(p^2)} &\stackrel{p^2\gg1}{\approx} \tilde Z_3 - \frac{3 g^2 N_c}{(8\pi)^2} \int_{p^2}^{\Lambda^2}\!dq^2 \,\frac{G(q^2)Z(q^2)}{q^2} \label{2_LG_ghDyson-Schwinger equation_ymax}\\
	\frac{1}{Z(p^2)} &\stackrel{p^2\gg1}{\approx}   Z_3 + \frac{g^2 N_c}{3(8 \pi)^2} \frac{1}{p^2} \int_{p^2}^{\Lambda^2}\!dq^2 \,\left(\frac{(\zeta-2)\,p^2 }{q^2} - (\zeta-4)\right)\,\left(G(q^2)\right)^2 \label{2_glDyson-Schwinger equation_ymax} \\
	& + \frac{g^2 N_c}{3(8 \pi)^2}  \frac{1}{p^2} \int_{p^2}^{\Lambda^2}\!dq^2 \left( \frac{7}{2} \frac{ p^4}{q^4}  - (\zeta +24)\,\frac{ p^2 }{q^2} + 6 (\zeta -4)  \right) \left(G(q^2)\right)^{-4\gamma_{3A}} \left(Z(q^2)\right)^{2-2\gamma_{3A}}\,. \nonumber
	\end{align}
    \end{subequations}
    From \eqref{2_olo_ymax} one can extract the structure of ultraviolet-divergences  of the ghost and gluon equations in the chosen truncation. While the ghost equation is logarithmically divergent, both terms in the gluon equation are quadratically divergent. These terms which generate the quadratic divergences are completely subtracted for $\zeta=4$ as expected. In perturbation theory quadratic divergences are absorbed by a mass-counterterm such that the renormalization condition \eqref{1_rc_Zm} is fulfilled. Moreover in gauge invariant regularization schemes as, e.g. dimensional regularization, this problem does not appear since the quadratic divergences of the ghost and gluon loops cancel \cite{Muta}.  One way to mimic the cancellation mechanism of perturbation theory is to construct a "counter-term" $\tau_{gl}$ such that the "regularized tensor structure" $\widetilde{\K}_{gl} = \K_{gl} - \tau_{gl}$ cancel the quadratic divergences of the ghost loop. Such a construction is 
obtained by setting $\tau_{gl} =- \frac{1}{q^2}\frac{5}{4}(4-\zeta)$. The resulting gluon equation \eqref{2_LG_glDyson-Schwinger equation} where the kernel ${\K}_{gl}$ is replaced by 
	\be  \widetilde{\K}_{gl}={\K}_{gl} + \frac{1}{q^2}\frac{5}{4}(4-\zeta) \ee
    is free of quadratic divergences for any value of the parameter $\zeta$.

    In an appropriate regularization the full Dyson-Schwinger equations are finite. The quadratic divergences considered here thus arise from the hard cutoff regularization but also the chosen truncation. The subtraction procedure presented here and as developed in \cite{Fischer:2002hna,Fischer:2003zc} can thus be interpreted as non-perturbative corrections to the tree-level vertices used in the truncation.

    The integrals in \eqref{2_olo_ymax} can be performed if the dressing functions are replaced by their logarithmic asymptotic behavior. Taking into account the ultraviolet-leading terms on both sides of the equations and absorbing the cutoff dependence with the renormalization constants $Z_3$ and $\tilde Z_3$ yields
        \begin{subequations}\label{2_olo_ultraviolet}
	\begin{align}
	 \left(1+ \beta_0 g^2\, \log{\frac{p^2}{\mu^2}} \right)^{-\gamma_c} &= \frac{g^2 N_c\,}{(8\pi)^2}\,3\, \frac{ \left( 1+ \beta_0 g^2 \,\log{\frac{p^2}{\mu ^2}}\right)^{1+\gamma_A+\gamma_c}}{ \beta_0 g^2\,(1+\gamma_A  +\gamma_c)} \label{2_LG_ghDyson-Schwinger equation_ultraviolet}\\
	\left(1+ \beta_0 g^2\, \log{\frac{p^2}{\mu^2}} \right)^{-\gamma_A} & =   \frac{g^2 N_c}{(8 \pi)^2}\,\frac{26}{3} \,\frac{\left( 1+ \beta_0 g^2  \,\log{\frac{p^2}{\mu ^2}}\right)^{1+2\gamma_c}}{ \beta_0 g^2\,(1+2 \gamma_c)} \label{2_glDyson-Schwinger equation_ultraviolet}\,.
	\end{align}
    \end{subequations}
    Self-consistency demands that the exponents and coefficients on both sides of the equations match. For the exponents one gets back \eqref{2_Taylor_gi}, which just states that Taylor's theorem was implemented consistently. For the coefficients to equalize on both sides of equations \eqref{2_olo_ultraviolet} one finds
	\be 1 = \frac{9}{44 \left(1+\gamma_A + \gamma_c \right)}\,, \qquad \text{and,}\qquad  1 = \frac{13}{22 \left(1+2 \gamma_c \right)} \,,\ee
    which is fulfilled by the perturbative anomalous dimensions \eqref{2_gAgc}. The relations \eqref{2_olo_ultraviolet} provide non-trivial checks for the used vertex models and show that the given truncation correctly reproduces resummed perturbation theory. In addition it is independent on the parameter $\zeta$, i.e. details of the subtraction of the quadratic divergences, in the ultraviolet.

    \subsection{IR-Analysis}
    The Dyson-Schwinger equations not only have to be solved self-consistently in the high-energy, but also in the low-energy regime. One possibility to gain such self-consistent solutions is to assume power-law behavior of the dressing functions in the deep-infrared,
	\be Z(p^2) \stackrel{p^2\ll1}{\approx} c_A (p^2)^{\delta_A}\,,\qquad\qquad   G(p^2) \stackrel{p^2\ll1}{\approx}c_c (p^2)^{\delta_c}\,, \label{2_powerlaw}\ee
    with the constant infrared-coefficients $c_i$ and the infrared-exponents $\delta_i$.

    The central integral for the infrared analysis with one loop  is given by \cite{Lerche:2002ep,Fischer:2003zc,Huber:2007kc}
	\begin{align}
	    I_{ol}(p^2,a,b,d) & := \int\!\dbar^dq\,\left(q^{2}\right)^a\left((p-q)^2 \right)^b   \label{2_infraredGamma} \\
			       &\hspace{-13mm} = (4\pi)^{-\frac{d}{2}} \left(p^2 \right)^{\frac{d}{2}+a+b} \frac{\Gamma\left(a+\frac{d}{2}\right)\Gamma\left(b+\frac{d}{2}\right)\Gamma\left(-\left(a+b+\frac{d}{2}\right)\right)}{\Gamma\left(-a\right)\Gamma\left(-b\right)\Gamma\left(a+b+d\right)} \,. \nonumber
	\end{align}
     For an analysis of the infrared asymptotics, one plugs the ans\"atze \eqref{2_powerlaw} into \eqref{2_olo} and uses the relation $2 p q\,z = p^2+q^2-k^2$ and the integral \eqref{2_infraredGamma}. Taking only the infrared-leading diagrams into account yields the two equations
    \begin{subequations}\label{2_olo_infrared}
     \begin{align}
	  \frac{1}{c_c (p^2)^{\delta_c}} & = c_A c_c\, \left(p^2\right)^{\delta_A+\delta_c}\,\frac{g^2 N_c}{ 32 \pi^2}\, \frac{3  \, \delta_A}{  \delta_A+\delta_c}\, \frac{ \Gamma\left(\delta_A\right) \Gamma\left(\delta_c+2\right) \Gamma\left(-\delta_A-\delta_c+1\right)}{ \Gamma\left(2-\delta_A\right) \Gamma\left(1-\delta_c\right)\Gamma\left(\delta_A+\delta_c+3\right)} \,,\\
	  \frac{1}{c_A (p^2)^{\delta_A}} & = c_c^2  \left(p^2\right)^{2 \delta_c} \frac{g^2 N_c}{32 \pi ^2 }\, \frac{ 3 (2-\zeta)+4\, \delta_c (1-\zeta)}{6\,\delta_c (1+2\, \delta_c) } \, \frac{\Gamma\left(1-2\, \delta_c\right) \Gamma\left(2+\delta_c\right)^2}{\Gamma\left(1-\delta_c\right)^2 \Gamma\left(4+2\, \delta_c\right)}\,.
     \end{align}
    \end{subequations}
    The algebra can be performed by an computer algebra system such as \form, \cite{Vermaseren:2000nd}. Self-consistency in the infrared demands that the powers in $p^2$ and the coefficients match on both sides of \eqref{2_olo_infrared}. The exponents give the scaling relation
	\be 0=2\delta_c + \delta_A \label{2_olo_scaling}\ee
    which uniquely relates the infrared exponents of ghost and gluon propagators. Equating both equations in \eqref{2_olo_infrared} and using \eqref{2_olo_scaling} yields,
      \be \frac{6\, \Gamma\left(-2\delta_c\right) \Gamma\left(\delta_c+2\right) \Gamma\left(\delta_c+1\right)}{ \Gamma\left(2+2\delta_c\right) \Gamma\left(1-\delta_c\right)\Gamma\left(3-\delta_c\right)} =  \frac{ 3 (2-\zeta)+4\, \delta_c (1-\zeta)}{6\,\delta_c (1+2\, \delta_c) } \, \frac{\Gamma\left(1-2\, \delta_c\right) \Gamma\left(2+\delta_c\right)^2}{\Gamma\left(1-\delta_c\right)^2 \Gamma\left(4+2\, \delta_c\right)} \,.\label{2_olo_kappa}\ee
    For the transverse projector, $\zeta = 1$, aboves equation uniquely determines the infrared exponents
	\be \kappa \equiv-\delta_c = \frac{1}{2}\delta_A  = \frac{93-\sqrt{1201}}{98} \approx 0.59535 \,.\ee

    Ghost and gluon equations can then equivalently be used to restrict the infrared coefficients, $c_A$ and $c_c$. The ghost equation yields the condition
      \be 1 = c_A c_c^2\,\,\frac{g^2 N_c}{ 32 \pi^2}\,  \frac{6\, \Gamma\left(2\kappa\right) \Gamma\left(2-\kappa\right) \Gamma\left(1-\kappa\right)}{ \Gamma\left(2-2\kappa\right) \Gamma\left(1+\kappa\right)\Gamma\left(3+\kappa\right)} \,.\label{2_olo_ircon} \ee
    
    \subsection{Renormalization and Results\label{sec_2_olo_ren}}

    In principle, Yang-Mills theory in the linear covariant gauge, \eqref{2_def_LG}, features eight renormalization constants,
      \be Z_3,\,Z_1,\,Z_4,\,\tilde Z_3,\,\tilde Z_1,\,Z_\xi,\,Z_m\text{ and }Z_g\,. \ee
    These renormalization constants are not independent but constrained by \eqref{1_def_Z1Z4}, \eqref{1_rc_Zm},\eqref{1_STI_lg} and \eqref{2_Taylor}, leaving two independent renormalization constants, i.e. two renormalization conditions to be imposed.

    After subtracting the quadratic divergences the integrals in \eqref{2_olo} are logarithmically divergent. A numerically tractable renormalization scheme which dispense with explicitly calculating the renormalization constants $Z_3$ and $\tilde Z_3$ is to subtract equations \eqref{2_olo} at some, not necessary identical scales $\sigma_{gh}^2$ and $\sigma_{gl}^2$. One then has to choose the boundary conditions $G(\sigma_{gh}^2)$ and $Z(\sigma_{gl}^2)$. While the gluon equation can be subtracted in the ultraviolet, the ghost equation has to be subtracted in the deep infrared, ideally at vanishing momentum. This is due to the fact, that the ghost-self energy shows a infrared-vanishing  asymptotic behavior $\propto (p^2)^{\kappa}$. The boundary condition corresponding to the scaling solution is $G(0)^{-1} = 0$. With the self-energies for ghost and gluon,
    \begin{subequations}\label{2_olo_Pi}
	  \begin{align}
	      \widetilde \Pi_{gh}^{olo}(p^2)& =  - g^2 N_c \int\!\dbar^4 q \,\frac{1-z_q^2}{k^4}\,G(q^2)Z(k^2) \,\,, \label{2_LG_ghSE}\\
	      \widetilde\Pi_{gl}^{olo}(p^2) &=  \frac{g^2 N_c}{3} \int\!\dbar^4 q \,\frac{q^2(1-\zeta\,z_q^2) - (1-\zeta) p\, q\, z_q}{p^2\,q^2\,k^2}\,G(q^2)G(k^2) \label{2_LG_glSE}\\ 
		& \qquad + \frac{g^2 N_c}{3}\, \int\!\dbar^4 q\,\widetilde{\K}_{gl}\,\left(Z(q^2) Z(k^2) \right)^{(1-\gamma_{3A})} \left(G(q^2) G(k^2) \right)^{-2\gamma_{3A}}\,,\nonumber
	  \end{align}     
    \end{subequations}
    the equations in a MOM-scheme, which are then solved numerically are given by
    \begin{subequations} \label{2_olo_MOM}
	  \begin{align}
	      \frac{1}{G(p^2)} & =  \widetilde\Pi_{gh}^{olo}(p^2)  - \widetilde\Pi_{gh}^{olo}(0)   \label{2_LG_ghDyson-Schwinger equation_MOM}\\
	       \frac{1}{Z(p^2)} &=  \frac{1}{Z(\sigma_{gl}^2)}  + \widetilde\Pi_{gl}^{olo}(p^2)  - \widetilde\Pi_{gl}^{olo}(\sigma_{gl}^2)  \label{2_LG_glDyson-Schwinger equation_MOM}\,.
	  \end{align}     
    \end{subequations}
    Numerical details can be found in  \appref{secA_YMtech}.

    In principle two methods of implying the renormalization conditions can be used. Traditionally, the renormalization condition for the gluon propagator is set in the MOM scheme by $Z(\sigma_{gl})$. However, the physical value of $\sigma_{gl}$ is not known. This is accieved by setting the scale via the condition that at the Z-boson mass the strong running coupling of the ghost-gluon vertex, 
	\be \alpha_{s}(p^2) = \frac{g^2}{4\pi}Z(p^2)G(p^2)^2 \,,\label{2_coupl}\ee
    takes the experimental value $ \alpha_{s}(M_Z^2) = 0.1184$ \cite{Beringer:1900zz}. This scale setting procedure ignores the fact that the experimental value has been obtained for five dynamical quarks which are absent in Yang-Mills theory. Still it is of some practical use.

    When comparing the calculated dressing functions to results from lattice gauge theory, another renormalization procedure is used. First the abscissa is fixed to the lattice-scale by identifying the location of the maxima in the gluon propagator dressing function. Then the ordinate is fixed to some value at the renormalization scale $\mu^2$. In this thesis $Z\left(\mu^2\right) = G\left(\mu^2\right) = 1$ with $\mu = 4\,\GeV$ is implied.

    \begin{figure}[t]
    \centering
    \includegraphics[width=.495\textwidth]{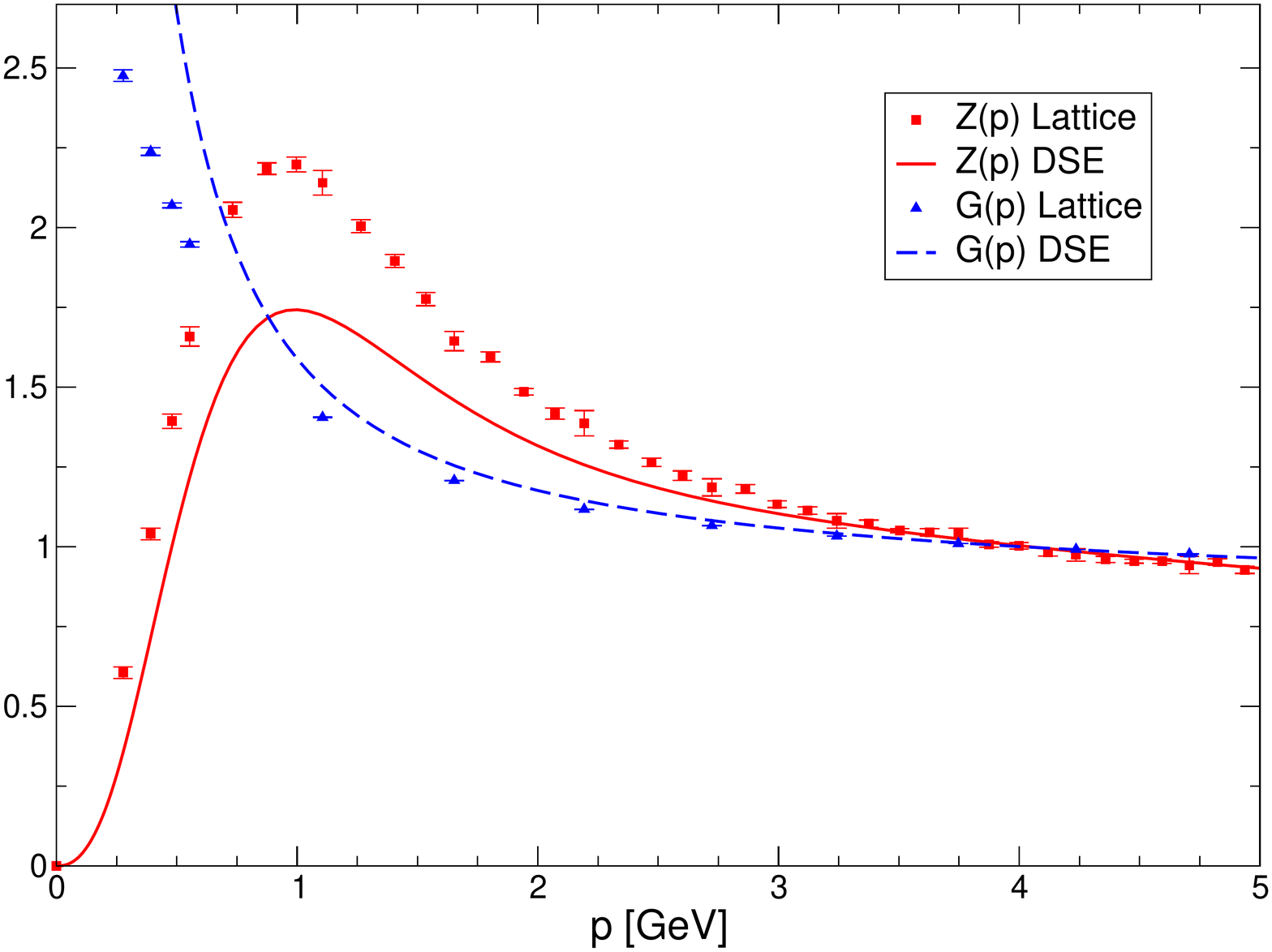}
    \includegraphics[width=.495\textwidth]{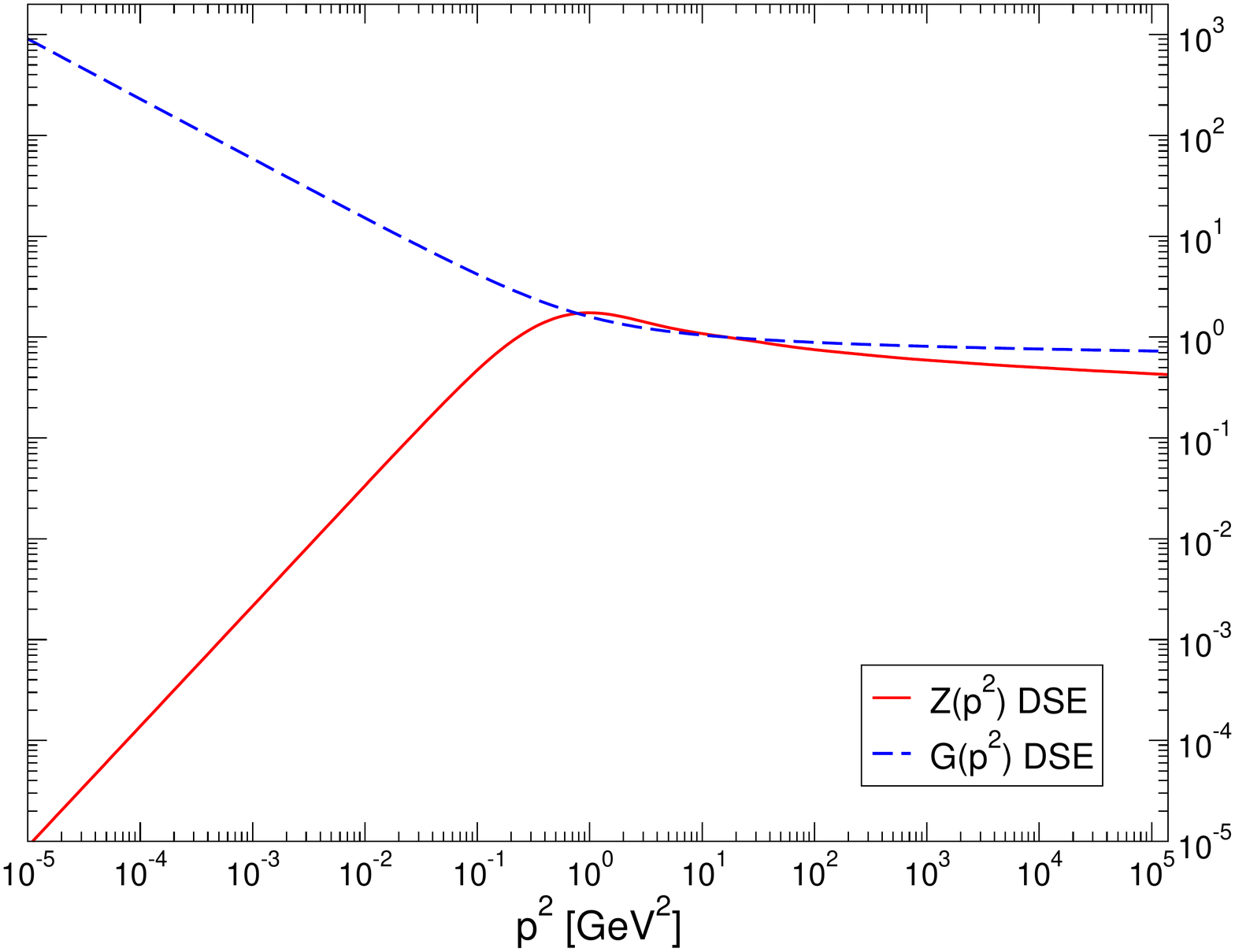}
    \caption{Gluon and ghost dressing functions $Z$ and $G$ as calculated from \eqref{2_olo_MOM} and compared to lattice results from \cite{Sternbeck:2006rd}. The dressing functions are renormalized to $1$ at $\mu = 4 \,\GeV$. The plot on the left shows the mismatch between lattice and continuum results in the mid and low momentum regime. The right plot shows the power-law behavior of the Dyson-Schwinger equation-solutions in the infrared.\label{fig_olo_latcomp}}
    \end{figure}
 
    In \figref{fig_olo_latcomp} results for the gluon and ghost dressing functions are shown and compared to lattice results from \cite{Sternbeck:2006rd}. While in the high-energy regime both methods agree, there is some disagreement for medium and low momenta. In the deep-infrared this does not come as a surprise as the lattice results represent a decoupling solution, while the Dyson-Schwinger results correspond to the scaling solution. The power-law behavior of the dressing functions can be seen on the double-logarithmic plot on the right-hand side of \figref{fig_olo_latcomp}. The mid-momentum  regime, however, is more worrying. While a more advanced three-gluon vertex model can absorb most of the mismatch \cite{Huber:2012kd}, one can also argue that the truncated two-loop diagrams could be the reason. This was also found when back-coupling the tree-level structure of the three-gluon vertex as calculated from its own Dyson-Schwinger equation \cite{Blum:2014gna}. In the next section it is investigated if 
the sunset-diagram can account for the mismatch of lattice and continuum results in the mid-momentum regime.

\section{Including the Sunset Diagram \label{sec_2_sun}}
    In this section the truncation is improved by introducing the sunset diagram into the calculation.  When calculating diagrams of order two and higher, the main complication is the appearance of overlapping divergences. In  perturbation theory the problem of overlapping divergences is solved by Zimmermann's "Forrest formula", e.g. \cite{Collins:1984}, for a pedagogical introduction see \cite{windisch:dipl}. For the sunset-diagram one has the fortunate situation that only a global quadratic divergence remains \cite{windisch:dipl}, which, in principle can be subtracted using a BPHZ scheme, \cite{Alkofer:2011di}. Here a method is employed which is more closely related to the calculations above. The sunset-diagram in the context of Dyson-Schwinger equations has been considered in \cite{Bloch:2003yu}, by approximating it with a set of one-loop integrals. 
    
    The sunset diagram introduces a Green function into the equations which has not been considered yet, the dressed four-gluon vertex. In \cite{Kellermann:2008iw} a semi-perturbative analysis identified the infrared-leading contributions. As guiding lines for the vertex construction one can use the asymptotic behavior as obtained from perturbation theory for high-energies and the infrared-scaling  \eqref{2_generalSR}. Along the lines of the one-loop-only calculations, the four-gluon vertex is modeled by its tree-level structure times a dressing functions which accounts for the ultraviolet and infrared behavior,
	  \begin{multline} \!\!\Gamma^{abcd}_{\nu\alpha\beta\gamma}(-p,p_1,p_2,p_3)\approx D_{4\Gamma}(-p,p_1,p_2,p_3) \Gamma^{(0)abcd}_{\nu\alpha\beta\gamma}=  \frac{1}{Z_4} \frac{\left[G(p_1^2)G(p_2^2)\right]^\alpha}
	 {Z(p_3^2)\,\left[Z(p_1^2)Z(p_2^2)\right]^{1-\beta}}\label{2_def_4g}\,.\!\!\end{multline}	
    The factor $1/Z(p_3^2)$ effectively undresses one of the legs inside the sunset diagram. This trick allows for faster calculations since one can effectively perform some angular integrals analytically.

    Given the scaling relation for all vertex dressing functions in Yang-Mills theory, \eqref{2_generalSR}, one finds that the dressing function of the four-gluon vertex should behave as $D_{4\Gamma}^{IR}\propto (p^2)^{-4\kappa}$ and thus
	  \be \beta = \frac{1+\alpha}{2} \,.\ee

    The high-energy behavior of the vertex model is constructed as such that it assumes the perturbative behavior which is dictated by $Z_4 = Z_3/\tilde Z_3^2$. Denoting the anomalous dimension of the four-gluon vertex by $\gamma_{4A}$ one finds
	  \be \gamma_{4A} = - \gamma_A  + 2\gamma_c \, \ee
    which yields $\alpha = -4\gamma_c$. 

    After projecting the gluon propagator Dyson-Schwinger equation \eqref{2_LG_glDyson-Schwinger equation_1} with the generalized Brown-Pennington projector, taking the normalized color trace and using the four-gluon vertex model \eqref{2_def_4g} the self-energy contribution from the sunset diagram reads,
	\be	\Pi_{sun}(p^2) = - \frac{g^4 N_c^2}{18\,p^2}  \int\! \dbar^4 q_1 \int\! \dbar^4 q_2\,\, \K_{sun}\, \frac{\left(Z(p_1^2)Z(p_2^2)\right)^{\frac{1}{2}-2\gamma_c}\left(G(p_1^2)G(p_2^2)\right)^{-4\gamma_c}}{p_1^2\, p_2^2 \, p_3^2} \label{2_LG_sun}\ee
    with the tensor structure of the sunset $\K_{sun}$ being given by
	\begin{align}
	    \K_{sun}=& \,\,9\,(5-\zeta) + 9\, z_{12}z_{13}z_{23} - 
	    3\,\zeta \bigl( z_{01}^2 + z_{02}^2 + z_{03}^2 + z_{12}^2 + z_{13}^2 
	    + z_{23}^2  \\
	    & \,\,  - z_{01}^2 z_{23}^2 - z_{02}^2 z_{13}^2 -z_{03}^2 z_{12}^2  + z_{01} z_{02} z_{13} z_{23} + z_{01} z_{03} z_{12} z_{23} 
	    + z_{02} z_{03} z_{12} z_{13}
	      \nonumber\\
	    & \,\,  - 3 \left( z_{01}z_{03}z_{13} + z_{01}z_{02}z_{12}  + z_{02}z_{03}z_{23} \right) \bigr) \,.\nonumber
	\end{align}
    The cosine $z_{ij}$ are defined via the normalized scalar product
	  \be z_{ij} = \frac{\p_i \cdot \p_j}{\sqrt{p_i^2\,p_j^2}} \,.\ee

    Before starting a detailed analysis of \eqref{2_LG_sun} it is advantageous to choose a specific momentum partitioning for the two internal and the external momentum. For the calculations in this section we choose, see \appref{secA_sunset} for details,
      \be \p_1 = \q_1\,, \qquad\quad \p_2 = \q_2 + \p\,,\qquad\quad\text{ and,}\quad\qquad \p_3 = -\q_1-\q_2 \,. \label{2_LG_sun_MP} \ee

\subsection{UV-Analysis}
      The analysis of the high-energy regime allows for the identification of quadratic and logarithmic divergent terms. For a two-loop term additional complifications arise due to the overlapping structure of the divergences which manifest themselves in a case-by-case analysis which has to be performed. For details see \appref{secA_sun_UV}. The quadratic divergences of the sunset are subtracted within the diagram itself, such that the cancellation mechanism of the one-loop graphs remains unaffected.

      After integrating out the trivial angles \eqref{2_LG_sun} reads
      	\begin{multline} \Pi_{sun}(p^2) = - \frac{g^4 N_c^2}{9(2\pi)^6\,p^2} \int_0^\Lambda dq_1 \int_0^\Lambda dq_2 \int_{-1}^1 \dzs_1\int_{-1}^1 \dzs_2 \int_{-1}^1 dy \\
	  \times\,\,\K_{sun} \, q_1q_2^3\,\frac{\left(Z(q_1^2)Z((p+q_2)^2)\right)^{\frac{1}{2}-2\gamma_c} \left(G(q_1^2)G((p+q_2)^2)\right)^{-4\gamma_c}}{ (p+q_2)^2 (-q_1-q_2)^2} \,.\label{2_LG_sun_2}
      \end{multline}
      where a factor of $\sqrt{1-z_i^2}$ is absorbed into the measure, 
	  \be \dzs_i \equiv dz_i\,\sqrt{1-z_i^2}\,. \ee
      The ultraviolet behavior is investigated by taking all momenta to be large and the dressing functions assume their logarithmic behavior. The integration momenta are larger than the external momentum since this parts contribute most to the integrals in the absence of poles as the cutoff may be arbitrary large. Since the logarithm is varying only slowly it is safe to assume
	  \be d_i\left((q_2+p)^2\right)  \stackrel{p^2,q_2^2\gg 1}{\longrightarrow} d_i(q_2^2)  \ee
      and one ends up with the integral
	  \begin{multline}  \Pi_{sun}^{ymax}(p^2) = - \frac{g^4 N_c^2}{9(2\pi)^6\,p^2} \int_p^\Lambda dq_1 \int_p^\Lambda dq_2 \\ \times\,\, \left(Z(q_1^2)Z(q_2^2)\right)^{\frac{1}{2}-2\gamma_c} \left(G(q_1^2)G(q_2^2)\right)^{-4\gamma_c} \,I_{sun}^{ang}(p,q_1,q_2) \label{2_LG_sun_ymax}\end{multline}
    where the angular integral was defined
	  \be I_{sun}^{ang}(p,q_1,q_2) = \int_{-1}^1 \dzs_1 \int_{-1}^1 \dzs_2 \int_{-1}^1 dy \,\frac{\K_{sun} \, q_1q_2^3}{ (p+q_2)^2 (-q_1-q_2 )^2}\,. \label{2_sun_angint}\ee
    In \tabref{A_tab_suntensor} the result of the angular integral analogous to \eqref{2_sun_angint} for any element of $\K_{sun}$ is given individually. Using these results one finds for the angular integral $I_{sun}^{ang}$ for $q_1>q_2$
      \begin{subequations}\label{2_sun_angint_out}
	  \be I_{sun}^{ang}(p,q_1,q_2) = - \frac{189\pi^2}{32} \frac{q_2}{q_1} (\zeta - 4) +\frac{9\pi^2}{32} \frac{q_2^3}{q_1^3} (\zeta - 4) - \frac{9\pi^2}{32} \frac{p^2q_2}{q_1^3} (\zeta - 1)\,, \ee
      while for the momentum configuration $q_2>q_1$ one gets
      \begin{multline}
	I_{sun}^{ang}(p,q_1,q_2) = - \frac{189\pi^2}{32} \frac{q_1}{q_2} (\zeta - 4) +\frac{9\pi^2}{32} \frac{q_1^3}{q_2^3} (\zeta - 4) - \frac{27\pi^2}{32} \frac{p^2q_1}{q_2^3} \\ + \frac{3\pi^2}{16} \frac{p^2q_1^3}{q_2^5} (6+\zeta) - \zeta \frac{15\pi^2}{32} \left( \frac{p^4q_1^3}{q_2^7} + \frac{p^2q_1^5}{q_2^7} - \frac{p^4q_1^5}{q_2^9}\right)\,.
      \end{multline}
      \end{subequations}

      The divergence structure of the sunset diagram can now be identified. The terms proportional to $\frac{q_<}{q_>}$ and $\frac{q_<^3}{q_>^3}$, where $q_> > q_<$, lead to quadratic divergences, while the terms proprtional to $p^2$ lead to logarithmic divergences. The quadratic divergences are proportional to $(\zeta-4)$ which is expected from general considerations of the generalized Brown-Pennington projector. This serves as cross-check for the analytical calculation.  Using the results given in \tabref{A_tab_suntensor} one can construct a term which subtracts the quadratic divergences of the integral for both momentum configurations. If in the self-energy contribution of the sunset diagram \eqref{2_LG_sun}  the integration kernel $\K_{sun}$ is exchanged by
	  \be \widetilde \K_{sun} = \K_{sun} - \left(\zeta-4 \right) \left( -\frac{225}{16}  + \frac{9}{4}\left(z_{12}^2 + 3 z_{01}^2\right)\right)\label{2_sun_Ktilde}\,,\ee
      no quadratic divergences appear in the integrals.

      A more detailed insight into the ultraviolet behavior of the sunset diagram can be obtained by performing the radial integrals in \eqref{2_LG_sun_ymax}. Therefore one approximates the dressing functions by their logarithmic behavior \eqref{2_def_gend} and uses the integral \eqref{A_sun_genultravioletint}. After subtracting the quadratic divergences, the high-energy asymptotics of the contributions of the sunset diagram to the gluon Dyson-Schwinger equation in leading order is given by
	  \begin{multline} \widetilde \Pi_{sun}^{ymax}(p^2) = - \frac{g^2 N_c }{(4\pi)^2} \,\frac{3}{176}\,  \frac{ 27 - 19 \zeta}{18\, \gamma_c } \\ \times \,\,  \left(\left(1+\beta_0 g^2\, \log{\frac{\Lambda^2}{\mu^2}} \right)^{2\gamma_c} -\left(1+\beta_0 g^2\, \log{\frac{p^2}{\mu^2}} \right)^{2\gamma_c}\right)\,, \label{2_sun_ultravioletas} 
	  \end{multline}
      The contributions of the sunset diagram are suppressed compared to the gluon and ghost loop contributions by one order in the logarithmic factor. It will thus not interfere with the ultraviolet-asymptotics and self-consistency relations of the gluon equation in \eqref{2_olo_ultraviolet}. In addition it is possible to absorb the logarithmic divergence into the renormalization constant $Z_3$.

\subsection{IR-Analysis}
      The general infrared analysis for the scaling solution as presented in \cite{Fischer:2006vf,Fischer:2009tn} allows to determine the infrared-behavior of a diagram in a Dyson-Schwinger equation if the scaling solution is used consistently. Since the four-gluon vertex model is consistent with the scaling solution \eqref{2_generalSR} one easily counts for the infrared exponent
	  \be \Pi_{sun}^{IR}(p^2) \propto \left(p^2\right)^{2\kappa} \ee
      and thus the sunset is suppressed in the infrared, in particular compared to the infrared-divergent ghost-loop contribution.
      It does therefore not interfere with the infrared-analysis of the one-loop-only truncation.
      
\subsection{Renormalization and Results}

      \begin{figure}[t]
      \centering
      \includegraphics[width=.495\textwidth,]{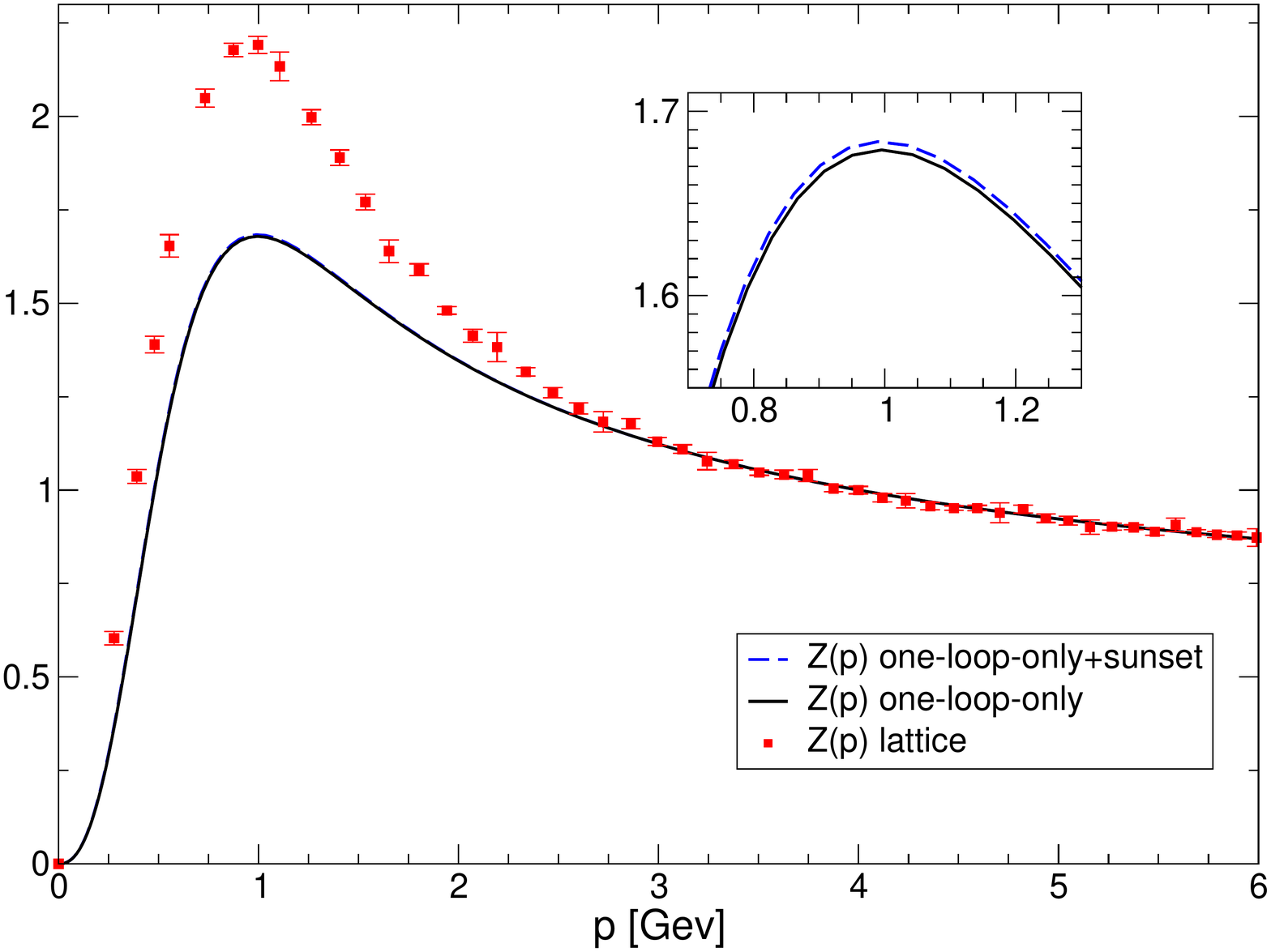}
      \includegraphics[width=0.495\textwidth,]{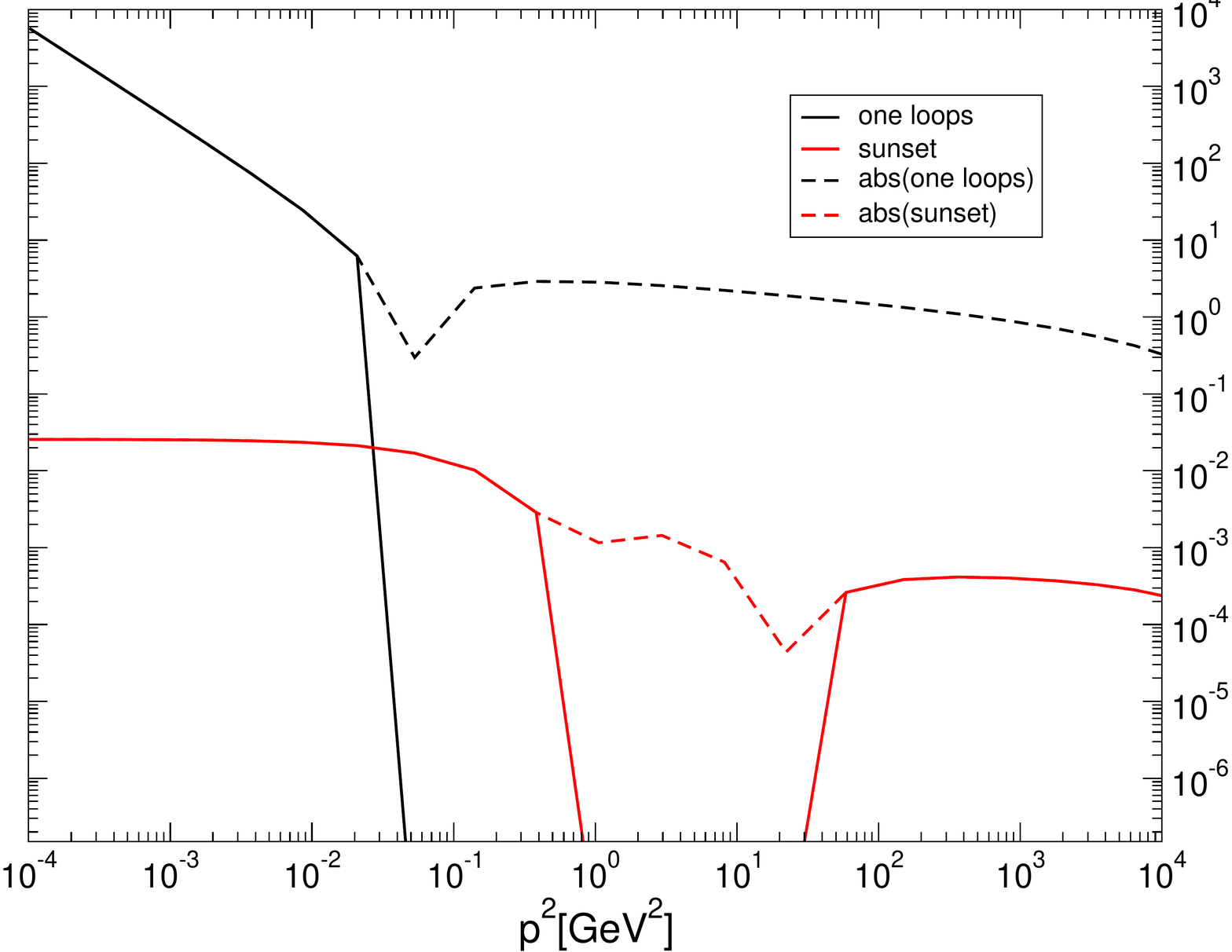}
      \caption{\label{fig_sun_LG} \emph{Left: }The Landau gauge 
      gluon dressing function with (blue, dashed) and without (black, solid) contributions
      from the sunset diagram compared to lattice data of \cite{Sternbeck:2006rd}. 
      \emph{ Right: } Contributions to the gluon self-energy of the one-loop diagrams and the sunset 
      after subtracting quadratic divergences.}
      \end{figure}

      Being subleading in the infrared and the ultraviolet, the sunset diagram can safely be included into the calculation without changing the self-consistency relations found in the last section. As described above, the overlapping quadratic divergences of the sunset diagram can be subtracted in the diagram itself. The residual logarithmic divergences can then be treated in a MOM-scheme. Using the one-loop self-energies \eqref{2_olo_Pi}, the gluon equation in a MOM-scheme with the sunset diagram included reads
	  \be \frac{1}{Z(p^2)} =  \frac{1}{Z(\sigma_{gl}^2)}  + \widetilde\Pi_{gl}^{olo}(p^2) + \widetilde \Pi_{sun}(p^2) - \widetilde\Pi_{gl}^{olo}(\sigma_{gl}^2)  - \widetilde \Pi_{sun}(\sigma_{gl}^2)  \,,\ee 
      where $\widetilde \Pi_{sun}(p^2)$ is $\Pi_{sun}(p^2)$ from \eqref{2_LG_sun} with the integration kernel $\K_{sun}$ exchanged by $\widetilde\K_{sun}$ from \eqref{2_sun_Ktilde}. The equation is solved in a coupled system with the ghost equation \eqref{2_LG_ghDyson-Schwinger equation_MOM}.

      The results for this calculation for the gluon dressing function are given in \figref{fig_sun_LG}. While the analytical considerations already showed, that the sunset diagram does not contribute in the far ultraviolet and deep infrared regions, it does not contribute significantly in the mid-momentum regime, too. In the right plot in \figref{fig_sun_LG} the contributions of the sunset diagram are displayed after convergence was obtained. Compared to the contributions from the one-loop diagrams they are suppressed by at least one order of magnitude.

\section{Summary}

      In this section the propagator Dyson-Schwinger equations of Yang-Mills theory in Landau gauge were presented. Two kinds of solutions, the unique scaling solution and the family of decoupling solutions, were introduced and the contemporary status of the discussion in the literature was summarized. The well-known one-loop-only truncation was introduced and treated in some detail. The results from Dyson-Schwinger equations in this truncation differ from the ones obtained in the lattice in the mid-momentum regime around $1 \GeV$. One possible reason for this mismatch are the two-loop terms, which are not accounted for in this truncation.

      Subsequently the truncation was improved by including the sunset diagram into the calculation. It could be shown that the overlapping quadratic divergences can be subtracted by an explicit counterterm construction. The effect of the sunset-diagram is negligibly small. Still, the inclusion of the sunset diagram is a technical advance, which serves useful in the next section. In the Maximal Abelian gauge sunset diagrams are known to be the dominant contributions in the infrared regime.

%% file: 3_MAG.tex
\chapter{Maximal Abelian Gauge \label{sec_MAG}}

    The Maximal Abelian Gauge (MAG) was developed to investigate the Dual-Super\-con\-duc\-tor Picture of confinement as described in \secref{sec_dsc} based on the 't Hooft's Abelian projection \cite{Hooft1981455}. In an effective model calculation it was found that the Abelian gauge fields dominate in the low-energy regime which lead to the "Hypothesis of Abelian Dominance", \cite{Ezawa:1982bf}. The main idea of the Maximal Abelian Gauge (MAG) is to split the color algebra into its Cartan subalgebra, the maximal set of commuting generators, and the corresponding coset algebra. On the lattice, the non-local gauge condition minimizes the norm of the coset link variables, \cite{Kronfeld:1987vd,Kronfeld:1987ri}. In the continuum this leads to a non-linear condition of the coset fields, e.g. \cite{Shinohara:2001cw,Fazio:2001rm}. 
    
    The literature on the MAG is vast, so here only some results can be summarized. Generally, in all studies on the Yang-Mills propagators, the hypothesis of Abelian dominance could be confirmed. The high-energy behavior of QCD in the MAG was calculated up to three-loop order for the propagators \cite{Gracey:2005vu} and two-loop order for the three-point functions \cite{Bell:2013xma}. On the lattice it could be shown, that the Abelian degrees of freedom carry the major contributions to the string tension \cite{Stack:2002sy}. When calculating $SU(3)$ gluon propagators on the lattice it is found that the Cartan gluon propagator surpasses the coset gluon propagator \cite{Gongyo:2012jb,Gongyo:2013sha} for all momenta. In an infrared analysis of Yang-Mills theory in the MAG a scaling-type of solution was found \cite{Huber:2009wh,Huber:PhD,Huber:2011fw}. While the Cartan gluon propagator is infrared enhanced, the coset gluon and ghosts are infrared suppressed.

    Gribov-copy effects and the Gribov horizon in the MAG were extensively studied in the literature. The Faddeev-Popov operator was constructed in \cite{Bruckmann:2000xd,Capri:2010an}. The Gribov region is bound in the coset directions, but unbound in the Cartan directions \cite{Capri:2010an}. It was even possible to explicitly construct Gribov copies \cite{Bruckmann:2000xd,Capri:2013vka} and to formulate a Gribov-Zwanziger like theory in the MAG \cite{Capri:2006cz,Gongyo:2013rua}. The Ward identities have been derived using algebraical renormalization resulting in the most general counter terms \cite{Fazio:2001rm,Dudal:2004rx}. 
    
    Several versions of interpolating gauges between the Landau gauge and the MAG where constructed and investigated, \cite{Hata:1992np,Capri:2005zj,Gracey:2005vu,Huber:PhD}.  In \cite{Hata:1992np} it was shown, however, that these gauges are not smoothly connected.

\section{Lagrangian, Gauge Fixing and General Aspects}

    After introducing new notations concerning the splitting of the gauge algebra into the Cartan subalgebra and the corresponding coset algebra, the equivariant BRST construction of gauge fixing, developed  for SU(2) in \cite{Schaden:1998hz,Schaden:1999ew}, is extended for general $SU(N)$ gauge groups. This method has several benefits compared to the standard BRST derivations. It clearly distinguishes between the Cartan and coset degrees of freedom by implementing gauge-fixing as a two-step process. First, the coset degrees of freedom are gauge fixed, breaking the local symmetry group to the Maximal Abelian subgroup. This residual Abelian symmetry is the fixed in a second step using usual Abelian BRST transformations. As a byproduct, the unpleasant non-local shift in the diagonal ghost is avoided, which usually is performed to decouple them from the theory, \cite{Capri:2005tj}. If one considers the equivariant construction for the coset space only, the gauge-fixing process 
    avoids Neuberger's $\frac{0}{0}$-problem \cite{Neuberger:1986vv}. The gauge fixing on the lattice is exact, resulting in a confining Abelian gauge theory. In addition this construction allows for a proper definition of the physical Hilbert-space in the MAG analogous to Landau gauge.

\subsection{Cartan Subalgebra and Color Splitting\label{sec_3_CS}}
    The Cartan subalgebra $\c$ is defined as a maximal set of commuting generators of an algebra,
      \be \c = \left\{T^i: \com{T^i}{T^j} = 0 \right\} \,.\label{3_def_Cartan}\ee
    For $su(N)-$algebras, the Cartan subalgebra is $(N-1)$ dimensional. In this case the Cartan subalgebra is the Abelian $u(1)^{N-1}$ algebra, i.e. the (maximal) "Abelian content" of the underlying $su(N)$ algebra. One usually chooses diagonal generators, e.g. $\{\frac{\tau_3}{2}\}$ for $su(2)$ and $\{\frac{\lambda_3}{2},\frac{\lambda_8}{2}\}$ for $su(3)$. Since any $SU(N)$ group possesses $N^2-1$ generators, the corresponding coset algebra is $N^2 - N$ dimensional. In the following the contributions for the Cartan subalgebra will be denoted by the indices
      \be i,j,k,\dots = 1,\dots,N-1 \,.\ee
    For the coset algebra the indices
      \be a,b,c,\dots = 1,\dots, N^2-N \ee
    will be used, while
      \be r,s,t,\dots = 1,\dots, N^2-1 \ee
    are kept for the full algebra. Gauge fields $A_\mu^r$ are elements of the gauge algebra. As such they can be split into parts which lie in the Cartan subalgebra, $A_\mu^i$, and parts in the coset algebra, $B_\mu^a$,
      \be A_\mu^r T^r = A_\mu^i T^i + B_\mu^a T^a \,. \ee
    This \emph{color-splitting} can be performed for any function of the gauge fields, in particular the field strength tensor, $F_{\mu\nu}^r T^r = f_{\mu\nu}^i T^i + F_{\mu\nu}^a T^a $ with
	\begin{align}
	  f_{\mu\nu}^i & = \partial_\mu A_\nu^i  - \partial_\nu A_\mu^i + \cp{B_\mu}{B_\nu}{i} \,, \label{3_def_fabel}\\
	  F_{\mu\nu}^a & = \D^{ab}_\mu B_\nu^b - \D_\nu^{ab} B_\mu^b + \cp{B_\mu}{B_\nu}{a} \,.\label{3_def_FnAbel}
	\end{align}
    In the last equation the covariant derivative with respect to the Cartan gluons only was introduced,
	\be \D_\mu^{ab} = \delta^{ab}\partial_\mu + g f^{aib}A_\mu^i \,. \ee
    The Yang-Mills Lagrangian, \eqref{1_def_LYM}, can be re-expressed in this color-split manner as
	  \begin{align} 
	      \LYM & = \frac{1}{4}f_{\mu\nu}^i f_{\mu\nu}^i +  \frac{1}{4}F_{\mu\nu}^a F_{\mu\nu}^a \label{3_def_LYM}\\
		   & = -\frac{1}{2} A_\mu^i\left(\delta_{\mu\nu}\partial^2 - \partial_\mu\partial_\nu \right)A_\nu^i + (\partial_\mu A_\nu^i) \cp{B_\mu}{B_\nu}{i} + \frac{1}{4} \cp{B_\mu}{B_\nu}{i}\cp{B_\mu}{B_\nu}{i}  \label{3_MAG_YM}\\
		   & -\frac{1}{2} B_\mu^a\left(\delta_{\mu\nu}\D^2 - \D_\mu\D_\nu \right)^{ab}B_\nu^b  + (\D_\mu B_\nu)^a \cp{B_\mu}{B_\nu}{a} + \frac{1}{4} \cp{B_\mu}{B_\nu}{a}\cp{B_\mu}{B_\nu}{a} \,. \nonumber
	  \end{align}

    The structure constants of the gauge group determine the commutator of the generators, \eqref{A_def_gen}, $\com{T^r}{T^s} = i f^{rst}T^t$. The Cartan subgroup \eqref{3_def_Cartan} implies the condition
	  \be  f^{ijk} = f^{ija} = 0 \,.\ee
    The non-vanishing structure constants have at least two indices in the coset space. The Jacobi-identity \eqref{A_Jacobi} splits into four equations,
	  \begin{subequations}\label{3_Jacobi}
	      \begin{align}
		  0 & = f^{abc}f^{ade} + f^{abd}f^{aec} + f^{abe}f^{acd} \,,\\
		  0 & = f^{iab}f^{icd} + f^{iac}f^{idb} + f^{iad}f^{ibc}\,, \\
		  0 & = f^{abc}f^{adi} + f^{abd}f^{aic} + f^{abi}f^{acd} \,,\\
		  0 & = f^{abj}f^{aci} + f^{abi}f^{ajc} \,.
	      \end{align}
	  \end{subequations}
    By construction the Lagrangian \eqref{3_def_LYM} is invariant under local $SU(N)-$transformations, \eqref{1_gtA_inf}. The local gauge parameter possesses contributions in the Cartan and coset sub-algebras
	\be \vartheta^r T^r = \vartheta^i T^i + \theta^a T^a \,.\ee
    The Cartan and coset gluons transform differently under gauge transformations in the Cartan and coset sub-algebras. The Cartan transformations, denoted by $\delta_C$, read
	\begin{align} 
	    \delta_C A_\mu^i  &= \partial_\mu \vartheta^i & \delta_C B_\mu^a &  = \cp{B_\mu}{\vartheta}{a} \label{3_GT_cart}\,.
    \intertext{While the Cartan gluon transforms as $N-1$ fold copies of an Abelian gauge boson the coset gluons transform like adjoint matter fields. The Cartan transformations are closed with respect to the Cartan subalgebra and the coset space. The coset transformations are denoted by $\delta_\eps$. They read}
	 \delta_\eps A_\eps^i & = \cp{B_\mu}{\theta}{i} & \delta_\eps B_\mu^a  & = \D_\mu^{ab} \theta^b +  \cp{B_\mu}{\theta}{a}  \,. \label{3_GT_coset}
    \end{align}
    The coset transformations are not closed. They mix the Cartan and coset fields.

    The gauge group $SU(2)$ is a special case in color-split theories since its Cartan subalgebra is one dimensional. The only non-vanishing structure constant is $f^{abi}$, which leads to significant simplifications as, e.g., the last two terms of \eqref{3_MAG_YM} vanish and the Jacobi-identities \eqref{3_Jacobi} boil down to the identity $ f^{iab} = - f^{iba}$. The gauge group $SU(2)$ in the MAG is also interesting for physical reasons as it allows to investigate relations between Yang-Mills theory and QED. These relations will be worked out in the following and offer insights when studying the Dyson-Schwinger equation for the gauge-boson propagator in different gauges and models in \secref{sec_saturation}.

\subsection{Gauge Fixing: the Equivariant BRST Construction \label{sec_3_lag}}

    The MAG rests on the idea to maximize the effect of the the Abelian part of the gauge fields, i.e.  elements of the Cartan subalgebra, with respect to the non-Abelian parts, i.e. elements of the coset algebra. The non-local interpretation of this condition is to minimize the norm of the coset gauge-fields, \cite{Shinohara:2001cw},
	\be R_{MAG} = \frac{1}{2} \int\!\! d^4x\, B_\mu^a B_\mu^a \,. \label{3_RMAG}\ee
    The functional $R_{MAG}$ shall be extremal with respect to coset transformations
	\be 0 \stackrel{!}{=}\delta_{\eps}  R_{MAG} = -  \int\!\! d^4x\,\, \theta^b\,\D_\mu^{ba} B_\mu^a \,, \label{3_RMAG_1}\ee
    which has to be fulfilled for any infinitesimal parameter $\theta^a$. The corresponding local condition reads
	\be \D_{\mu}^{ab}B_\mu^b(x) = 0, \quad \forall x \,.\label{3_localGC}\ee
    Another coset variation of $R_{MAG}$  yields
	\be \delta_{\eps}^2  R_{MAG} =  \int\!\! d^4x\!\int\!\! d^4y\,\,\theta^a(y) \,\M^{ab}_{MAG}(x,y)\, \theta^b(x) \ee
    with the Faddeev-Popov operator of the MAG, \cite{Bruckmann:2000xd,Capri:2010an}, 
	\be  \M^{ab}_{MAG}(x,y) = \left(-\D_\mu^{ac}\D_\mu^{cb} - g f^{adc} B_\mu^d \D_\mu^{cb} - g^2 f^{iac}f^{idb}B_\mu^c B_\mu^d \right)\delta(x-y)\,, \ee
    which is always positive in the first Gribov region. \eqref{3_localGC} really characterizes a minimum of the functional $R_{MAG}$. When comparing the MAG gauge fixing condition to Landau gauge, \eqref{1_def_LG}, one finds two significant differences. While in Landau gauge all color directions are treated equivalently, the MAG gauge fixing condition breaks color symmetry explicitly. This is an immediate result of the physical idea to investigate the influence of the Cartan subalgebra. In addition Landau gauge has in total $N^2-1$ conditions for the $N^2-1$ gauge fields, while the MAG condition \eqref{3_localGC} only contains $N^2-N$ conditions. MAG is such under-constrained with respect to Landau gauge. This is a consequence of the functional \eqref{3_RMAG} only fixing the coset gluons but not the Cartan ones. To fully fix the gauge one has to imply additional constrains. In this thesis the Cartan gluons are fixed to the Abelian Landau gauge,
	\be \partial_\mu A_\mu^i = 0 \,. \label{3_localGC_cart}\ee

    The two gauge-fixing conditions \eqref{3_localGC} and \eqref{3_localGC_cart} suggest that the gauge-fixing Lagrangian also consist of two parts,
	\be \Lgf^{MAG} = \Lgf^{\eps} + \Lgf^{C} \,.\ee
   The first term $\Lgf^{\eps}$ fixes the coset parts only and breaks the gauge group from $SU(N)$ down to $U(1)^{N-1}$. Such a partial (equivariant) gauge fixing construction was developed in \cite{Schaden:1998hz,Schaden:1999ew}. Only recently a complete treatment of a general equivariant gauge-fixing construction in the continuum in terms of cohomology theory was performed \cite{Ferrari:2013aza}. 
   
   For the partially gauge fixed Yang-Mills theory 
	\be \L^{abel} = \L_{YM}  +   \Lgf^{\eps}\label{3_def_Label}\ee
    one expects the usual properties of non-Abelian gauge theories as asymptotic freedom in the high-energy sector and Confinement in the infrared. On the other hand,  $\L^{abel}$ is an Abelian gauge theory with gauge group $U(1)^{N-1}$, as such one expects a simple structure of, e.g., Ward-identities. In particular for the gauge group $SU(2)$ the gauge group of $\L^{abel}$ is the same gauge group as QED. The Cartan gluon in this case really is the photon. The difference between these two theories is the matter sector only. While in QED the photon is coupled to some Dirac fermions in the fundamental representation of $U(1)$, in $\L^{abel}$ the photon couples to some vector boson and fermions which transform according to \eqref{3_GT_cart}. Considering $\L^{abel}$ as an Abelian gauge theory one thus finds that Confinement is not a property of non-Abelian gauge theories only, but it also depends on the matter content of the theory. A similar statement is known for a long time for QCD where there exists a 
critical number of fermion flavors which leads to a breakdown of asymptotic freedom. 
   
   Based on the coset gauge-transformation \eqref{3_GT_coset} one defines the equivariant BRST transformations as
	\begin{subequations} 
	\label{eBRST}
	    \begin{align}
		\se A_\mu^i & = \cp{B_\mu}{c}{i}, && \se B^a_\mu  = (\D_\mu c)^a +\cp{B_\mu}{c}{a} \,,\\
		\se c^a & = -\frac{1}{2}\cp{c}{c}{a}, && \se \bar c^a = b^a - \frac{1}{2} \cp{\bar c}{c}{a}\,,
	      \end{align}\vspace{-4mm}
	    \be \hspace{-15mm}  \se b^a = -\frac{1}{2}\cp{c}{b}{a} + \frac{1}{8} \cp{\bar c}{\cp{c}{c}{b}}{a} +\frac{1}{2} \cp{\bar c}{\cp{c}{c}{i}}{a} \,,\ee
	\end{subequations}
    where for uniqueness the inner cross-products are denoted with their corresponding index. Accordingly one defines the equivariant anti-BRST transformations
	\begin{subequations} 
	\label{anti_eBRST}
	    \begin{align}
		\seb A_\mu^i & = \cp{B_\mu}{\bar c}{i}, & \seb B^a_\mu & = (\D_\mu \bar c)^a +\cp{B_\mu}{\bar c}{a} \,,\\
		\seb \bar c^a & = -\frac{1}{2}\cp{\bar c}{\bar c}{a}, & \seb  c^a &=- b^a - \frac{1}{2} \cp{\bar c}{c}{a}\,,
	      \end{align}\vspace{-4mm}
	    \be \hspace{-15mm}  \seb b^a = -\frac{1}{2}\cp{\bar c}{b}{a} + \frac{1}{8} \cp{\cp{\bar c}{\bar c}{b}}{ c}{a} +\frac{1}{2}  \cp{\cp{\bar c}{\bar c}{i}}{ c}{a}\,. \ee
	\end{subequations}
    By construction these transformations are not nilpotent, but their square yield a Cartan transformation with the parameter $\theta = \frac 1 2 \cp{c}{c}{i}, \frac 1 2 \cp{\bar c}{\bar c}{i},  \frac 1 2 \cp{\bar c}{ c}{i}$ for the transformations $\se^2,\seb^2$ and $\frac{1}{2}\com{\se}{\seb}$, respectively. By construction the ghost $c^a$, antighost $\bar c^a$ and NL-field $b^a$ take values in the coset-space only. Using the equivariant (anti-) BRST transformations, the coset gauge-fixing Lagrangian is defined as 
    \begin{align}
      \Lgf^\eps & = \frac{i}{2}\,\se\seb\left(B_\mu^a B_\mu^a - i \xi \bar c^a c^a   \right) = \se\left(\bar c^a \left(\frac{\xi}{2} b^a - i (\D_\mu B_\mu)^a \right)  \right)  \label{3_Leps}\\
		  & = \frac{\xi}{2}\,b^2 - i b^a (\D_\mu B_\mu)^a + i \bar c^a (\D_\mu\D_\mu c)^a + i \cp{B_\mu}{\bar c}{i}\cp{B_\mu}{c}{i} \\
		  & \,\,\,+ \frac{i}{2} \, B^a_\mu \left( \cp{\D_\mu\bar c}{c}{a} - \cp{\bar c}{\D_\mu c}{a} \right) + \frac{\xi}{8}\,\cp{\bar c}{c}{a}\cp{\bar c}{c}{a} + \frac{\xi}{2} \cp{\bar c}{c}{i}\cp{\bar c}{c}{i} \,.\nonumber
    \end{align}
    In the literature the quartic ghost interactions usually have been addressed to renormalizability issues in the MAG, e.g. \cite{Min:1985bx}. In the equivariant construction, however, these terms appear naturally at tree level and are required by gauge-invariance \cite{Ferrari:2013aza}. By construction the gauge-fixing Lagrangian is invariant under $U(1)^{N-1}-$ and equivariant BRST transformations, $\se \Lgf^\eps = 0$.

    To complete the gauge fixing process one needs to introduce BRST transformation with respect to the Abelian subgroup which leave $\L^{abel}$ invariant. Based one the Abelian transformations \eqref{3_GT_cart} and introducing Cartan ghost $\omega^i$, the corresponding antighost $\bar \omega^i$ and NL-field $\eta^i$, the Cartan BRST-transformations read
    \begin{subequations}\label{3_def_cartanBRST}
	\begin{align}
	    \sC A_\mu^i & = \partial_\mu \omega^i, &  \sC \omega^i & = 0, & \sC \bar \omega^i &= \eta^i, & \sC  \eta^i &= 0, \\
	    \sC B^a_\mu  &= \cp{B_\mu}{\omega}{a}, & \sC c^a &= -\cp{c}{\omega}{a}, & \sC \bar c^a&= -\cp{\bar c}{\omega}{a},& \sC b^a &= \cp{b}{\omega}{a}.	    
	\end{align}
    \end{subequations}
    The Cartan gluon, ghost, antighost and NL fields transform as a $(N-1)-$fold copy of the corresponding field in QED, while the Coset gluons, ghost, antighost and NL fields transformation as matter fields. For completeness one can also define the Cartan anti-BRST transformations
    \begin{subequations}\label{3_def_cartanABRST}
	\begin{align}
	    \sCb A_\mu^i & = \partial_\mu  \bar\omega^i, &  \sCb \bar \omega^i & = 0, & \sCb \omega^i &=- \eta^i, & \sCb  \eta^i &= 0, \\
	    \sCb B^a_\mu  &= \cp{B_\mu}{\bar \omega}{a}, & \sCb c^a &= -\cp{c}{\bar\omega}{a}, & \sCb \bar c^a&= -\cp{\bar c}{\bar \omega}{a},& \sCb b^a &= \cp{b}{\bar\omega}{a}.	    
	\end{align}
    \end{subequations}
    Both sets of BRST-transformations are nilpotent and leave $\L^{abel}$ invariant. They anticommute, $\acom{\sC}{\sCb} = 0$. The gauge fixing Lagrangian implementing the Cartan gauge condition \eqref{3_localGC_cart} can be written as a Cartan BRST exact expression, 
	\begin{multline} \label{3_Lgf_Cartan}
	    \Lgf^C   = \frac{i}{2} \sC\sCb\left( A_\mu^iA_\mu^i - i \lambda \,\bar \omega^i \omega^i \right)   =\sC\left(\bar \omega^i \left(\frac{\lambda}{2} \eta^i - i \partial_\mu A_\mu^i \right)  \right)  \\ 
	    = \frac{\lambda}{2}\,\eta^2 - i \eta^i \,\partial_\mu A_\mu^i + i \bar \omega^i \partial^2 \omega^i \,,
	\end{multline}
    where the gauge parameter $\lambda$ was introduced. As in any Abelian gauge theory the Abelian ghosts decouple from the theory and can be integrated out trivially. The Abelian ghost, antighost and NL-field are taken to be invariant under the equivariant (anti-) BRST transformations,
	\be \se \omega^i = \se \bar \omega^i = \se \eta^i = \seb \omega^i = \seb \bar \omega^i = \seb \eta^i=0 \,.\ee
    The full gauge-fixing Lagrangian for the MAG is then given as
	\begin{align}
	  \Lgf^{MAG} & = \frac{\xi}{2}\,b^2 + \frac{\lambda}{2}\,\eta^2 - i b^a (\D_\mu B_\mu)^a - i \eta^i \,\partial_\mu A_\mu^i  + i \bar c^a (\D_\mu\D_\mu c)^a   \nonumber\\
	  & \,\,\, + i \cp{B_\mu}{\bar c}{i}\cp{B_\mu}{c}{i} + \frac{i}{2} \, B^a_\mu \left( \cp{\D_\mu\bar c}{c}{a} - \cp{\bar c}{\D_\mu c}{a} \right) \label{3_Lgf_compl}\\
		      & \,\,\, + \frac{\xi}{8}\,\cp{\bar c}{c}{a}\cp{\bar c}{c}{a} + \frac{\xi}{2} \cp{\bar c}{c}{i}\cp{\bar c}{c}{i} \,.\nonumber
	\end{align}
    It is important to note that the Abelian gauge fixing not only breaks local $\uone^{N-1}$-symmetry, but also the global equivariant BRST symmetry since $\se \Lgf^C \neq 0$. The symmetries left are the respective invariance under Cartan BRST, Cartan anti-BRST and global Cartan color transformations.

    \subsection{The Renormalized Lagrangian}

    In the Dyson-Schwinger equations used below, it is of technical advance to avoid mixed propagators. Therefore one usually integrates out the NL-fields, which yields
	\begin{align}
	    \L^{MAG}_{gf} &  = -\frac{1}{2 \lambda} A_\mu^i \partial_\mu \partial_\nu A_\nu^i -\frac{1}{2\xi} B_\mu^a (\D_\mu\D_\nu)^{ab} B_\nu^b + i \bar c^a (\D^2)^{ab} c^b \nonumber \\ 
	    &\qquad - \frac{i}{2}  (\D_\mu\bar c)^a \cp{B_\mu}{c}{a}  - \frac{i}{2} \cp{B_\mu}{\bar c}{a} (\D_\mu c)^a \label{3_Lgf_intout}\\
	    &\qquad+ i \cp{B_\mu}{\bar c}{i} \cp{B_\mu}{c}{i} + \frac{\xi}{2}\,\cp{\bar c}{c}{i}\cp{\bar c}{c}{i}  + \frac{\xi}{8}\,\cp{\bar c}{c}{a} \cp{\bar c}{c}{a} \nonumber\,.
	\end{align}
    By construction in the MAG the global color symmetry is broken and thus it can not be taken for granted that all gluon fields renormalize identically. It has to be distinguished between the coset and the Cartan gluons. In Yang-Mills theory with the gauge fixing Lagrangian \eqref{3_Lgf_intout} the following terms are subject to renormalization, \cite{Gracey:2005vu},
	\begin{align}
	     B_{0\,\mu}^{a} &= \sqrt{Z_B}\,  B_{R\,\mu}^{a} & A_{0\,\mu}^i &= \sqrt{Z_A}\,  A_{R\,\mu}^{i} & c_0^a & = \sqrt{Z_c}\,c^a_R & \bar c_0^a & = \sqrt{Z_c}\,\bar c^{a}_R \label{3_def_Z_fund}\\
	    g_0 &= Z_g\,g_{R} & \xi_0 & = Z_\xi \xi_{R} & \lambda_0 & = Z_\lambda \lambda_{R} \nonumber
	\end{align}
    where the indices $0$ and $R$ denote bare and renormalized fields, respectively.  The gauge part of QED in the linear covariant gauge features only one renormalization constant due to the Ward identities for the longitudinal photon propagator and for the electron-photon vertex \cite{Bohm:2001yx}. The latter identity states that the product of the renormalization constants of the electric coupling and the one of the photon wave function is finite. Since Yang-Mills theory in the MAG can be interpreted as the free Maxwell theory coupled to some adjoint matter, a similar behavior can also be expected for this theory. 
    
    Performing exactly the same considerations that lead to \eqref{1_STI_lg} constrain the renormalization of the Cartan gauge parameter to 
	    \be Z_\lambda = Z_A \,.\label{3_WI_cartan}\ee	
    Note that such an equation can not be given for the coset gluons, since the equivariant BRST in general is not preserved. The corresponding equation for the gauge coupling can be argued for in a similar way using the $\uone^{N-1}$ gauge symmetry, \cite{Kondo:1997pc,Quandt:1997rw}. It is given by
	    \be Z_g \sqrt{Z_A} = 1  \,,\label{3_diagghost}\ee
    which is excacly the same relation as in QED \cite{Bohm:2001yx}. In a more general treatment of the MAG using algebraic renormalization one finds that \eqref{3_diagghost} is a corollary of a more general relation controlling the dependence of the theory on the diagonal ghost, the so-called  \emph{diagonal ghost equation}, \cite{Fazio:2001rm}. It can be interpreted as a non-renormalization theorem similar to the Taylor's theorem in Landau gauge as the rescaled field strength, $\mathcal A_\mu^i = g A_\mu^i$, does not renormalize. An immediate corollary of this is that the covariant derivative does not contain any renormalization constant
	  \be \D^{ab}_\mu = \delta^{ab}\partial_\mu + g_0 f^{aib}A_{0\,\mu}^i  = \delta^{ab}\partial_\mu +  g_R f^{aib} A_{R\,\mu}^{i} \,.\ee
     {\allowdisplaybreaks
   Changing from bare to renormalized fields, thereby dropping the index $R$, and expanding all cross-products, the renormalized Lagrangian $ \L_{MAG} = \LYM + \L^{MAG}_{gf}$ reads
	  \begin{align}
	      \L_{MAG}    &= -\frac{Z_A}{2} \delta^{ij}A_\mu^i\left(\delta_{\mu\nu}\partial^2 - \left(1 - \frac{1}{Z_A\,\lambda}\right) \partial_\mu\partial_\nu \right)A_\nu^j \label{3_LMAG}\\
	      &\hspace{-10mm} -\frac{Z_B}{2} \delta^{ab} B_\mu^a\left(\delta_{\mu\nu}\partial^2 - \left(1-\frac{1}{Z_\xi\, \xi}\right) \partial_\mu\partial_\nu \right)B_\nu^b  + Z_c\, i \bar c^a  \partial^2  c^a\nonumber \\
		& \hspace{-10mm}  + Z_{ABB} gf^{iab} \left(1-\frac{1}{Z_\xi \xi} \right) A_\nu^i (\partial_\mu B_\mu)^a B_\nu^b +  Z_{ABB} gf^{iab} A_\mu^i B_\nu^a (\partial_\mu B_\nu)^b   \nonumber\\
		& \hspace{-10mm}  + Z_{ABB} gf^{iab} (\partial_\mu A_\nu^i)  B_\mu^a B_\nu^b    + Z_{3B}\, gf^{abc} (\partial_\mu B_\nu)^a B_\mu^b B_\nu^c  + Z_{Ac\bar c}\,ig f^{abi} A_\mu^i (\partial_\mu \bar c^a)  c^b\nonumber\\
		& \hspace{-10mm}   - Z_{Ac\bar c}\,ig f^{abi} A_\mu^i \bar c^a \partial_\mu c^b + \frac{Z_{Bc\bar c}\,ig}{2}\,f^{abc} B_\mu^a ( \partial_\mu\bar c^b)  c^c  - \frac{Z_{Bc\bar c}\,ig}{2}\,f^{abc} B_\mu^a \bar c^b \partial_\mu c^c\nonumber\\
		& \hspace{-10mm}  +  \frac{Z_{AABB}\,g^2}{2}f^{aic}f^{aje} A_\mu^i  A_\mu^j B_\nu^c B_\nu^e -  \frac{Z_{AABB}\,g^2}{2}\left(1-\frac{1}{Z_\xi \xi}\right) f^{aic}f^{ajd} A_\mu^i  A_\nu^j B_\mu^c B_\nu^d  \nonumber\\
		 & \hspace{-10mm}+  Z_{ABBB}\,g^2 f^{aic}f^{ade} A_\mu^i B_\nu^c B_\mu^d B_\nu^e   + \frac{Z_{4B}\,g^2}{4}f^{abi}f^{cdi} B_\mu^a B_\nu^b B_\mu^c B_\nu^d  \nonumber\\
		& \hspace{-10mm}  +  \frac{Z_{4B}\,g^2}{4}f^{abc}f^{ade} B_\mu^bB_\nu^c B_\mu^d B_\nu^e  - Z_{AAc\bar c}\, ig^2 f^{aib}f^{ajc} A_\mu^i A_\mu^j \bar c^b c^c \nonumber\\
		& \hspace{-10mm}- \frac{Z_{ABc\bar c}\,ig^2}{2}\,f^{abc}f^{aid}A^i_\mu B_\mu^b \bar c^d  c^c  - \frac{Z_{ABc\bar c}\,ig^2}{2}\,f^{abc}f^{aid} A^i_\mu B_\mu^b \bar c^c  c^d \nonumber\\
		& \hspace{-10mm} + Z_{BBc\bar c}\,ig^2f^{iab}f^{icd} B_\mu^a  B_\mu^c \bar c^b  c^d - \frac{Z_{4c}\,\xi g^2}{2}\,f^{iab}f^{icd} \bar c^a \bar c^c c^b  c^d  - \frac{Z_{4c}\,\xi g^2}{8}\,\,f^{abc}f^{ade}\bar c^b \bar c^d c^c c^e\,. \nonumber
	  \end{align}
    Implying the MAG non-renormalization theorem, the vertex renormalization constants and field strength renormalization constants are related by the following identities
	  \begin{align}
	    Z_{ABB} &= Z_B\,, & Z_{3B} &= Z_g Z_B^{3/2}\,,\label{3_def_Z}\\
	    Z_{AABB} & = Z_B  \,, & Z_{ABBB} & =Z_g Z_B^{3/2} \,, & Z_{4B} &= Z_g^2Z_B^2\,, \nonumber \\
	    Z_{Ac\bar c} & = Z_c \,, &Z_{Bc\bar c} &= Z_g Z_c Z_B^{1/2}\,,& Z_{4c} &= Z_g^2 Z_\xi Z_c^2\,,\nonumber\\
	    Z_{AAc\bar c} &= Z_c \,, & Z_{ABc\bar c} &=  Z_g Z_c Z_B^{1/2}\,, &  Z_{BBc\bar c} &= Z_g^2Z_BZ_c\,.\nonumber
	  \end{align}
    For completeness and for later reference, the Feynman rules for the Cartan gluon and ghost propagators and their 3- and 4-point functions are given,
    \begin{align}
      D^{ik\,(0)}_{\iota\kappa} (p)\equiv \Fvev{A_\iota^i\,A_\kappa^k}^{tl} & = \delta^{ik} \,\frac{1}{Z_A\,p^2} \left(\delta_{\iota\kappa} - (1-Z_A \lambda)\, \frac{p_\iota p_\kappa}{p^2} \right) \,,\label{3_FR_tl_AA}\\
	D^{ab\,(0)}_{\alpha\beta} (p)\equiv \Fvev{B_\alpha^a\,B_\beta^b}^{tl}& = \delta^{ab} \,\frac{1}{Z_B\,p^2} \left(\delta_{\alpha\beta} - (1-Z_{\xi}\xi)\, \frac{p_\alpha p_\beta}{p^2} \right) \,,\label{3_FR_tl_BB}\\
    D^{ab\,(0)} (p)\equiv \Fvev{c^a \, \bar c^b}^{tl} &= \delta^{ab}\,\frac{i}{Z_c\,p^2}\,, \label{3_FR_tl_cc}\\
    \Gamma^{iab\,(0)}_\iota(p_A,p_c,p_{\bar c})\equiv \Fvev{A_\iota^i\,c^a \bar c^b}^{tl} &= Z_{Ac\bar c}\, g f^{iab} \left(p_c + p_{\bar c} \right)_\iota \,,  \label{3_FR_tl_Acc} \\
    \Gamma^{ikab\,(0)}_{\iota\kappa}(p_{A_1},p_{A_2},p_c,p_{\bar c}) \equiv \Fvev{A_\iota^i\,A_\kappa^k\,c^a \bar c^b}^{tl} &= -Z_{AAc\bar c}\,2ig^2 f^{cia}f^{ckb} \delta_{\iota\kappa}  \label{3_FR_tl_AAcc} \,.
    \end{align}
    }
    There are six independent renormalization constants in \eqref{3_def_Z_fund}, which are reduced to four by the identities \eqref{3_WI_cartan} and \eqref{3_diagghost}. In \cite{Fazio:2001rm} it was, however, found that the most general counter term in the MAG consist of five free-parameters. The equivariant construction employed here features the additional Faddeev-Popov conjugation symmetry which is not apparent in \cite{Fazio:2001rm} and which relates the renormalization constants of the coset ghost and antighost. This symmetry is part of the even larger equivariant algebra presented in \secref{sec_3_algebra}.

\subsection{Strong Running Coupling}
    The MAG non-renormalization theorem \eqref{3_diagghost} states that the gauge coupling renormalization is directly related to the renormalization of the two-point function of the Cartan gluon. In other words, to calculate the gauge coupling renormalization, i.e. the \emph{QCD $\beta$-function}, one only needs to calculate the renormalization of the Cartan gluon propagator, which has explicitly been checked in \cite{Gracey:2005vu}. This behavior can be considered as "Abelian dominance in the high energy regime" \cite{Quandt:1997rw}. In addition it allows for a  simple non-perturbative definition of the strong running coupling $\alpha_S$ in the MAG analogous to the ghost-gluon vertex coupling in Landau gauge \cite{vonSmekal:1997vx, Fischer:2003zc}. Therefore one observes that the strong running coupling renormalizes as the square of the gauge coupling,
	\be \alpha_{s}(\Lambda^2) = \frac{g^2_0(\Lambda^2)}{4\pi} =  \frac{Z_g^2(\Lambda^2,\mu^2) g^2_R(\mu^2)}{4\pi} = Z_g^2(\Lambda^2,\mu^2) \alpha_{s}(\mu^2)\,, \ee
    where $\Lambda$ denotes the cutoff and $\mu$ the renormalization scale. Now consider the tree-level structure of the diagonal gluon propagator \eqref{3_FR_tl_AA} to be dressed with a renormalized function $a(p^2,\mu^2)$, for a proper definition see \eqref{3_FR_AA} below. It is connected to its bare counterpart by the renormalization constant $Z_A$, $a(p^2;\mu^2) = Z_A^{-1}(\Lambda^2,\mu^2)  a(p^2;\Lambda^2)$. Expressing the renormalization constants by the renormalized and unrenormalized dressing functions, the MAG non-renormalization theorem implies the  renormalization group invariant identity 
	\be 1 = \frac{\alpha_{s}(\Lambda^2)\,a(p^2;\Lambda^2) }{ \alpha_{s}(\mu^2)\, a(p^2;\mu^2)} \,,\ee
    which allows for a definition of the strong running coupling in the MAG
	  \be  \alpha_s(p^2) \equiv \alpha_s(\mu^2)a(p^2;\mu^2) \,. \label{3_alphas}\ee
    The definition \eqref{3_alphas} implies the renormalization condition $a(\mu^2;\mu^2)=1$. The strong running coupling in the MAG is given by the dressing function of the Cartan gluon only.
    
    If the scaling solution \eqref{3_IR_prop} is realized, the non-perturbative running coupling in the MAG is infrared diverging. This behavior is in stark contrast to Landau gauge where the strong running coupling is infrared-constant for the scaling solution and infrared-vanishing for decoupling-type of solutions \cite{Fischer:2008uz}. This seems contradicting if one encumbers the strong running coupling with some physical reality. The strong running coupling, however, is not an observable itself but \textit{"rather a quantity defined in the context of perturbation theory, which enters predictions for experimentally measurable observables"} \cite{Beringer:1900zz}. As such it can be gauge dependent. The lowest energy for which the strong running coupling is currently extracted from experiment is the mass of the $\tau$-lepton, $M_\tau \approx 1.8 \,\GeV$.

    \subsection{Hilbert-Space and Quartet Mechanism}

    In Landau gauge a physical Hilbert-space can by constructed by the BRST cohomology \eqref{1_KO_HS}. In the MAG such a constructions is not possible a priori because of the two-step gauge fixing process. As such one needs to define the Hilbert-space in two steps. Denoting the conserved Cartan BRST charge generating the transformations $\sC$ by $Q_C$, analogously to \eqref{1_KO_HS}, one defines the Hilbert-space for the Abelian theory $\L^{abel}$,
      \be \H_{abel} = \ker{Q_C}\!/\, \im {Q_C}\,,\ee
    or analogously,
      \be \H_{abel} = \left\{ \left[\ket{\psi}\right] : \sC\ket{\psi}=0\,\text{ and } \ket{\psi'}\sim\ket{\psi} \Leftrightarrow \ket{\psi'} - \ket{\psi} = \sC\left(\ket{\chi}\right) \,;\forall \ket{\chi} \right\} \,.\ee

    Operators whose Cartan BRST transformation vanishes have the property that the vacuum expectation value of their equivariant BRST transformation also vanishes. Using the general Ward identity \eqref{1_def_WI} one finds
    \begin{align} 
	\vev{\se \O(y)} &= \vev{\O(y)\,\left(-i \eta^i \partial_\mu \cp{B_\mu}{c}{i}\right)(x)} = \vev{\O(y)\,\sC \left(\left(-i \bar\omega^i \partial_\mu \cp{B_\mu}{c}{i}\right)(x)\right)} \nonumber\\
	    &   = \sC \vev{\O(y)\,\left(-i \bar\omega^i \partial_\mu \cp{B_\mu}{c}{i}\right)(x)}   = 0 \,, \label{3_appB}
    \end{align}
    for all $\O$ with $\sC \O = 0$ and if the Cartan BRST symmetry is conserved. This means that on the space of Cartan BRST invariant operators, one can use the equivariant BRST charge $Q_\eps$ to define the cohomology of the $SU(N)-$theory
	\be   \H = \ker{Q_\eps}\!/\, \im {Q_\eps}\Bigr\vert_{\H_{abel}}\,. \ee
    In terms of equivalence classes of states, the Hilbert space for the full theory reads
	\begin{multline}  \H = \bigl\{ \left[\ket{\psi}\right] : \se\ket{\psi}=0\,\text{ and } \\ \ket{\psi'}\sim\ket{\psi} \Leftrightarrow \ket{\psi'} - \ket{\psi} = \se \left(\ket{\chi}\right) \,;\forall \ket{\chi}  \text{ and } \ket{\psi},\ket{\psi'} \in \H_{abel}\bigr\} \,.\end{multline}

    This construction ensures that all asymptotic states in the physical Hilbert space $\H$ are colorless. A dynamical explanation of the extinction of colored degrees of freedom, as gluons and ghosts, in the asymptotic state space is the quartet mechanism as worked out in \secref{sec_KO}. A set of operators $\{\chi,\beta,\gamma,\bar \gamma \}$ which are metric partners, \eqref{1_KO_MP}, and fulfill specific BRST transformation relations, \eqref{1_KO_quart}, are called a quartet. In this equivariant construction with two kinds of BRST transformation, also the quartet mechanism acts on two levels. The first level are the quartets with respect to the Cartan BRST transformations. Among them is the elementary quartet of the Abelian theory, the longitudinal Cartan gluon, and the Cartan ghost, antighost and NL-field. They compose $(N-1)$ copies of the Abelian elementary quartet. 
    The coset gluon, ghost, antighost and NL-field form non-perturbative bound-states with the Cartan ghosts and antighosts as do transverse gluons and quarks in Landau gauge \cite{Alkofer:2011pe}. What is not captured by the quartet mechanism with respect to the Cartan BRST transformations is the transverse Cartan gluon since it is invariant under these transformations, 
	\be \sC A_\mu^{i\,T} = T_{\mu\nu}\sC(A_\nu^i) = 0\,.\ee
    This is understandable from the point of view of the Abelian gauge theory, since there are free photons in the Coulomb-phase of such a theory. In a confining non-Abelian gauge theory, however, it seems contradictory to have a free gluon. This "flaw" is cured by the quartets of the equivariant BRST transformations. Here the transverse Cartan gluon forms a quartet with the operators $\gamma^i = \cp{B_\mu}{c}{i}, \bar \gamma^i = \cp{B_\mu}{\bar c}{i}$ and $\beta^i = \se\cp{B_\mu}{\bar c}{i}$. Such a quartet structure exists for any $U(1)^{N-1}$ invariant operator which is not $SU(N)$ invariant. The physical operators of the theory are the one which are invariant under Cartan and equivariant BRST transformations.

    \subsection{The Equivariant Algebra \label{sec_3_algebra}}

      \begin{figure}[!b]
	  \begin{center}
	  \begin{minipage}[h]{.6\textwidth} 
		    \includegraphics[scale=.47]{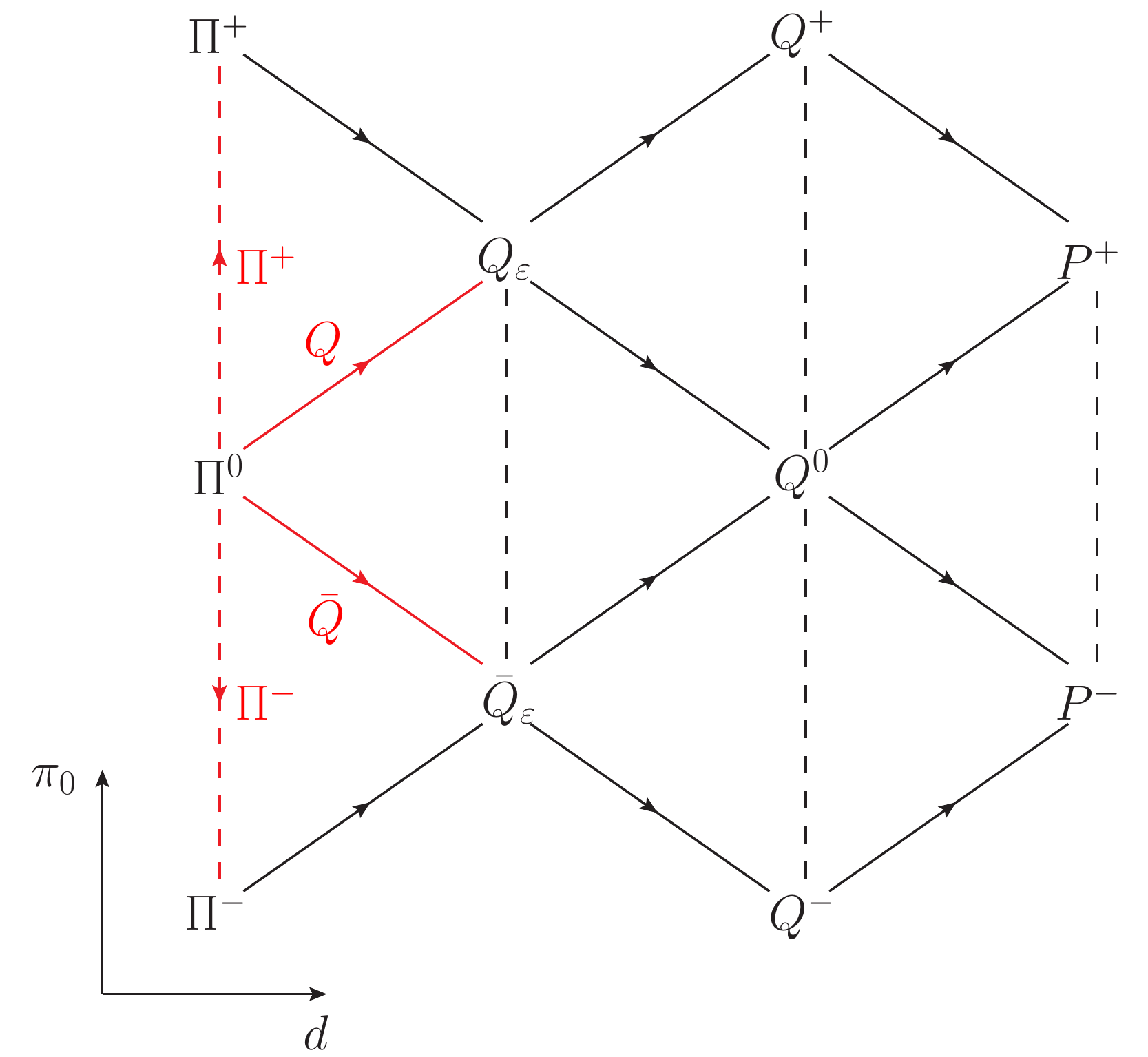} \\
	  \end{minipage}
	  \begin{minipage}[h]{.2\textwidth} 
		      \begin{tabular}{l | c | c }
			    Op. & $\pi^0$ & $d$ \\ \hline
			    $\Pi^+ $& 2 & 0 \\ \hline
			    $\Pi^0$ & 0 & 0 \\ \hline
			    $\Pi^-$ & -2 & 0 \\ \hline
			   $ Q_\eps$ & 1 & 1 \\ \hline
			    $\bar Q_\eps$ & -1 & 1 \\ \hline
			    $Q^+$ & 2 & 2 \\ \hline
			    $Q^0$ & 0 & 2 \\ \hline
			    $Q^-$ & -2 & 2 \\ \hline
			    $P^+$ & 1 & 3 \\ \hline
			    $P^-$ & -1 & 3 \\ 
			  \end{tabular}
	  \end{minipage}
	  \end{center}\vspace{-10mm}
	    \caption{The adjoint representation of the equivariant algebra $\mathbf E$. The red lines denote the ladder operators. The dimensionality increases from left to right, the ghost number from down to up.\label{fig_3_algebra}}
      \end{figure}

    Yang-Mills theory in partially gauge-fixed theory $\L^{abel}$, \eqref{3_def_Label}, features an equivariant algebra which is analogous to the deformed superalgebra of the massive Curci-Ferrari gauges \cite{ThierryMieg:1985yv,NakanishiOjima:1990}. By construction the theory is invariant under equivariant BRST and anti-BRST transformations and under Faddeev-Popov conjugation generated by the operators
      \be \Pi^+ = -i \int d^4x \,c^a(x)\var{\bar c^a(x)} \,,\qquad \text{and,}\qquad  \Pi^- = -i \int d^4x\, \bar c^a(x)\var{c^a(x)}\,. \ee
    The BRST and anti-BRST transformations are generated by the charges $ Q_\eps$ and $\bar  Q_\eps$ which are related via the commutation relations
	\be i\com{\Pi^-}{Q_\eps} = \bar Q_\eps  \,,\qquad \text{and,}\qquad i\com{\Pi^+}{\bar Q_\eps} = Q_\eps \,.\ee
    The commutator of the operators $\Pi^+$ and $\Pi^-$ yields the generator of the ghost number,
	\be \Pi^0 = i\com{\Pi^+}{\Pi^-} =  -i\int d^4x \,\left( c^a(x)\var{c^a(x)} -  \bar c^a(x)\var{\bar c^a(x)} \right)\,,\ee
    which also is a symmetry of the theory. The operators $\{Q_\eps,\bar Q_\eps,\Pi^+,\Pi^0,\Pi^-\}$ are not closed in the algebraic sense since the equivariant BRST transformations are not nilpotent but generate $U(1)^{N-1}$ transformations. These additional transformations with the transformation parameters
	\be \vartheta^i \in \{\cp{c}{c}{i},\cp{\bar c}{c}{i},\cp{\bar c}{\bar c}{i},\cp{b}{c}{i},\cp{b}{\bar c}{i} \} \ee
    are respectively performed by the operators $\{Q^+,Q^0,Q^-,P^+,P^-\}$. The operators are explicitly given in \eqref{B_algeb_op}.

      The (graded) algebra of the symmetry operators 
	\be \mathbf E = \{ Q_\eps,\bar Q_\eps,\Pi^+,\Pi^0,\Pi^-,Q^+,Q^0,Q^-,P^+,P^-\} \ee
    is closed and the non-vanishing commutation relations  are given in \eqref{B_algeb}. The equivariant algebra $\mathbf E$ is represented by its adjoint multiplet depicted in \figref{fig_3_algebra}. Assuming a symmetric division of the canonical dimension, $d$, and ghost-number, $\pi^0$, among ghost and anti-ghost,
	\be d(c^a)=d(\bar c ^a) = 1\,,\qquad d(b^a)=2\,,\qquad \pi^0(c) = 1\,,\quad \pi^0(\bar c) = -1 \,,\ee
    every operator can uniquely be specified by its ghost-number $\pi^0$ and canonical dimension $d$. Operators with the same canonical dimension have a difference of $2$ in the ghost number. There are two operators for any ghost number from $-2$ to $2$, respectively. There are four ladder operators, $Q_\eps,\bar Q_\eps,\Pi^+$ and $\Pi^-$, of which only two are needed to generate all operators from one "seed-operator". Interestingly there is no operator which goes from the right to the left in the adjoint multiplet, i.e. which reduces the dimensionality. There is nothing like a "BRST-annihilation" operator. In contrast, the operator $\Pi^+$ raises the ghost number by $2$, while $\Pi^-$ lowers the ghost number by $2$.

\section{The Dyson-Schwinger Equations}

    \begin{figure}[tp]
      \centering
      \includegraphics[width=\textwidth]{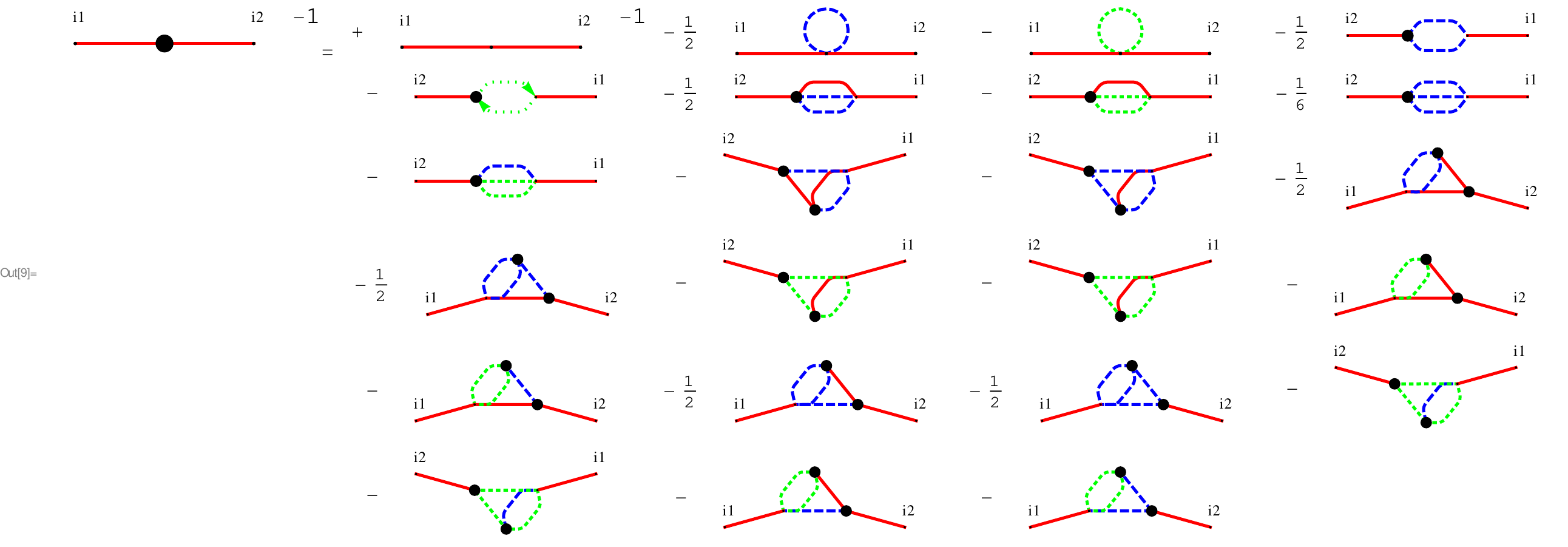}
      \caption{The DSE of the Cartan gluon propagator. Color coding as in \figref{3_fig:cDSEfull}.  \label{3_fig:ADSEfull}}
      \vspace{7mm}
	\includegraphics[width=\textwidth]{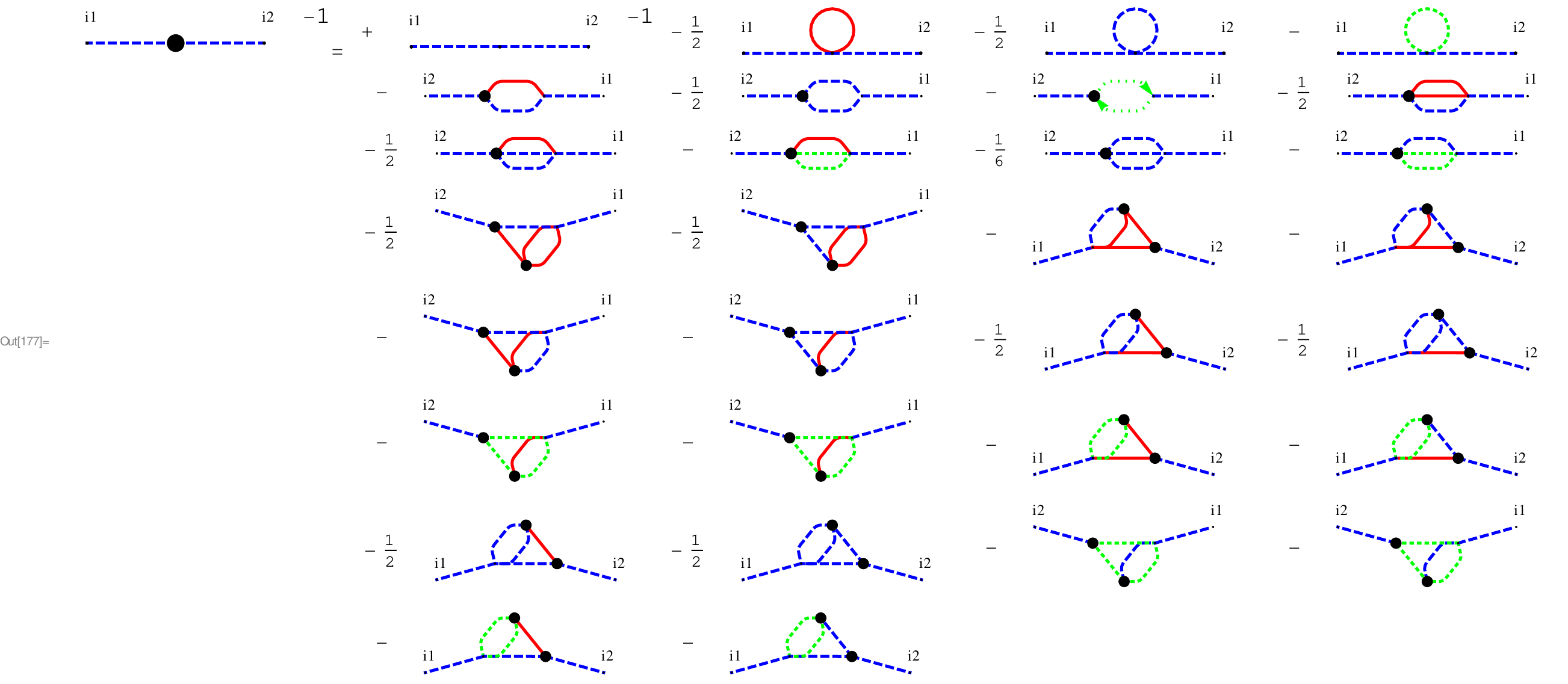}
	\caption{The DSE of the coset gluon propagator. Color coding as in \figref{3_fig:cDSEfull}. \label{3_fig:BDSEfull}}
      \vspace{7mm}
	\includegraphics[width=\textwidth]{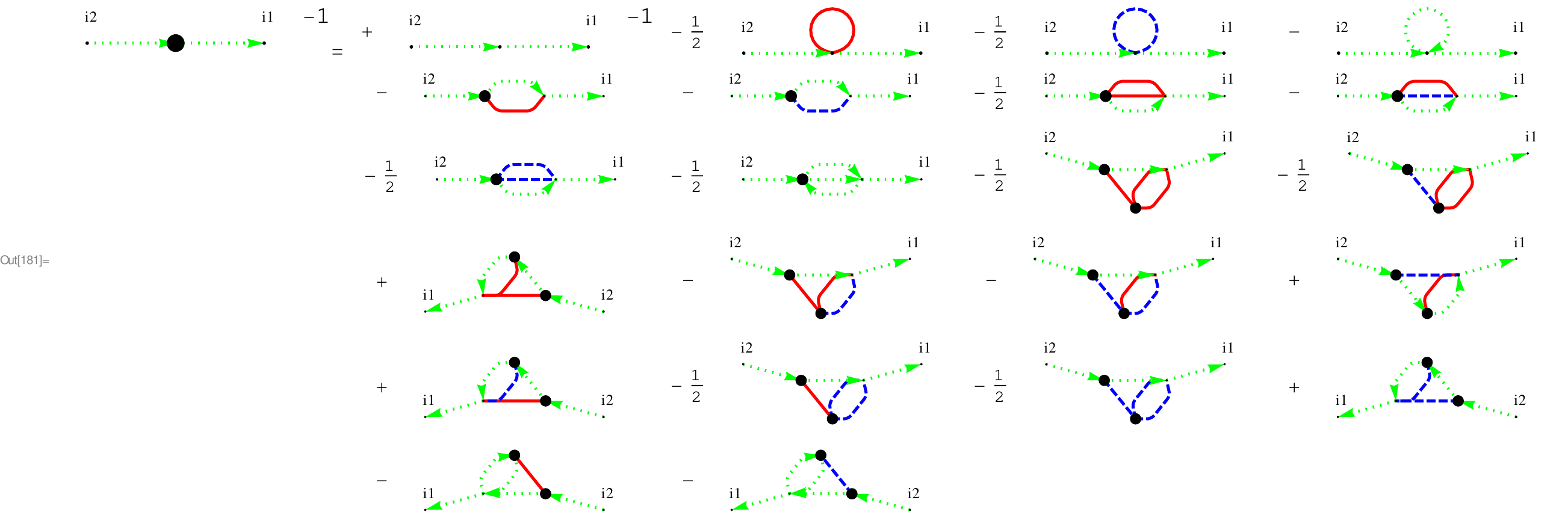}
	\caption{The DSE of the coset ghost propagator. The Cartan gluon is represented by the red solid line, the coset gluon by the blue dashed line and the ghost by the green dotted line. \label{3_fig:cDSEfull}}
    \end{figure}

    It is not much known about the Dyson-Schwinger equations in the MAG, in particular if one compares with the situation in Landau gauge. A first exploratory study of the Yang-Mills propagator system up to one-loop was performed in \cite{Shinohara:2003mx}. The main problem is that a derivation already of only the propagator Dyson-Schwinger equations by hand is a very tedious task due to the missing color symmetry and the additional interactions in the Lagrangian \eqref{3_LMAG}. Such a calculation is the perfect task for a computer algebra system and with the invention of the \mathematica-package \emph{DoDSE} \cite{Alkofer:2008nt} which is now part of the enlarged package \emph{DoFUN} \cite{Huber:2011qr} this derivation became a calculation of minutes. The Dyson-Schwinger equation for the Cartan gluon propagator is depicted in \figref{3_fig:ADSEfull}, for the coset gluon propagator in \figref{3_fig:BDSEfull} and the coset ghost propagator in \figref{3_fig:cDSEfull}. They were first derived in 
    \cite{Huber:2009wh}. 

    An infrared analysis of the full-set of Dyson-Schwinger equations and Exact Renormalization Group equations of the MAG was presented in \cite{Huber:2009wh,Huber:PhD}. A scaling-type of solution was found. It confirms the hypothesis of Abelian dominance: The Cartan gluons are infrared enhanced, while the coset gluons and ghosts are infrared suppressed.  A peculiarity in the MAG is that the infrared exponents of the four-point functions vanishes, i.e. the four-point functions of the MAG become infrared constant. The behavior of the three-point functions cannot be determined unambiguously.  With the definition of the full propagators of the Cartan gluon, the coset gluon and the coset ghost,
	\begin{align}
	D^{ik}_{\iota\kappa} (p) \equiv \Fvev{A_\iota^i\,A_\kappa^k} &= \delta^{ik}\,\frac{a(p^2)}{p^2} \left(\delta_{\iota\kappa} -  \frac{p_\iota p_\kappa}{p^2} \right) \,, \label{3_FR_AA}\\
	D^{ab}_{\alpha\beta} (p)\equiv \Fvev{B_\alpha^a\,B_\beta^b}&= \delta^{ab} \,\frac{b(p^2)}{p^2} \left(\delta_{\alpha\beta} - \left(1-Z_\xi \xi\right)\, \frac{p_\alpha p_\beta}{p^2} \right)\,, \label{3_FR_BB}\\
	    D^{ab} (p)\equiv \Fvev{c^a \, \bar c^b}& = i\delta^{ab}\,\frac{c(p^2)}{p^2}\,,  \label{3_FR_cc}
	\end{align}
    the infrared power-law solution in the MAG reads
	\be a(p^2)\rightarrow c_a\, \left(p^2\right)^{-\kappa_{MAG}}\,,\quad b(p^2)\rightarrow c_b\,\left(p^2\right)^{\kappa_{MAG}}\,,\quad\text{and,}\quad c(p^2)\rightarrow c_c\,\left(p^2\right)^{\kappa_{MAG}}\,, \label{3_IR_prop}\ee
    The infrared coefficients $c_a,c_b$ and $c_c$ stay unconstrained by the infrared analysis. The MAG-infrared exponent is a real positive number, $\kappa_{MAG}\approx 0.74$ \cite{Huber:PhD}.
  
    In \cite{Huber:2009wh,Huber:PhD} it was found, that the two-loop diagrams are the leading diagrams in the infrared regime of the Dyson-Schwinger equations. While for the sunset diagrams this is definite, the squint diagram might or might not be of leading order. This ambiguity is due to the ambiguity in the infrared behavior of the three-point functions. In fact this study was the motivation to study the sunset diagram in the Landau gauge in the first place. 

    \subsection{A Maximal Truncation}
\begin{figure}[t]
    \centering
    \includegraphics[width=\textwidth]{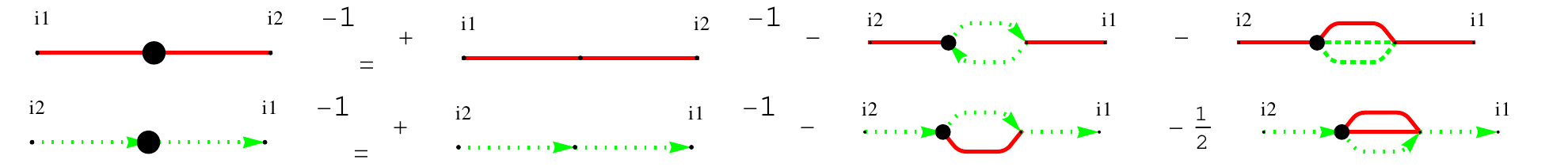}
    \caption{The maximal truncation of the propagator Dyson-Schwinger equations of the MAG used in this work. It contains respectively one infrared and one ultraviolet-leading of the Cartan gluon and the coset ghost Dyson-Schwinger equation. Color coding as in \figref{3_fig:cDSEfull}.\label{3_fig:DSE_trunc}}
    \end{figure}

    The most truncated set of equations which still entails a UV and IR leading diagram and does not contain the coset-gluon is depicted in \figref{3_fig:DSE_trunc}. The coset gluons have been dismissed due to the simpler Lorentz structure of the coset ghosts. In the deep infrared they obey the same power law behavior. This set of equations contains the tree-level and full Cartan gluon and ghost propagators and the tree-level and full $Ac\bar c-$ and $AAc\bar c-$vertices. While the propagators are solutions of the equations and are to be calculated self-consistently, the vertices have to be modeled. In this work the vertex models consist of the tree-level structure times some dressing function which arrange for the correct behavior in the UV and IR, 
	\begin{align}
	\Gamma^{iab}_\iota(p_A,p_c,p_{\bar c})\equiv \vev{A_\iota^i\,c^a \bar c^b}& \approx g f^{iab} \left(p_c + p_{\bar c} \right)_\iota \,D_{3\Gamma}(p_A^2,p_c^2,p_{\bar c}^2) \,, \label{3_FR_Acc}\\
	\Gamma^{ikab}_{\iota\kappa}(p_{A_1},p_{A_2},p_c,p_{\bar c}) \equiv \vev{A_\iota^i\,A_\kappa^k\,c^a \bar c^b} &\approx -2ig^2 f^{cia}f^{ckb} \delta_{\iota\kappa} \, D_{4\Gamma}(p_{A_1}^2,p_{A_2}^2,p_c^2,p_{\bar c}^2)  \label{3_FR_AAcc} \,.
	\end{align}

   The vertex dressing functions are constrained in their high and low energy asymptotics. While the $AAc\bar c$-vertex is infrared constant, the $Ac\bar c$-vertex has some ambiguity in its infrared power law, \cite{Huber:PhD},
	\be D_{3\Gamma}(p^2,p^2,p^2) \propto \left(p^2\right)^{\kappa_{MAG}\,\lambda_{3\Gamma}}, \qquad\text{with } -\frac{1}{2}\leq\lambda_{3\Gamma}\leq 0\,.\ee 
   In the high energy-regime they are determined by their anomalous dimensions, $\gamma_{Ac\bar c}$ and $\gamma_{AAc\bar c}$. 
   The vertex dressing functions are modeled as
	\begin{align}
	    D_{3\Gamma}(p_1^2,p_2^2,p^2) & = \frac{1}{Z_{Ac\bar c}}\, \left( a(p_1^2) a(p_2^2)  \right)^{\beta^{3\Gamma}_a} \left( c(p_1^2) c(p_2^2)  \right)^{\beta^{3\Gamma}_c} \\
	    D_{4\Gamma}(p_1^2,p_2^2,p_3^2,p^2)  & = \frac{1}{Z_{AAc\bar c}}\, \left(a(p_1^2)a(p_2^2)c(p_1^2)c(p_2^2) \right)^{\beta^{4\Gamma}}
	\end{align}
    where the exponents are given by
	\be \beta^{3\Gamma}_a = -\left(1+\frac{\lambda_{3\Gamma}}{2}\right) \frac{\gamma_c}{\gamma_A+\gamma_c}\,,\qquad \beta^{3\Gamma}_c =\beta^{3\Gamma}_a + \frac{\lambda_{3\Gamma}}{2} \,,\qquad
	\beta^{4\Gamma}_a = \frac{ \gamma_{AAc\bar c}}{\gamma_A+\gamma_c}  \,.
	\ee
    The parameter $\lambda_{3\Gamma}$ is a real number whose exact value will be fixed later.

\begin{figure}[t]
      \centering
      \includegraphics[width=.27\textwidth]{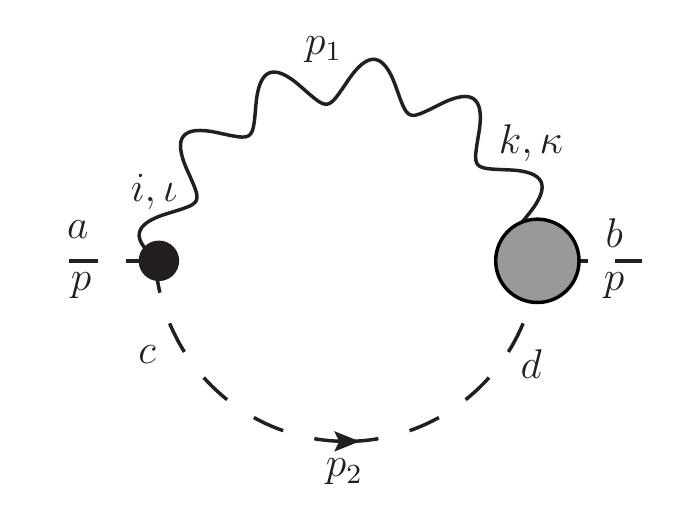}\hspace{-5.5mm}
      \includegraphics[width=.27\textwidth]{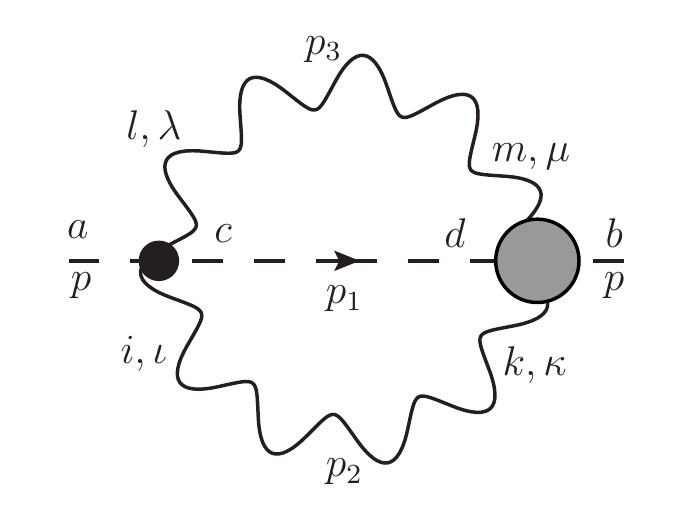} \hspace{-6.5mm}
      \includegraphics[width=.27\textwidth]{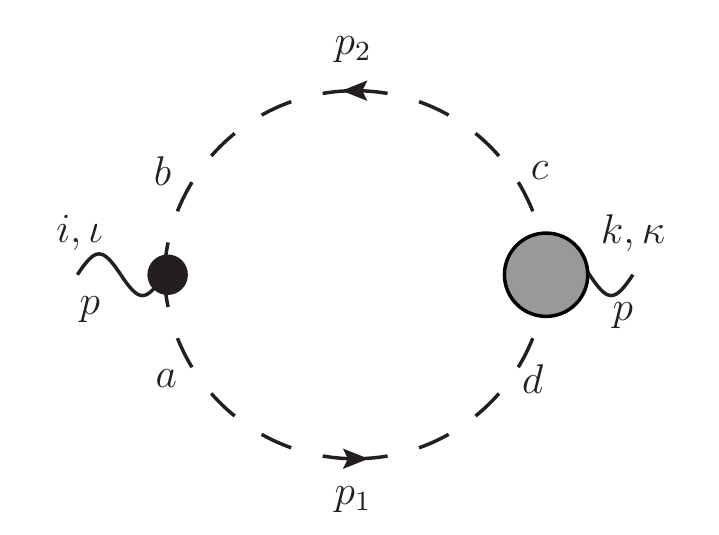}\hspace{-4.5mm}
      \includegraphics[width=.27\textwidth]{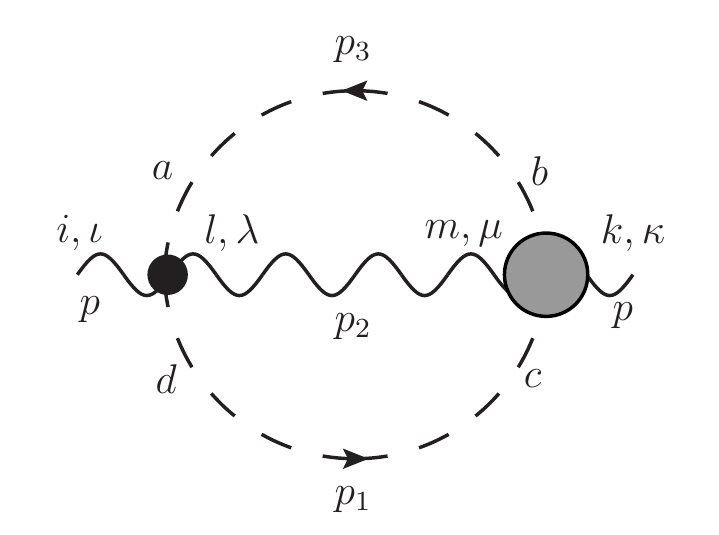}
      \caption{The full index dependences of the one-loop and sunset diagram appearing in the ghost and gluon equations, \eqref{3_ghostDSE1} and \eqref{3_gluonDSE1}. The Cartan gluon is depicted by the wiggly line, the ghost by the dashed line. \label{3_fig:ghostglue}}
      \end{figure}

    The mathematical expressions for the truncated equations in \figref{3_fig:DSE_trunc} read
	\begin{subequations}\label{3_DSE_1}
	\begin{align}
	\left[ D^{ab} (\p)\right]^{-1} & = \left[ D^{ab\,0} (\p)\right]^{-1} - \Gamma^{iac\,(0)}_\iota(-\p_1,\p,\p_2) \Gamma^{kdb}_\kappa(\p_1,\p_2,\p) D^{cd} (\p_2)D^{ik}_{\iota\kappa} (\p_1) \label{3_ghostDSE1}\\ 
	& \quad + \frac{1}{2} \Gamma^{ilac\,(0)}_{\iota\lambda}(-\p_2,-\p_3,\p,\p_1)  \Gamma^{kmdb}_{\kappa\mu}(\p_2,\p_3,\p_1,\p) D^{cd} (\p_1)D^{ik}_{\iota\kappa} (\p_2) D^{lm}_{\lambda\mu} (\p_3) \nonumber\\
	\left[ D^{ik}_{\iota\kappa} (\p)\right]^{-1} & = \left[ D^{ik\,(0)}_{\iota\kappa} (\p)\right]^{-1} + \Gamma^{iba\,(0)}_\iota(\p,\p_2,\p_1) \Gamma^{kdc}_\kappa(-\p,\p_1,\p_2) D^{ad}(\p_1)D^{cb}(\p_2) \label{3_gluonDSE1}\\
	& -  \Gamma^{ilad\,(0)}_{\iota\lambda}(\p,-\p_2,\p_3,\p_1) \Gamma^{kmcb}_{\kappa\mu}(-\p,\p_2,\p_1,\p_3) D^{dc}(\p_1)D^{ba}(\p_3) D^{lm}_{\lambda\mu}(\p_2) \nonumber \,.
	\end{align}
	\end{subequations}
  For the definition of the indices and the momenta see \figref{3_fig:ghostglue}. In the following $N_c$ denotes the number of colors, $N_d$ the number of Cartan (diagonal) and $N_o$ the number of coset (off-diagonal) generators respectively. For the one-loop terms an asymmetric momentum partitioning is chosen
	\be
	  \p_1 = \q\,,  \qquad \p_2 = \p-\q\,. \label{3_MR_ol}
	\ee
   To be able to transport the result of the infrared analysis of the sunset diagram of \cite{Huber:PhD} to this calculations, the same momentum routing is used
	\be \p_1 = \q_1\,,\qquad \p_2 = \q_2\,, \qquad \p_3 = \p-\q_1-\q_2 \,.\label{MR_tl} \ee
    Further technical details of the investigation of the sunset diagram are shifted to the appendix \appref{secA_sunset}.

    Inserting the 1PI Green functions as defined above into the truncated Dyson-Schwinger equations \eqref{3_DSE_1},  taking the normalized color trace and projecting the gluon equation with the generalized Brown-Pennington projector, \eqref{LG_def_BP}, yields
	{\allowdisplaybreaks
      \begin{subequations}\label{3_DSE_2}
	\begin{align}
	    \frac{1}{ c(p^2)}   =  Z_c & -g^2 N_c \, \frac{N_d}{N_o} \, 4 \,\int\!\dbar^4q \, \frac{ \T_c^{ol}}{p_1^2\,p_2^2}\,a(p_1^2)c(p_2^2)\left(a(p_1^2) a(p_2^2)\right)^{\beta_a^{3\Gamma}} \left(c(p_1^2) c(p_2^2)\right)^{\beta_c^{3\Gamma}}\nonumber\\ 
	    &  + \frac{2(g^2 N_c)^2 }{p^2} \left(\frac{N_d}{N_o}\right)^2\int\!\dbar^4q_1 \int\!\dbar^4q_2  \label{3_ghostDSE2}\\
	    & \hspace{15mm}\times\,\T_c^{sun}\frac{a(p_2^2)a(p_3^2)c(p_1^2)}{p_1^2\,p_2^2\,p_3^2} \,\left(a(p_1^2)a(p_2^2)c(p_1^2)c(p_2^2) \right)^{\beta^{4\Gamma}} \nonumber\,, \\ 
	    \frac{1}{a(p^2)}  = Z_A & + \frac{g^2N_c}{3\,p^2}\,\int\dbar^4q \,\frac{ T_A^{ol}}{p_1^2 \,p_2^2}\,\left(c(p_1^2) c(p_2^2)\right)^{1+\beta_c^{3\Gamma}} \left(a(p_1^2) a(p_2^2)\right)^{\beta_a^{3\Gamma}}\label{3_gluonDSE2}\\
	    &  +  \frac{4 \left(g^2N_c\right)^2}{3\,p^2} \frac{N_d}{N_o}  \int\!\dbar^4q_1 \int\!\dbar^4q_2 \nonumber\\
	    & \hspace{15mm}\times\,T_A^{sun} \frac{a(p_2^2)c(p_3^2)c(p_1^2)}{p_1^2\,p_2^2\,p_3^2} \, \left(a(p_1^2)a(p_2^2)c(p_1^2)c(p_2^2) \right)^{\beta^{4\Gamma}}\,.\nonumber
	\end{align}
	\end{subequations}
     For later reference the independent angles are integrated out,
	\begin{subequations}\label{3_DSE_3}\begin{align}
	\frac{1}{ c(p^2)}   =  Z_c &-\, \frac{g^2 N_c }{2\pi^3}\, \frac{N_d}{N_o}  \int\!dq^2\!\int\!\dzs\,\,  \frac{\T_c^{ol}\,q^2}{p_1^2\,p_2^2}\,a(p_1^2)c(p_2^2)\left(a(p_1^2) a(p_2^2)\right)^{\beta_a^{3\Gamma}} \left(c(p_1^2) c(p_2^2)\right)^{\beta_c^{3\Gamma}}\nonumber\\ 
	  &  + \frac{(g^2 N_c)^2 }{16\pi^6\,p^2} \left(\frac{N_d}{N_o}\right)^2\int\!dq_1 \!\int\! dq_2\!\int\! \dzs_1\!\int\! \dzs_2\!\int\! dy\label{3_ghostDSE3}\\
	    & \hspace{15mm} \times q_1^3\,q_2^3 \,\T_c^{sun} \,\frac{a(p_2^2)a(p_3^2)c(p_1^2)}{p_1^2\,p_2^2\,p_3^2} \,\left(a(p_1^2)a(p_2^2)c(p_1^2)c(p_2^2) \right)^{\beta^{4\Gamma}}\,, \nonumber\\ 
	\frac{1}{a(p^2)}  = Z_A & +\frac{g^2N_c}{24 \pi^3\,p^2}\int\!dq^2\!\int\!\dzs\, \frac{T_A^{ol}  q^2 }{p_1^2 \,p_2^2}\,\left(c(p_1^2) c(p_2^2)\right)^{1+\beta_c^{3\Gamma}}\left(a(p_1^2) a(p_2^2)\right)^{\beta_a^{3\Gamma}}\label{3_gluonDSE3}\\
	& + \frac{\left(g^2N_c\right)^2}{24\pi^6\,p^2} \frac{N_d}{N_o} \int\!dq_1 \!\int\! dq_2\!\int\! \dzs_1\!\int\! \dzs_2\!\int\! dy\nonumber\\
	& \hspace{15mm}\times q_1^3\,q_2^3\,T_A^{sun}\, \frac{a(p_2^2)c(p_3^2)c(p_1^2)}{p_1^2\,p_2^2\,p_3^2} \,\left(a(p_1^2)a(p_2^2)c(p_1^2)c(p_2^2) \right)^{\beta^{4\Gamma}}\,.\nonumber
	\end{align}\end{subequations}
      }

    The dimensionless tensor structures read
	\begin{align}
	    \T_c^{ol} & = 1-z_q^2\,,  &  T_A^{ol} &= 4 q^2 (1-\zeta z_q^2)+(1-\zeta)\left(p^2+4 p q z_q \right)\,,\label{3_tensor} \\
	    \T_c^{sun} & = 2+z_{23}^2\,,  &  T_A^{sun} &= 3-\zeta+\zeta z_{02}^2\,.                 \nonumber
	\end{align}

    Analogous to the considerations in \secref{sec_LG} an investigation of the high and low energy asymptotic of the Eqs.~\ref{3_DSE_3} follows. As there is, respectively, one one-loop and one two-loop term in each equation, possible quadratic divergences are subtracted within the diagrams themselves. A counter term construction $\tau_i$ is added to the tensor structures in \eqref{3_tensor} such that the integral with the corresponding regularized tensor  $\widetilde\T_i = \T_i - \tau_i$ is free of quadratic divergences.

\subsection{UV-Analysis}
    
    The high-energy regime of Yang-Mills theory can be investigated using usual perturbation theory. The logarithmic scaling of the primitively divergent Green functions is encoded in the renormalization constants \eqref{3_def_Z_fund} and \eqref{3_def_Z}. An immediate consequence of the MAG renormalization theorem, \eqref{3_diagghost}, is $Z_{Ac\bar c} = Z_{AAc\bar c} = Z_c$  and thus
	\be \gamma_{Ac\bar c} = \gamma_{AAc\bar c} = -\gamma_c  \,.\label{3_AD_vert}\ee
    Due to the simple vertex models employed in this thesis it will not be able to keep all of the relations \eqref{3_AD_vert}, which will be seen below. Fortunatelly, the numerical error by breaking this relation is small.
  
    For the MAG the anomalous dimensions of the propagators where calculated up to three-loops in \cite{Gracey:2005vu}. For $SU(3)$ gauge theory, the one-loop resummed anomalous dimensions read
	\be \gamma_A = -1\,,\qquad \gamma_B = -\frac{5}{11}\,,\qquad\text{and,}\qquad \gamma_c =- \frac{15}{44}\,. \label{3_aD_gracey} \ee
    While the anomalous dimensions of the coset fields somewhat depend on the $SU(N)$-gauge group, but are negative in any case, the anomalous dimension of the Cartan gluon is fixed for any gauge group. In fact it is a corollary of the MAG non-renormalization theorem. However, it was not put as an input into the calculation \cite{Gracey:2005vu} and thus this result can be interpreted as an explicit confirmation of \eqref{3_diagghost}. The high-energy behaviour of the vertex functions was calculated only recently in \cite{Bell:2013xma}. In the following the anomalous dimensions are considered unfixed. It has to be seen in how far the analytical values can be reproduced.

    \subsubsection{One-loop diagrams}
    The UV-analysis of one-loop integrals follows the one in Landau gauge presented in \secref{sec_2_UV_olo}. Performing the same approximation that lead to \eqref{2_olo_ymax} and subsuming the color indices into the prefactors $C_c^{ol} = -\frac{g^2N_cN_d}{2\pi^3N_o}$ and $ C_A^{ol} = \frac{g^2N_c}{24\pi^3}$, the one-loop contribution to the truncated ghost and gluon DSE, \eqref{3_DSE_3}, read 
	\begin{subequations}
	\begin{align}
	I_c^{ol,UV} & = C_c^{ol}\,\int_{p^2}^{\Lambda^2}\!\!dq^2\, \left(a(q^2)\right)^{1+2\beta_a^{3\Gamma}} \left(c(q^2)\right)^{1+2\beta_c^{3\Gamma}}  \int_{-1}^1\!\dzs \, \frac{\T_c^{ol}}{(p-q)^2}   \,, \\
	  I_A^{ol,UV} & = \frac{C_A^{ol}}{p^2}\,\int_{p^2}^{\Lambda^2}\!\!dq^2\, \left(a(q^2)\right)^{2\beta_a^{3\Gamma}} \left(c(q^2)\right)^{2+2\beta_c^{3\Gamma}}   \int_{-1}^1\!\dzs \, \frac{q^2}{(p-q)^2}  \,\T^{ol}_A\,\,,
	\end{align}\end{subequations}
    Integrating out the angular variables yields
    \begin{subequations}\begin{align}
    I_c^{ol,UV} & = C_c^{ol}\,\int_{p^2}^{\Lambda^2}\!\!dq^2\, \left(a(q^2)\right)^{1+2\beta_a^{3\Gamma}} \left(c(q^2)\right)^{1+2\beta_c^{3\Gamma}}\,\frac{\pi }{8 q^2}  \left(3 - \frac{p^2}{q^2}\right) \,,  \label{3_UV_ghost_angint}\\
      I_A^{ol,UV} & = \frac{C_A^{ol}}{p^2}\,\int_{p^2}^{\Lambda^2}\!\!dq^2\,  \left(a(q^2)\right)^{2\beta_a^{3\Gamma}} \left(c(q^2)\right)^{2+2\beta_c^{3\Gamma}} \, \frac{\pi}{2} \left((4-\zeta) + (3-4 \zeta ) \,\frac{p^2}{q^2}\right) \label{3_UV_glue_angint}
    \end{align}\end{subequations}
    The one-loop contribution to the ghost self-energy is only logarithmically divergent, so no manipulations are needed. The one-loop contribution to the gluon self-energy is quadratically divergent. A possible counterterm which resembles the Brown-Pennington case in the limit $\zeta\rightarrow 4$ is given by $\tau_{A}^{ol} = (4-\zeta)$. The regularized tensor structure then reads
	\begin{align}
	\widetilde \T^{ol}_A &  =\,\zeta\left(1- 4 \, z_q^2\right) + \frac{1-\zeta}{q^2}\left(p^2 + 4\,pqz_q\right) \,.\label{3_ol_glue_Treg}
	\end{align}
    For all zeta the quadratic divergences in \eqref{3_UV_glue_angint} disappear and only logarithmic divergences remain, 
	\be \widetilde I_A^{ol,UV}  = \frac{C_A^{ol}\,\pi }{2}\,\left(3- 4 \zeta \right)\,\,\int_{p^2}^{\Lambda^2}\!dq^2\, \left(a(q^2)\right)^{2\beta_a^{3\Gamma}} \left(c(q^2)\right)^{2+2\beta_c^{3\Gamma}} \, \frac{1}{q^2} \,. \label{3_UV_glue_angint_reg}  \ee

    Having only logarithmic divergences left and assuming that the dressing functions take on their logarithmic scaling one can perform the integrations in \eqref{3_UV_ghost_angint} and \eqref{3_UV_glue_angint_reg} to give in leading order
	\begin{subequations}\label{3_IolUV}\begin{align}
		I^{ol,UV}_c & = -\frac{9}{11}\,\frac{N_d}{N_o}\, \frac{1}{1+\gamma_A+\gamma_c+2\gamma_{Ac\bar c}}\, \Biggl(\left(\beta_0g^2 \log{\frac{\Lambda^2}{\mu^2}}+1\right)^{1+\gamma_A+\gamma_c+2\gamma_{Ac\bar c}} \label{3_UV_ghost_ol} \\
		  & \hspace{60mm} -\left( \beta_0g^2\log{\frac{p^2}{\mu^2}}+1\right)^{1+\gamma_A+\gamma_c+2\gamma_{Ac\bar c}} \Biggr) \nonumber\\
		\widetilde I^{ol,UV}_A & = \frac{3- 4 \zeta}{11}\,\frac{1}{1+2\gamma_c+2\gamma_{Ac\bar c}}  \label{3_UV_gluon_ol}  \Biggl( \left(\beta_0g^2 \log{\frac{\Lambda^2}{\mu^2}}+1\right)^{1+2\gamma_c+2\gamma_{Ac\bar c}} \\ 
		 & \hspace{60mm}-\left( \beta_0g^2\log{\frac{p^2}{\mu^2}}+1\right)^{1+2\gamma_c+2\gamma_{Ac\bar c}} \Biggr)\,. \nonumber
	\end{align}\end{subequations}

    \subsubsection{Sunset Diagrams}

    Neglecting the angular dependence of the arguments in the dressing functions , the angular integrals of the sunset diagrams in \eqref{3_DSE_3} are given by their tree-level structure
	\be \label{3_Isunang}
	    I^{sun,ang}_{A,c}(p,q_1,q_2)  = q_1^3q_2^3\int\!\dzs_1 \int\!\dzs_2 \int\!dy \,\,\frac{T_{A,c}^{sun}}{p_1^2\,p_2^2\,p_3^2}\,. 
	\ee
    Further details on the UV-analysis of the sunset diagrams are given in \appref{secA_sun_UV}. The integrals $ I^{sun,ang}_{A,c}$ can be evaluated using the results given in \tabref{A_tab_MAG_sun}. The corresponding radial integrals are quadratically divergent. The overlapping divergences manifest themselves in a case-by-case analysis if some momentum in the integrals is larger than another one. Since the tensor structure in the sunset diagrams in \eqref{3_DSE_3} is relatively puny, the counter term construction effects the result of the integral considerably. A proper construction scheme for the counter term is thus in order. The guiding line chosen here is to take only these tensor structures into account, which could have been generated by exchanging legs within the diagram, i.e. changing a ghost line with a gluon line. For the sunset diagram in the ghost equation this are $\{z_{12},z_{13},z_{23}\}$. The gluon line in the sunset diagram of the gluon equation is contracted with the external projector and thus the 
possible tensor structures are $\{z_{01},z_{02},z_{03}\}$.
    The counter term for the ghost equation is defined uniquely by aboves argument and is given by
    \be \tau_c^{sun} = z_{13}^2 + 2 z_{23}^2 + 3 z_{12}^2\,,\qquad\text{which yields,}\qquad \widetilde\T_c^{sun} = 2- z_{13}^2 - z_{23}^2 - 3 z_{12}^2\,. \label{3_tl_ghost_Treg}\ee
    
    In the gluon equation there is an ambiguity in constructing this counter term. Here two possibilities are presented. One which continuously interpolates to the Brown-Pennington case, the other one which removes the dependence on the parameter $\zeta$ and is maximally symmetric,
      \begin{subequations}\label{3_tensreg}
      \begin{align}
	  \tau_{A,1}^{sun} & = \frac{3}{4} \,\left( 4-\zeta \right)\,, &  \widetilde \T_{A,1}^{sun} & = -\frac{\zeta}{4}\left(1-4 z_{02}^2 \right) \,, \\
	  \tau_{A,2}^{sun} & = (1-\zeta) + \frac{8}{3}\left( z_{01}^2 + z_{02}^2 + z_{03}^2 \right) + \zeta z_{02}^2\,, & \widetilde \T_{A,2}^{sun} & = 2- \frac{8}{3}\left( z_{01}^2 + z_{02}^2 + z_{03}^2 \right)\,.
      \end{align}
      \end{subequations}
    Inserting the regularized tensor structures $\widetilde T_i$ into the integrals \eqref{3_Isunang} only leaves terms proportional to $p^2$ and higher, i.e. terms which correspond to logarithmic divergences and lower. Approximating the dressing functions by their perturbative behavior and using \eqref{A_sun_genultravioletint} yields
      \begin{subequations} \label{3_IsunUV}
      \begin{align}
	I_c^{sun,UV} & = \frac{g^2 N_c}{(4\pi)^2} \frac{N_d^2}{N_o^2}\,\frac{9}{11}\,\frac{1}{1+2\gamma_A+ \gamma_c+2\tilde\gamma_{AAc\bar c}}  \\ 
	& \hspace{-5mm}\times\left( \left(1+\beta_0g^2 \log{\frac{\Lambda^2}{\mu^2}}\right)^{1+2\gamma_A+ \gamma_c+2\tilde\gamma_{AAc\bar c}}- \left(1+\beta_0g^2 \log{\frac{p^2}{\mu^2}}\right)^{1+2\gamma_A+ \gamma_c+2\tilde\gamma_{AAc\bar c}} \right) \nonumber\\
	I_A^{sun,UV} & = -\frac{g^2 N_c}{(4\pi)^2} \frac{N_d}{N_o}\,\frac{t_i}{11}\,\frac{1}{1+\gamma_A+2\gamma_c+2\tilde\gamma_{AAc\bar c}} \\
	&\hspace{-5mm}\times\left( \left(1+\beta_0g^2 \log{\frac{\Lambda^2}{\mu^2}}\right)^{1+\gamma_A+2\gamma_c+2\tilde\gamma_{AAc\bar c}}- \left(1+\beta_0g^2 \log{\frac{p^2}{\mu^2}}\right)^{1+\gamma_A+2\gamma_c+2\tilde\gamma_{AAc\bar c}} \right)  \nonumber
      \end{align}
      \end{subequations}
    The factor $t_i$ represents a numerical factor for the different regularized tensor structures \eqref{3_tensreg}, $ t_1 = \zeta$ and $ t_2 = \frac{16}{3}$. If the relations \eqref{3_AD_vert} are fulfilled the rhs of the gluon equation diverges. Therefore one is not able to keep $\gamma_{AAc\bar c} = -\gamma_c$. This divergence is a model artifact, since the vertex \eqref{3_FR_AAcc} pulls the complete anomalous dimension of the four-point vertex into the integral. In the full theory there would be some contribution from the external gluon line. The anomalous dimension of the four gluon vertex is thus modeled as $\gamma_{AAc\bar c}\approx 0.9 \gamma_c$. Since the sunset diagrams are subleading in the UV the error induced by this modeling can be considered to be small.
    
    \subsubsection{Self-Consistency in the UV}

    When comparing the logarithmic running of the one-loop \eqref{3_IolUV} and sunset diagrams \eqref{3_IsunUV} one finds that the one-loop terms are leading in the ultraviolet. Plugging these results into the Dyson-Schwinger equations \eqref{3_DSE_3}, taking the UV leading terms on both sides and absorbing all dependence on the cut-off into the renormalization constants $Z_A$ and $Z_c$ yields
    \begin{subequations}\begin{align}
    \left(1+\beta_0g^2 \log{\frac{p^2}{\mu^2}} \right)^{-\gamma_c} & = \frac{9}{11}\,\frac{N_d}{N_o}\, \frac{\left(\beta_0g^2 \log{\frac{p^2}{\mu^2}}+1\right)^{1+\gamma_A+\gamma_c+2\gamma_{Ac\bar c}}}{1+\gamma_A+\gamma_c+2\gamma_{Ac\bar c}}\,,\label{UV_sc_c}\\ 
    \left(1+\beta_0g^2 \log{\frac{p^2}{\mu^2}}  \right)^{-\gamma_A} & = - \frac{3- 4 \zeta}{11}\,\frac{ \left(1+ \beta_0g^2\log{\frac{p^2}{\mu^2}}\right)^{1+2\gamma_c+2\gamma_{Ac\bar c}} }{1+2\gamma_c+2\gamma_{Ac\bar c}}\,. \label{UV_sc_A} 
    \end{align}\end{subequations}
    Self consistency in the high-energy regime thus constrains the anomalous dimensions
	\begin{align}
	  1+\gamma_a+2\gamma_c + 2\gamma_{Ac\bar c} &= 0 \,,\\
	  1+\gamma_a+\gamma_c + 2\gamma_{Ac\bar c} &= \frac{9\,N_d}{11\,N_o}\,, \\
	  1+2\gamma_c + 2\gamma_{Ac\bar c} &=-\frac{3-4\zeta}{11}\,,
	\end{align}
    which yields
	\be  \gamma_A = \frac{3-4\zeta}{11}\,,\qquad \gamma_c = -\frac{9\,N_d}{11\,N_o}\,,\qquad\text{and,}\qquad\gamma_{Ac\bar c} = \frac{-7+2\zeta+9 \frac{N_d}{N_o}}{11}\,. \ee

    Of all the relations presented in the preamble to this subsection the equation $\gamma_A = - 1$ seems to be the strongest, since it directly related to the MAG renormalization theorem. Keeping this immediately fixes the parameter $\zeta \equiv \frac{7}{2}$. The parameter $\zeta$ stems from the Brown-Pennington projector. It is unfixed in the range $\zeta\in\,[1,4]$ and it does not harm if it is fixed to any value in the interval since the tensor structures \eqref{3_tensreg} are constructed such that the quadratic divergences are subtracted for any value of $\zeta$. Imposing $\zeta = \frac{7}{2}$ yields
	  \be \gamma_c = -\gamma_{Ac\bar c} = - \frac{9\,N_d}{11\,N_o}\,, \ee
    and thus the relevant relation in \eqref{3_AD_vert} also holds.

    To summarize, in the self-consistent solution of the truncation considered it is not possible to meet the analytical values of the anomalous dimensions \eqref{3_aD_gracey}. However, the main features are captured. The MAG non-renormalization theorem is kept and the anomalous dimensions of the propagators are negative. The relation between the anomalous dimensions of the coset ghost propagator and the $Ac\bar c$-vertex can be fulfilled.

\subsection{IR-Analysis}
    For the infrared analysis the dressing functions in \eqref{3_ghostDSE2} and \eqref{3_gluonDSE2} are approximated by their power-law behavior and the tensor structure are replaced by their regulated counterparts with all quadratic divergences removed.
    
    \subsubsection{One-Loop Diagrams}
      The one-loop integrals in \eqref{3_ghostDSE2} and \eqref{3_gluonDSE2} are solved using \eqref{2_infraredGamma}. They yield the expected subleading behavior. The one-loop contribution to the ghost equation is proportional to the scaling of the Cartan-gluon--ghost vertex, $\left(p^2\right)^{\kappa\lambda_{3\Gamma}}$, the one-loop contribution to the gluon equation is infrared finite $\propto \left(p^2\right)^{\kappa(2+\lambda_{3\Gamma})} $. The Gamma functions yield constrains on the vertex parameter $\lambda_{3\Gamma}$ in terms of the infrared exponent $\kappa_{MAG}$, which are, however, not very severe. The most relevant constrain is that this parameter is not allowed to vanish, $\lambda_{3\Gamma}\neq 0$. In the numerical solution the parameter is easily adjusted to fulfill this constraint.
    
    \subsubsection{Sunset Diagrams}
      The general integral for the infrared analysis of the sunset-diagram was obtained in \cite{Huber:PhD} and reads
      \begin{align} 
	      I_{sun}(p^2;a,b,c,d,e,f) & :=\int\!\dbar^d q_1\int\!\dbar^d q_2 \,\,(p_1^2)^a\, (p_2^2)^b\, (p_3^2)^c \,z_{01}^e \,z_{02}^f \label{3_def_IsunIR} \\
	      &\hspace{-29mm}=  \left(4 \pi \right)^{-d} (p^2)^{a+b+c+d}  \sum_{n_1,n_2,n_3,n_4,n_5=0}^{\max\{e,f\}} \frac{(-2)^{n_3}}{n_1!n_2!n_3!n_4!n_5!} \nonumber\\
	      &\hspace{-29mm} \times\left(a+b+c+\frac{3d+e+f}{2},-2(a+b+c)-\frac{5d}{2}-n_1-n_2-n_3-n_4-n_5\right)_P \nonumber\\
	      &\hspace{-29mm} \times \left(-a+\frac{e}{2},2a+\frac{d}{2}-n_1-n_2+n_5\right)_P \left(-b+\frac{f}{2},2b+\frac{d}{2}+n_2-n_4-n_5\right)_P \nonumber \\
	      &\hspace{-29mm} \times \left(-c,2c+\frac{d}{2}+n_1+n_4\right)_P P\left(e,n_1+n_2,n_3\right)\,P\left(f,n_4+n_5,n_3\right)\,,	\nonumber
      \end{align}
      with the Pochhammer symbols $ \left(a,b \right)_P = \frac{\Gamma\left(a+b\right)}{\Gamma\left(a\right)}$ and the $P$-symbols defined as
      \be  P(a,b,c) = \left(-\frac{a}{2},b+\frac{c}{2} \right)_P \left(\frac{1}{2}-\frac{a}{2},b+\frac{c}{2} \right)_P \,.\ee
      More details are found in \appref{secA_sun_IR}. \eqref{3_def_IsunIR} is sufficient to evaluate any tensor structure appearing in the calculation. With \eqref{3_def_IsunIR}, the infrared asymptotics of the sunset terms in \eqref{3_DSE_2} with the tensor structures replaced by their regularized counterparts reads
	  \begin{subequations}\label{3_IRsun}
	  \begin{align}
	    I_c^{sun} & =\frac{\left(g^2 N_c\right)^2}{(4\pi)^4} \frac{N_d^2}{N_o^2}\,(p^2)^{-\kappa}\,c_a^{2+2\beta^{4\Gamma}}c_c^{1+2\beta^{4\Gamma}}\,\frac{6\kappa}{(2-\kappa)(1-\kappa)^2(1+\kappa)^2} \,,\label{3_IRsun_c} \\
	    I_{A,1}^{sun} & =\frac{\left(g^2 N_c\right)^2}{(4\pi)^4} \frac{N_d}{N_o}\,(p^2)^{\kappa}\,c_a^{1+2\beta^{4\Gamma}}c_c^{2+2\beta^{4\Gamma}}\,\frac{\zeta (-1+\kappa)}{\kappa (1+\kappa)^2(2+\kappa)(3+\kappa)} \,, \label{3_IRsun_A}\\
	    I_{A,2}^{sun} & =\frac{\left(g^2 N_c\right)^2}{(4\pi)^4} \frac{N_d}{N_o}\,(p^2)^{\kappa}\,c_a^{1+2\beta^{4\Gamma}}c_c^{2+2\beta^{4\Gamma}}\, \frac{8}{9}\,\frac{9\kappa^3+10\kappa^2+9\kappa-4}{(1-\kappa)\kappa^2(1+\kappa)^2(2+\kappa)(3+\kappa)} \,,
	  \end{align}
	  \end{subequations}
    where $I_{A,1}^{sun}$ corresponds to the regularized tensor structure $\widetilde\T_{A,1}^{sun}$ and  $I_{A,2}^{sun}$ to $\widetilde\T_{A,2}^{sun}$ in \eqref{3_tensreg}.

    \subsubsection{Self-Consistency in the IR}
    When comparing the scaling of the sunset diagrams and the one-loop terms in the deep-infrared with the momentum $p^2$ one finds, as expected, that the sunset diagrams are dominant in the infrared. When plugging the infrared asymptotics of the sunset diagrams \eqref{3_IRsun} into the Dyson-Schwinger equations \eqref{3_DSE_2} and taking the IR-leading terms into account, one yields relations as
	  \be 1 = \frac{\left(g^2 N_c\right)^2}{(4\pi)^4} \frac{N_d}{N_o}\, c_a^{2+2\beta^{4\Gamma}}c_c^{2+2\beta^{4\Gamma}}\,f_{i} \label{3_IRrel}\ee
    with the functions $f_{i}$  are the ratios of polynomials and $\Gamma$-functions in \eqref{3_IRsun}. The function in the ghost equation $f_{c}$ picks up an additional color factor. If one is able to fulfill respectively one relation \eqref{3_IRrel} from the ghost and gluon equations for one specific value of $\kappa$ a self-consistent solution in the infrared is possible.

    \begin{figure}[t]
    \centering
    \includegraphics[width=.7\textwidth]{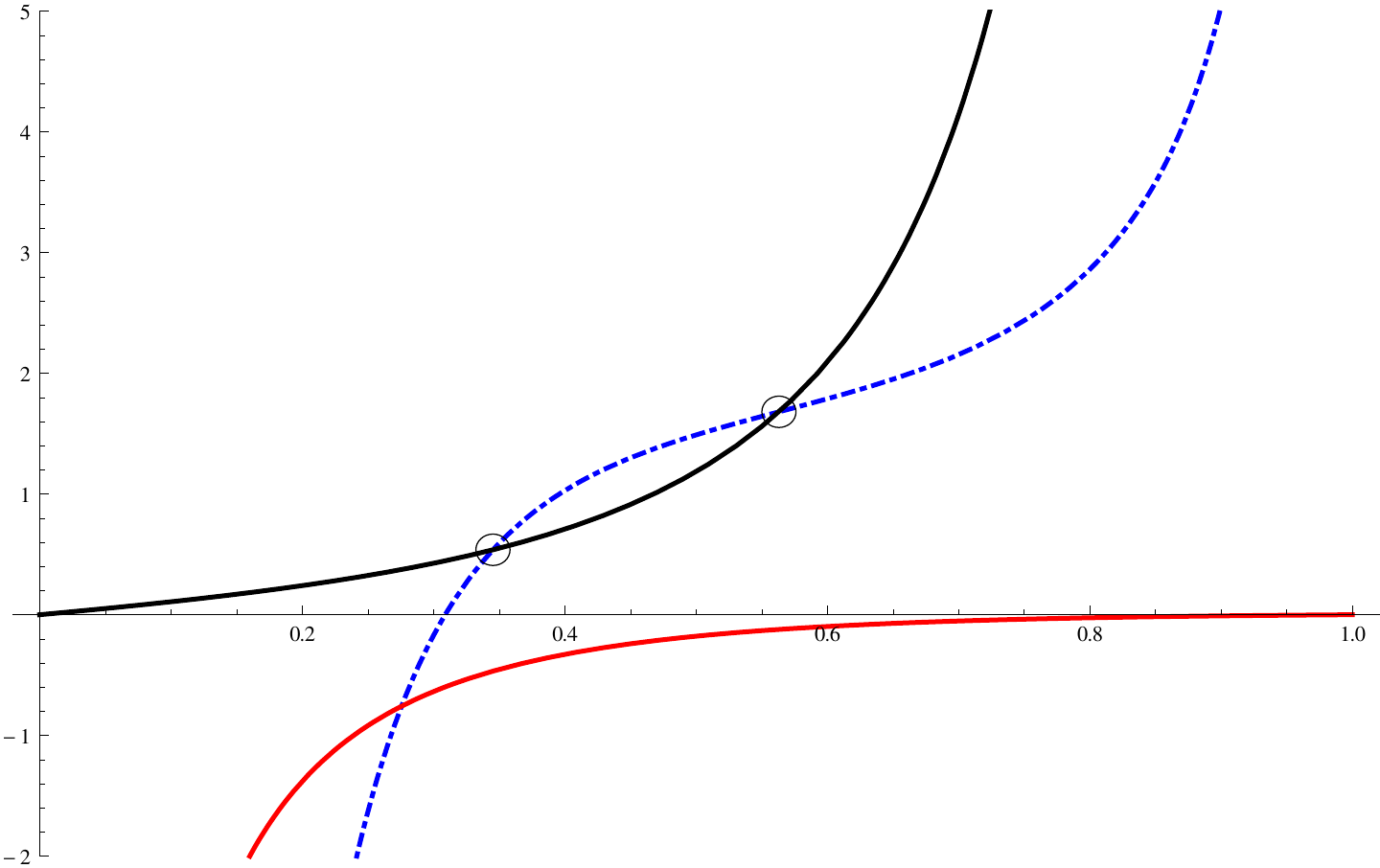}
    \caption{The functions $f_i$ as described in the text. The color coding is: $f_{c}\rightarrow$black-solid, $f_{A,1}\rightarrow $ red-solid, $f_{A,2}\rightarrow$ blue--dash-dotted.   \label{fig:3_IRsc}}
    \end{figure}

    A graphical representation of the functions  $f_i$ is given in \figref{fig:3_IRsc}. One searches for a self-consistent solution in the first quadrant, since the calculated value of $\kappa_{MAG}$ is real and positive and a negative value of the integral would lead to a sign change in the dressing function. One finds a drastic influence of the different tensor structures in the gluon equation. While there is no self consistent solution for the tensor structure $\widetilde T_{A,1}^{sun}$ there are self-consistent solutions possible for the tensor structure $\widetilde T_{A,2}^{sun}$ for the values  
	\be  \kappa \in \{ 0.345328,0.563162\}\,.\ee

\subsection{Results}

\begin{figure}[th]
 \centering
 \includegraphics[width=.6\textwidth]{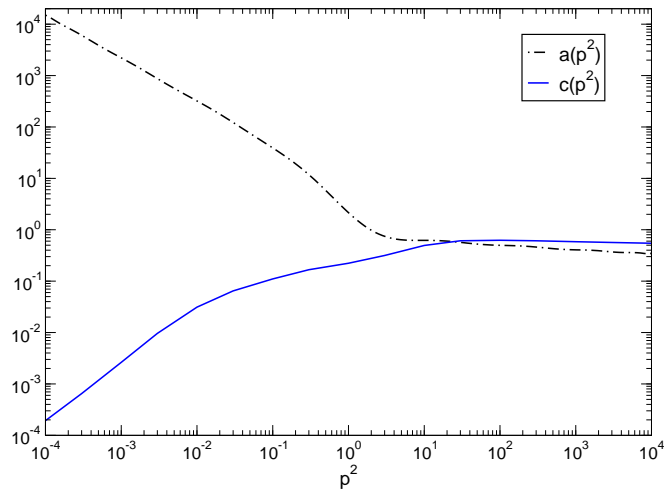}
 \caption{The solution of the ghost equation $c(p^2)$, blue solid, as explained in the text. The gluon dressing function $a(p^2)$, black dash-dotted, is used as fixed input. \label{fig:3_solution}}
\end{figure}

    The ghost equation \eqref{3_ghostDSE3} is solved with the tensor structure replaced by their regularized counter part for $\kappa=0.563162$ with $\lambda_{3\Gamma}=-0.1$ and the renormalized coupling $g^2=4\pi$. The logarithmic divergences are subtracted in a MOM scheme as developed in the last section. The ghost equation is iterated in a fixed point iteration, leaving the gluon dressing function as fixed input. It is modeled according to \eqref{B_olo_ansatz} with $a=2$, $b=1$, $s=1$ and $c_A=7$. 
    
    The double-logarithmic plot in \figref{fig:3_solution} nicely shows the obtained power-law solution in the infrared which is due to the contribution of the sunset diagram. The ultraviolet behavior is driven by the one-loop terms. The mid-momentum regime seems a little bit wiggly, which might be due to problems when numerically resolving the complicated structure of infrared enhancement and suppression in the angular integrals. No renormalization has been applied so no direct interpretation of the momentum scale is possible. The perturbative behavior changes over to the infrared regime at around $p^2\approx 1$ which can approximately be identified with $\Lambda_{QCD}$.

\section{Summary}

    In this section the Maximal Abelian gauge of Yang-Mills theory and its propagator Dyson-Schwinger equations were presented. In the first part of the section the gauge fixing Lagrangian was introduced by a generalization of the equivariant gauge fixing procedure to $SU(N)$. Subsequently several terms where introduced, which have not been treated in the literature so far in the MAG: The vertex renormalization constants were defined, a non-perturbative definition of the Yang-Mills strong-running coupling was given, a Hilbert space of physical states was constructed and the equivariant algebra of symmetries of the partially gauge fixed theory was presented.

    In the second part the Dyson-Schwinger equations of the MAG and a maximal truncation of them was presented. The truncation is the minimal subset of equations containing respectively one infrared and one ultraviolet leading diagram without (the need of) taking into account coset gluons. An analysis of the high- and low-momentum asymptotics of this set of equations was given and the quadratic divergences where subtracted successfully. The employed subtraction method, however, has significant impact on the infrared behavior. As such the analytic value of the MAG-infrared exponent cannot be reproduced.

    Obviously, the numerical solution of the Dyson-Schwinger equations in the MAG presented here is only one step on a long way of a thorough investigation of the MAG Green functions. More technical effort is needed for a self-consistent solution of the coupled set of equations. Still, the presented calculation is the first solution of a Dyson-Schwinger equation with a two-loop diagram as infrared leading term.

%% file: 4_ccquartet.tex
\chapter{The Quartet Mechanism in Generalized Covariant Gauge \label{sec_cc}}

	 In the Landau gauge limit of the linear covariant gauge, Yang-Mills theory is invariant under Faddeev-Popov conjugation (FPC),
	    \be  c^r\FP \bar c^r\,,\qquad \bar c^r\FP - c^r\,,\qquad b^r\FP b^r + \cp{\bar c}{c}{r}\,. \label{4_FPtrafo_LG} \ee 
	These transformations also relate the BRST and anti-BRST transformations as given in \eqref{1_def_BRST} and \eqref{1_def_aBRST} for any value of $\xi$,
	    \be s \FP \bar s \,,\quad\qquad\text{and, } \qquad\quad \bar s \FP s \,. \ee

	In the Kugo-Ojima scenario, as described in \secref{sec_KO}, ghost and antighost fields are treated differently. While, e.g., in the channel $\cp{A_\mu}{\bar c}{}$ a bound state has to be assumed to account for the function $u(p^2)$ in \eqref{1_def_u}, for $\cp{A_\mu}{c}{}$ such a nonperturbative contribution is excluded. The Kugo-Ojima scenario thus breaks the FPC symmetry of Landau gauge. Moreover, the construction of the physical Hilbert-space, the quartet mechanism and the Kugo-Ojima confinement criterion do not make any use of the anti-BRST symmetry, but only use invariance under BRST transformations. The physical relevance of the anti-BRST symmetry is thus unclear \cite{NakanishiOjima:1990}. This situation becomes even more unclear when realizing that the BRST exact part of the conserved color current, \eqref{1_KO_current}, can be rewritten as an anti-BRST exact expression
	    \be  \N^r_\mu =  i s(D_\mu\bar c)^r  = i s \bar s A_\mu^r = -i\bar s ( D_\mu c)^r \,. \label{4_KO_current}\ee

	For the current $\N^r_\mu$ not to couple to massless modes, i.e. the color charge to exist, one introduces a function $u(p^2)$ such that for the asymptotic field of $(D_\mu \bar c)^r$ holds
	    \be \left[(D_\mu\bar c)^r\right]^{as.} =  \left(\one + u\right) \partial_\mu\bar \gamma^r \,\,\pto \,\,0 \,.\label{4_Dcbas_1}\ee
	Generally, a symmetry transformation and the asymptotic limit can be interchanged if this symmetry is unbroken \cite{Kugo:1979gm}. If BRST symmetry is assumed to hold also in the asymptotic regime one has $s \bar \gamma^r = \beta^r$ and thus
	    \be [\N^r_\mu]^{as.} =  (\one + u) \partial_\mu \beta^r\,. \label{4_KO_currAS} \ee
	Either anti-BRST symmetry is broken and $[\bar s (D_\mu c)^r]^{as} = \partial_\mu \beta^r+ f $ with an unknown function $f$, or anti-BRST symmetry is preserved also in the asymptotic regime. If it is unbroken, comparing \eqref{4_KO_current} and \eqref{4_KO_currAS} yields that anti-BRST tranformations pick up a factor of $(\one + u)$ as , e.g.,
	    \be  s (\gamma^r) = (\one + u)\beta^r\,.  \ee
	This factor vanishes for in the deep infrared if the confinement criterion \eqref{4_Dcbas_1} holds.

	Several questions arise in the context of the Kugo-Ojima scenario with respect to FPC and anti-BRST symmetry: Is it possible to keep FPC in Landau gauge? What is the role of anti-BRST symmetry? Are there any contradictions if anti-BRST symmetry is preserved/broken?

	In this section the generalized covariant gauge is introduced, which extends the linear covariant gauge by a new gauge parameter which continuously interpolates between the ghost and antighost fields. It is then investigated with respect to the role of FPC symmetry and anti-BRST symmetry.

	The question of consistency of the Kugo-Ojima mechanism with BRST and anti-BRST symmetry has already been investigated in \cite{Shintani:1983ar,Shintani:1984cv} with negative result. However, in this study the limit $p^2\rightarrow 0$ has been performed before constructing correlation functions. In this work here, first the corresponding correlation functions are constructed and then their asymptotic limit is investigated. This is considered to be more transparent with respect to the formulation of a quantum field theory in the path-integral formalism. Intermediate results for currents and equations of motion are given in \appref{A_curr}.

	\section{Lagrangian, Symmetries and Charges\label{sec_4_lag}}

	The generalized covariant gauge is defined by the gauge-fixing Lagrangian, \cite{Baulieu:1981sb,ThierryMieg:1985yv,Alkofer:2003jr},
	      \be \Lgf^{c\bar c} = \frac{\xi}{2} b^2 - i b^r\, \partial_\mu A_\mu^r - i \alpha (D_\mu \bar c)^r \partial_\mu c^r - i \bar \alpha \, \partial_\mu \bar c ^r (D_\mu c)^r + \frac{\alpha \bar \alpha \xi}{2} \,\cp{\bar c}{c}{2} \label{4_def_Lgf}\,. \ee
	The two gauge parameters $\alpha$ and $\bar \alpha$ are restricted to the range $0\leq \alpha,\bar \alpha \leq 1$. They are not independent but reduce to effectively one gauge parameter by the condition $\alpha + \bar \alpha = 1$. The parameter $\alpha$  interpolates between the linear covariant gauge, $\alpha=0$, its conjugate, $\alpha=1$, and a Curci-Ferrari gauge, $\alpha = \frac{1}{2}$.  In the Landau-gauge limit, $\xi\rightarrow 0$ for any $\alpha$, and in the Curci-Ferrari gauge, $\alpha  = \frac{1}{2}$ for any $\xi$, the Lagrangian $ \Lgf^{c\bar c}$ is invariant under the $\alpha-$depend FPC,
	    \be  c^r\FP \bar c^r\,,\qquad \bar c^r\FP - c^r\,,\qquad b^r \FP b^r + \left( \bar \alpha-\alpha\right)\cp{\bar c}{c}{r}\, \label{4_FPtrafo} \,.\ee 
	The BRST transformations gain a dependence on the new gauge-parameter
	    \begin{align}
		\sa A_\mu^r &= (D_\mu c)^r & \sa c^r &= -\frac{1}{2} \cp{c}{c}{r}  \label{4_def_BRST}\\
		\sa \bar c^r &= b^r-\alpha \cp{\bar c}{c}{r} & \sa b^r & = -\alpha\cp{c}{b}{r} + \frac{\alpha\bar\alpha}{2}\cp{\bar c}{\cp{c}{c}{}}{r}  \nonumber
	    \end{align}
	as do the anti-BRST transformations
	    \begin{align}
		\sab A_\mu^r &= (D_\mu \bar c)^r & \sab \bar  c^r &= -\frac{1}{2} \cp{\bar c}{\bar c}{r}  \label{4_def_aBRST}\\
		\sab c^r &= - b^r-\bar \alpha \cp{\bar c}{c}{r} & \sab b^r & = -\bar \alpha\cp{\bar c}{b}{r} + \frac{\alpha\bar\alpha}{2}\cp{\cp{\bar c}{\bar c}{}}{c}{r} \,. \nonumber
	    \end{align}
      The BRST and anti-BRST transformations are nilpotent. They anticommute and are related by FPC for any $\alpha$ and $\xi$
	    \be \sa^2 = \sab^2 = \acom{\sa}{\sab} = 0\,,\qquad \sa \FP \sab\,, \qquad \sab \FP -\sa \label{4_BRSTFP} \,.\ee
      Invariance under BRST and anti-BRST transformations can most easily be seen when expressing the gauge-fixing Lagrangian as BRST and anti-BRST variation, 
	    \be \Lgf^{c\bar c} = \sa \left(\bar c^r \left(\frac{\xi}{2}b^r - i \partial_\mu A_\mu^r  \right) \right) = - \sab \left(c^r \left(\frac{\xi}{2}b^r - i \partial_\mu A_\mu^r  \right) \right) \,.\label{4_Lag_1}\ee
      In comparison to \eqref{1_eq_gengauge} it is instructive to write
	    \be  \Lgf^{c\bar c} = \sa\sab \left( \frac{i}{2} \,A_\mu^rA_\mu^r - \xi\,\alpha  \,c^r \bar c^r   \right) + \xi \left(\frac{1}{2}-\alpha \right)  \sa(\bar c^r \sa \bar c^r)\,.\label{4_Lag_2}\ee
      While the first term in the last equation is invariant under FPC, the second term is not. It vanishes and thus FPC symmetry is restored only for $\alpha = \frac{1}{2}$ or $\xi = 0$. 

      The Kugo-Ojima confinement scenario rests on a proper definition of the global color charge corresponding to the global color symmetry \eqref{1_KO_gc}. The generalized covariant gauge is globally color symmetric and the corresponding Noether current is given by
	    \be J^{\alpha\,r}_\mu = \partial_\nu F_{\mu\nu}^r + i \bar \alpha \,\sa (D_\mu \bar c)^r - i \alpha\, \sab  (D_\mu  c)^r \,. \ee
      Not surprisingly, the BRST exact contribution to the conserved color current gains a dependence on the new gauge parameter $\alpha$. For better comparison with other results it is instructive to rewritte the conserved color current as
	    \begin{align} 
		  \N^{\alpha\,r}_\mu & =  \,i\sa\sab A_\mu^r \, =\,  i \bar \alpha \,\sa (D_\mu \bar c)^r - i \alpha\, \sab  (D_\mu  c)^r  \\ 
		  & = \,i(D_\mu b)^r +i \bar \alpha \cp{\bar c}{D_\mu c}{r} - i\alpha \cp{c}{D_\mu \bar c}{r} \,.\nonumber 
	    \end{align}
      The ghost charge is the conserved charge with respect to the invariance of the Lagrangian under rescaling of the gauge-fields, $\delta_\lambda c^r = \lambda c^r$ and $\delta_\lambda \bar c^r = -\lambda\bar c^r,\, \lambda\in \complC$,
	    \begin{multline} 
		Q_c^{\alpha} =i \int\!\!d^3 x \left( -\alpha\, (D_0\bar c)^r c^r -\bar \alpha\, (\partial_0 \bar c^r) c^r + \bar \alpha\, \bar c^r (D_0 c)^r + \alpha \,\bar c^r \partial_0c^r\right) \\  
			= i \int\!\!d^3 x \left(\bar c^r \partial_0 c^r - (D_0\bar c)^r c^r \right) = i \int\!\!d^3 x \left( \bar c^r (D_0 c)^r - (\partial_0 \bar c^r)  c^r\right) \,. 
	    \end{multline}
      The ghost charge operator is independent on the gauge-parameter $\alpha$. Denoting the action in generalized covariant gauge by $\S^{c\bar c} = \int\,\! d^4x \left( \LYM + \Lgf^{c\bar c}\right)$ the Dyson-Schwinger equations for the ghost and antighost read
	  \begin{subequations}\label{4_DSEs}
	      \begin{align}
		  -\delta^{rs}\delta(x-y) & = \vev{\bar c^s(y)\,\var[\S^{c\bar c}]{\bar c^r(x)}} =  \vev{\bar c^s(y)\,\,\sa\! \left(i\partial_\mu A_\mu^r - \xi b^r \right)\!(x)} \label{4_ghostDSE}\\
			& = \vev{\bar c^s(y)\,\left( i \partial_\mu D_\mu c - i  \alpha \cp{\partial_\mu A_\mu}{c}{} - \frac{\alpha \bar\alpha\xi}{2} \cp{\bar c}{\cp{c}{c}{}}{} \right)^r\!\!(x)} \nonumber \\
		   \delta^{rs}\delta(x-y) & = \vev{\var[\S^{c\bar c}]{c^r(x)}\,c^s(y)} =  \vev{-\sab\! \left(i\partial_\mu A_\mu^r - \xi b^r \right)\!(x)\,\,c^s(y)}\label{4_aghostDSE} \\
			& = \vev{\left(- i \partial_\mu D_\mu \bar c + i \bar \alpha \cp{\partial_\mu A_\mu}{\bar c}{} + \frac{\alpha \bar\alpha\xi}{2} \cp{\cp{\bar c}{\bar c}{}}{c}{} \right)^r\!\!(x)\,\,c^s(y)} \nonumber 
	      \end{align}
	  \end{subequations}
   For later reference, the Dyson-Schwinger equation of the NL field is rewritten as,
	  \be 0 =  \Fvev{\var[\sa\bar c^r]{b^s}} - \Fvev{\sa(\bar c^r)\,\var[\S^{c\bar c}]{b^s}} = \Fvev{\var[\sab c^r]{b^s}}  -\Fvev{\sab( c^r)\,\var[\S^{c\bar c}]{b^s}} \label{4_WI}\,,\ee
    with
	  \be  \Fvev{\var[\sa\bar c^r]{b^s}}  =  - \Fvev{\var[\sab c^r]{b^s}} = \Fvev{\var[b^r]{b^s}}  = \delta^{rs}  \label{4_WI_2}\,.\ee

    \section{FPC-Invariant Asymptotic States}
	In this subsection the FPC invariant gauges, Landau gauge and Curci-Ferrari gauge, are investigated with respect to FPC invariant asymptotic states. An assignment of the asymptotic states which respects FPC symmetry is given by
	\begin{subequations}\label{4_AS_FPs}
	    \begin{align}
	      D_\mu^{rs}c^s &\as (\one + v)\,\partial_\mu\gamma^r, & c^r &\as(\one + v)^{-1} \,\gamma^r \\
	      D_\mu^{rs}\bar c^s &\as (\one + v)\,\partial_\mu\bar \gamma^r, & \bar c^r &\as(\one + v)^{-1} \,\bar \gamma^r \\
	      b^r &\as \beta^r & A_\mu^r &\as \partial_\mu\chi^r
	    \end{align}
	\end{subequations}
	To proceed it is now assumed that BRST and anti-BRST symmetry are unbroken, which then allows to change the asymptotic limit and the (anti-)BRST transformations. It is then investigated whether this treatment leads to contradictions. The BRST and anti-BRST transformations of the asymptotic fields read
	      \begin{align}
		 \sa \chi^r &=  (\one + v) \,\gamma^r\,, & \sa \gamma^r & = 0\,, & \sa \bar \gamma^r &=  (\one + v)\, \beta^r\,, & \sa \beta^r &= 0 \,,\label{4_sAS_FPs}\\
		 \sab \chi^r &=  (\one + v) \,\bar \gamma^r\,, & \sab \bar \gamma^r & = 0\,, & \sab \gamma^r &= - (\one + v)\, \beta^r\,, & \sab \beta^r &= 0 \,.\label{4_sbAS_FPs}
	      \end{align}
	First, the confinement criterion \eqref{4_Dcbas_1} under the assignment \eqref{4_AS_FPs} is checked. The asymptotic contribution to the current $\mathcal N_\mu^{\alpha\,r}$, for any $\{\xi,\alpha\}$ is given by
	      \be \N^{\alpha\,r}_\mu =  i\sa\sab A_\mu^r \as i (\one + v)^2\,\partial_\mu \beta^r\,.\ee
	The corresponding charge is well-defined if $v(p^2)\pto -1$. A comparison with the original Kugo-Ojima construction yields $(\one+v)^2 = (\one+u)$. It is thus possible to transport the Kugo-Ojima confinement criterion to the FPC-symmetric case as well. It is now checked if the asymptotic stated \eqref{4_sAS_FPs} and \eqref{4_sbAS_FPs} indeed form quartets.

    \subsubsection{Landau Gauge}
	The Dyson-Schwinger equations provide relations between the Green functions of a quantum field theory which have to be fulfilled on the microscopic level as in the asymptotic regime. Using the assignment \eqref{4_AS_FPs}, the asymptotic regime of the antighost Dyson-Schwinger equation \eqref{4_aghostDSE} in Landau gauge is given by 
	      \be  -\delta^{rs} = \Fvev{\bar c^r(y)\,\left( i \partial_\mu D_\mu c \right)^s\!(x) } \as \Fvev{\bar \gamma^r(y)\,\, i \partial^2 \gamma^s(x) } \,. \ee 
      From this equation one extracts the massless correlation function
	     \be \Fvev{\bar \gamma^r\, \gamma^s }  =  \frac{- i\,\delta^{rs}}{p^2}\,. \label{4_LG_q1}\ee 
      An analogous relation can be obtained from \eqref{4_ghostDSE}. The asymptotic regime of \eqref{4_WI} reads in Landau gauge
	      \be \delta^{rs} = -\Fvev{\sa(\bar c^r)(y)\,\,i\partial_\mu A_\mu^s(x)} \pto -\Fvev{\beta^r(y)\,\,i\partial^2\chi^s(x)}\,, \ee
      which yields the massless correlation function of the asymptotic fields
	      \be \Fvev{\beta^r\,\chi^s}= -\frac{ i \delta^{rs}}{p^2}\,. \label{4_LG_q2}\ee
      One thus gets back the identity
	      \be 0 =  \Fvev{\beta^r\,\chi^s} - \Fvev{\bar \gamma^r\, \gamma^s } = \frac{1}{\one+v} \sa \Fvev{\bar \gamma^r\,\chi^s} \,.\label{4_LG_quartet}\ee
      The same argumentation can be applied for anti-BRST to obtain the corresponding equation
	      \be  0 = -\frac{1}{1+v} \sab \Fvev{ \chi^r\,\gamma^s}  \label{4_LG_aquartet}\ee

      \eqref{4_LG_quartet} and \eqref{4_LG_aquartet} state that BRST and anti-BRST symmetry are unbroken in this particular channel. Both symmetries were assumed to get to \eqref{4_sAS_FPs} and \eqref{4_sbAS_FPs}. It has thus been shown that a FPC symmetric assignment of the asymptotic states is possible without contradictions. The seemingly mysterious factors of $(\one+v)$ cancel each other. \eqref{4_LG_quartet} and \eqref{4_LG_aquartet} hold for any value of $p^2$.

	\subsubsection{Curci-Ferrari gauge}
	
	If $\xi\neq 0$, the Lagrangian $\Lgf^{c\bar c}$ is invariant under FPC only for $\alpha = \bar \alpha = \frac 1 2$, which is assumed for this subsection. Using the assignment \eqref{4_AS_FPs}, the Dyson-Schwinger equations for ghost and antighost, \eqref{4_DSEs}, in the asymptotic regime read,
	  \begin{subequations}\label{4_DSEs_CF_as}
	      \begin{align}
		  \delta^{rs} & = i p^2 \Fvev{\bar \gamma^s\,\gamma^r} + \frac{1}{1+v} \Fvev{\bar \gamma^s\, \left[ \frac{i}{2} \cp{\partial_\mu A_\mu}{c}{r} + \frac{\xi}{8} \cp{\bar c}{\cp{c}{c}{}}{r} \right]^{as}}\,, \label{4_ghostDSE_CF_as}\\
		   \delta^{rs} & =  i p^2 \Fvev{\bar \gamma^s\,\gamma^r} + \frac{1}{1+v} \Fvev{ \left[ \frac{i}{2} \cp{\partial_\mu A_\mu}{\bar c}{s} + \frac{\xi}{8} \cp{\cp{\bar c}{\bar c}{}}{c}{s} \right]^{as}\,\,\gamma^r}\,.
	      \end{align}
	  \end{subequations}	    
      The contributions of the respectively last terms in \eqref{4_DSEs_CF_as} are equal due to FPC-symmetry. Their contribution can be described by the function $w(p^2)$ defined as 
	  \begin{multline} \delta^{rs} w(p^2) =  \frac{1}{1+v} \Fvev{\bar \gamma^s\, \left[ \frac{i}{2} \cp{\partial_\mu A_\mu}{c}{r} + \frac{\xi}{8} \cp{\bar c}{\cp{c}{c}{}}{r} \right]^{as}} \\ = \frac{1}{1+v} \Fvev{ \left[ \frac{i}{2} \cp{\partial_\mu A_\mu}{\bar c}{s} + \frac{\xi}{8} \cp{\cp{\bar c}{\bar c}{}}{c}{s} \right]^{as}\,\,\gamma^r}\,,\end{multline}
      which yields
	  \be \Fvev{\bar \gamma^s \,\gamma^r} = \frac{-i\delta^{rs}}{p^2}\,\left(1-w(p^2)\right)\,. \ee
      In the asymptotic limit, contributions to $w(p^2)$ stem from ghost-antighost--bound-states or bound-states of the longitudinal gluon with ghost or antighost fields. Bound states of ghost and antighost, or the corresponding condensate, play a crucial role in the non-perturbative mass generation mechanisms as e.g. in \cite{Schaden:1999ew}. In the current investigation it cannot be decided if these states exist. However, if they do not exist, then $w(p^2) \pto 0$ and thus one gains a massless correlation function $\vev{\bar \gamma^s\,\gamma^r}$. Using the BRST symmetry transformations \eqref{4_sAS_FPs} one finds another massless correlation function $\vev{\beta^s \chi^r}$ and thus a massless elementary quartet $\{\beta^s, \chi^s,\bar \gamma^s,\gamma^r\}$ which fulfill the relations
	  \be 0 = \frac{1}{(\one+v)}\sa\Fvev{\bar\gamma^r\,\chi^s} = - \frac{1}{(\one+v)}\sab \Fvev{\gamma^r\,\chi^s} \,.\ee
      If, however, such bound states exist one has $w(p^2\rightarrow 0)\neq 0$. Even more, if these bound states saturate the Dyson-Schwinger equations \eqref{4_DSEs_CF_as}, i.e. $w(p^2\rightarrow 0)= 1$, one has that 
	  \be p^2\vev{\bar \gamma^s\,\gamma^r} \pto 0 \,.\ee
      The correlation function $\vev{\bar \gamma^s\,\gamma^r} $ diverges less than a massless pole, or even is finite or vanishes. This means that in this case there is no simple massless elementary quartet as expected in the original Kugo-Ojima scenario. The consequences are not clear from the point of this work. It has to be noted, however, that the mass of the correlation functions of the quartet members $\{\chi^r,\beta^r,\gamma^r,\bar \gamma^r\}$ is not essential as long as the conditions \eqref{1_KO_MP} and \eqref{1_KO_quart} are fulfilled.

      Also in the Curci-Ferrari gauge it is possible to construct the asymptotic states of the microscopic field in a FPC-invariant manner. The assumption of BRST and anti-BRST symmetry does not yield any contradiction.

      \section{Rotation in the Ghost Fields\label{sec_4_etagh}}
	  The gauge  parameter $\alpha$ interpolates between the ghost and antighost fields. In this subsection it is investigated how far one can get back to a linear-covariant--like gauge by rewriting the general covariant gauge.

	  New ghost and antighost fields, $\eta$ and $\bar \eta$, as linear combinations of the original fields of the general linear covariant gauge are defined as
	      \be \eta^r = \sqrt{\bar \alpha}\, c + \sqrt{\alpha}\,\bar c\,,\qquad\text{and,}\qquad \bar \eta^r = -\sqrt{ \alpha}\, c + \sqrt{\bar \alpha}\,\bar c\,.\ee
	  Since the Jacobian of this transformation is $1$, it leaves the measure of the path integral invariant. The transformation of the gluon field yields the new BRST and anti-BRST transformations,
	      \be \set = \sqrt{\bar \alpha}\, \sa + \sqrt{\alpha}\,\sab\,,\qquad\text{and,}\qquad \setb = -\sqrt{ \alpha}\, \sa + \sqrt{\bar \alpha}\,\sab\,.\ee
	  Nilpotency and anticommutivity are transported from the original transformations,
	      \be \set^2 = \acom{\set}{\setb} = \setb^2 = 0\,. \ee
	  Written in the new ghosts $\eta^r$ and $\bar \eta^r$ and introducing also a new NL field, 
	      \be  b^r_\eta = b^r + \frac{\sqrt{\alpha\bar\alpha}}{2}\left(\cp{c}{c}{r}-\cp{\bar c}{\bar c}{r} \right)\,,\ee
	the BRST and anti-BRST transformations \eqref{4_def_BRST} and \eqref{4_def_aBRST} read
	      \begin{align}
	       \set A_\mu^r &= (D_\mu\eta)^r\,, & \set \eta^r & = -\frac{1}{2} \cp{\eta}{\eta}{r}\,, \\
		\set \bar \eta^r &= b_\eta^r\,, & \set b_\eta^r & = 0 \nonumber\,, \\
	       \setb A_\mu^r &= (D_\mu\bar \eta)^r\,, & \setb \bar \eta^r & = -\frac{1}{2} \cp{\bar\eta}{\bar\eta}{r}\,, \\
		\setb  \eta^r &= - b_\eta^r-\cp{\bar \eta}{\eta}{r}\,, & \set b_\eta^r & = -\cp{\bar \eta}{b_\eta}{r} \nonumber\,.
	      \end{align}
	  By construction, these expressions recover the BRST and anti-BRST transformations of the linear covariant gauge on a formal level. The gauge fixing Lagrangian $\Lgf^{c\bar c}$ in these new fields reads,
	      \begin{multline} 
		  \Lgf^{c\bar c} = \Lgf^{\eta} = \frac{\xi}{2} b^2_\eta  - i(\partial_\mu\bar\eta^r)(D_\mu\eta)^r +\xi\alpha\bar\alpha\,\cp{\eta}{\bar \eta}{2}  \\ - b^r_\eta\left(i \partial_\mu A_\mu^r + \xi \frac{\sqrt{\alpha \bar\alpha}}{2} \left((\bar\alpha-\alpha)\left(\cp{\eta}{\eta}{}-\cp{\bar\eta}{\bar\eta}{} \right)^r - 4\sqrt{\alpha\bar\alpha}\cp{\eta}{\bar\eta}{r} \right) \right)\,.
	      \end{multline}
	  In the Landau gauge limit one formally comes back to the Landau gauge limit of the linear covariant gauge,
		\be \Lgf^{\eta} =  -i b^r_\eta \partial_\mu A_\mu^r - i(\partial_\mu\bar\eta^r)(D_\mu\eta)^r.\ee
	  In the symmetric limit, $\alpha = \bar \alpha = \frac{1}{2}$, one has
	      \be \Lgf^{\eta} = \frac{\xi}{2}(b'_\eta)^2 - i {b_\eta'}^r \,\partial_\mu A_\mu^r - \frac{i}{2}\, (\partial_\mu\bar\eta)^r( D_\mu\eta)^r - \frac{i}{2}\,(D_\mu\bar\eta)^r\partial_\mu \eta^r+ \frac{\xi}{8}\cp{\eta}{\bar \eta}{2}\,, \label{4_eta_L3}\ee
	  where the NL field was shifted again by $b_\eta^r = {b_\eta'}^r - \frac{1}{2} \cp{\eta}{\bar \eta}{r}$. \eqref{4_eta_L3} is essentially the same Lagrangian as one gets for the ghost and antighost fields $c^r,\bar c^r$ for $\alpha=\bar\alpha = \frac{1}{2}$. 

	  One thus finds that, for $\xi = 0$, the dependence of the parameter $\alpha$ can be removed from the general covariant gauge. In this case the general covariant gauge \emph{is equivalent} to the linear covariant gauge. For $\alpha = \frac 1 2$ there is another invariance. The theory does not change if it is written in the fields $c^r,\bar c^r$ or $\eta^r,\bar \eta^r$. Thus there is a symmetry under a rotation of $\frac \pi 4$ in field space.

	  Since $\sa\sab = \set\setb$, the BRST exact part of the global color current reads
		  \be \N ^r_\mu = \set\setb A_\mu^r \,,\ee
	  which has to become massive for vanishing momentum. As shown above, one can either choose a FPC invariant assignment of the asymptotic states as \eqref{4_AS_FPs} o an asymmetric one as \eqref{1_KO_elq}. The physics is independent on the specific assignment.

	  \section{Generalized Quartets}
	  The two defining properties of BRST quartets $\{\chi,\beta,\gamma,\bar\gamma\}$ are their metric relations,  \eqref{1_KO_MP},
	      \be \vev{\bar\gamma\,\gamma} = \vev{\beta\,\chi} = \one\,, \label{4_quart_MR}\ee
	  and the BRST transformation properties, \eqref{1_KO_quart},
	      \be \com{Q_B}{\chi} = \gamma \,,\qquad\text{and,}\qquad \acom{Q_B}{\bar\gamma} = \beta\,. \label{4_quart_BRST}\ee
	  While it is relatively easy to find objects which obey the definitions \eqref{4_quart_BRST}, the problem is to find metric relations as \eqref{4_quart_MR}. Usually therefore general identities for Green functions such as Dyson-Schwinger equations or Ward identities  are employed. The elementary quartet, considered up to now in this thesis is related via the ghost Dyson-Schwinger equation and the Slavnov-Taylor identity for the longitudinal gluon. In the special case of Landau gauge, however, one can also use the Dyson-Schwinger equation of the NL field \eqref{4_WI}. 

	  The particular form of the ghost and antighost Dyson-Schwinger equations \eqref{4_DSEs} and the BRST transformations allow for a more general quartet, which obeys the defining equations \eqref{4_quart_MR} and \eqref{4_quart_BRST} for all values of the gauge parameters $\alpha$ and $\xi$. Denoting the equation of motion of the NL field by 
	      \be \f^r = \var[\S^{c\bar c}]{b^r} = \xi b^r - i \partial_\mu A_\mu^r \,,\ee
	  the Dyson-Schwinger equations for the ghost and antighost, \eqref{4_DSEs}, read
	      \be \delta^{rs}\delta(x-y) =   \vev{\bar c^s(y)\,\,\sa\left(\f^r\!(x)\right)} = \vev{\sab\left(\f^r(x)\right)\,\,c^s(y)} \,.\ee
	  The identities \eqref{4_WI} can be rewritten as
	      \be \delta^{rs}\delta(x-y) =  \vev{\sa(\left(\bar c^s(y)\right)\,\,\f^r(x)} = -\vev{\sab\left(c^r(y)\right)\,\f^s(x)}\,.\ee
	  Combining the last two equations yields the relations
	      \begin{subequations} \label{4_STI}
	      \be 0 = \vev{\sa(\bar c^s)\,\f^r} -  \vev{\bar c^s\,\sa(\f^r)} = \sa \vev{\bar c^s\,\f^r} \ee
	  and
	      \be 0 = \vev{\sab(\f^r)\,c^s} + \vev{\f^r\,\sab(c^s)} = \sab \vev{\f^r\,c^s}\,.\ee
	      \end{subequations}
	  It is important to note that neither BRST nor anti-BRST symmetry are assumed to obtain Eqs.~\ref{4_STI}. The only input into these equations are the Dyson-Schwinger equations, \eqref{4_DSEs} and \eqref{4_WI}. Assigning the asymptotic fields according to
	      \be \f^r\as \chi^r\,,\quad \bar c^r\as\bar\gamma^r\,,\quad\sa\f^r\as\gamma^r\,,\quad\sa\bar c^r\as\beta^r \ee
	  and given the explicit representation of the unity operator $\one = \delta^{rs}\delta(x-y)$, all requirements for a BRST quartet, \eqref{4_quart_MR} and \eqref{4_quart_BRST}, are fulfilled. These quartets are defined only by the equations of motion and are independent on a spontaneous breaking of BRST or anti-BRST symmetry. This general quartet mechanism holds for any gauge which obeys the relations
	      \be \var[s(\bar c^r(x))]{b^s(y)} = \delta^{rs}\delta(x-y)\,,\qquad\text{and,}\qquad  \var[\S]{\bar c^s(y)} = s\left(\var[\S]{b^s(y)}\right) \,.\label{4_genQ_con}\ee
	  For Landau gauge one recovers the original BRST quartet up to factors of $p^2$.

	  \section{Summary}
	  In this section the Kugo-Ojima scenario in the generalized covariant gauge has been investigated. Special attention has been drawn on the FPC symmetric gauges, the Landau and Curci-Ferrari gauges which both are a one parameter family of gauges. It could be shown that a FPC symmetric assignment of the asymptotic states is possible. No contradictions were found if BRST and anti-BRST were assumed. The quartet structure of linear covariant gauge also exists in the FPC invariant gauges.
	  
	  In \secref{sec_4_etagh} new ghost fields have been introduced which absorb the dependence on the gauge parameter $\alpha$. In the case of Landau gauge it could be shown that all Landau gauges are equivalent to the Landau gauge of linear covariant gauge, i.e., $\xi=\alpha=0$. In the Curci-Ferrari gauge, FPC is enlarged by an additional rotation-symmetry in field-space around $\frac \pi 4$. The BRST exact contribution to the conserved color current can always be made massive, either with an FPC symmetric or asymmetric assignment of the asymptotic fields. 

	  In the last subsection four operators could be identified, which build up a quartet without the assumption of BRST or anti-BRST symmetry. The only input into this considerations are the quantum equations of motion. This quartet generally exist in any gauge which fulfills conditions \eqref{4_genQ_con}. In the Landau gauge this generalized quartet recovers the original elementary quartet up to factors of $p^2$.

%% file: 5_saturation.tex
\chapter{Generalization of the Kugo-Ojima Confinement Criterion\label{sec_saturation}}

    The Kugo-Ojima confinement scenario is defined in the linear covariant gauge. As physical results do not depend on the gauge, if the Kugo-Ojima scenario correctly describes confinement, then some characteristics of it should be found in other gauges as well. A literal translation of the Kugo-Ojima scenario to Abelian gauges, however, is not possible \cite{Suzuki:1983cg,Hata:1992np}. 
    
    In \cite{Suzuki:1983cg} it was found that in Abelian gauges the color current is not well defined in the Abelian directions, which was then interpreted as a breakdown of the Kugo-Ojima scenario. A renormalization group analysis of interpolating gauges including Abelian ones, \cite{Hata:1992np}, showed that the Abelian gauges form an infrared unstable invariant subspace. They are considerably distinct to usual covariant gauges. In addition, in any Abelian theory, the global (color-)charge is not well-defined. It is therefore plausible, that the generalization of the Kugo-Ojima scenario to Abelian gauges employed in \cite{Suzuki:1983cg} is too stringent. In the following investigation a corollary of the Kugo-Ojima confinement criterion is found which, indeed, can be generalized to Abelian and a variety of other gauges.
    
    This corollary of the Kugo-Ojima confinement criterion is found in terms of the quantum equation of motion of the gluon,
	\be \delta_{\sigma\mu}\delta^{rs}\delta(x-y) = \vev{A_\sigma^s(y)\,\,\var[(\SYM+\S_{gf})]{A_\mu^r(x)}} \,.\label{5_DSE_general}\ee
    This equation describes transport of adjoint color charges between the two spacetime-points $x$ and $y$. Color confinement means that there should be no such transport measurable for large distances, or equivalently, low momenta.
    
    Being formally the same in any gauge, the question arises whether it is possible to find gauge-invariant signatures of confinement in the low-energy regime of \eqref{5_DSE_general}. The only difference between different gauges is the specific gauge-fixing action $\S_{gf}$. It is found that, based on the Kugo-Ojima scenario, indeed it is possible to transport a generalized confinement criterion to other gauges and models. After a first test of this criterion in the generalized covariant gauge, the dependence on color symmetry is investigated by checking the generalized criterion in the Maximal Abelian gauge. The dependence on Lorentz covariance is investigated using Coulomb gauge. The horizon condition, which implements non-perturbative gauge-fixing but breaks BRST symmetry, is introduced in the Gribov-Zwanziger theory. 

    The gauge-boson Dyson-Schwinger equation is not only formally identical between different gauges of Yang-Mills theory but also between Abelian and non-Abelian gauge theory. It is thus interesting to check the gluon Dyson-Schwinger equation against the photon Dyson-Schwinger equation of QED in linear covariant gauge and also the Abelian Higgs model. 
    The comparison between the gauge-boson Dyson-Schwinger equation of Abelian and non-Abelian theories also allows for a more detailed understanding of the significant difference of Abelian and usual covariant gauges. This is particularly true for $SU(2)-$Yang-Mills theory in the Maximal Abelian gauge with equivariant gauge fixing. This is an Abelian theory in linear covariant gauge which is confining. The only difference between this theory, ordinary QED and the Abelian-Higgs model is the matter content of the theory.

    It is long known that in gauge-Higgs models the confining and Higgs phases are analytically connected \cite{Fradkin:1978dv,Lang:1981qg} and there is no gauge invariant order parameter distinguishing the two phases \cite{Frohlich:1981yi}. In principle there is only one Higgs/confinement phase. Still, the gauge-dependent propagators vary in the two regimes \cite{Capri:2012ah,Capri:2013oja,Maas:2013aia}. In this study here the theories can only be either in the Higgs or the confining phase, since we do not study a model which possesses both phases. The non-Abelian Higgs theory, which is a natural candidate to investigate the similarities/differences of the confining and Higgs-phases, is left for future work.

    \section{General Considerations}
  
    In the following investigation universal aspects of the gauge-boson Dyson-Schwinger equation are investigated. The first finding is already that, in all models investigated, the gauge-boson Dyson-Schwinger equation can be written as
	\be \delta^{rs}\delta_{\sigma\mu} = \Fvev{A_{\sigma}^s(y)\,\left(-\partial_\nu F_{\nu\mu}^r - j_\mu^{r} + i s\xi^r_\mu \right)(x) } \label{5_genDyson-Schwinger equation} \ee
    with the Noether current of the conserved global gauge symmetry $j_\mu^r$ and some function $\xi_\mu^r$. In most of the cases investigated, the BRST exact term can be written as $s\xi^r = s\bar s A_\mu^r$, as the BRST exact contribution in \eqref{1_KO_current}. Only in the Gribov-Zwanziger theory additional terms arise due to the auxiliary ghosts implementing the horizon function. On the lhs of \eqref{5_genDyson-Schwinger equation} there is essentially a \one-operator in color and Lorentz-space, which is independent on the momentum. The question asked in this investigation is: \emph{How is the Dyson-Schwinger equation of the gauge boson saturated for vanishing momentum?} In other words: which terms dominate the rhs of \eqref{5_genDyson-Schwinger equation} in the low energy regime such that the equation is fulfilled. The first contribution on the rhs is transverse due to the asymmetry of the field-strength tensor and thus features one Lorentz scalar function
	\be -\Fvev{A_{\sigma}^s(y)\,\partial_\nu F_{\nu\mu}^r(x)} = \delta^{rs} T_{\sigma\mu}\,\, f(p^2)\,. \label{5_deff}\ee
    Where the transverse projector $T_{\mu\nu}$ and the longitudinal projector $L_{\mu\nu}$ are defined via
	\be T_{\mu\nu} = \delta_{\mu\nu} - L_{\mu\nu}\,,\qquad\text{and,}\qquad L_{\mu\nu} = \frac{p_\mu p_\nu}{p^2}\,.\ee
    The function $f$ yields the criterion distinguishes between the Coulomb and the Higgs~/~confinement phase of the model. The infrared limit $f(0)\neq 0$ implies a pole at $p^2=0$ due to a massless transverse vector boson in the correlator
	\be\label{5_masslessvb}
	    \Fvev{ A^s_\sigma(y)\,  F^r_{\nu\mu}(x)}=- i (\delta_{\sigma\mu} p_\nu-	\delta_{\sigma\nu}p_\mu) \frac{f(p^2)}{p^2}\delta^{rs}\, .
	\ee
	A model with $f(0)\neq 0$ thus has a massless photon and describes a Coulomb phase. For $f(0)= 0$ there is no massless gauge-boson and the theory is in the Higgs~/~confinement phase. \eqref{5_masslessvb} follows from \eqref{5_deff} due to Poincar{\'e} invariance and the  antisymmetry of the field strength tensor $F_{\mu\nu}$, only.

    Classically the current $j_\mu^r$ is conserved. However in a quantum field theory the corresponding symmetry can be broken spontaneously. Then its contribution picks up a longitudinal part to define the two functions
	  \be- \Fvev{A_\sigma^s(y)\,j_\mu^r(x) } = \delta^{rs} \left( T_{\sigma\mu}\,\, i(p^2) +  L_{\sigma\mu}\,\,  j(p^2)\right)\,. \ee
    The BRST exact contribution is not restricted by general arguments and defines the functions
	  \be i \Fvev{A_{\sigma}^s(y)\, s\xi^r_\mu (x) } = -\delta^{rs}\left( T_{\sigma\mu}\,\, u(p^2) +  L_{\sigma\mu}\,\,  v(p^2)\right)\,.  \ee
    Transverse and longitudinal projection of the Dyson-Schwinger equation \eqref{5_genDyson-Schwinger equation} yields the relations
	  \be f(p^2) + i(p^2) - u(p^2) = 1\,,\qquad\text{and,}\qquad  j(p^2) - v(p^2) = 1\,.\label{5_gen_sat}\ee
    The first equation of \eqref{5_gen_sat}, in particular the way it is saturated in the limit of vanishing momentum, differentiates between Coulomb and Higgs~/~confinement phase of the theory. The second equation, describing the longitudinal contributions to the Dyson-Schwinger equation, cannot be used to distinguish between the different phases. The longitudinal contributions are always unphysical.

    With exception of the Gribov-Zwanziger theory in \secref{sec_5_GZ}, BRST and anti-BRST symmetry is assumed to be unbroken. The corresponding conserved charges are denoted by $Q_B$ and $\bar Q_B$. The ghost number $Q_c$ is always conserved. The subset $\P$ of physical operators can then be defined by, \cite{Becchi:1975nq},   
	\begin{align}\label{5_physOP}
	\P & = \{\O ; \,[Q_{B},\O]=0,\text{ and }[Q_c,\O ] = 0 \} / \{\O: \O= [Q_{B},\O']\}\ .
	\end{align}   
    Note that for the various cases below the BRST transformations and thus also $Q_B$ varies. The same is true for the ghost-fields and $Q_c$. The content of \eqref{5_physOP} is gauge-dependent. The set of physical operators, however, is not.

\section{Kugo-Ojima revisited in Landau gauge\label{sec_5_lc}}

    The Kugo-Ojima confinement scenario was introduced in \secref{sec_KO}. It is defined in the linear covariant gauge \eqref{1_KO_lag}. The crucial point in this criterion is that both contributions of the current
	\be J_\mu^r =   - \partial_\nu F_{\nu\mu}^r + i s(D_\mu\bar c)^r  \label{5_lc_curr}\ee
    individually do not couple to a massless mode. In particular this means that they do not couple to massless gluon. Defining the function $f_{lc}(p^2)$ by
        \be \Fvev{ A^s_\sigma(y)\,  F^r_{\nu\mu}(x)} =- i \delta^{rs}\,\left(\delta_{\sigma\mu} p_\nu - \delta_{\sigma\nu} p_\mu \right) \frac{f_{lc}(p^2)}{p^2}\,, \label{5_lc_deff}\ee
    the absence of a coupling between $- \partial_\nu F_{\nu\mu}^r$ and $A_\sigma^s$ for vanishing momenta implies $f_{lc}(p^2) \pto 0$.
    
     The absence of a massless modes in the BRST exact term in the current \eqref{5_lc_curr} motivates the introduction of a function $u(p^2)$, \eqref{1_def_u}, which is defined via the correlation function \cite{Kugo:1995km},
	\be i\Fvev{(D_\sigma c)^r(y)\,(D_\mu \bar c )^s(x) } = \delta^{rs}\left( T_{\sigma\mu}\,u_{lc}(p^2)  +   L_{\sigma\mu}v_{lc}(p^2)\right)\, , \label{5_lc_defu} \ee
    and assumes the infrared value $u_{lc}(p^2)\pto -1$, \eqref{1_KO_crit}. The longitudinal part of aboves equation is restricted to $v_{lc}(p^2) = -1$ via
	\begin{multline} \delta^{rs} i p_\nu\, v_{lc}(p^2) = i p_\mu  \Fvev{(iD_\mu c)^r(y)\,(D_\nu \bar c )^s(x) }  \\=  \Fvev{(i\partial_\mu D_\mu c)^r(y)\,(D_\nu \bar c )^s(x) } = \Fvev{\var[\S_{lc}]{\bar c^r(y)}\,(D_\nu \bar c )^s(x)} = -\delta^{rs}i p_\nu \label{5_lc_long}\end{multline}
    The crucial point is to realize that the Kugo-Ojima criterion $u_{lc}(p^2)\pto -1$ implies this correlation function to yield unity,
	\be i\Fvev{(D_\sigma c)^r(y)\,(D_\mu \bar c )^s(x) } \pto -\delta_{\sigma\mu}\delta^{rs}\,.\label{5_lc_DcDcb}\ee
    The current \eqref{5_lc_curr} is equivalent to the  Noether current of global color symmetry,
	\be j_\mu^{lc\,r} = \cp{A_\nu}{(F_{\nu\mu} + i \delta_{\mu\nu} b)}{r}  - i\cp{c}{\partial_\mu\bar c}{r} + i\cp{\bar c}{D_\mu c}{r} \, , \label{5_lcg_current} \ee
    up to the equation of motion of the gluon. In turn this equation of motion can be rewritten and plugged into the Dyson-Schinger equation of the gluon
      \be \delta^{rs}\delta_{\mu\sigma} = \Fvev{A_{\sigma}^s(y)\,\var[S_{lc}]{A_{\mu}^r(x)} }  = \Fvev{A_{\sigma}^s(y)\,\left(-\partial_\nu F_{\nu\mu}^r - j_\mu^{lc\,r} + i s\bar s A_\mu^r\right)(x) } \,. \label{5_lc_dse} \ee
    If BRST symmetry is unbroken one has $\vev{A_\sigma^s\,s\bar s (A_\mu^r)} = - \vev{s(A_\sigma^s)\,\bar s (A_\mu^r)}$ and thus, if $u_{lc}(p^2)\pto -1$, the gluon Dyson-Schwinger equation is saturated by the BRST exact contributions, \eqref{5_lc_DcDcb}, only.
    
    Any physical state is invariant under BRST transformations \eqref{1_KO_subs}. Thus, if the Kugo-Ojima confinement criterion is fulfilled, i.e.
	\be\label{5_lc_KOcrit}  u_{lc}(0)=-1\qquad \text{and}\qquad f_{lc}(0)=0\,,\ee
    then the Dyson-Schwinger equation of the gluon is saturated by unphysical degrees of freedom only. The gluon is only of finite range but the conserved color current $ j_\mu^{lc\,r}$ does not contribute in the infrared.

    Confinement is not compatible with physical states in the adjoint color representation.  An alternative confinement criterion which resembles the Kugo-Ojima criterion in the linear covariant gauge, but which is transportable to other gauges, is that \emph{for vanishing momentum the Dyson-Schwinger equation of the gauge boson propagator is saturated by unphysical degrees of freedom only}, without any contributions from the conserved color current.

\section{Phases of Abelian Gauge Theories\label{sec_5_Abel}}

    Before comparing this reinterpreted version of the Kugo-Ojima criterion with different gauges of Yang-Mills theory, it is interesting to investigate the gauge-boson Dyson-Schwinger equation of Abelian gauge theories. With the Abelian field strength tensor $F_{\mu\nu} = \partial_\mu A_\nu - \partial_\nu A_\mu$ the Lagrangian of a general Abelian gauge theory in a linear covariant gauge reads

    \begin{align}\label{5_A_L}
	\L_\uone&= \frac{1}{4} F_{\mu\nu} F_{\mu\nu} + \LM +s\left(\bar c\left(\frac \xi 2 b  -i\partial_\mu A_\mu +i \gamma(\phi,\dots)\right)\right)\\
		&= \frac{1}{4} F_{\mu\nu} F_{\mu\nu}  + \LM +\frac \xi 2 b^2  -ib\partial_\mu A_\mu+i b\gamma(\phi,\dots) + i\bar c\partial^2 c-i\bar c s\gamma(\phi,\dots )\, .\nonumber
	\end{align}

	The gauge parameter $\xi$ and the NL, ghost and antighost fields, $b,c,$ and $\bar c$, are analogous to their non-Abelian counterparts. The local function $\gamma(\dots)$ of canonical dimension $2$ and vanishing ghost number is a polynomial of matter fields $\phi$ that
	does not depend on the gauge field $A_\mu$ or the NL field $b$. The matter part, \LM, is invariant under  Abelian gauge transformations. It includes covariantly coupled fermions and/or bosons.  
	
	The Abelian BRST and anti-BRST transformations are given by
	\begin{align}
	s A_\mu &=\partial_\mu c \, , & s c&=0\, ,& s \bar c &=b \, , & s b &=0\, ,\label{5_A_BRST}\\
	\bar s A_\mu &=\partial_\mu \bar c \, , & \bar s \bar c&=0 \, ,& \bar s c &=- b \, , & \bar s b &=0\, .\label{5_A_aBRST}
	\end{align}
	The matter fields transform under BRST and anti-BRST variations as under infinitesimal gauge transformations with the ghost or respectively the antighost as gauge parameter $ s\phi = c\phi$, and, $ \bar s \phi = \bar c \phi$. By construction  BRST and anti-BRST transformations are nilpotent and they anticommute,  
	\be s^2=\bar s^2=\acom{s}{\bar s}=0\,.\ee

	In \secref{sec_KO}, it was shown that negative norm states associated with asymptotic BRST quartets are unphysical. The elementary quartet consisting of longitudinal photon, ghost, antighost and NL field, thus is not observable. Contrarily, transversely polarized photons are physical. 

	The Dyson-Schwinger equation for the Abelian gauge boson in linear covariant gauges in momentum space reads,
	\be
	\delta_{\sigma\mu}  = -\Fvev{ A_\sigma(y)\,  \partial_\nu F_{\nu\mu}(x)} -\Fvev{A_\sigma(y)\,
	    {j_\mu}^{\hspace{-.4em}\uone}(x)} + i\vev{ A_\sigma(y)\, s \bar s A_\mu(x)}  \, , \label{5_qed_dse}
	\ee
	where the conserved global \uone-current, $j_\mu^\uone$, is obtained from the matter part of the action alone, $ {j_\mu}^{\hspace{-.4em}\uone}(x)=-\frac{\delta \S_M}{\delta A_\mu(x)}\, ,\label{5_U1-current} $ with $\S_M=\int d^4x \LM$. The last term on the rhs in \eqref{5_qed_dse} arises from the linear covariant gauge fixing in \eqref{5_A_L}. In contrast to the non-Abelian case it is purely longitudinal. Longitudinal and transverse  projection of \eqref{5_qed_dse} give the identities,
	\begin{subequations}\label{5_A_decompose}
	\begin{align}
	\label{5_A_longitudinal}
	L_{\sigma\mu}&=\Fvev{A_\sigma(y)\, i \partial_\mu s\bar c(x)}- 
	L_{\mu\rho}\Fvev{A_\sigma(y)\, {j_\rho}^{\hspace{-.4em}\uone}(x)} \, ,
	\\
	\label{5_A_transverse}
	T_{\sigma\mu}  &= -\Fvev{ A_\sigma(y)\, \partial_\nu F_{\nu\mu} } 
	-T_{\mu\rho}\Fvev{ A_\sigma(y)\,{j_\rho}^{\hspace{-.4em}\uone}(x)}\, .
	\end{align}
	\end{subequations}

	Using the equation of motion of the NL field, \eqref{5_A_longitudinal} yields the Ward identity for the longitudinal photon propagator,
	\be \xi\, p_\sigma= p^2 p_\nu \,\Fvev{A_\sigma(y)\,A_\nu(x)} -ip^2\Fvev{A_\sigma(y)\,\gamma(x)}-i\xi\Fvev{A_\sigma(y)\,\partial_\nu {j_\nu}^{\hspace{-.4em}\uone}(x)} \,, \label{5_A_wi} \ee
	where $\gamma(x)=\gamma(\phi(x),\dots)$ is the local function of the fields in the BRST exact term of \eqref{5_A_L}.

	In the Abelian case, $f_\uone(p^2)$ defined by
	\be T_{\sigma\mu}\, f_\uone(p^2)=-\Fvev{ A_\sigma(y)\,  \partial_\nu F_{\nu\mu}(x)}\ ,\label{5_A_transdef} 
	\ee	
	determines the transverse part of the vector boson propagator,
	\be\label{5_A_photontransverse}
		T_{\mu\nu} \, \Fvev{A_\sigma(y)\,A_\nu(x)}=\frac{f_\uone(p^2)}{p^2} T_{\sigma\mu}\,.
	\ee
	The photon is massless if $f_\uone(0)>0$. In terms of \eqref{5_gen_sat} one has $u_\uone(p^2)=0$ due to the gauge-fixing contributions being longitudinal only and thus immediately $i_\uone(p^2) + f_\uone(p^2)=1$. If the current saturates the transverse part of the Dyson-Schwinger equation, \eqref{5_A_transverse}, in the infrared,
	      \be\label{5_A_higgscond}
	      -T_{\mu\nu}\Fvev{ A_\sigma(y)\,{j_\nu}^{\hspace{-.4em}\uone}(x)}\stackrel{p^2\rightarrow 0}
	      {\longrightarrow}T_{\sigma\mu}\ ,
	      \ee
	one has $i_\uone(p^2) = 1$ and thus $f_\uone(0)=0$.
 
	These relations hold for any Abelian gauge theory in linear covariant gauges. Next Abelian gauge theories in the Coulomb and Higgs phase are examined in more detail. In Sec.~\ref{sec_5_mag} a confining Abelian gauge theory is investigated, Yang-Mills theory in the MAG.

\subsection{The Coulomb Phase\label{sec_acoulomb}}
	To investigate the Coulomb phase of the theory in some more detail consider usual QED in the general covariant gauge, i.e. $ \LM = \bar \psi( -\slashed D + m ) \psi$ and $\gamma = 0$. The Coulomb phase features a massless photon which translates into a non-vanishing function $f_\uone$ for vanishing momentum. Since the Abelian gauge symmetry is unbroken, the correlation function  \vev{ A_\sigma(y)\,{j_\nu}^{\hspace{-.4em}\uone}(x)} is transverse in any covariant gauge. The Ward identity \eqref{5_A_wi} obtains the familiar form
	      \be \label{5_A_longCoulomb} \xi\frac{p_\sigma}{p^2}= p_\nu \,\Fvev{A_\sigma(y)\,A_\nu(x)}\,,\ee
	which, together with \eqref{5_A_photontransverse} fully determines the photon propagator,
	\be \label{5_A_propCoul}
	      \Fvev{A_\mu(y)\,A_\nu(x)} = \frac{f_\uone(p^2)}{p^2} T_{\mu\nu}+ \frac{\xi }{p^2} L_{\mu\nu}\ . 
	\ee
	The photon is massless with $f_\uone(0)>0$. There are now two interpretations of the same mathematical fact, either one states: $f_\uone(0)>0$ because \eqref{5_A_higgscond} does not hold, or: because of $f_\uone(0)>0$, \eqref{5_A_higgscond} does not hold. Either ways, the non-vanishing of the function $f_\uone$ in the infrared and the failure of \eqref{5_A_higgscond} are intrinsically related.

	The longitudinal part of the photon Dyson-Schwinger equation \eqref{5_A_longitudinal} and the ghost Dyson-Schwinger equation yield that the elementary quartet is free and massless:
	      \be\label{5_A_propquartet}
		  \Fvev{b(y)\,A_\mu(x)}=\Fvev{\bar c(y)\, \partial_\mu c(x)}=-\frac{p_\mu}{p^2}  \,.
	      \ee	
	It is interesting to note that in the canonical formalism $f(0)\neq 0$ implies that the electromagnetic charge operator is not well defined. Up to terms proportional to the photon equation of motion, this charge is equivalent to
	  \be\label{5_A_Qfalse}  Q=\int d^3x (i\partial_0 b(x)-\partial_\nu F_{\nu 0}(x))  = \int d^3x i\partial_0 b(x)+\int_{S_\infty} d{\mathbf \sigma}_i F_{0i} \equiv \N+\mathcal G\,. \ee
	  Due to the antisymmetry of the field strength tensor the current $-\partial_\nu F_{\nu\mu}$, and thus the charge $\mathcal G$, is conserved. The charge $\N$ can be written as BRST exact expression. With $\mathcal C = i\int d^3x \bar c$ one has $\N = \acom{Q_B}{\mathcal C}$. Furthermore, the equal time commutator of $\mathcal G$ with any \emph{local} physical operator $\Phi(x)\in\P$ vanishes,
		\be\label{5_A_causloc}
		[\Phi(x),\mathcal G]=0\  \text{ for all \emph{local}}\  \Phi(x)\in\P\ ,
		\ee
	  because causality requires operators with spatial separation to commute. One thus has for the electric charge of any local operator  $\Phi(x)\in\P$
	      \begin{multline}\label{5_A_chargecomm}
	      \com{Q}{\Phi(x)} = \com{\N+\mathcal G}{\Phi(x)}=\com{\acom{\mathcal C}{Q_{BRST}}}
	      {\Phi(x)} \\
	      = \acom{\mathcal C}{\com{Q_{BRST}}{\Phi(x)}}+\acom{Q_{BRST}}{\com{\mathcal C}{\Phi(x)}} = \acom{Q_{BRST}}{\com{\mathcal C}{\Phi(x)}} \,.
	      \end{multline}
	 The electric charge of a local operator in QED is unphysical. All \emph{local physical} operators $\Phi(x)\in \P$ are uncharged. Physical operators creating charged particles like the electron necessarily are not local, \cite{Bagan:1999jf}. Such non-local states are only possible since the photon is massless, i.e. of infinite range.

	 \subsection{The Abelian Higgs Phase \label{sec_higgs}}
	  A ``spontaneously broken'' Abelian gauge theory in the Higgs phase satisfies the Dyson-Schwinger equation of \eqref{5_qed_dse} differently. From the general discussion one expects that  $f(0)=0$, the vector boson is massive and the current saturates the transverse Dyson-Schwinger equation at low momenta, i.e.,  \eqref{5_A_higgscond} holds. Also in the Higgs phase one expects (unphysical) massless excitations. This scenario is verified explicitly in the Abelian Higgs model with quartic coupling $\lambda$ and a negative quadratic term,
	  \begin{align} 
	      \LM^\text{Higgs}& = \frac{1}{2} (D_\mu \Phi)^* (D_\mu \Phi) + \lambda \left( \lvert\Phi\rvert^2  - v^2\right)^2 	  \label{5_A_LHiggs}\\ 
		    &= \frac 1 2 \left( (\partial_\mu \pp)^2 + (\partial_\mu\pm)^2 \right) + \frac {m^2}{2}\,A_\mu^2 + m\, \phi_- \partial_\mu A_\mu  + g \, A_\mu \left( \pm \partial_\mu \pp-\pp \partial_\mu \pm  \right) 
			   \nonumber \\& + gm \, A_\mu^2 \pp  + \frac{g^2}{2}\,A_\mu^2 \left( \pp^2 + \pm^2 \right) + 
		  \lambda \left( \pp^2 + \pm^2 + 2 \pp v \right)^2  \,.  \nonumber
	  \end{align}
	  In the Higgs phase with $v>0$ the fields are parametrized by,  $\Phi = \phi + v ,\, \pp = {\frac 1 2} (\phi^*+\phi) ,\, \pm  ={\frac i 2}( \phi^*-\phi) $. The tree level photon mass is $m = gv$. The massless Goldstone-mode $\pm$ couples to the longitudinal photon. The matter fields transform under the Abelian gauge transformations as $\delta\Phi = i g \theta \Phi ,\, \delta\pp = - g \theta \pm ,\, \delta \pm  =  g \theta \left( \pp +  v \right)$. Replacing $\theta(x)$ by the anti-commuting ghost field one arrives at the BRST variations of the reparametrized matter fields
		\be \label{5_A_matter} s\pp  = -gc \,\pm  \, ,\qquad s \pm  =  g c \left( \pp + v \right)  \, .\ee

	  A convenient gauge that eliminates the  bilinear coupling of $\pm$ to the longitudinal photon is given by the 't~Hooft gauge,  \cite{'tHooft:1971rn},
	    \begin{align}\label{5_A_tHooftgf}
		\Lgf^\text{'t~Hooft} & = s\left(\bar c \left(  \frac \xi 2 b -i \partial_\mu A_\mu 
		+ i\xi m \pm \right) \right)\\
			& = \frac \xi 2 b'^2 - i b' \partial_\mu A_\mu - m \, \pm \partial_\mu A_\mu 
			+ \frac{\xi m^2}{2}\,\pm^2 + i \bar c \left( \partial^2 - gm\xi \pp - m^2 \xi \right) c  
			\,,\nonumber
	    \end{align}
	    where in the second expression the NL field has been shifted: $b = b' -i m \pm$. In the following the Lagrangian \eqref{5_A_L} is investigated with $\LM = \LM^\text{Higgs}$ and $\gamma = \xi m\pm$. The BRST exact term $\Lgf^\text{'t~Hooft}$ breaks not only local but also global \uone-symmetry explicitly. The gauge invariant and classically  conserved matter current is given by
	    \begin{align}
		{j_\mu}^{\hspace{-.4em}\uone}&=  \delta \Phi \,\var[\LM^\text{Higgs}]{\partial_\mu \Phi} 
		+ \delta \Phi^* \,\var[\LM^\text{Higgs}]{\partial_\mu \Phi^*} \label{5_A_higgs_curr} \\
		&= g \left( \pp \partial_\mu \pm-\pm \partial_\mu \pp  \right) + m \partial_\mu \pm  -  A_\mu \left( g^2( \pp^2 + \pm^2)+m^2 +2 mg \pp \right)\,. \nonumber
	    \end{align}
	    The current is BRST invariant, and its divergence is unphysical because  the global gauge invariance of the model is broken by BRST exact terms only. In fact, the divergence $\partial_\mu{j_\mu}^{\hspace{-.4em}\uone}$  is BRST exact up to equations of motion. Using the equation of motion of the field $\pm$ and the NL field $b$, up to tree-level one has
	    \be \label{5_A_divj} \partial_\mu{j_\mu}^{\hspace{-.4em}\uone}\approx m\partial^2\pm-m^2 \partial_\mu A_\mu= m^2(\xi m\pm - \partial_\mu A_\mu) = i m^2\xi s\bar c\,.\ee

	    In the broken phase,  the current contribution to \eqref{5_A_longitudinal} does not vanish and in fact saturates  the longitudinal Dyson-Schwinger equation at low momenta. Since the divergence of the current is BRST exact up to  equations of motion, it does not create physically observable Goldstone bosons. The unphysical field  $\pm$ does not contribute to the Ward identity at tree level which is a feature of the 'tHooft gauge. \eqref{5_A_wi} gives the tree-level longitudinal propagator in the Higgs phase:
	      \be
		p_\nu \,\Fvev{A_\sigma(y)\,A_\nu(x)} = \frac{\xi p_\sigma}{p^2 + \xi m^2}  \label{higgs_wi}  
	      \ee
	      which may be directly verified from the quadratic terms of the action \eqref{5_A_LHiggs}. 
	      The tree-level correlation functions of the elementary quartet are given by
		    \be  \Fvev{b(y) A_{\sigma}(x)} = - \frac{p_\sigma}{p^2+\xi m^2}\,,\qquad\text{and,}\qquad\Fvev{\bar c(y)\,c(x)} = - \frac{i}{p^2+\xi m^2}\,.\ee
	      For $\xi\neq 0$ one thus has a massive behaviour in the longitudinal contributions of the Dyson-Schwinger equation \eqref{5_A_longitudinal}. For vanishing momentum the current contributions saturate this equation. The propagator of the longitudinal gluon and the elementary quartet become massive. In the tranverse case, $\xi\rightarrow 0$, however, the picture is different. The longitudinal photon propagator vanishes and the elementary quartet becomes massless. Since in this case there is no explicit breaking of the global gauge symmetry the current $j_\mu^\uone$ is classically conserved. 
	     
	      In contrast, the transverse Dyson-Schwinger equation yields information of the phase-structure independent on the gauge parameter. In tree-level approximation the function $f_\uone(p^2)$ defined by \eqref{5_A_transdef}, in the Higgs phase is 
		    \be f_\uone(p^2) \approx \frac{p^2}{p^2 + m^2} . \label{5_A_higgs_sat} \ee
	      Since $f_\uone(0)=0$ the transverse vector boson is short ranged in this phase,  
		  \be  T_{\mu\nu}  \Fvev{A_{\sigma}(y)\,A_\nu(x)} \approx \frac{1}{p^2 + m^2}T_{\sigma\mu}  \,.  \label{5_A_higgs_witrans}\ee
	      The current of \eqref{5_A_higgs_curr} thus also saturates the transverse Dyson-Schwinger equation at low momenta, and \eqref{5_A_higgscond} holds.    

	      These examples illustrate (at tree level) the characteristics that distinguish the unbroken Coulomb and ``spontaneously broken'' Higgs phases of Abelian gauge theories. If  the current contribution saturates the transverse Dyson-Schwinger equation of the photon at low momenta, the model is in a Higgs phase. If the current contribution fails to saturate the transverse Dyson-Schwinger equation at low momenta, the Abelian gauge theory describes a Coulomb phase with a massless vector particle.  The (conserved) transverse part of the Abelian current in our examples is BRST invariant and does not include BRST exact terms.  It apparently creates physical particles only. 

\section{Generalized Kugo-Ojima Criteria in Non-Abelian Gauge Theories}
	      In this section the above found generalization of the Kugo-Ojima confinement criterion is tested in various gauges of Yang-Mills theory. First the already introduced generalized covariant gauges and the Maximal Abelian gauges are investigated. The the Lorentz non-covariant Coulomb gauge and the minimal Landau gauge in the Gribov-Zwanziger framework are introduced and investigated with respect of the generalized confinement criterion.

\subsection{Generalized Linear Covariant Gauge}
	  The generalized linear covariant  gauge, its Lagrangian and the corresponding BRST and anti-BRST transformations are introduced in \secref{sec_4_lag}. The gauge-fixing Lagrangian is given in \eqref{4_def_Lgf} which defines Yang-Mills theory in the generalized covariant gauge,
		  \be \L_{GCG} = \LYM + \frac{\xi}{2} b^2 - i b^r\, \partial_\mu A_\mu^r - i \alpha (D_\mu \bar c)^r \partial_\mu c^r - i \bar \alpha \, \partial_\mu \bar c ^r (D_\mu c)^r + \frac{\alpha \bar \alpha \xi}{2} \,\cp{\bar c}{c}{2} \label{5_cc_Lag}\,.\ee
	  As the Lagrangian \eqref{5_cc_Lag} is invariant under global color transformations, the corresponding Noether current is conserved but depends on the gauge parameter $\alpha$,
	  \begin{multline}
	      j_\mu^{GCG\,r} = \cp{A_\nu}{(F_{\nu\mu} + i b \delta_{\mu\nu})}{r} 
		- i \cp{c}{(\alpha D_\mu\bar c + \bar \alpha \partial_\mu \bar c)}{r} \\+ 
		i \cp{\bar c}{(\bar \alpha D_\mu c + \alpha \partial_\mu c )}{r}\,. \label{cc_cur}
	  \end{multline}
	  This conserved current again is part of the variation of the action and the gluon Dyson-Schwinger equation
	  takes the form
	      \be  \delta^{rs}\delta_{\mu\sigma} = \Fvev{A_{\sigma}^r(y)\,(-\partial_\nu F_{\nu\mu} - j_{\mu}^{GCG})^s(x) } + \vev{A_{\sigma}^r(y)\,(i\sa\sab A_\mu )^s(x) }\,. \label{5_cc_dse}\ee
	  The last term in \eqref{5_cc_dse} again involves only unphysical excitations. It is of the same  form as in the linear covariant gauge studied above,
	  \begin{multline} 
	      \Fvev{A_{\sigma}^r(y)\,(i\sa\sab A_\mu )^s(x)} = -i\Fvev{(D_{\sigma}c)^r(y)\,
	      (D_\mu \bar c)^s(x)} \\ = \delta^{rs}\left( L_{\sigma\mu}-T_{\sigma\mu}\,u_{GCG}(p^2) 
	      \right)\,.\label{5_cc_sat}
	  \end{multline}
	  Unlike in linear covariant gauge, the equation of motion of the ghost by itself does not suffice to determine the longitudinal part of \eqref{5_cc_sat}. Instead one has, \eqref{B_ghostEOM},
	    \be\label{5_cc_longcc}
		\frac{\delta S_{GCG}}{\delta c^r} = - i(\partial_\mu D_\mu \bar c)^r+ \frac{\xi \bar \alpha}{2} s (\bar c\times\bar c)^r\,.
	    \ee
	  Using nilpotency of the BRST transformations, an analogous calculation to \eqref{5_lc_long} determines the  longitudinal part of the correlation function in \eqref{5_cc_sat}. As in linear covariant gauge, unphysical degrees of  freedom saturate the longitudinal part of the Dyson-Schwinger equation in \eqref{5_cc_dse}, and the current matrix  element is transverse.

	  The form factor, $f_{GCG}(p^2)$, is defined as in \eqref{5_lc_deff}, and the same discussion as in \secref{sec_5_lc}  applies.  So, provided that 
	  \be f_{GCG}(0)=0\,,\qquad\text{and,}\qquad u_{GCG}(0)=-1\,,\ee
	  the transverse Dyson-Schwinger equation is   saturated by unphysical degrees  of freedom in the infrared. One formally has the same confinement criterion as in linear covariant gauge. However, although the  unphysical correlation functions in \eqref{5_cc_sat} and \eqref{5_lc_defu} are formally similar,  the unphysical sectors differ, and $u_{GCG}(p^2)\neq u_{lc}(p^2)$ if $\alpha\neq 0$. However, the confinement criterion stated above asserts that these functions coincide at $p^2=0$ in any gauge parametrized by $(\alpha,\xi)$.

\subsection{Maximal Abelian Gauge\label{sec_5_mag}}

	The Maximal Abelian gauge was introduced in \secref{sec_MAG}. It breaks color symmetry by differentiating between the gluons in the Cartan subalgebra and the corresponding coset space. An equivariant gauge fixing construction fixes the coset degrees of freedom breaks the gauge symmetry from $SU(N)$ down to $U(1)^{N-1}$ and only an additional gauge fixing of the Abelian theory completely fixes the gauge, \secref{sec_3_lag}. In principle one has thus a $(N-1)$-fold copy of an Abelian gauge theory as \eqref{5_A_L}, where the matter content is given by the Coset degrees of freedom. Although a more general discussion is possible, for simplicity the gauge group $SU(2)$ is considered in the following only. It illustrates the main points and  directly connects to the considerations in \secref{sec_5_Abel}. The Cartan subalgebra in this case is one dimensional and the corresponding index is suppressed in the following. The coset space is two dimensional with components $a=1,2$. Given the equivariant coset gauge 
fixing, \eqref{3_Leps},
	    \begin{multline} \L_{gf}^\eps  = \frac{\xi}{2}\,b^2 - i b^a (\D_\mu B_\mu)^a + i \bar c^a (\D_\mu\D_\mu c)^a + i \cp{B_\mu}{\bar c}{}\cp{B_\mu}{c}{}  + \frac{\xi}{2} \cp{\bar c}{c}{}\cp{\bar c}{c}{} \,,
	    \end{multline}
	and the gauge fixing Lagrangian of the Abelian Cartan algebra, \eqref{3_Lgf_Cartan}
	  \be 
	    \Lgf^C  = \frac{\lambda}{2}\,\eta^2 - i \eta \,\partial_\mu A_\mu  \,,
	  \ee
	in the following $SU(2)-$ Yang-Mills theory in MAG,
	      \be  \L_{MAG} = \LYM + \Lgf^\eps + \Lgf^C\,,\label{5_mag_Lag}\ee    
	specified by the action $\S_{MAG} = \int\!d^4x \L_{MAG}(x)$ is considered. 
	
	The $\uone$ gauge fixing $\Lgf^C$ of \eqref{5_mag_Lag} not only explicitly breaks the local \uone-gauge symmetry but global symmetries as well. $\Lgf^C$ is symmetric under global \uone-transformations, but breaks the global equivariant BRST, anti-BRST and $SU(2)$ symmetries explicitly. For any, not necessarily local, operator $\O$ one has the Ward identities,
	    \begin{subequations} \label{5_mag_wardId}
		\begin{align}
		  \label{5_U1Ward}
		    \vev{\delta_x \O}=\vev{\O\,\delta_x S_{MAG}}&=i\vev{\O \ \partial^2 \eta(x)}\,,\\
		  \label{5_seWard}
		    \vev{\se\O}=\vev{\O\,\se S_{MAG}}&=i\vev{\O\  \int\!\!d^4y \left(\partial_\mu \eta(y)\right)\se A_\mu(y)}\,,\\
		  \label{5_sebWard}
		    \vev{\seb\O}=\vev{\O\,\seb S_{MAG} }&=i\vev{\O\  \int\!\!d^4y \left(\partial_\mu \eta(y)\right)\seb A_\mu(y)}\,, 
		\end{align}
	    \end{subequations}
	where the generator of the local \uone-transformations is given by
	    \begin{multline}
		    \label{5_MAG_delta}
		      \delta_x = \partial_\mu\var{A_\mu(x)} +  \cp{B_\mu(x)}{\var{B_\mu(x)}}{} + \cp{c(x)}{\var{c(x)}}{} \\
			+ \cp{\bar c(x)}{\var{\bar c(x)}}{} + \cp{b(x)}{\var{b(x)}}{}\,.
	    \end{multline}
	Defining the set $\W$ of \uone-invariant operators, $\O\in \W\Leftrightarrow \delta_x \O=0$, one finds that  $\vev{\se\O}=\vev{\seb\O}=0\ \ \text{for all  } \O\in\W\,$, \eqref{3_appB}. On the set $\W$ of $\uone$ invariant operators  of the equivalent Abelian gauge theory one thus recovers $\se$ and $\seb$ as nilpotent BRST symmetries. One can then define the set of physical operators of the underlying non-Abelian $SU(2)$ gauge theory as a subset of $\W$, $\P\subseteq\W$ ,
	\begin{align}\label{5_mag_physOP}
	\P & = \{\O \in \W ; \,[Q_{\eps},\O]=0,\text{ and }[Q_c,\O ] = 0 \} / \{[Q_{\eps},\O];\  [Q_c,\O]=-\O \text{ and } \O \in\W\}\ .
	\end{align} 
	The conserved $\uone$ current of the Cartan subalgebra of $SU(2)$ in MAG reads
	\be  j^{MAG}_\mu  = i \com{\se}{\seb}A_\mu + \cp{B_\nu}{( F_{\nu\mu} - i \delta_{\mu\nu} \,b)}{} \label{5_mag_curr_brst}\,. \ee
	Each term in \eqref{5_mag_curr_brst} separately is an element of $\W$, but $ i \com{\se}{\seb}A_\mu$ does not create physical transverse states.

	The Dyson-Schwinger equation of the Cartan gluon depends on the conserved Abelian Noether current \eqref{5_mag_curr_brst} in the same way as in any Abelian gauge theory in linear covariant gauge studied in \secref{sec_5_Abel},
	    \be \delta_{\sigma\mu}\delta(x-y) = \vev{A_\sigma(y)\,\left( - \partial_\nu f_{\nu\mu} - j^{MAG}_\mu \right)(x)} + \vev{A_\sigma(y)\,i\sC\sCb A_\mu(x)} \,.\label{5_mag_dse}
	    \ee
	As for an unbroken Abelian gauge theory,  the last term of \eqref{5_mag_dse} saturates the longitudinal part of the Dyson-Schwinger equation due to the Abelian Ward identity of \eqref{5_U1Ward},
	    \be \label{5_satlongMAG} 
	    \partial_\sigma^x \delta(x-y)=i\vev{A_\sigma(y)\, \partial^2 \eta(x)}
		    \Rightarrow \Fvev{A_\sigma(y)\, i\partial_\mu \eta(x)} = L_{\sigma\mu}\ .
	    \ee

	The first correlator in \eqref{5_mag_dse} is transverse due to the anti-symmetry of $f_{\mu\nu}$ and the current matrix element thus is transverse as well,
	    \be\label{5_jmagsat}
		\Fvev{A_\sigma(y)\,j^{MAG}_\mu(x)}=(f_{MAG}(p^2)-1)T_{\mu\nu}\ .
	    \ee
	Defining the functions $f_{MAG},\, u_{MAG},\, h_{MAG},$ and $\ell_{MAG}$ from the correlators
      \begin{subequations}
	  \begin{align}
	      -\Fvev{A_\sigma(y)\, \partial_\nu f_{\nu\mu} (x)} &  = f_{MAG}(p^2) T_{\sigma\mu}  \label{5_mag_deff} \\
	      i\Fvev{A_\sigma(y)\, \com{\se}{\seb}A_\mu (x)} & = u_{MAG}(p^2) T_{\sigma\mu} + \ell_{MAG}(p^2) L_{\sigma\mu} \label{5_mag_defi} \\
	      -\Fvev{A_\sigma(y)\, \cp{B_\nu}{(F_{\nu\mu} - \delta_{\mu\nu} i b)}{}} & = h_{MAG}(p^2) T_{\sigma\mu} +\ell_{MAG}(p^2) L_{\sigma\mu} \,,
	  \end{align}
	  \end{subequations}
      the transverse part of \eqref{5_mag_dse} yields the constraint , 
	  \be\label{5_MAGcond} f_{MAG}(p^2) + h_{MAG}(p^2) = 1+ u_{MAG}(p^2)\,.  \ee
      As in linear covariant gauge and the generalized linear covariant gauge, the transverse Dyson-Schwinger equation is saturated by unphysical degrees of freedom and the Cartan color charge of physical states vanishes if one requires   the Kugo-Ojima-like criterion
      \be\label{5_MAG_sat} f_{MAG}(0)=0 \ \text{and}\ \ u_{MAG}(0)=-1\, .      \ee

      The conditions~(\ref{5_MAG_sat}) guarantee saturation of the gluon  Dyson-Schwinger equation in the infrared by unphysical degrees of freedom in MAG. It implies that physical states are colorless. However, in contrast to general covariant gauges,   \eqref{5_jmagsat} together with \eqref{5_MAG_sat} show that a Kugo-Ojima  scenario  can only be realized in MAG if unphysical degrees of freedom created by the  conserved Abelian current $j_\mu^{MAG}$ saturate the Abelian Dyson-Schwinger equation at low momenta.  From the point of view of the Abelian gauge theory, saturation of the transverse Dyson-Schwinger equation at  low momenta  in confinement and Higgs phases are similar.  The Abelian current saturates the Dyson-Schwinger equation at low energies in the Abelian Higgs phase as well as in the 
      confinement phase of the $SU(2)$  gauge theory in Maximal Abelian gauge. The only difference is that whereas \emph{physical} degrees of freedom contribute to the Dyson-Schwinger equation in the Higgs phase, only \emph{unphysical}  states saturate it in the confinement phase.

      In this context it is interesting to consider the condition $f_{MAG}(0) = 0$ more closely. If the scaling solution \eqref{3_IR_prop} is realized, the Abelian propagator is enhanced in the infrared \cite{Huber:2009wh}.  This scenario can be reconciled with the criterion of \eqref{5_MAG_sat} due to the definition of the Abelian field strength tensor $f_{\mu\nu}$, \eqref{3_def_fabel}.  \eqref{5_mag_deff} implies  
      \be  f_{MAG}(p^2) T_{\sigma\mu} = p^2 T_{\mu\nu} \Fvev{A_\sigma(y) \,A_\nu(x)} - 
      \Fvev{A_\sigma(y) \partial_\nu\cp{B_\nu}{B_\mu}{}\!(x)} \,. \label{5_mag_satf} \ee
      Although a massive Abelian vector boson allows one to fulfill $f_{MAG}(0)=0$, the last correlator of \eqref{5_mag_satf} prohibits one from asserting that the diagonal gluon propagator has to be suppressed at low momenta. 
      Introducing the function $a_{MAG}(p^2)$, 
      \be\label{5_MAG_enhanced}
      \Fvev{A_\sigma(y)  \cp{B_\nu}{B_\mu}{}\!(x)} = -i (\delta_{\sigma\nu} p_\mu
      -\delta_{\sigma\mu}p_\nu) a_{MAG}(p^2) ,
      \ee
      \eqref{5_mag_satf} states that,
      \be f_{MAG}(p^2)  = p^2 ({\textstyle \frac{1}{3}} T_{\mu\nu} \Fvev{A_\mu(y) \,A_\nu(x)} - a_{MAG}(p^2)) \,. \label{5_mag_satf1}
      \ee
      If the Cartan gluon  correlator is infrared enhanced, \eqref{5_mag_satf1} determines only the infrared singular behavior of  $a_{MAG}(p^2)$  when $f_{MAG}(0)=0$.

      To gain some more information about the functions defined above, we define a $\uone$ invariant transverse field strength,
      \be\label{5_EM} \W\ni G_{\mu\nu}=\partial_\mu A_\nu-\partial_\nu A_\mu\,.
      \ee
      It is not an invariant of the equivariant BRST (or anti-BRST) and,  in contrast to the $\uone$ gauge theory considered in \secref{sec_5_Abel}, is not a physical operator of the $SU(2)$ gauge theory. The function $u_{MAG}(p^2)$ defined in \eqref{5_mag_defi} also describes the correlation functions,
      \begin{multline}
	    \label{5_uG}
	    u_{MAG}(p^2)(\delta_{\rho\mu}p_\sigma-\delta_{\sigma\mu}p_\rho) =  \Fvev{G_{\rho\sigma}(y)\, 
	    \com{\se}{\seb}A_\mu (x)} \\ =\Fvev{\seb G_{\rho\sigma}(y)\, \se A_\mu (x)} - \Fvev{\se G_{\rho\sigma}(y)\, 
	    \seb A_\mu (x)} \, .
      \end{multline}
      In close analogy to general covariant gauges consider
	    \begin{multline}\label{5_MAG_cc}
	    u_{MAG}(p^2) T_{\mu\nu}+v_{MAG}(p^2) L_{\mu\nu} = 2i\Fvev{\seb A_\nu(y)\,\se A_\mu(x)} \\ =2i\Fvev{ \cp{B_\nu}{\bar c}{}(y)\, \cp{B_\mu}{c}{}(x)}\,,
	    \end{multline}
      The definition of the function $u_{MAG}(p^2)$ in MAG apparently differs  by a factor of $-2$ from that of general linear covariant gauge given by \eqref{5_cc_sat}. These functions are form factors of gauge dependent correlation functions that are not required to coincide in two different gauges. In \cite{Hata:1992np} it was found that Landau gauge and the MAG are not smoothly connected. Thus it is even expected that the functions saturating the transverse Dyson-Schwinger equation with unphysical degrees of freedom cannot be related by a continuous parameter. Still, the physical implications remain unchanged.
        
      \subsection{Coulomb Gauge \label{sec_CG}} 
	Coulomb gauge breaks manifest Lorentz covariance by treating timelike and spacelike gluons differently. It is described by the Lagrangian,
	      \be \L_C = \LYM - i b^r \partial_i A_i^r - i \partial_i\bar c^r (D_ic)^r \,, \label{5_c_lag} \ee
	where Latin indices denote spatial components of a Lorentz-vector, $i,j,\dots = 1,2,3$. The BRST transformations are the same as in \eqref{1_def_BRST}, 
	      \be\label{5_C_BRSTtrafo}	
		    s A_0^r  = (D_0 c)^r, \quad s A_i^r  = (D_i c)^r, \quad sc^r  = -\frac{1}{2} \cp{c}{c}{r}, \quad s \bar c ^r  = b^r, \quad  sb^r = 0 \,.
	      \ee
	The BRST charge in Coulomb gauge can be written in terms of Gauss's law, \cite{Zwanziger:1998ez},
	  \be Q_{B} = - \int\! d^3x\, c^r (D_i F_{i0})^r = \int\! d^3x \,c^r \var[S_C]{A_0^r} \,.\label{5_C_BRST}
	  \ee
	The anti-BRST transformations and corresponding charge may be defined analogously and the set of physical operators is defined via the BRST cohomology, \eqref{5_physOP}. Coulomb gauge  manifestly preserves global color symmetry and the color currents 
	    \begin{subequations}
	      \begin{align}
		    \label{5_temporalj}
		    {j^C_0}^r  &= \cp{A_i}{F_{i0}}{r},\\ 
		    {j^C_i}^r &= \cp{A_0}{F_{0i}}{r} + \cp{A_j}{F_{ji}}{r} + \cp{A_i}{b}{r}  - i \cp{c}{\partial_i \bar c}{r} + i \cp{\bar c}{D_i c}{r} \, ,
		    \label{5_spatialj}
	      \end{align}
	      \end{subequations}
	are conserved. The absence of manifest Lorentz invariance in Coulomb gauge implies two distinct gluonic equations of motion, the time component of which is
	      \begin{align} 
	      \delta^{rs} & = -\Fvev{A_0^s(y)\,D_iF_{i0}^r(x) }\,. \label{5_C_temporaleom}
	      \end{align}
	Since all physical states satisfy Gauss's Law in Coulomb gauge, this equation of motion is saturated by unphysical states only, whether the model confines or not. To see that all states contributing to \eqref{5_C_temporaleom} are unphysical note that physical states $\ket{\Psi_{phys}}$  are created by physical operators defined in \eqref{5_physOP}. They have vanishing ghost number and are annihilated by the ``Gauss-BRST" charge \eqref{5_C_BRST},
	      \be Q_{B} \ket{\Psi_{phys}} = 0\,. \label{5_C_subs} \ee
	The ghost field $c$ does not annihilate $\ket{\Psi_{phys}}$, as it contains ghost creation and annihilation contributions. \eqref{5_C_subs} thus has to be ensured by gluonic contributions only, and one gets back Gauss's law as the subsidiary condition condition
		\be \var[S_C]{A_0^r(x)}\, \ket{\Psi_{phys}} = 0 \qquad \forall\,x\,. \label{5_C_subs2} \ee
	Any non-vanishing contribution to \eqref{5_C_temporaleom} thus must be due to unphysical $\ket{\psi} \not\in \{\ket{\Psi_{phys}}\}$. It even can be shown that the temporal Dyson-Schwinger equation is saturated by instantanous contributions only, \cite{Schaden:2013ffa}.

	The discussion of the spatial components of the DSE is very similar to the one in linear covariant gauge. The equation of motion for the spatial part of the gluon propagator is given by
		\be
		\delta^{rs}\delta_{ij}= \Fvev{A_j^s(y)\, \left( - \partial_\nu F_{\nu i}^r(x) - j^r_i(x) \right) } 
		    + i \Fvev{A_j^s(y)\,s\bar s A_i ^r}\, . \label{5_coul_spatialeom} 
		\ee
	The first matrix element necessarily is spatially transverse in Coulomb gauge with $\partial_i A_i=0$.  The Faddeev-Popov operator of Coulomb gauge is instantaneous,
		\begin{align}
			M^{rs}(x,y) &=- \partial_iD_i^{rs}\delta(\textbf{y} - \textbf{x} ) \delta(y_0-x_0) := \delta(y_0-x_0) M^{rs}(\textbf x,\textbf y )\ , 
		  \end{align}
	and so the contribution of the last term in \eqref{5_coul_spatialeom} is instantaneous,   
	      \begin{multline}
	      i \vev{A_j^s(y)\,s\bar s A_i^r(x)} = -i \vev{(D_j c)^s(y)\,(D_i\bar c)^r(x)} \\ = 
	      -i \vev{(D_j^{bc}(y) \,(D_i^{ad}(x)\left[M^{-1}(\textbf{y, \textbf x}) \right]^{cd}} 
	      \delta(y_0-x_0)\label{coul_inst} 
	      \end{multline}
	and its Fourier-transform depends on spatial momenta only. The equation of motion of the ghost determines the longitudinal part of the correlation function
	      \be  -i \Fvev{(D_j c)^s(y)\,(D_i\bar c)^r(x)} = - t_{ij} u_C(\textbf p ^2) + l_{ij}\,,\ee
	where $t_{ij}$ and $l_{ij}$ are the longitudinal and transverse spatial projectors. The confinement criterion reads,
	      \be\label{5_Coulconf} \lim_{\textbf p ^2\rightarrow 0} u_C(\textbf p ^2) = -1 \qquad    \text{ and} \qquad \lim_{\textbf p ^2\rightarrow 0} f_C(p_0,\textbf p ) = 0,   \ee
	where the function $f(p_0,\textbf p)$ is defined by
		\be -\Fvev{A_k^s(y)\,\partial_\nu F_{\nu j}^r(x)} = \delta^{rs}t_{kj} f(p_0,\textbf p)\,,\ee
	which ensures that the spatial DSE is saturated by unphysical degrees of freedom in Coulomb gauge.

\subsection{Covariant Gribov-Zwanziger Model (Minimal Landau Gauge)\label{sec_5_GZ}}	  
	
	As last example for checking the generalized confinement criterion the Gribov-Zwanziger framework is investigated. It was introduced in \secref{sec_gz}. The gauge fixing which includes the horizon condition as non-perturbative gauge-fixing condition is given in \eqref{1_gz_lag}. The corresponding BRST transformations in terms of the shifted fields are given in \eqref{1_gz_shiftBRST}.
	
	The auxiliary ghost-fields, $\vph^{a}_{\mu b},\bar\vph^{a}_{\mu b},\omega^{a}_{\mu b}, \text{and } \bar \omega^{a}_{\mu b}$ are vector fields with two color indices that transform under the adjoint representation of the global color group. These color indices transform independently as  
	    \be \delta\Psi^{a}_{\mu b} = g f^{acd} \Psi_{\mu b}^{c} \delta\vartheta^d +  g f^{bcd}\Psi_{\mu c}^{a}\delta\vartheta^d \label{5_gz_traf2} \,\ee
	for any $\Psi^{a}_{\mu b} \in \{\vph^{a}_{\mu b},\bar\vph^{a}_{\mu b},\omega^{a}_{\mu b},\bar \omega^{a}_{\mu b}\}$. The equation of motion of the gluon implied by the Lagrangian $\L_{GZ} = \LYM + \L_{GZ}^{gf}$ is given by
	      \begin{multline}\label{5_gz_eom1}
		  \var[S_{GZ}]{A_\mu^r}  = - \partial_\nu F_{\nu\mu}^r - j_\mu^{lc\,r} + s(D_\mu \bar c)^r 
		    + \cp{\vph}{\partial_\mu \bar \vph}{r} + \cp{\omega}{\partial_\mu\bar\omega}{r} \\
		    - \cp{c}{\cp{\partial_\mu \omega}{\vph}{}}{r}  - \gamma^{1/2} \cp{c}{\tr{\omega_\mu}}{r} + \gamma^{1/2} \tr[r]{\vph_\mu-\bar 
		    \vph_\mu} \,,
	      \end{multline}
	where all indices that are summed over are suppressed. Contractions with structure constants in the ``covariant" and ``contravariant" color indices are denoted by
	    \be \cp{\Psi}{\Omega}{r} = g f^{rst} \Psi^{s}_{\mu b} \Omega^{t}_{\mu b} \,,\qquad\text{and,}\qquad \cpt{\Psi}{\Omega}{r} = g f^{rst} \Psi^{b}_{\mu s} \Omega^{b}_{\mu t} \,. \ee
	\eqref{5_gz_eom1} includes the global color current $j_\mu^{lc\,r}$ of linear covariant gauge given in \eqref{5_lcg_current}. However,  $j_\mu^{lc\,r}$ is not the conserved color current of the Gribov-Zwanziger action since the auxiliary fields transform according to \eqref{5_gz_traf2}. The corresponding conserved color current is
	      \begin{align} \label{5_gz_cur}
		  j_\mu^{GZ\,r} & = j_\mu^{lc\,r} + \cp{c}{\cp{\partial_\mu \bar \omega}{\vph}{}}{r} + 
		  \gamma^{1/2}\cp{c}{\tr{\bar\omega_\mu}}{r}   - \cp{\vph}{\partial_\mu \bar \vph}{r} - \cpt{\vph}{\partial_\mu \bar \vph}{r} \\
		   &-  \cp{\bar \vph}{D_\mu \vph}{r}  - \cpt{\bar \vph}{D_\mu \vph}{r} - \cp{\omega}{\partial_\mu\bar \omega}{r} -
			  \cpt{\omega}{\partial_\mu\bar \omega}{r}\label{gz_cur}  + \cp{\bar \omega}{D_\mu \omega}{r} 
			  \nonumber \\
			  & + \cpt{\bar \omega}{D_\mu \omega}{r}   + \cp{\bar\omega}{\cp{D_\mu c}{\vph}{}}{r} + 
			  \cpt{\bar\omega}{\cp{D_\mu c}{\vph}{}}{r} \,. \nonumber
	      \end{align}
	  Using \eqref{5_gz_cur} the equation of motion \eqref{5_gz_eom1} may be  rewritten as, 
	      \be \var[S_{GZ}]{A_\mu^r} = - \partial_\nu F_{\nu\mu}^r - j_\mu^{GZ\,a} + s\chi^r_\mu \,, \label{5_gz_eom2} \ee
	  with 
	      \be  \chi^r_\mu = (D_\mu \bar c)^r - \cp{\bar \omega}{D_\mu \vph}{r}  - 
	      \cpt{\bar \omega}{D_\mu \vph}{r} - \cpt{\vph}{\partial_\mu \bar \omega}{r} - \gamma^{1/2} \tr[r]{\bar\omega_\mu} \,. \ee
	  The gluon Dyson-Schwinger equation of the Gribov-Zwanziger action therefore has the now already familiar form
	      \be \delta^{rs}\delta_{\mu\sigma} = -\Fvev{A_{\sigma}^r(y)\,(\partial_\nu F_{\nu\mu}^s +j_\mu^{GZ\,s})(x) }  + i\Fvev{A_{\sigma}^r(y)\,s\chi^s_\mu(x) }\,.  \label{5_gz_dse} \ee
	  Color transport is short-ranged and the current matrix element does not contribute in the infrared if the functions $f_{GZ}(p^2)$ and $u_{GZ}(p^2)$ defined by
		\begin{align}
			-\Fvev{A_{\sigma}^r(y)\,\partial_\nu F_{\nu\mu}^s (x) } &=\delta^{rs}T_{\sigma\mu}\, f_{GZ}(p^2) \,,\label{5_fGZ}\\
			i\Fvev{A_{\sigma}^r(y)\,s\chi^s_\mu(x)} &=\delta^{rs}\left(L_{\sigma\mu}-T_{\sigma\mu}\,u_{GZ}(p^2)\right)\,,\label{5_uGZ}
		\end{align}
	  satisfy the criteria
		\be\label{5_confGZ} f_{GZ}(0)=0\qquad\text{and}\qquad u_{GZ}(0)=-1\ . \ee
	  However, in this case of a spontaneously broken BRST symmetry it is not entirely clear that,
		\be \label{5_gz_quartet}
		      0=\vev{s(A_{\sigma}^r(y)\, \chi^s_\mu(x))}  =\vev{A_{\sigma}^r(y)\, s(\chi^s_\mu(x))}+\vev{(D_\sigma c)^r(y)\, \chi^s_\mu(x)}\ ,
		 \ee
	  which would imply that only (unphysical) quartet states contribute to  the matrix element of  \eqref{5_uGZ}. Due to the equations of motion of the anti-ghost $\bar c$ and of the NL field $b^r$ the longitudinal part of \eqref{5_gz_quartet} is satisfied.  Although a proof is lacking, it therefore is at least plausible that the transverse part of \eqref{5_gz_quartet} also holds.
  
	  The Gribov-Zwanziger action incorporates non-perturbative features and in fact satisfies the criteria (\ref{5_confGZ}) already at tree-level. Expanding the gluon Dyson-Schwinger equation to tree level yields, 
	      \begin{multline}
		  \delta^{rs}\delta_{\sigma\mu} \approx p^2 T_{\mu\nu} \Fvev{A_\sigma^s(y)\, A_\nu^r(x)} +  \Fvev{A_\sigma^s(y) \,i\partial_\mu b^r(x)}\\ + \gamma^{1/2} g f^{acd} \Fvev{A_\sigma^s(y) \, (\vph^{c}_{\mu d} -  \bar \vph^{c}_{\mu d})(x)} \,. \label{5_gz_eom_tl}
	      \end{multline} 
	    One again has that the longitudinal part of the gluon propagator is saturated by  the NL field as in the foregoing investigations. The transverse part of \eqref{5_gz_eom_tl} is satisfied  by the  tree-level propagators, given for example in \cite{Vandersickel:2012tz, Gracey:2009mj}, 
	      \be  \vev{A_\sigma^s(y)\, A_\mu^r(x)}_{FT} \approx \delta^{ab} \,T_{\sigma\mu} \,\frac{p^2}{p^4 +  \lambda^4} \ee
	      and,
	      \be \vev{A_\sigma^s(y) \, (\vph^{c,d}_\mu -  \bar \vph^{c,d}_\mu)(x)}_{FT}  \approx  f^{bcd} \,T_{\sigma\nu}\, \frac{2g \gamma^{1/2}}{p^4 + \lambda^4} \,.\ee
	    The parameter $\lambda$ is defined via the relation $\lambda^4 = 2 N_c g^2 \gamma$. The Gribov-Zwanziger gluon propagator vanishes in the infrared, and $f_{GZ}(0)=0$.  In addition the last term in \eqref{5_gz_eom_tl}, derived entirely from the BRST exact term in  \eqref{5_gz_eom2}, saturates the  transverse part of the gluon Dyson-Schwinger equation at tree-level for vanishing momenta, 
	      \be \delta^{ab}T_{\sigma\mu} =  \delta^{ab}T_{\sigma\mu} \frac{p^4}{p^4+\lambda^4} + \delta^{ab}T_{\sigma\mu} \frac{\lambda^4}{p^4+\lambda^4} \,. \ee
	    Quite strikingly, both criteria of \eqref{5_confGZ} for a confining phase are thus satisfied  by the Gribov-Zwanziger Lagrangian already at tree level. Perturbative calculations to one- and two-loop order in three \cite{Gracey:2010df} and four \cite{Ford:2009ar,Gracey:2009zz,Gracey:2009mj} dimensions as well as a non-perturbative infrared analysis show that in the infrared the gluon propagator remains suppressed beyond tree-level. The  ghost propagator diverges more strongly than a massless pole \cite{Zwanziger:1992qr} and the Dyson-Schwinger equations as the corresponding infrared exponents of the scaling solution remain unaffected, \cite{Zwanziger:2001kw,Huber:2009tx}. 	  
	  
\section{Summary}

	In this section, a generalization of the Kugo-Ojima confinement criterium was formulated that can be transported to the generalized covariant gauge, MAG, Coulomb gauge, as well as the Gribov-Zwanziger framework. It was found that the gluon Dyson-Schwinger equation can be saturated by BRST-exact terms only in the infrared.

	In the considered cases the Dyson-Schwinger equation of the gauge boson reads
	    \be\label{5_DSEG}
	    \delta_{\sigma\mu}\delta^{rs} = -\Fvev{A^r_\sigma(y)\, \partial_\nu F^s_{\nu\mu}(x)} -\Fvev{A^r_\sigma(y)\, \tilde j^s_\mu (x)} \,.
	    \ee
	The current $\tilde j_\mu^r(x)$ is given by the canonical Noether current $j^r_\mu (x)$ of the global gauge symmetry plus a BRST-exact contribution,
	\be\label{5_tj}
	\tilde j_\mu^r(x)=j_\mu^r(x) +s \xi_\mu ^r(x) \ .
	\ee
	The phases are distinguished by the manner in which the transverse projection of \eqref{5_DSEG} is saturated in the infrared limit.

	The Coulomb phase is characterized by $f(p^2\rightarrow 0)>0$, i.e. contributions of the first term in \eqref{5_DSEG} in the infrared, which yields physical transverse gauge bosons of infinite range. 
	
	If $f(p^2\rightarrow 0)=0$ the theory is either in the Higgs-phase or the confining phase, i.e. the gauge boson is only of finite range.
	
	In the study here in principle we find a difference between the saturation of the gauge-boson Dyson-Schwinger equation in the infrared between the Higgs and the confining phases. This difference manifests itself in the contributions of the current \eqref{5_tj} in the infrared. In the Higgs phase the transverse part of \eqref{5_DSEG} is saturated by the contribution of the physical gauge boson mass in $j_\mu^r(x)$. Thus physical contributions saturate \eqref{5_DSEG} in the infrared in the absence of a massless gauge boson. If this is not the case, then \eqref{5_DSEG} has to be saturated by the unphysical (BRST-exact) contributions to \eqref{5_tj}. This pattern is then recognized as the confining phase of a gauge theory. The influence of the gauge fixing aspects on the Higgs-confinement transition is beyond the scope of the study presented here. An analysis of the non-Abelian Higgs model is desirable to investigate this interrelation. In pure gauge theory there is only the confining phase.

%% file: 6_conclusio.tex
\chapter{Summary and Outlook \label{sec_conclusio}}

In this thesis the Dyson-Schwinger equation of gauge bosons were investigated with respect to signals of confinement. 

In the first part the  Dyson-Schwinger equation of the gluon propagator in Landau gauge, \secref{sec_LG}, and Maximal Abelian gauge, \secref{sec_MAG}, was investigated under respective approximations. The level of approximation in the Landau gauge was improved compared to foregoing studies by the first full inclusion of the sunset diagram into this calculation. The main technical obstacle introduced by two-loop diagrams are overlapping quadratic divergences. These divergences could be subtracted via an explicit counter term construction which can be interpreted as mimicking the non-perturbative tensor structure of the truncated n-point functions. Even though the overlapping divergences have different structures in different momentum configurations, it was possible to subtract all quadratic divergences with one additional term in the tensor structure. 

The method developed in this work is applicable to any kind of two-loop term in the Dyson-Schwinger equations as, e.g. the squint-diagram. Still more advanced subtraction techniques are desirable and can be developed on basis of the results presented here. A self-consistent solution of the Dyson-Schwinger equations of the ghost and gluon propagators with all terms included, is the next step, which should be performed. Such a calculation is also needed to properly calculate back-coupling effects of the three-gluon vertex. A calculation which takes into account all terms in the equations for the propagators and which closes one the level of three-point functions would be a major step which now seems to be possible.

The knowledge gained in the Landau gauge calculations was then transported to the MAG, where the first solution of the Dyson-Schwinger equation of the ghost propagator was performed. The sunset diagram here is the dominant contribution in the low-energy regime. 

In the MAG the investigation presented here is only one step further on the long way of a detailed investigation of the QCD Green functions in this gauge. More work needs to be done in the gauge sector but also matter-gauge sector. Compared to Landau gauge the Dyson-Schwinger equations in the MAG are mostly unknown territory. Beside the gluon and ghost propagators the most interesting Green functions seem to be the quark-gluon vertices. In this studies, however, one has to take into account that the needed three and four-point functions are even less explored than in Landau gauge.

In the second part of this thesis attempts where made to transport aspects of the Kugo-Ojima confinement scenario to various gauges. In  \secref{sec_cc} focus was put on the quartet mechanism with respect to Faddeev-Popov conjugation invariant gauges in the two-parameter family of generalized covariant gauges. The original Kugo-Ojima formalism suffers from the shortcoming that ghost and antighost fields are treated differently and as such breaks Faddeev-Popov conjugation symmetry of Landau gauge. In this work it was shown that with an Faddeev-Popov conjugation symmetric assignment of the asymptotic fields, one can keep this symmetry without contradictions in gauges with this symmetry. The Kugo-Ojima scenario can be generalized to these gauges without contradiction. The role of the anti-BRST symmetry stays unclear. Still it is the first known generalization of the Kugo-Ojima scenario to other gauges and can lead the way to further studies in other Faddeev-Popov conjugation symmetric gauges.

By interpreting the Kugo-Ojima confinement criterion in terms of correlation functions in the Dyson-Schwinger equation of the gluon propagator, in \secref{sec_saturation}, a generalization of this criterion was found which can be transported to other gauges as well. If the Kugo-Ojima confinement criterion is fulfilled, then the gluon Dyson-Schwinger equation in Landau gauge in the deep infrared is saturated by unphysical degrees of freedom only. Practically this is just a rephrasing of the long-known ghost-dominance in this gauge. Analogous unphysical contributions could be found in all other gauges investigated in Yang-Mills theory and the Gribov-Zwanziger model, but not in QED or the Abelian-Higgs model. While the longitudinal sector is always saturated by unphysical contribution, the transverse sector of the Dyson-Schwinger equation must be saturated by physical degrees of freedom in the Abelian case. 

Further checks of the generalized confinement criterion in other gauges, e.g. axial gauge or light-cone gauge, are interesting. In particular as the definition of the physical state-space in these gauges might differ from the ones used in this work. A more detailed investigation of the Higgs/confinement transition can be performed in Non-Abelian--Higgs models. Interesting gauges for this model are the Landau gauge, as it directly connects to the Kugo-Ojima scenario, the 't Hooft gauge, as it connects to the Abelian-Higgs model presented in this work, and the MAG, as in this gauge the Higgs/confinement transition might be visible most clearly.

Generally, the studies in \secref{sec_cc} and \secref{sec_saturation} show that it is possible to generalize existing confinement criteria from one gauge to other ones. The ambition of this kind of investigations is a gauge-invariant definition of confinement in gauge-fixed studies.

%% file: A1_conventions.tex
\chapter{Conventions\label{secA_con}}

\section{Gauge Theory in Euclidean Spacetime\label{secA_gt}}
    Consider a matter field $\phi(x)$ which transforms under a gauge transformations $U(x)$ as 
	\be \phi(x) \rightarrow \phi'(x) = U(x)\phi(x) = e^{ig\vartheta^r(x)T^r} \phi \approx  \phi + i g \vartheta^r(x) T^r \phi  \ee
    with the $N$ generators of the gauge group $\mathcal G$ given in some representation, $T^r$, $r=1,\dots,N$, the spacetime dependent real gauge parameters $\vartheta^r(x)$ and the gauge coupling $g$. The gauge coupling $g$ is a real constant parameter which in principle can be absorbed into the gauge parameter by redefinition. The generators are considered to be traceless and Hermitian and fulfill the commutation and normalization relations
	\be \com{T^r}{T^s} = i f^{rst}T^t\,, \qquad \text{and, }\qquad \tr{T^rT^s} = \frac{\delta^{rs}}{2}\,,  \label{A_def_gen}\ee
    where the real structure constants of the gauge group $f^{rst}$ were defined. The structure constants are totally anti-symmetric and fulfill the Jacobi-identity
	\be f^{rst}f^{ruv} + f^{rsu}f^{rvt} + f^{rsv}f^{rtu} = 0 \,. \label{A_Jacobi}\ee

    By construction, the covariant derivative shall transform covariantly under gauge transformations,
	\be D_\mu\phi(x)\rightarrow \left(D_\mu\phi(x)\right)'=U(x)(D_\mu\phi(x)) \label{A_trafo_covder}\,.\ee
    This particular transformation behavior can be achieved by defining the covariant derivate as
	\be D_\mu\phi =  \partial_\mu - i g A_\mu^r(x) T^r\,,  \ee
   with the gauge fields $A_\mu^r(x)$ transforming as
	\begin{align} 
	    A^r_\mu(x) T^r  \rightarrow \left(A^r_\mu(x)T^r\right)^U & = U(x)A^r(x)T^rU^{-1}(x) - \frac{i}{g} \left(\partial_\mu U(x)\right) U^{-1}(x)\\
					     & \approx A^r_\mu(x)T^r + \left(\partial_\mu \vartheta^r(x) - g \vartheta^s(x)A_\mu^t(x) f^{rst}\right)T^r 
	\end{align}
    Using the generators of the adjoint representation of the gauge group, $(T^r)_{st} = i f^{srt} $, one finds for the infinitesimal variation of the gauge fields $A_\mu^r(x)$
	\be \delta_\vartheta A_\mu^r(x) = D_\mu^{rs}\vartheta^s(x)\,,\label{A_trafo_A} \ee
    with the covariant derivative in the adjoint representation given explicitly as
	\be D_\mu^{rs} = \delta^{rs}\partial_\mu + g f^{rts} A_\mu^t \,. \ee
    In the course of this thesis the notion of the adjoint cross-product is used frequently. It is defined as the contraction of two fields with the structure constants of the group,
	\be \cp{\phi}{\psi}{r} = gf^{rst}\phi^s\psi^t \label{A_def_cp} \,.\ee
    It can be used to rewrite the covariant derivative in the adjoint representation as
	\be D_\mu^{rs}\phi^s = \partial_\mu\phi^r + \cp{A_\mu}{\phi}{r} \label{A_def_covder}\,.\ee
    The fundamental representation of the $SU(N)$ group is generated by the generalization of the Gell-Mann matrices, $(T^r)_{st} = \tfrac 1 2 \lambda^r_{st}$. The covariant derivative in the fundamental representation then reads6
	\be (D_\mu)_{rs}\phi_s = \partial_\mu\phi_r - igA_\mu^t \frac{\lambda^t_{rs}}{2} \phi_s \label{A_def_fundcovder}\,.\ee

    The field strength tensor can be defined via the commutator of the covariant derivative with itself,
	\be  T^rF^r_{\mu\nu} = \frac{i}{g}\com{D_\mu}{D_\nu} = T^r\left(\partial_\mu A_\nu^r - \partial_\nu A_\mu^r + \cp{A_\mu}{A_\nu}{r} \right)\,.\label{A_def_fieldstrength}\ee
    The Lagrangian of Yang-Mills theory \cite{Yang:1954ek} is proportional to the square of the field strength tensor
	\begin{align} 
	      \LYM &= \frac{1}{4}F_{\mu\nu}^r F_{\mu\nu}^r \label{A_def_LYM}\\
		   & = -\frac{1}{2} \delta^{rs} A_\mu^r\left(\delta_{\mu\nu}\partial^2 - \partial_\mu\partial_\nu \right)A_\nu^s+ (\partial_\mu A_\nu^r) \cp{A_\mu}{A_\nu}{r} + \frac{1}{4} \cp{A_\mu}{A_\nu}{r}\cp{A_\mu}{A_\nu}{r}\,.\nonumber	 
	\end{align}

    An Abelian gauge theory is a special case of the aboves considerations. In this case all structure constants in \eqref{A_def_gen} vanish. The gauge transformations simplify to
	\be
	    A_\mu(x)   \rightarrow A_\mu^U(x)  =  - \frac{i}{g} \left(\partial_\mu U(x)\right) U^{-1}(x) \approx A_\mu(x) + \partial_\mu \vartheta(x)  
	\ee
    and thus $\delta A_\mu(x) = \partial_\mu \theta$. The covariant derivative in the Abelian case reads 
	\be D_\mu\phi =  \partial_\mu - i g A_\mu(x) \ee
    which yields the Abelian field strength tensor $F_{\mu\nu} = \partial_\mu A_\nu - \partial_\nu A_\mu$ and the Lagrangian of the Maxwell theory
	\begin{align} 
	      \LM = \frac{1}{4}F_{\mu\nu} F_{\mu\nu} = -\frac{1}{2} A_\mu\left(\delta_{\mu\nu}\partial^2 - \partial_\mu\partial_\nu \right)A_\nu\,.
	\end{align}

    \subsection*{Fourier Transform\label{app_FT}}
    The Fourier transform of a function $f(x)$ is defined as
	\be \mathcal {FT}[f(x)] = \int \!d^4 x\, e^{-i\, p\cdot x} f(x) \equiv \F{f}(p) \ee
    whereas the back transformation is defined as
	\be \mathcal {FT}^{-1}[\F{f}(p^2)] = \int \!\dbar^4 p\, e^{i\, p\cdot x} \F{f}(p) = f(x)\,.\ee
    The derivative operator transforms into the momentum $\partial_\mu^x \leftrightarrow ip_\mu$. The vacuum expectation value of an operator is defined as $\vev{\O} = \mathcal{N} \int\![dA] \,\O \, e^{-\S}$. If the operator is local, $\O = \O(x)$, then its Fourier-transform is defined as above,
	  \be \Fvev{\O(x)} =  \int\!\! d^4 x\,\, e^{-ip\cdot x} \, \vev{\O(x)} = \mathcal{N} \int\![dA] \, \left( \int\!\! d^4 x\,\, e^{-ip\cdot x} \,\O(x) \right)\,e^{-\S}\,.\ee
    Is the operator non-local, however, the situation is ambiguous, as it is not clear which position space coordinate should be integrated over. In this work, in particular in \secref{sec_saturation}, the following definition is used. Given a non-local operator at two spacetime points $\O = \O_1(y) \O_2(x)$. Then the Fourier transform of the vacuum expectation value $\vev{\O}$ of the operator is given by
	  \be \Fvev{\O} = \Fvev{\O_1(y)\, \O_2(x)} \equiv    \int\!\! d^4\xi\,\, e^{-ip\cdot\xi}  \vev{\O_1(x+\xi)\, \O_2(x)} \,,\ee
    with $\xi = y-x$. Depending on the spacetime point the derivative operator now picks up a sign,
	  \be \partial_\mu^y \leftrightarrow ip_\mu \,,\qquad\text{while,}\qquad \partial_\mu^x \leftrightarrow -ip_\mu \,.\ee

    \subsection*{Integral Measure}
    A general four vector in Euclidean space can be expressed in spherical coordinates as
    \begin{equation}\label{A_kdef}
    \k =   \begin{pmatrix} \k_0 \\ \k_1 \\ \k_2 \\ \k_3  \end{pmatrix} = k \begin{pmatrix} \sin\psi \sin\theta  \sin{\varphi}\\ \sin\psi \sin\theta \cos{\varphi} \\ \sin\psi  \cos \theta \\ \cos \psi    \end{pmatrix}   = k \begin{pmatrix} \sqrt{1-z^2} \sqrt{1-y^2}\,  \sin{\varphi}\\ \sqrt{1-z^2} \sqrt{1-y^2}\,  \cos{\varphi} \\ \sqrt{1-z^2}\,  y \\ z    \end{pmatrix} \,,
    \end{equation}
    with $k = \sqrt{(\k\cdot\k)} =  \lvert\k\rvert$. The corresponding integral measure in momentum space reads,
    \begin{multline}
    \int\!\dbar^4k = \frac{1}{(2\pi)^4} \int_0^\infty\!\!dk^2\, \frac{k^2}{2} \, \int_{-1}^1\!\! dz \,\sqrt{1-z^2}\, \int_{-1}^1\!\! dy\, \int_{-\pi}^\pi\!\! d\varphi \\ =  \frac{1}{(2\pi)^4} \int_0^\infty\!\!dk\,k^3\, \int_{-1}^1\!\! dz \,\sqrt{1-z^2}\, \int_{-1}^1\!\! dy\, \int_{-\pi}^\pi\!\! d\varphi    \label{A_measure_ol}\,.
    \end{multline}
    For brevity in this thesis the following notation is used,
	    \be  dz \,\sqrt{1-z^2} \equiv \dzs \,. \label{a_def_dzsq}\ee

\section{Currents and Charges in Classical and Quantum Theories\label{app_currents}}

    An alternative to the path-integral quantization of a quantum field theory is the canonical formalism. Even though it is explained in many textbooks, e.g. \cite{NakanishiOjima:1990,PeskinSchroeder}, the main formulas relevant for this thesis are collected here to make it self-contained and to fix the conventions. In this section all sums are denoted explicitly.

\subsubsection*{Action, Lagrangian and Hammilton}
    A classical field theory of the set of fields $\phi = \{\phi_i\}$ in four spacetime dimensions is defined the action $\S$
	\be \S = \int\!\!d^4x\,\L(\phi,\partial \phi) \,.\ee
    where $\L$ represents the Lagrangian of the theory. The action is a functional of fields, which vanish at infinity. Its variation is given by
	\be \delta \S =  \int\!\!d^4x\,\sum_i\,\delta\phi_i\,\left( \var[\L]{\phi_i} - \partial_\mu \var[\L]{\partial_\mu\phi_i} \right)\,.\ee
    The principle of least action then immediately yields the Euler-Lagrange equations for the field $\phi_i$
	\be 0 = \var[\S]{\phi_i} =  \var[\L]{\phi_i} - \partial_\mu \var[\L]{\partial_\mu\phi_i} \,.\ee
    The Hammiltonian of the theory is a function of the fields and their canonical momenta $\pi = \{\pi_i\}$ and is defined via the Legendre transform of the Lagrangian
	\be \mathcal H (\phi,\pi) = -\L (\phi,\partial_\mu \phi) + \sum_i (\partial_0 \phi_i) \pi_i\,.\ee
    Since the Hammiltonian does not depend on the derivatives of the fields one immediately gets for the canonical momenta of the field $\phi_i$
	\be \pi_i = \var[\L]{\partial_0 \phi_i}\,. \ee

\subsubsection*{Currents and Charges in Classical Theories}
    Given the theory is invariant under the continuous transformations $\delta_\Lambda$ of the fields, $\delta_\Lambda\S = 0$, i.e. $\delta_\Lambda$ represents a continuous symmetry of the theory, there exists a current $J_\mu^{\Lambda}$ which is conserved, $\partial_\mu J_\mu^{\Lambda} = 0$. This current is given by
	\be J_\mu^{\Lambda} = \sum_i\,  \delta_\Lambda \phi_i \,\var[\L]{\partial_\mu \phi_i}\,.\label{A_noethercurr}\ee
    To any conserved current there exists a conserved charge as the 3-integral over the time component,
	\be Q^{\Lambda} =\int\!\!d^3x \,  J^\Lambda_0 = \sum_i\int\!\!d^3x \,  \delta_\Lambda \phi_i \,\var[\L]{\partial_0 \phi_i} =\sum_i \int\!\!d^3x \,  \delta_\Lambda \phi_i \,\pi_i \,. \label{A_noethercharge}\ee
    The charge $Q^{\Lambda}$ is conserved. It is independent on the time $x_0$.

\subsubsection*{Currents and Charges in Quantum Theories}
    In canonical quantization one implies commutation relation onto the canonical variables of the theory. For a field theory this are the fields and their canonical momenta. In Euclidean spacetime the canonical commutation relations read
	\be \com{\phi_i(x)}{\pi_k(y)} = i \delta^{(4)}(x-y) \delta_{ik} \,.\label{A_cancomrel}\ee
    The momentum operator in quantum mechanics in position-space representation is given by $\hat{\vec p} \rightarrow -i \vec \nabla $. In quantum field theory this language can be transported into a "field-space representation". If
	\be \pi_i \rightarrow -i \var{\phi_i} \ee
    the canonical commutation relations are fulfilled. Using this language, any classical Noether charge can be represented by the corresponding charge operator
	\be Q^{\Lambda} \rightarrow -i \sum_i \int\!\!d^4x \,  \delta_\Lambda \phi_i \,\var{\phi_i}\,.\label{A_charge_quantum}\ee
   This charge operator is the generator of the symmetry transformations,
	\be \com{iQ^{\Lambda}}{\phi_k} = \delta_\Lambda \phi_k\,.  \ee

%% file: A2_technicalities.tex
\chapter{Technicalities}

\section{Numerical Solution of the Propagator Equations\label{secA_YMtech}}
    In this appendix numerical details on solution of the coupled system of ghost and gluon Dyson-Schwinger equations in Landau gauge are presented. The same techniques can be transported to other systems of Dyson-Schwinger equations. The main ideas presented here were developed in \cite{Atkinson:1997tu,Fischer:2003zc}. For a calculation using GPUs see \cite{Hopfer:2012ht}.

    \subsection*{Chebychev Expansion and Input Functions}
    Before the equations itself can be considered the ghost and gluon propagator dressing functions have to be prepared for the numerical treatement. From analysis of the low and high-energy asymptotics of the ghost and gluon dressing functions as performed in \secref{sec_2_olo}, one knows that in the deep infrared the dressing functions obtain a power-law behavior, \eqref{2_powerlaw}, while in the ultraviolet they resemble the logarithmic scaling of one-loop perturbation theory, \eqref{2_def_gend}.  A simple ansatz for a start function $d_i\in\{Z(p^2),G(p^2)\}$ which has been proven successful reads
	\be d_i(p^2) = c_i \left(p^2\right)^{\delta_i} \,\left(\frac{s}{s+p^2} \right)^a + e_i \left(\frac{p^2}{s+p^2} \right)^b \left(g^2 \beta_0 \log{\frac{p^2}{\sigma_{gl}}+1}+1 \right)^{\gamma_i}\label{B_olo_ansatz}\ee
    The parameter $s$ sets the scale of the transition from the IR to the UV regime, the parameters $a$ and $b$ can be used to model this transition smoothly. Usual values are $s=1$ and $a,b\in\{1,2,3\}$. The infrared exponents $\delta_i$ and the anomalous dimensions $\gamma_i$ have been obtained in \secref{sec_LG}. The values of $e_i$ are taken to be of order $1$ and the infrared coefficients $c_i$ are related via \eqref{2_olo_ircon}. One of them is choosen of order $1$. If one only recalculates the gluon and ghost system in Landau gauge in one-loop-only truncation one can also use the fits given in \cite{Fischer:2003zc} to gain faster convergence. 

    The Dyson-Schwinger equations are solved for some functions $Z(p^2)$ and $G(p^2)$. Since in a numerical calculations these functions will always be known only on a finite set of external momenta $\{p^2_k\}$ one has to use appropriate interpolation and extrapolation techniques, as e.g. splines or Pad\'e approximations. In this work a Chebychev expansion is used as explained in detail in \cite{NumRecF77}. Therefore the dressing functions are approximated by $N_x$ Chebyshev polynomials. The two sets of coefficients are denoted by $\{a_k\}$ for the gluon dressing function, $Z(\{a_k\};p^2)$, and $\{b_k\}$ for the ghost dressing function $G(\{b_k\};p^2)$ with $k=1,\dots,N_x$. The $N_x$ roots $y_k$ of the highest Chebyshev polynomial are mapped from the interval $(-1,1)$ onto an interval $(\eps,\lambda)$ with the mapping
	\be x_k = \sqrt{\lambda\,\eps} \left(\frac{\lambda}{\eps}\right)^{\frac{y_k}{2}} \,.\label{B_olo_extgrid}\ee
    The parameter $\eps$ is chosen to be in the deep-infrared regime, while $\lambda$ is in the perturbative high-energy regime. In the course of the calculation the dressing functions will also be tested outside the interval $(\eps,\lambda)$. In the infrared regime $(0,\eps)$ one assumed the power law behavior
	\be Z(\{a_k\};p^2) \stackrel{p^2<\eps}{=} c_A (p^2)^{\delta_A}\,,\qquad\text{and}\qquad G(\{b_k\};p^2) \stackrel{p^2<\eps}{=} c_c (p^2)^{\delta_c}\,,\ee
    while for high energies the perturbative asymptotics is assumed,
	\begin{align} 
	    Z(\{a_k\};p^2) & \stackrel{p^2>\lambda}{=} Z(\{a_k\};\sigma_{gl}^2) \left( \frac{11 N_c}{12\pi} \alpha_s(\{a_k\},\{b_k\};\sigma_{gl}^2) \,\log{\frac{p^2}{\sigma_{gl}^2}}+1\right)^{\gamma_A} \\
	    G(\{b_k\};p^2) & \stackrel{p^2>\lambda}{=} G(\{b_k\};\sigma_{gl}^2) \left( \frac{11 N_c}{12\pi} \alpha_s(\{a_k\},\{b_k\};\sigma_{gl}^2) \,\log{\frac{p^2}{\sigma_{gl}^2}}+1\right)^{\gamma_c}
	\end{align}
    with the strong running coupling of the ghost-gluon vertex as defined in \eqref{2_coupl},
	\be \alpha_s(\{a_k\},\{b_k\};p^2) = \frac{g^2}{4\pi}\,Z(\{a_k\};p^2)G(\{b_k\};p^2)^2 \,.\ee
    The value $\sigma_{gl}$ is the subtraction point of the gluon equation $\sigma_{gl}<\lambda$.

    \subsection*{Self-Energies and Integrals}
    The equations to be solved are given in \eqref{2_olo_MOM}. The self-energies \eqref{2_olo_Pi} read after integrating out the trivial angles
	  \begin{subequations}\label{A_LG_SE}
	  \begin{align}
	     \widetilde \Pi_{gh}^{olo}(\{a_k\},\{b_k\};p^2) & =  C_{ghost} \int_0^{\Lambda^2}\!\!dq^2\!\!\int_{-1}^{1}\! dz_q^{\sqrt{}} \,\,\K_{ghse} (p^2,q^2,z_q)\,D_{ghse}(\{a_k\},\{b_k\};q^2,k^2)\, \label{A_LG_ghDyson-Schwinger equation_MOM}\\
	       \widetilde\Pi_{gl}^{olo}((\{a_k\},\{b_k\};p^2) &=  C_{gluon} \int_0^{\Lambda^2}\!\!dq^2\!\!\int_{-1}^{1}\! dz_q^{\sqrt{}} \Bigl(\K_{gh} (p^2,q^2,z_q;\zeta)\,D_{gh}(\{a_k\},\{b_k\};q^2,k^2)\nonumber  
	      \\ &\hspace{20mm} + \widetilde{\K}_{gl}(p^2,q^2,z_q;\zeta)\,D_{gl}(\{a_k\},\{b_k\};q^2,k^2) \Bigr)\,. \label{A_LG_glSE}
	  \end{align} 
	  \end{subequations}
    Now and in the following the notation $k^2 = k^2(p^2,q^2,z_q) = (\p-\q)^2 = p^2+q^2 - 2\,z_q\,\sqrt{p^2\,q^2}$ will be used. The measure $\dzs$ is an abbreviation for $ \dzs =  dz\,\sqrt{1-z^2}$, \eqref{a_def_dzsq}.
    The integrals \eqref{A_LG_SE}  feature the prefactors,
	    \be C_{ghost} = -\frac{g^2 N_c}{(2\pi)^3}\,,\qquad\text{and,}\qquad C_{gluon} = \frac{g^2 N_c}{3(2\pi)^3}\,,\ee
    the integration kernels,
	    \begin{subequations}\label{A_LG_kernels}
	    \begin{align}
	      \K_{ghse}(p^2,q^2,z_q) &= \frac{q^2\left(1-z_q^2\right)}{k^4}\,,\\
	      \K_{gh} (p^2,q^2,z_q;\zeta) &= \frac{q^2(1-\zeta\,z_q^2) - (1-\zeta) \sqrt{p^2\,q^2}\, z_q}{p^2\,q^2\,k^2}\,, \\
	      \widetilde{\K}_{gl}(p^2,q^2,z_q;\zeta) & = \frac{1}{q^2} \Biggl[ \frac{10-3\zeta}{4}+ \frac{1}{k^2}\Biggl(q^2 \left(-\frac{9}{2} + 5\zeta \,z_q^2\right) + p^2 \left(\frac{\zeta-10}{2} + \frac{9+\zeta}{2} \,z_q^2\right) \label{A_def_kerQ}\\ 
		&\hspace{5mm} - \sqrt{p^2\,q^2} \,z_q\,\left(2+4\zeta+\zeta z_q^2\right) \Biggr) +\frac{1}{k^4}\Biggl(q^4 \left(\frac{2-\zeta}{2} + \zeta z_q^2\right) + p^4 \left(\frac{1}{2} +  z_q^2\right) \nonumber   \\
		&\hspace{5mm}+ p^2q^2\,\frac{z_q^2}{2}(\zeta - 1)- 3\, p^2q^2 + \sqrt{p^2\,q^2} \, z_q \left( q^2 \left(1-\zeta z_q^2 \right) + p^2 \left(1- z_q^2 \right) \right)\Biggr) \Biggr] \nonumber\,,
	    \end{align}
	    \end{subequations}
    and the dressing functions,
	    \begin{subequations}
	     \begin{align} 
		D_{ghse}(\{a_k\},\{b_k\};q^2,k^2) &= G(\{b_k\};q^2)Z(\{a_k\};k^2) \,,\\
		D_{gh}(\{a_k\},\{b_k\};q^2,k^2) & =  G(\{b_k\};q^2)G(\{b_k\};k^2)\,,\\      
		D_{gl}(\{a_k\},\{b_k\};q^2,k^2) & =  \left(Z(\{a_k\};q^2) Z(\{a_k\};k^2) \right)^{(1-\gamma_{3A})} \left(G(\{b_k\};q^2) G(\{b_k\};k^2) \right)^{-2\gamma_{3A}}\,.
	     \end{align}
	    \end{subequations}

    The integrals are solved using standard numerical methods, \cite{NumRecF77,NumRecF90}. The prefactor in the angular integral suggests Chebyshev integration weights of second kind. The radial integral is split into three integration ranges, $(0,\eps),(\eps,p^2),(p^2,\Lambda^2)$. One possible integration method is to use Gauss-Legendre integration, where the nodes are mapped such that they ideally are equally distributed on a logarithmic scale. In this work the following mapping is used \cite{privcom:HopfHor}. Given the Gauss-Legendre nodes $x^{(0,1)}$ in the interval $(0,1)$ and the corresponding weights $w^{(0,1)}$ one maps them onto the nodes $x^{(A,B)}$ and weights $w^{(A,B)}$ in the interval $(A,B)$ via 
      \begin{subequations} \label{A_mapping_radial}
	  \begin{align} 
	  x^{(A,B)} &= A + s \, \frac{ \left( 1 + \frac{B-A}{s}  \right) ^{ x^{(0,1)} } - 1 } { 1 + e - e^{x^{(0,1)}} }\,,\\
	    w^{(A,B)} &=  w^{(0,1)} \, \frac{ s\,\log{ 1 + \frac{B-A}{s} } \, \left( 1 + \frac{B-A}{s}  \right) ^{ x^{(0,1)} } + (x^{(A,B)}-A)\, e^{x^{(0,1)}} }  { 1 + e - e^{x^{(0,1)}} }\,.
	  \end{align}
      \end{subequations}
    The parameter $s$ allows for some control on the distribution of the nodes in the interval $(A,B)$. A choice which has been prooven useful is $s=A$.

    \subsection*{Newton-Method}
    In \secref{sec_2_olo} the equations have been treated in a MOM-scheme, which is an elegant way to get rid of the logarithmic divergences and the correspoding renormalization constants $Z_3$ and $\tilde Z_3$.  Defining the functions $f\left(\{a_k\},\{b_k\};p^2\right)$ and $g\left(\{a_k\},\{b_k\};p^2\right)$,
	\begin{subequations}
	\begin{align}
	   g\left(\{a_k\},\{b_k\};p^2\right) & \equiv -\frac{1}{G\left(\{b_k\};p^2\right)} + \widetilde \Pi_{gh}^{olo}\left(\{a_k\},\{b_k\};p^2\right)-\widetilde \Pi_{gh}^{olo}\left(\{a_k\},\{b_k\};0\right)  \\
	  f\left(\{a_k\},\{b_k\};p^2\right)  & \equiv -\frac{1}{Z\left(\{a_k\};p^2\right)}+ \frac{1}{Z(\sigma_{gl}^2)} \label{B_MOM_f1} \\
	  &\hspace{30mm} + \widetilde \Pi_{gl}^{olo}\left(\{a_k\},\{b_k\};p^2\right)-\widetilde \Pi_{gl}^{olo}\left(\{a_k\},\{b_k\};\sigma_{gl}^2\right) \,, \nonumber 
	\end{align}
	\end{subequations}
    the ghost and gluon Dyson-Schwinger equations in the MOM-scheme, \eqref{2_olo_MOM}, read
	\begin{subequations}\label{B_MOM_cond}
	\begin{align}
	   0 &= g\left(\{a_k\},\{b_k\};p^2\right) \,, \\
	  0 &= f\left(\{a_k\},\{b_k\};p^2\right) \,. \label{B_MOM_f2}
	\end{align}
	\end{subequations}
    The self-consistent solution of the Dyson-Schwinger equations numerically boils down to finding the root of the functions $f$ and $g$ for every external momentum $p^2$ with respect to the functions $Z(p^2)$ and $G(p^2)$, i.e. their expansion coefficients $\{a_k\}$ and $\{b_k\}$. One possible technique is the Newton method \cite{Atkinson:1997tu}. The $j$-th coefficient in the $n$-th and $(n+1)$-th iteration steps are related by
	  \be a_j^n = a_j^{n+1} + \Delta a_j^{n+1} \,,\qquad \text{and,}\qquad  b_j^n = b_j^{n+1} + \Delta b_j^{n+1}\,.\ee
   In one dimension a Newton step is given by $\Delta x_{n+1} =  \frac{f(x_n)}{f'(x_n)}$ which vanishes at the root. In more than one dimensions the Newton method implies to solve the set of linear equations $A_{n}\Delta x_{n+1} = B_n$ with the $(2N_x,2N_x)$-matrix
	\be 
	    A_n = 
	    \begin{pmatrix}
		  \var[f\left(\{a^n_k\},\{b^n_k\};p^2\right)]{a_j} & \var[f\left(\{a^n_k\},\{b^n_k\};p^2\right)]{b_j} \\
		  \var[g\left(\{a^n_k\},\{b^n_k\};p^2\right)]{a_j} & \var[g\left(\{a^n_k\},\{b^n_k\};p^2\right)]{b_j} 
	    \end{pmatrix}\,,
	\ee
    the solution vector $\Delta x_{n+1} = \left(\Delta a_j^{n+1},\Delta b_j^{n+1} \right)^T$ and the inhomogeneity 
	\be B_n = \left(f\left(\{a^n_k\},\{b^n_k\};p^2\right),g\left(\{a^n_k\},\{b^n_k\};p^2\right) \right)^T\,. \ee
    As external momentum grid one uses $\{p^2_k\} = \{x_k\}$ with the definition \eqref{B_olo_extgrid}. The variations in the Newton-matrix are calculated numerically as, e.g., 
	\begin{multline} \var[f\left(\{a^n_k\},\{b^n_k\};p^2\right)]{a_j} = \frac{1}{2\,h^n_j} \Bigl( f\left(\{a^n_1,\dots,a^n_j+h^n_j,\dots,a^n_{N_x}\},\{b^n_k\};p^2\right) \\- f\left(\{a^n_1,\dots,a^n_j-h^n_j,\dots,a^n_{N_x}\},\{b^n_k\};p^2\right) \Bigr) \label{B_deriv} \end{multline}
    with $h_j^{n} = h + \sqrt{h} \,a_j^{n}$ with some minimum stepsize $h$. The set of linear equations is solved using standard libraries which yields the new coefficients
	\be a_j^{n+1} = a_j^n - \omega \Delta a_j^{n+1} \,,\qquad \text{and,}\qquad b_j^{n+1} = b_j^n - \omega \Delta b_j^{n+1}\,. \label{B_olo_relax}\ee
   In \eqref{B_olo_relax} a relaxation factor $\omega$ was introduced, which can be used to improve convergence properties.  
   For the infrared fit of the dressing functions one needs the explicit values of $c_A$ and $c_c$. They are not independent since they are related via \eqref{2_olo_ircon}. One thus extracts one of the values directly from the calculation and calculates the other one via \eqref{2_olo_ircon}. This procedure is iterated until convergence is reached.

    \subsection*{Setting the Parameters}
    The calculation is performed on some external grid $\{p^2\}=\{x_k\}$. Before one sets the scale in the renormalization procedure the values $\{x_k\}$ do not have any physical meaning. Still it is advantageous to choose values which are of a similar order of magnitude as the physical results will be. The external momentum grid is completely determined by the parameter $\eps$ and $\lambda$ in \eqref{B_olo_extgrid}, of which in this work are chosen in the range
	\be \eps = 10^{-4\dots-6}\,,\qquad\text{and,}\qquad \lambda = 10^{4\dots 6}\,. \ee
    The final result must be independent on the cutoff $\Lambda$ which is tested in the range $\Lambda^2 = (10\dots 100) \lambda$. The subtraction point of the gluon equation is chosen to be in the high-energy regime inside the region of Chebychev approximation, $\sigma_{gl}^2 = (0.8\dots0.95)\lambda$.

    The number of Chebyshev polynomials $N_x$ is one of the most significant numbers with respect to the speed of the calculation, since the size of the Matrix $A_n$ and thus the number of integrals to be evaluated scale with $N_x^2$. If it is too low, the calculation is too crude and there maybe even will be no solution, if it is too high, the calculation is too expensive. A good compromise on standard desktop computers is $N_x = 20 \dots 25$. For the radial integration $N_q = 64$ integration points have been used per integration range. The angular integration is sufficiently exact with $N_z = 32$ nodes. The derivatives \eqref{B_deriv} have to be independent on the stepsize $h$, however if $h$ is too small on runs into numerical instabilities, in contrast if it is too big finite-size effects kick in. In this work $h=10^{-6}$ was used.

    \subsection*{Setting the Scales: Renormalization}
    
    In Landau gauge there are two renormalization condition to be fulfilled. This fits the two free parameters left in the calculation above. The renormalized coupling $g$ and the value of the gluon dressing function at the subtraction scale $Z(\sigma_{gl}^2)$. One can now take two points of view. The more traditional point of view is to fix these values from the beginning of the calculation, which then uniquely defines the theory one works in. However, then one cannot be sure if one works in the physically realized theory until the very end of the calculation. One then has to choose the correct value for $g^2$, which in general changes if some other parameters of the calculation are changed. Another point of view, which is less stringent but leaves more flexibility is to interpret these values as boundary conditions. The renormalization process is then shifted to a later step in the calculation. 
    The drawback of this method is that one knows the physical values of the parameters, such as the cutoff or the subtraction point, only at the very end of the calculation. One does not have direct control onto this parameters in physical units. 

    In this work the latter point  of view is taken. For the coupling constant one takes a value of order $1$, e.g. a convenient choice is $g^2 = 4\pi$. The value $Z(\sigma_{gl}^2)$ is chosen to avoid zero crossings in \eqref{B_MOM_f1}, which is obtained in this setup for $Z(\sigma_{gl}^2) = 0.258$. The renormalization procedure is then performed at the very end of the calculation as described in \secref{sec_2_olo_ren}.

\section{The Sunset Diagram \label{secA_sunset}}

\subsection{Kinematics\label{secA_sun_kin}}
    Being a two-loop diagram the sunset possesses two integration momenta, which extends the kinematical considerations compared to the one-loop case. The external momentum defines a coordinate system in which the integral momenta can be expressed as
	\begin{align}
	  \p&= p\begin{pmatrix} 0 \\ 0 \\ 0 \\ 1 \end{pmatrix} \,,&
	  \q_1&= q_1\begin{pmatrix} 0 \\ 0 \\ \sqrt{1-z_1^2} \\ z_1 \end{pmatrix} \,,&
	  \q_2 &= q_2\begin{pmatrix} 0 \\ \sqrt{1-y^2}\, \sqrt{1-z_2^2} \\ y\, \sqrt{1-z_2^2} \\ z_2 \end{pmatrix}\,.
	\end{align}
    The system features six independent Lorentz invariant scalars, the three moduli, 
    \begin{subequations}\label{B_sun_indepLI}
	\be p = \sqrt{\p\cdot \p} \,, \qquad q_1 = \sqrt{\q_1\cdot \q_1}\,,\qquad q_2 = \sqrt{\q_2 \cdot \q_2} \,,  \ee
    and three angles which characterize the scalar products
	\begin{multline}
	\p \cdot \q_1  = p\, k\, z_1\,, \qquad  \p\cdot \q_2  = p\, q_2\, z_2\,,\qquad\text{and,}\\
	 \q_1\cdot \q_2   = q_1 q_2\, \left(z_1 z_2 + y \sqrt{1-z_1^2}\sqrt{1-z_2^2}\right) \equiv q_1 q_2\, y_{12}\,.
	\end{multline}
    \end{subequations}
    The cosine $y_{12} =  z_1 z_2 + y \sqrt{1-z_1^2}\sqrt{1-z_2^2}$ is defined for later convenience.

      \begin{figure}[t]
      \centering
      \includegraphics[width=.4\textwidth]{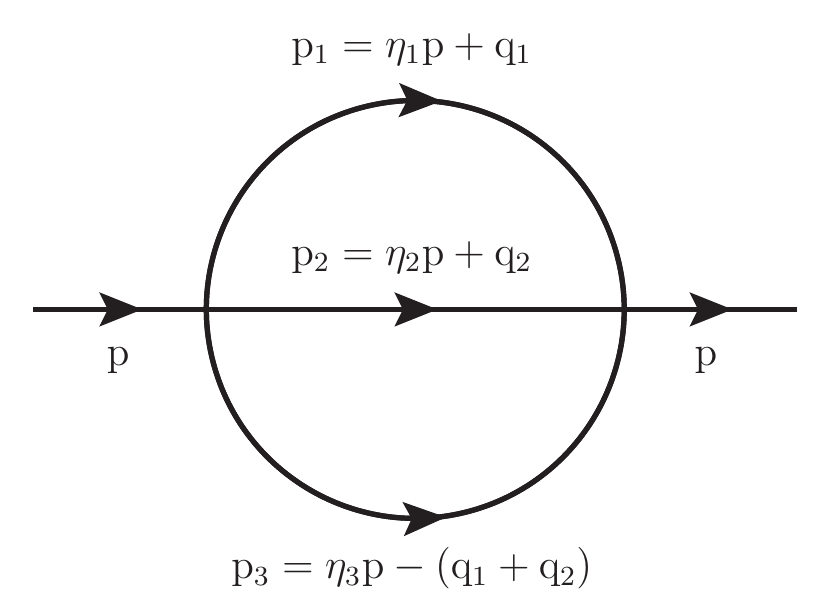}
      \caption{The kinematical configuration of the sunset diagram.}
      \label{fig:sun_momenta}
      \end{figure}

    The sunset diagram features three internal momenta, $\p_1, \p_2$ and $\p_3$, which are linear combinations of the external momentum $\p$ and the two integration momenta $\q_1$ and $\q_2$. In principle, the external momentum can be split arbitrarily onto the respective internal momenta which is accounted for by the introduction of momentum partitioning parameters $\{\eta_1,\eta_2,\eta_3\}$. They fulfill the relation $\eta_1 + \eta_2 + \eta_3 = 1$. The internal momenta are then defined as, see \figref{fig:sun_momenta},
	\begin{align}
	\p_1&= \eta_1\, \p + \q_1\,,	& \p_2&= \eta_2 \,\p + \q_2\,,& \p_3&= \eta_3\, \p - (\q_1+\q_2)\,.
	\end{align}

    In practical calculations of the Dyson-Schwinger equations there will be contractions of the three internal momenta with each other, but also of the internal momenta with the external momentum. They form the Lorentz scalars
	\begin{subequations}
	    \begin{align}
	      p_1^2&= \eta_1^2 \,p^2 + q_1^2 + 2 \,\eta_1\, p\, q_1 \,z_1 \\
	      p_2^2&= \eta_2^2\, p^2 + q_2^2 + 2\, \eta_2\, p \,q_2 \,z_2 \\
	      p_3^2&= \eta_3^2 \,p^2 + q_1^2 + q_2^2 + 2\, q_1 q_2 \,y_{12} - 2\, \eta_3\, p \,q_1\, z_1 - 2\, \eta_3\, p\,  q_2\, z_2  \\
	    \p \cdot \p_1&= \eta_1 p^2 + p \,q_1 \,z_1 \\
	    \p \cdot \p_2&= \eta_2 p^2 + p \,q_2 \,z_2 \\
	    \p \cdot \p_3&= \eta_3 p^2 - p \,q_1 \,z_1 - p \,q_2 \,z_2 \\
	      \p_1 \cdot \p_2 &= \eta_1\eta_2\, p^2 + \eta_1\,p \,q_2 \,z_2 + \eta_2 \,p\, q_1\,z_1 + q_1 q_2 \,y_{12} \\
	      \p_1 \cdot \p_3 &= \eta_1\eta_3 \,p^2 + (\eta_3-\eta_1)\, p\, q_1 \,z_1 - \eta_1 \,p \,q_2\, z_2 - q_1^2 -  q_1 q_2 \,y_{12} \\
	      \p_2 \cdot \p_3 &= \eta_2\eta_3 \,p^2 + (\eta_3 - \eta_2)\, p\,q_2 \,z_2 - \eta_2 \,p \,q_1 \,z_1 -q_2^2 - q_1 q_2 \,y_{12}  \,.
	    \end{align}
	\end{subequations}

    All together there are six angles described by the cosine, $\{z_{01},z_{02},$ $z_{03},z_{12},z_{13},z_{23}\}$, which are defined by
	\be z_{ij} = \frac{\p_i\cdot \p_j}{\sqrt{p_i^2 p_j^2}}. \ee
    The external momentum is denoted by $i=0$.

    \subsubsection{Landau Gauge}
    In the Landau gauge calculations in \secref{sec_2_sun} including the sunset diagram the momentum partitioning
	\be \eta_2=1\,,\qquad\eta_1=\eta_3=0 \ee
    is applied. The internal momenta are then given by
	\begin{align}
	\p_1&=  q_1 	& \p_2&= \p + \q_2	& \p_3&=  - (\q_1+\q_2)\,, \label{A_MP_LG}
	\end{align}
    which form the Lorentz invariants
	\begin{align}
	  p_1^2&=  q_1^2\,,  & p_2^2&=  p^2 + q_2^2 + 2 p q_2 z_2\,, &  p_3^2&=  q_1^2 + q_2^2 + 2 q_1 q_2 y_{12} \,,  \nonumber\\
	  z_{01} &= z_1\,, & z_{02} &= \frac{ p + q_2 z_2 }{p_2}\,,& z_{03} &= - \frac{ q_1 z_1 + q_2 z_2 }{p_3}\,,\\
	z_{12} &= \frac{p z_1 + q_2 y_{12}}{p_2}\,, & z_{13} &= -\frac{q_1 + q_2 y_{12}}{p_3}\,,& z_{23} &= -\frac{p\, q_2 z_2 +  p \,q_1 z_1 + q_2^2 + q_1 q_2 y_{12}}{p_2\,p_3} \,. \nonumber
	\end{align}

    \subsubsection{Maximal Abelian Gauge}
    In the calculations in Maximal Abelian gauge it is required to compare to known results for the infrared analysis from  \cite{Huber:PhD}. Therefore the momentum partitioning
	\be \eta_3=1\,,\qquad \eta_1=\eta_2=0 \ee
    is used. The internal momenta then read
    \begin{align}
    \p_1&= \q_1 	& \p_2&= \q_2	& \p_3&=  \p - (\q_1+\q_2)\,. \label{A_MP_MAG}
    \end{align}
    They form the following Lorentz invariants
    \begin{align}
      p_1^2&=  q_1^2\,, &  p_2^2&=  q_2^2\,, & p_3^2&=  p^2 + q_1^2 + q_2^2 + 2\, q_1 q_2 y_{12} - 2\, pq_1 z_1 - 2 \, pq_2 z_2 \,, \nonumber \\
    z_{01} &= z_1\,,& z_{02} &= z_2\,,& z_{03} &= \frac{p - q_1 z_1 - q_2 z_2}{p_3}\,, \label{A_MAG_LS}\\
    z_{12} &= y_{12}\,,& z_{13} &= \frac{p\, z_1  - q_1 - q_2 y_{12}}{p_3}\,, &z_{23} &= \frac{ p\, z_2 - q_2 - q_1 y_{12} }{p_3}  \,. \nonumber
    \end{align}
    Note that there are only five linear independent Lorentz invariants inside the sunset diagram. In addition there is the external momentum squared. All Lorentz scalars in \eqref{A_MAG_LS} can be expressed by the linearly independent set
      \be \left\{p^2,p_1^2,p_2^2,p_3^3,z_{01}, z_{02} \right\} \label{A_MAG_indLS}\,. \ee

\subsection{UV Analysis \label{secA_sun_UV}}

    The integrals of the sunset diagram read,
	\be I_{sun} = \int \dbar^4 q_1  \int \dbar^4 q_2   \frac{\T}{p_1^2\, p_2^2 \, p_3^2}\,D(p_1,p_2,p_3,p)\ee
    with some dimensionless tensor structure $\T$, which is a function of the cosine $z_{ij}$, and the combined dressing function $D$ wherein all dressing functions are absorbed. Integrating out the trivial angles yields
	\begin{multline}
	I_{sun} = \frac{2}{(2\pi)^6}  \int_0^\Lambda \! dq_1  \int_0^\Lambda\! dq_2  \int_{-1}^1\!dz_1^{\sqrt{}}  \int_{-1}^1\!dz_2^{\sqrt{}}   \int_{-1}^1\!dy\,\,  q_1^3q_2^3 \,\, \frac{\T}{p_1^2\, p_2^2 \, p_3^2}\,D(p_1^2,p_2^2,p_3^2,p^2)\,. \label{A_sunset}
	\end{multline}
 
    For an analysis of the high-energy behavior one usually integrates out the tree-level structure of the angular integrals. Therefore several approximations have to be performed. Since the cutoff $\Lambda$ can be arbitrary large and in general the integrals are quadratically or logarithmically divergent for any finite external momentum the main contributions to the integrals stem from the momentum regime $p<q_{1,2}<\Lambda$. When all momenta are large, i.e. excluding exceptional momenta, the dressing functions adapt their perturbative behavior and all angular dependence can be neglected. All these approximations allow to pull the dressing function in front of the angular integral
	\be I_{sun} \approx  \frac{2}{(2\pi)^6}  \int_p^\Lambda \! dq_1  \int_p^\Lambda\! dq_2\,  D(q_1^2,q_2^2) I_{sun}^{ang} \,, \label{A_sunset_ymax}\ee
    with the angular integral
	\be I_{sun}^{ang}(\T) = \int_{-1}^1\!dz_1^{\sqrt{}}   \int_{-1}^1\!dz_2^{\sqrt{}}   \int_{-1}^1\!dy \,q_1^3 \,q_2^3\,\, \frac{\T}{p_1^2\, p_2^2 \, p_3^2} \,. \label{A_sunset_angint}\ee
    
    The result of $I_{sun}^{ang}$ depends on the chosen momentum partitioning. In Landau gauge the momentum partitioning \eqref{A_MP_LG} is used. The main advantage of this momentum partitioning scheme is that in the UV analysis the different momenta are not entangled and one only needs to discriminate the cases $q_1> q_2$ and $q_2>q_1$ when solving the integral \eqref{A_sunset_angint}. The results for this calculation for all tensor structures appearing in the Landau gauge calculation are given in \tabref{A_tab_suntensor}. The integrals were solved using \mathematica, \cite{mathematica8}, where the limits have to be taken with some care to obtain the correct result.

\begin{table}[pt] \linespread{2}
  \begin{flushright}
    \begin{tabular}{|r | c | c|}
    \hline
	$\T$ & $q_1 > q_2>p$   &  $q_2 > q_1>p$   \\
    \hline
	$1$ & $\frac{\pi^2}{2} \,\frac{q_2}{q_1}$   &  $\frac{\pi^2}{2} \,\frac{q_1}{q_2}$  \\
	$z_{01}^2$ & $\frac{\pi^2}{8} \left(\frac{q_2}{q_1} + \frac{1}{3} \frac{p^2 q_2}{q_1^3}  \right)$  &    $\frac{\pi^2}{8} \left( \frac{q_1}{q_2} + \frac{1}{3}\,\frac{p^2 q_1^3}{q_2^5} \right)$  \\
	$z_{02}^2$ & $\frac{\pi^2}{8} \, \frac{q_2}{q_1} $  & $\frac{\pi^2}{8}  \frac{q_1}{q_2}$\\
	$z_{03}^2$ & $\frac{\pi^2}{8} \, \frac{q_2}{q_1}$ &  $\frac{\pi^2}{8}  \left(  \frac{q_1}{q_2}  + \frac{p^2 q_1}{q_2^3} - \frac{p^2 q_1^3}{q_2^5} \right)$   \\
	$z_{12}^2$ &  $\frac{\pi^2}{8} \left( \frac{q_2}{q_1} + \frac{q_2^3}{q_1^3} - \frac{p^2 q_2}{q_1^3}\right)$  &  $\frac{\pi^2}{8}  \left(  \frac{q_1}{q_2}  + \frac{ q_1^3}{q_2^3} - \frac{p^2 q_1^3}{q_2^5} \right)$ \\
	$z_{13}^2$ &  $\frac{\pi^2}{2} \left( \frac{q_2}{q_1} - \frac{3}{4} \,\frac{q_2^3}{q_1^3}\right)$  & $\frac{\pi^2}{8}  \frac{q_1}{q_2}$ \\
	$z_{23}^2$ &   $\frac{\pi^2}{8} \, \frac{q_2}{q_1}$   &  $\frac{\pi^2}{2}  \left(  \frac{q_1}{q_2}  - \frac{3}{4}\, \frac{q_1^3}{q_2^3} - \frac{3}{4}\, \frac{p^2 q_1}{q_2^3} + \frac{3}{4}\, \frac{p^2 q_1^3}{q_2^5} \right)$ \\
	$z_{01}^2 z_{23}^2$ & $\frac{\pi^2}{32} \left( \frac{q_2}{q_1} + \frac{1}{6}\, \frac{p^2q_2}{q_1^3}\right) $  &  $\frac{\pi^2}{8}  \left(  \frac{q_1}{q_2}  - \frac{3}{4}\, \frac{q_1^3}{q_2^3} - \frac{3}{4}\, \frac{p^2 q_1}{q_2^3} + \frac{13}{6}\, \frac{p^2 q_1^3}{q_2^5} - \frac{5}{4} \left( \frac{p^2 q_1^5}{q_2^7}+  \frac{p^4 q_1^3}{q_2^7} - \frac{p^4 q_1^5}{q^9}  \right) \right)$ \\
	$z_{02}^2 z_{13}^2$ & $\frac{\pi^2}{8} \left( \frac{q_2}{q_1} - \frac{3}{4} \, \frac{q_2^3}{q_1^3}\right)$    &  $\frac{\pi^2}{32}  \frac{q_1}{q_2}$ \\
	$z_{03}^2 z_{12}^2$ & $\frac{\pi^2}{32} \left( \frac{q_2}{q_1} +  \frac{q_2^3}{q_1^3} - \frac{2}{3} \, \frac{p^2 q_2}{q_1^3}\right)$    &  $\frac{\pi^2}{32}  \left(  \frac{q_1}{q_2}  + \frac{q_1^3}{q_2^3} + \frac{p^2 q_1}{q_2^3} + \frac{10}{3}\, \frac{p^2 q_1^3}{q_2^5} - 5 \left( \frac{p^2 q_1^5}{q_2^7}+  \frac{p^4 q_1^3}{q_2^7} - \frac{p^4 q_1^5}{q^9}  \right) \right)$ \\
	$z_{01} z_{02} z_{12}$ & $\frac{\pi^2}{32} \left( \frac{q_2}{q_1} +  \frac{q_2^3}{q_1^3} - \frac{4}{3} \, \frac{p^2 q_2}{q_1^3}\right)$   &  $\frac{\pi^2}{32}  \left(  \frac{q_1}{q_2}  + \frac{q_1^3}{q_2^3} - \frac{4}{3}\, \frac{p^2 q_1^3}{q_2^5} \right)$ \\
	$z_{01} z_{03} z_{13}$ & $\frac{\pi^2}{8} \left( \frac{q_2}{q_1} - \frac{3}{4}\,  \frac{q_2^3}{q_1^3} - \frac{1}{12} \, \frac{p^2 q_2}{q_1^3}\right)$   &  $\frac{\pi^2}{32}  \left(  \frac{q_1}{q_2}  + \frac{p^2q_1}{q_2^3} - \frac{4}{3}\, \frac{p^2 q_1^3}{q_2^5} \right)$ \\
	$z_{02} z_{03} z_{23} $  & $\frac{\pi^2}{32}  \frac{q_2}{q_1}$   &  $\frac{\pi^2}{8}  \left(  \frac{q_1}{q_2}  - \frac{3}{4}\, \frac{q_1^3}{q_2^3} - \frac{p^2 q_1}{q_2^3} + \frac{p^2 q_1^3}{q_2^5} \right)$ \\
	$z_{12} z_{13} z_{23} $  & $\frac{\pi^2}{8} \left( \frac{q_2}{q_1} -   \frac{q_2^3}{q_1^3} + \frac{1}{4} \, \frac{p^2 q_2}{q_1^3}\right)$    &  $\frac{\pi^2}{8}  \left(  \frac{q_1}{q_2}  -  \frac{q_1^3}{q_2^3} - \frac{3}{4}\, \frac{p^2 q_1}{q_2^3} +  \frac{p^2 q_1^3}{q_2^5} \right)$ \\
	$ z_{01} z_{02} z_{13} z_{23} $  & $\frac{\pi^2}{32} \left( \frac{q_2}{q_1} -   \frac{q_2^3}{q_1^3} + \frac{1}{3} \, \frac{p^2 q_2}{q_1^3}\right)$    &  $\frac{\pi^2}{32}  \left(  \frac{q_1}{q_2}  -  \frac{q_1^3}{q_2^3} -  \frac{p^2 q_1}{q_2^3} + \frac{4}{3}\,  \frac{p^2 q_1^3}{q_2^5} \right)$ \\
	$ z_{01} z_{03} z_{12} z_{23} $  & $\frac{\pi^2}{32} \left( \frac{q_2}{q_1} -   \frac{q_2^3}{q_1^3} \right) $   &   $\frac{\pi^2}{32}  \left(  \frac{q_1}{q_2}  -  \frac{q_1^3}{q_2^3} -  \frac{p^2 q_1}{q_2^3} + 6\,  \frac{p^2 q_1^3}{q_2^5} + 5 \left( \frac{p^2 q_1^5}{q_2^7}+  \frac{p^4 q_1^3}{q_2^7} - \frac{p^4 q_1^5}{q^9}  \right) \right) $ \\
	$ z_{02} z_{03} z_{12} z_{13} $  & $\frac{\pi^2}{32} \left( \frac{q_2}{q_1} -   \frac{q_2^3}{q_1^3} + \frac{1}{3} \, \frac{p^2 q_2}{q_1^3}\right)$   &  $\frac{\pi^2}{32}  \left(  \frac{q_1}{q_2}  -  \frac{q_1^3}{q_2^3} -  \frac{p^2 q_1}{q_2^3} + \frac{4}{3}\,  \frac{p^2 q_1^3}{q_2^5} \right)$  \\
    \hline
    \end{tabular} 
  \end{flushright}
 \linespread{1.2}
\caption{Results the angular integral $I^\T$ as defined in \eqref{A_sunset_angint} for the momentum partitioning uses in the Landau gauge calculations, \eqref{A_MP_LG}.\label{A_tab_suntensor}}
\vspace{7mm}
\linespread{2}
  \centering
    \begin{tabular}{|r | c | c|}
    \hline
	$\T$ & $q_1^2 > (\p-\q_2)^2$   & $ (\p-\q_2)^2 > q_1^2 $ \\
    \hline
	$1$ & $\frac{\pi^2}{2} \,\frac{q_2}{q_1}$   &  $\frac{\pi^2}{2} \,\frac{q_1}{q_2}$  \\
	$z_{01}^2$ & $\frac{\pi^2}{8} \,\frac{q_2}{q_1} \, \frac{p^2+q_1^2}{q_1^2}$   &    $\frac{\pi^2}{8}\,  \frac{q_1}{q_2}$  \\
	$z_{02}^2$ & $\frac{\pi^2}{8} \,\frac{q_2}{q_1}$ & $\frac{\pi^2}{8} \, \frac{q_1}{q_2}\,\frac{p^2+q_2^2}{q_2^2}$\\
	$z_{03}^2$ & $\frac{\pi^2}{8} \, \frac{q_2}{q_1} $  &  $\frac{\pi^2}{8}  \,  \frac{q_1}{q_2} $   \\
	$z_{12}^2$ & $\frac{\pi^2}{8} \left(\frac{q_2}{q_1}+\frac{q_2^3}{q_1^3} \right)$   &  $\frac{\pi^2}{8}  \left(  \frac{q_1}{q_2}  + \frac{q_1^3}{q_2^3} \right)$ \\
	$z_{13}^2$ & $\frac{\pi^2}{8} \left(4 \,\frac{q_2}{q_1} - 3\,\frac{q_2^3}{q_1^3} -3\, \frac{p^2 q_2}{q_1^3}\right)$   & $\frac{\pi^2}{8}  \frac{q_1}{q_2}$ \\
	$z_{23}^2$ & $\frac{\pi^2}{8} \frac{q_2}{q_1} $    &  $\frac{\pi^2}{8}  \left( 4\, \frac{q_1}{q_2}  - 3\, \frac{q_1^3}{q_2^3} - 3\, \frac{p^2 q_1}{q_2^3} \right)$ \\ \hline
    \end{tabular} 
 \linespread{1.2}
\caption{Results the angular integral $I^\T$ as defined in \eqref{A_sunset_angint} for the momentum partitioning used in the MAG calculations, \eqref{A_MP_MAG}. \label{A_tab_MAG_sun} }
\end{table}

    The results for the calculation in Maximal Abelian gauge are given in \tabref{A_tab_MAG_sun}. There the momentum partitioning \eqref{A_MP_MAG} was used. The drawback of this momentum configuration is that the momenta do not decouple. The case by case analysis has to be performed for $q_1^2 \gtrless (\p-\q_2)^2$ or $q_2^2 \gtrless (\p-\q_1)^2$, depending on the order of integration. The crucial point is that this case-by-case analysis is angular dependent which complicates the numerical calculations as the boundaries have to be adjusted. On the other hand this momentum partitioning is symmetric under exchange of the indices $1\leftrightarrow 2$. For both tables, \tabref{A_tab_suntensor} and \tabref{A_tab_MAG_sun}, applies that the terms independent on $p^2$ generate quadratic divergences, while the terms proportional to $p^2$ generate logarithmic divergences.

    The sunset diagrams considered in this thesis are generally quadratically divergent. This divergences are subtracted by constructing a tensor structure $\widetilde{\T}$ such that the integral in \eqref{A_sunset} with $\T$ replaced by $\widetilde{\T}$ is free of quadratic divergences. To proceed with the UV-analysis, the regularized angular integral $I_{sun}^{ang}(\widetilde{\T})$ is then put into the approximation \eqref{A_sunset_ymax},
	  \be \widetilde I_{sun} =  \frac{2}{(2\pi)^6}  \int_p^\Lambda \! dq_1  \int_p^\Lambda\! dq_2\,  D(q_1^2,q_2^2) I_{sun}^{ang}(\widetilde{\T}) \,, \label{A_sunset_ymax_tilde}\ee
    which is only logarithmic divergent and thus carries a factor of $p^2$. In the perturbative regime the dressing functions show logarithmic scaling with some exponent. When dressing function $D$ depends on the momentum $p_3^2$ in the momentum routing of the MAG after the angular approximations above, there appears a dressing function as $d_i(p_3^2)\approx d_i(q_1^2+q_2^2)$, which hinders further analytical calculations. A further approximation is thus given by $d_i(p_3^2)\approx d_i(q_>^2)$. Then the dressing functions factorize and the leading terms in \eqref{A_sunset_ymax_tilde} are described by
	  \begin{align} 
	      &\int_p^\Lambda \!\! dk\int_p^{k}\!\! dl \,\, \frac{l^\alpha}{k^{2+\alpha}} \left(1+\omega \, \log{\frac{k^2}{s}} \right)^{\lambda_1}\left(1+\omega\, \log{\frac{l^2}{s}} \right)^{\lambda_2} \label{A_sun_genultravioletint}\\
	      & \approx  \frac{1}{2\omega\,(1+\alpha)\left(1+\lambda_1+\lambda_2\right)} \left(\left(1+\omega\, \log{\frac{\Lambda^2}{s}} \right)^{1+\lambda_1+\lambda_2} -\left(1+\omega\, \log{\frac{p^2}{s}} \right)^{1+\lambda_1+\lambda_2} \right)\,. \nonumber
	  \end{align}
    \eqref{A_sun_genultravioletint} was obtained using the series expansion of the incomplete Gamma function, $ \Gamma(a,z) \approx z^{a-1}e^{-z} $, \cite{Abramowitz},  and taking only the ultraviolet-leading terms into account.

\subsection{IR Analysis\label{secA_sun_IR}}
      In the infrared analysis the dressing functions are replaced by their power-law behavior. For the one-loop calculations the corresponding integral is textbook knowledge, \cite{Muta}, and was introduced into the Dyson-Schwinger context in \cite{Lerche:2002ep,Fischer:2003zc}. 
	\begin{align}
	    I_{ol}(p^2,a,b,d) & := \int\!\dbar^dq\,\left(q^{2}\right)^a\left((p-q)^2 \right)^b   \label{A_infraredGamma} \\
			       & = (4\pi)^{-\frac{d}{2}} \left(p^2 \right)^{\frac{d}{2}+a+b} \frac{\Gamma\left(a+\frac{d}{2}\right)\Gamma\left(b+\frac{d}{2}\right)\Gamma\left(-\left(a+b+\frac{d}{2}\right)\right)}{\Gamma\left(-a\right)\Gamma\left(-b\right)\Gamma\left(a+b+d\right)}\,. \nonumber 
	\end{align}
      For a comparison to the solution of the sunset diagram it is interesting to rewrite the one-loop integral as
	\begin{multline}
	    I_{ol}(p^2,a,b,d)  =  (4\pi)^{-\frac{d}{2}} \left(p^2 \right)^{\frac{d}{2}+a+b} \left(-a,2a+\frac{d}{2} \right)_{\!\!P}\left(-b,2b+\frac{d}{2} \right)_{\!\!P} \\ \times\quad\left(a+b+d,-2(a+b)-\frac{3d}{2} \right)_{\!\!P} \,,\label{A_oli_PS}
	\end{multline}
      with the Pochhammer symbols,
	\be \left(a,b \right)_P = \frac{\Gamma\left(a+b\right)}{\Gamma\left(a\right)} \,. \label{A_def_PS}\ee
      The solution for the corresponding integral of the sunset diagram is given in App.~D of \cite{Huber:PhD}. The calculation is complicated by additional tensor structures compared to the one-loop case. Using the momentum routing \eqref{A_MP_MAG} one finds that there are six independent Lorentz scalars in the system, \eqref{A_MAG_indLS}, of which the external momentum can be pulled in front of the integral. For the sunset-diagrams it is thus sufficient to give the solution of the following integral,
      \begin{align} 
	      I_{sun}(p^2;a,b,c,d,e,f) & :=\int\!\dbar^d q_1\int\!\dbar^d q_2 \,\,(p_1^2)^a\, (p_2^2)^b\, (p_3^2)^c \,z_{01}^e \,z_{02}^f \label{A_def_IsunIR} \\
	      &\hspace{-29mm}=  \left(4 \pi \right)^{-d} (p^2)^{a+b+c+d}  \sum_{n_1,n_2,n_3,n_4,n_5=0}^{\max\{e,f\}} \frac{(-2)^{n_3}}{n_1!n_2!n_3!n_4!n_5!} \nonumber\\
	      &\hspace{-29mm} \times\left(a+b+c+\frac{3d+e+f}{2},-2(a+b+c)-\frac{5d}{2}-n_1-n_2-n_3-n_4-n_5\right)_P \nonumber\\
	      &\hspace{-29mm} \times \left(-a+\frac{e}{2},2a+\frac{d}{2}-n_1-n_2+n_5\right)_P \left(-b+\frac{f}{2},2b+\frac{d}{2}+n_2-n_4-n_5\right)_P \nonumber \\
	      &\hspace{-29mm} \times \left(-c,2c+\frac{d}{2}+n_1+n_4\right)_P P\left(e,n_1+n_2,n_3\right)\,P\left(f,n_4+n_5,n_3\right)\,.	\nonumber
      \end{align}
      The Pochhammer symbols are defined in \eqref{A_def_PS} and the $P$-symbols are defined as
      \be  P(a,b,c) = \left(-\frac{a}{2},b+\frac{c}{2} \right)_P \left(\frac{1}{2}-\frac{a}{2},b+\frac{c}{2} \right)_P \,.\ee
      Evaluation of \eqref{A_def_IsunIR} yields a long sum of products of Gamma-functions which is restricted by the order of the angular variables $z_{01}$ and $z_{02}$. In addition, the $P-$symbols truncate this sum heavily. Since in this work $\max\{e,f\} = 2$ the needed $P-$Symbols can be given explicitly. With $b,c\geq 0$ one has
      \begin{align}
      P(0,b,c) &= \frac{\Gamma\left(b+\frac{c}{2}\right)\Gamma\left(b+\frac{c}{2}+\frac{1}{2}\right)}{\Gamma\left(0\right)\Gamma\left(\frac{1}{2}\right)} = \begin{cases} 1 &\mbox{if } b=c= 0 \\
      0 & \mbox{else }  \end{cases} \\
      P(1,b,c) &= \frac{\Gamma\left(b+\frac{c}{2}-\frac{1}{2}\right)\Gamma\left(b+\frac{c}{2}\right)}{\Gamma\left(-\frac{1}{2}\right)\Gamma\left(0\right)} = \begin{cases} 1 &\mbox{if } b=c= 0 \\
      0 & \mbox{else }  \end{cases} \\
      P(2,b,c) &= \frac{\Gamma\left(b+\frac{c}{2}-1\right)\Gamma\left(b+\frac{c}{2}-\frac{1}{2}\right)}{\Gamma\left(-1\right)\Gamma\left(-\frac{1}{2}\right)} = \begin{cases} 1 &\mbox{if } b=c= 0 \\ - 1 &\mbox{if } b=0,c=1 \\  \frac{1}{2} &\mbox{if } b=0, c= 2 \mbox{ or }  b=1, c=0\\
      0 & \mbox{else }  \end{cases}\,.
      \end{align}
      Still the algebra to evaluate the sum in \eqref{A_def_IsunIR} is tedious. The computer algebra system \form, \cite{Vermaseren:2000nd}, is very useful in this respect. 

      Some helpful formulas when dealing with Pochhammer-symbols, which can be validated by explicit calculation, are
	\be  \frac{(a,b)_P}{(a,c)_P} = (a+c,-c+b)_P \,,\qquad \text{and,}\qquad (a,-n)_P = \frac{(-1)^n}{(1-a,n)_P}\,,\quad \forall n \in \mathbb{Z}\,.
	\ee

\subsection{Numerical Implementation}

    In the Landau gauge the task is to numerically implement the integral, \eqref{2_LG_sun_2},
	\begin{multline} I_{sun}(p^2) = c_{sun} \int_0^\Lambda dq_1 \int_0^\Lambda dq_2 \int_{-1}^1 dz_1^{\sqrt{}}\int_{-1}^1 dz_2^{\sqrt{}} \int_{-1}^1 dy \\
	  \times\,\,\widetilde{\K}_{sun} \, q_1q_2^3\,\frac{\left(Z(q_1^2)Z((\p+\q_2)^2)\right)^{\lambda_1} \left(G(q_1^2)G((\p+\q_2)^2)\right)^{\lambda_2}}{ (\p+\q_2)^2 (\q_1+\q_2)^2} \label{A_sun_LG}\end{multline}
    with $c_{sun} = - \frac{g^4 N_c^2}{9(2\pi)^6\,p^2}$, the regularized integration kernel \eqref{2_sun_Ktilde}, $\lambda_1 = \frac{1}{2}-2\gamma_c$ and $\lambda_2 = -4\gamma_c$. 

   \begin{wrapfigure}[13]{r}{.5\textwidth}
    \centering
    \includegraphics[width=.4\textwidth]{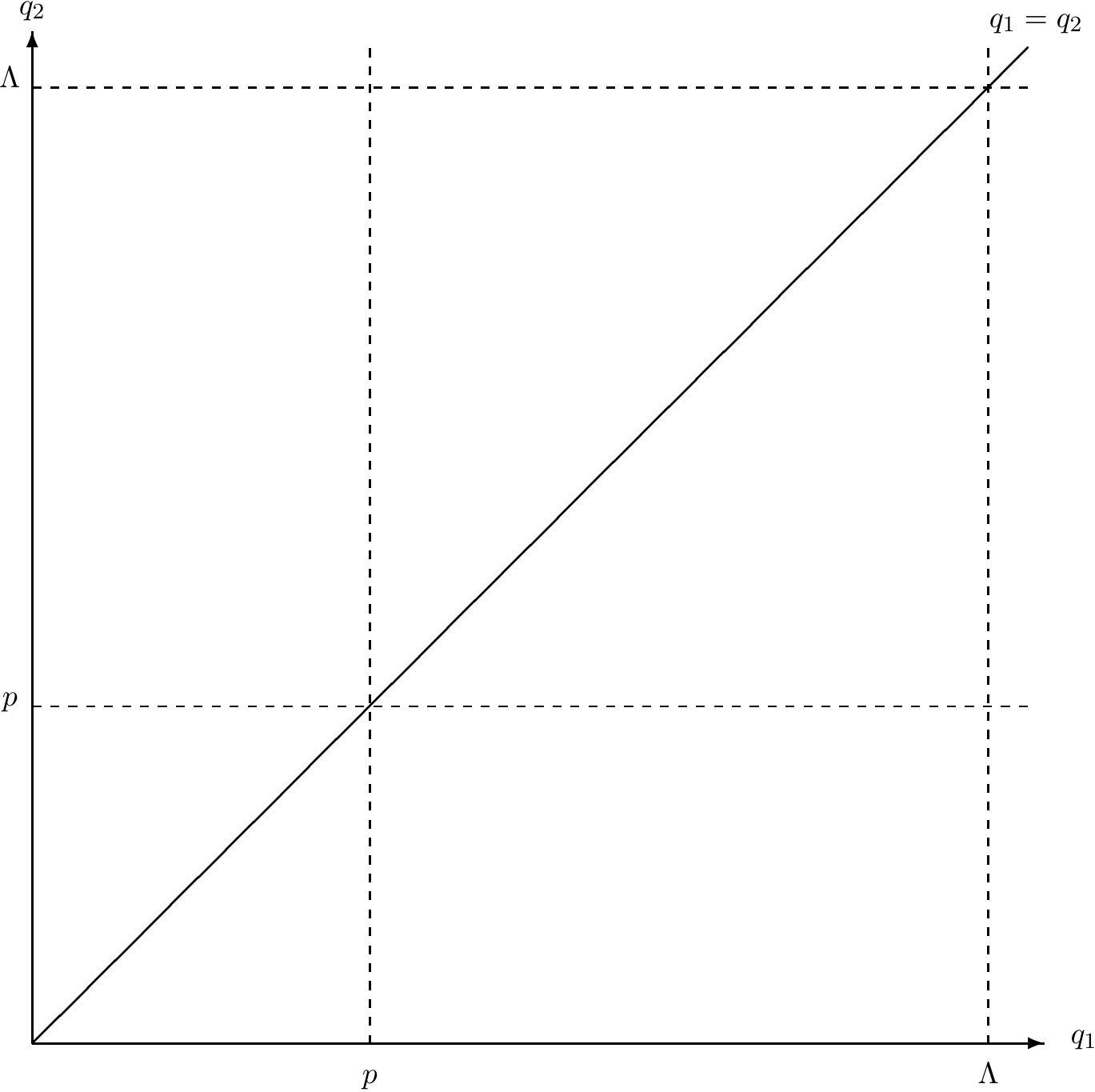}
    \caption{The intervals of integration.\label{fig:mom_twoloop}}
    \end{wrapfigure}

    In the last section it was found that the radial integral strongly depends on the momentum configuration, in particular the cases $q_1\gtrless q_2 $ have to be distinguished. In addition the external momentum represents a scale. The $(q_1,q_2)$-plane is divided into the intervals, $\{q_1;q_2\}$,
     \begin{multline} 
	  \{(0,p);(0,q_1)\},\,\{(p,\Lambda);(0,p)\},\\\{(p,\Lambda);(p,q_1)\},\,\{(0,q_2);(0,p)\},\\
	  \{(0,p);(p,\Lambda)\},\,\{(p,q_2);(p,\Lambda)\} \,. \nonumber
      \end{multline}

    These intervals are calculated on $32^2$ integration points using the mapping \eqref{A_mapping_radial}.

    By construction, in \eqref{A_sun_LG} one propagator is undressed, i.e. the momentum $p_3^2$ only appears in a tree-level structure. This allows to analytically integrate the angles $y$ and $z_1$ and there is only one angular integral left for numerical evaluation. For $(\q_1+\q_2)^2 = 0$ the integrand in \eqref{A_sun_LG} develops a pole. Therefore its vicinity has to be integrated with some care. For a ratio $\frac{q_<}{q_>} \leq 0.8$ $N_z=128$ points are used for the angular integral, while for $\frac{q_<}{q_>} > 0.8$ $N_z=512$ integration nodes have been used to get reliable results.

    In the MAG in principle the same techniques are applied. The only change is that here the singularity is not at $q_1=q_2$ but depends on the cosine $z_2$ as it sits at $q_1^2 = q_2^2 + p^2 + 2\,p q_2 z_2$ due to the different momentum partitioning. Also one has to take all angular integrations into account as also the momentum $\p_3$ carries a dressing function. This increases the calculational costs considerably.

\section{Conjugate Momenta, Currents and Charges \label{A_curr}}

    In this appendix field variations and the equations of motion of the fields in the generalized covariant gauge are given. As this gauge includes the general covariant gauge, these results can also be used for this case.

    \subsection{Yang-Mills Theory}
    The Lagrangian for Yang-Mills theory is given in \eqref{A_def_LYM}. The variations with respect to the gluon fields are given by
	\be	\var[\LYM]{ A^r_\rho}= \cp{F_{\sigma\rho}}{A_\sigma}{r} \,,\qquad\text{and,}\qquad \var[\LYM]{\partial_\sigma A^r_\rho}= F_{\sigma\rho}^{r} \ee
    which yields the equation of motion of classical Yang-Mills theory
	\be \var[\SYM]{A_\rho^r} = -D_{\sigma}^{rs}F_{\sigma\rho}^s\,. \ee

    \subsection{Generalized Linear Covariant Gauge}
      The Lagrangian of the generalized covariant gauge is given in \eqref{4_def_Lgf} and reads,
	  \be \L_{gf}^{c\bar c} = \frac{\xi}{2} b^2 - i b^r\, \partial_\mu A_\mu^r - i \alpha (D_\mu \bar c)^r \partial_\mu c^r - i \bar \alpha \, \partial_\mu \bar c ^r (D_\mu c)^r + \frac{\alpha \bar \alpha \xi}{2} \,\cp{\bar c}{c}{2} \,, \ee
      with $\alpha+\bar \alpha = 1$. The general covariant gauge is given by the case $\alpha = 0$.
      
      The variations of the gauge-fixing Lagrangian  with respect to the fields are given by,
      \begin{subequations}
	  \begin{align}
	    \var[\L_{gf}^{c\bar c}]{A^r_\rho}&=  i \bar \alpha \cp{\partial_\rho \bar c}{c}{r} - i \alpha  \cp{ \bar c}{\partial_\rho c}{r}\,, & \var[\L_{gf}^{c\bar c}]{ \partial_\sigma A^r_\rho}&=- i b^{r} \delta_{\sigma\rho}\,,\\
	    \var[\L_{gf}^{c\bar c}]{ c^r}&=  -i \bar \alpha \cp{A_\sigma}{\partial_\sigma \bar c}{r}  +\alpha \bar\alpha\xi \cp{\bar c}{\cp{\bar c}{c}{}}{r}\,,&
	    \var[\L_{gf}^{c\bar c}]{\partial_\rho c^r}&= i \alpha \left(D_\rho \bar c \right)^r +i \bar \alpha \,\partial_\rho \bar c^r\,,    \\
	     \var[\L_{gf}^{c\bar c}]{ \bar c^r}&=  i\alpha \cp{A_\sigma}{\partial_\sigma c}{r}  - \alpha \bar\alpha\xi \cp{\cp{\bar c}{c}{}}{c}{r}\,, &  \var[\L_{gf}^{c\bar c}]{ \partial_\rho \bar c^r}&= -i \bar \alpha \left(D_\rho c \right)^r -i\alpha\, \partial_\rho c^r\,,\\
	    \var[\L_{gf}^{c\bar c}]{ b^r}& = \xi b^r - i \partial_\sigma A^r_\sigma\,,
	  \end{align}
    \end{subequations}
      where always left-derivatives have been employed. The equations of motion are then given by
    \begin{subequations}  
	\begin{align} 
	    \var[\S^{c\bar c}]{A_\rho^r} &=-D_{\sigma}^{rs}F_{\sigma\rho}^s + i \bar \alpha \cp{\partial_\rho \bar c}{c}{r} - i \alpha  \cp{ \bar c}{\partial_\rho c}{r}+ i \partial_\rho b^{a} \,,\\   
	    \var[\S^{c\bar c}]{c^r} &= - i \left(\partial_\sigma D_\sigma \bar c\right)^r+ i \bar \alpha \cp{\partial_\sigma A_\sigma}{\bar c}{r} + \frac{\alpha \bar\alpha\xi}{2} \cp{\cp{\bar c}{\bar c}{}}{c}{r}= -\sab\left(  i \partial_\sigma A^r_\sigma - \xi b^r \right) \,,\label{B_ghostEOM}\\   
	    \var[\S^{c\bar c}]{\bar c^r} &=i \left(\partial_\sigma D_\sigma c\right)^r - i \alpha \cp{\partial_\sigma A_\sigma}{ c}{r}  - \frac{\alpha \bar\alpha\xi}{2} \cp{\bar c}{\cp{c}{c}{}}{r} = \sa \left(  i \partial_\sigma A^r_\sigma - \xi b^r \right)\,,\\
	    \var[\S^{c\bar c}]{b^r}& = \xi b^r - i \partial_\sigma A^r_\sigma \,.
      \end{align}   
    \end{subequations}                                   
    The $\alpha-$ dependent BRST and anti-BRST transformations are given in \eqref{4_def_BRST} and \eqref{4_def_aBRST}. They give the identity 
      \be b^r = \bar \alpha\, \sa \bar c^r - \alpha \sab\,c^r \,.\ee
    Using the Jacobi identity one can show the following identity, which serves useful in many calculations in this thesis,
	  \be \cp{\cp{\phi}{\psi}{}}{\psi}{a} = \frac 1 2 \cp{\phi}{\cp{\psi}{\psi}{}}{a}  \ee
      for any fermionic $\psi \in \{c^r,\bar c^r\}$ and any $\phi \in\{A_\mu^r,b^r,c^r,\bar c^r\}$.

\section{Equivariant BRST Algebra}
    The equivariant algebra described in \secref{sec_3_algebra} consists of the following operators. An integration over position space is implicit. Open indices in nested cross-products are not summed over, but just denote the index of the inner cross-product, e.g. $\cp{B_\mu}{\cp{c}{c}{i}}{a} = g^2 f^{abi}f^{icd}B_\mu^b c^c c^d$.
      \begin{subequations} \label{B_algeb_op}
      \begin{align}
       iQ_\eps &= \cp{B_\mu}{c} {i} \var{A_\mu^i } +\left((D_\mu c)^a + \cp{B_\mu}{c}{a}  \right) \var{B_\mu^a}-\frac{1}{2}\cp{c}{c}{a} \var{c^a} \\ 
	& \quad + \left( b^a - \frac{1}{2} \cp{\bar c}{c}{a}\right)\var{\bar c ^a} +  \left( -\frac{1}{2}\cp{c}{b}{a} + \frac{1}{8} \cp{\bar c}{\cp{c}{c}{b}}{a} +\frac{1}{2} \cp{\bar c}{\cp{c}{c}{i}}{a} \right)\var{b^a} \nonumber\\
       i\bar Q_\eps &= \cp{B_\mu}{\bar c} {i} \var{A_\mu^i } + \left( (D_\mu \bar c)^a + \cp{B_\mu}{\bar c}{a} \right) \var{B_\mu^a} + \left( -b^a - \frac{1}{2} \cp{\bar c}{c}{a}\right)\var{c ^a} \\  
	& \quad - \frac{1}{2}\cp{\bar c}{\bar c}{a} \var{\bar  c ^a} + \left( -\frac{1}{2}\cp{\bar c}{b}{a} + \frac{1}{8} \cp{\cp{\bar c}{\bar c}{a}}{c}{a} +\frac{1}{2} \cp{\cp{\bar c}{\bar c}{i}}{c}{a} \right)\var{b^a} \nonumber\\
       i\Pi^+ &= c^a \var{\bar c^a} \\
       i\Pi^0 &= c^a \var{ c^a} - \bar c^a \var{\bar c^a} \\
       i\Pi^- &=  \bar c^a \var{ c^a} \\
 	iQ^+ &= \partial_\mu\cp{c}{c} {i} \var{A_\mu^i } + \cp{B_\mu}{\cp{c}{c}{i}}{a} \var{B_\mu ^a} + \cp{\bar c}{\cp{c}{c}{i}}{a} \var{\bar c ^a}\\
	& \quad  \nonumber + \cp{b}{\cp{c}{c}{i}}{a} \var{b^a}\\
 	iQ^0 &= \partial_\mu\cp{\bar c}{c} {i} \var{A_\mu^i } + \cp{B_\mu}{\cp{\bar c}{c}{i}}{a} \var{B_\mu^a}+ \cp{c}{\cp{\bar c}{c}{i}}{a} \var{ c ^a} \\ 
	& \quad  \nonumber + \cp{\bar c}{\cp{\bar c}{c}{i}}{a} \var{\bar c ^a} + \cp{b}{\cp{\bar c}{c}{i}}{a} \var{b^a}\\
 	iQ^- &= \partial_\mu\cp{\bar c}{\bar c} {i} \var{A_\mu ^i} + \cp{B_\mu}{\cp{\bar c}{\bar c}{i}}{a} \var{B_\mu^a} + \cp{ c}{\cp{\bar c}{\bar c}{i}}{a} \var{c^a}\\
	& \quad \nonumber + \cp{b}{\cp{\bar c}{\bar c}{i}}{a} \var{b^a}\\
 	iP^+ &= \partial_\mu\cp{b}{c} {i} \var{A_\mu^i } + \cp{B_\mu}{\cp{b}{c}{i}}{a} \var{B_\mu^a}- \cp{c}{\cp{b}{c}{i}}{a} \var{c^a} \\ 
	& \quad \nonumber- \cp{\bar c}{\cp{b}{c}{i}}{a} \var{\bar c^a} + \cp{b}{\cp{b}{c}{i}}{a} \var{b^a} \\
 	iP^- &= \partial_\mu\cp{b}{\bar c} {i} \var{A_\mu^i } + \cp{B_\mu}{\cp{b}{\bar c}{i}}{a} \var{B_\mu^a}- \cp{c}{\cp{b}{\bar c}{i}}{a} \var{c^a} \\ 
	  & \quad \nonumber - \cp{\bar c}{\cp{b}{\bar c}{i}}{a} \var{\bar c^a} + \cp{b}{\cp{b}{\bar c}{i}}{a} \var{b^a} \,.
      \end{align}
      \end{subequations}
      These operators form the algebra $ \mathbf E = \{ Q_\eps,\bar Q_\eps,\Pi^+$ $,\Pi^0,\Pi^-,Q^+,Q^0$ $,Q^-,P^+,P^-\} $
      with the non-vanishing commutation relations
	  \begin{subequations} \label{B_algeb}
	  \begin{align}
	  i\com{ \Pi^+}{ \Pi^-} &= \Pi^0 \, & i\com{ \Pi^0}{ \Pi^+} & = 2 \Pi^+ \, & i\com{ \Pi^0 }{\Pi^- }& = -2 \Pi^- \, \\ 
	    i\acom{ Q_\eps}{ Q_\eps} &= Q^+ \, & i\acom{ Q_\eps}{ \bar Q_\eps} & = Q^0 \, & i\acom{ \bar Q_\eps}{\bar Q_\eps} & = Q^- \, \\
	    i\com{\Pi^+}{\bar Q_\eps} &= Q_\eps \, & i\com{ \Pi^-}{Q_\eps} & = \bar Q_\eps \, & i\com{\Pi^0}{Q_\eps} & = Q_\eps \, \\
	    i\com{ \Pi^0}{\bar Q_\eps} &= - \bar Q_\eps \, & i\com{ \Pi^+}{Q^0} &= Q^+ \, & i\com{\Pi^+}{Q^-}& = 2 Q^0\, \\
	      i\com{\Pi^0}{Q^-} &= -2 Q^- \, & i\com{\Pi^-}{Q^+} & = 2Q^0 \, & i\com{\Pi^0}{Q^+} & = 2 Q^+ \, \\
	    i\com{Q_\eps}{Q^0} &= P^+ \, & i\com{Q_\eps}{Q^-}  & = 2 P^- \, & i\com{\Pi^-}{Q^0} & = Q^- \,,\, \\
	    i\com{\bar Q_\eps}{Q^+} & = -2 P^- \, & i\com{\bar Q_\eps}{Q^0} & = -P^- \, & i\com{\Pi^0}{P^-} &= -P^-\,,\, \\
	  i\com{\Pi^0}{P^+} &= P^+ \, & i\com{\Pi^-}{P^+} & = P^- \, & i\com{\Pi^+}{P^-} & = P^+  \,.
	  \end{align}
	  \end{subequations}